\documentclass[aps,pra,twocolumn,amssymb,amsmath,amsfonts,floatfix]{revtex4-1}
\usepackage{CJK}
\pdfoutput=1
\usepackage{subfig}
\usepackage{bm}
\usepackage{hyperref}
\usepackage{color}
\usepackage{psfrag,euscript,eufrak,mathrsfs}
\usepackage{graphicx}
\usepackage{epsfig}
\usepackage{amssymb}
\def\Xint#1{\mathchoice
    {\XXint\displaystyle\textstyle{#1}}%
    {\XXint\textstyle\scriptstyle{#1}}%
    {\XXint\scriptstyle\scriptscriptstyle{#1}}%
    {\XXint\scriptscriptstyle\scriptscriptstyle{#1}}%
    \!\int}
\def\XXint#1#2#3{{\setbox0=\hbox{$#1{#2#3}{\int}$}
    \vcenter{\hbox{$#2#3$}}\kern-.5\wd0}}

\def\dashint{\Xint-}
\definecolor{dgreen}{rgb}{0.2 ,0.54, 0.2}

\begin{document}
\begin{CJK*}{GB}{} 
\title{Spinon confinement in the gapped antiferromagnetic XXZ spin-1/2 chain}
\author{{Sergei~B.~Rutkevich}}
\affiliation{Fakult\"at f\"ur Mathematik und Naturwissenschaften, Bergische Universit\"at Wuppertal, 42097 Wuppertal, Germany.}
\maketitle
\end{CJK*}
\date{October 7, 2022}
\begin{abstract}
The infinite  Heisenberg  XXZ spin-(1/2) chain in the gapped antiferromagnetic regime has two degenerate vacua and kink topological excitations (which are also called spinons) interpolating between these vacua as elementary excitations. Application of an arbitrary weak staggered longitudinal magnetic field $h$ induces a long-range  attractive potential between two adjacent spinons leading to their confinement into  'meson' bound states. Exploiting the integrability of the XXZ model in the deconfined phase $h=0$, we perform perturbative calculations of the energy spectra of the two-spinon bound states in the weak confinement regime at $h\to+0$, using the strength of the staggered magnetic field $h$ as a small parameter.
Both transverse and longitudinal dynamical structure factors of the local spin operators are calculated as well in the 
two-spinon approximation in the weak confinement regime to the leading order in $h$.
  \end{abstract}
\maketitle

\section{Introduction}
The notion of confinement plays a significant role in modern physics. This 
phenomenon occurs if the constituents of compound particles cannot be separated 
from each other and therefore cannot be observed directly. A famous  example
is the confinement of quarks in hadrons \cite{Nar04}, whose theoretical description 
remains a long-standing open problem of  Quantum Chromodynamics (QCD). 

It is remarkable that confinement of particles finds its realization 
not only in high-energy physics but also in condensed matter systems. 
In certain
quasi-one-dimensional crystals the 
confinement of topological kink magnetic excitations becomes
experimentally observable and demands for precise theoretical
predictions. An example is the compound $\mathrm{Co}\mathrm{Nb}_2\mathrm{O}_6$
in which kink confinement can be seen e.g.\ in neutron-scattering 
experiments \cite{Coldea10} or high-resolution terahertz spectroscopy
\cite{Mor14}.  The magnetic
structure of this compound \cite{Coldea10,Coldea20} can be described by the one-dimensional quantum Ising spin-chain
model, which is a paradigmatic model in the theory of quantum phase
transitions \cite{Sach99}. Kink confinement has  been  recently also
experimentally studied in the quasi-1D antiferromagnetic 
compounds  ${\rm {SrCo_2V_2O_8}}$ \cite{Wang15,Bera17} and ${\rm{BaCo_2V_2O_8}}$ 
\cite{Gr15,Faur17,Wang19}. 

The theoretical study of  confinement in condense-matter systems was started 
more than fourty years ago in the pioneering work of McCoy and Wu
\cite{McCoy78}, in which they examined the effect of the external symmetry-breaking 
magnetic field $h$ on the analytical structure of the two-point 
function in the Ising Field Theory (IFT) in the ferromagnetic phase. McCoy and Wu demonstrated 
that the square-root branch cut located at the imaginary axis in the   momentum complex plane, which is present in the two-point function
in the ordered phase at $h=0$, breaks up into a sequence of poles at any $h>0$. This change in the analytic structure of the
two-point function was associated in [\onlinecite{McCoy78}] with the confinement transition: fermions, which were  free  particles in the 
IFT at $h=0$, attract one-another and form bound states at $h>0$.

It becomes clear later \cite{Del96,DelMus98,Mus11}, that the mechanism of confinement discovered by McCoy and Wu
in IFT is quite general. It can be realized in  many 
 one-dimensional  Quantum Field Theories (QFTs) and spin-chain models, which are invariant under a
discrete symmetry group and  display a continuous order-to-disorder phase
transition. If  the system has two degenerate 
vacua $|vac\rangle^{(\mu)}$, $\mu=0,1$ in the ordered phase
due to a spontaneous breaking of the $\mathbb{Z}_2$-symmetry, the particle sector
of the theory should contain kinks $K_{\mu\nu}$, $\mu, \nu=0,1$ that interpolate between these vacua. The application
of the symmetry-breaking field, that shifts the  energy of  the vacuum 
$|vac\rangle^{(1)}$ to a lower value, lifts the degeneracy between the vacua.
As a result, the vacuum $|vac\rangle^{(1)}$ transforms into the true ground
state, whereas the state $|vac\rangle^{(0)}$  
turns into the unstable false vacuum. The energy difference between the true 
and false vacuum induces a long-range attractive interaction between kinks,
which, in turn, leads to their confinement: isolated kinks do not exist
anymore in the system, and the kinks bind into compound particles. 
In recent
decades particular realizations of this scenario in different 
one-dimensional QFT  and spin-chain models 
 have attracted much theoretical interest
\cite{Shiba80,DelMus98,FonZam2003,FZ06,Rut05,Del08,MusTak09,Mus11,Kor16,Rob19,Lagnese_2020,Lagnese_2022,Ram20,Mus22}.
Note, that due to analogy with QCD,  the kinks and two-kink bound states  in the confinement regime are 
often referred to as `quarks' and `mesons', respectively.

In the simplest phenomenological approach to   confinement in one-dimension, 
originating from the work of  McCoy and Wu
 \cite{McCoy78}, the two kinks are treated as
 quantum particles with the quadratic dispersion law
\begin{equation}\label{omquad}
\omega(p)=m_0+\frac{p^2}{2m},
\end{equation}
moving on the line and attracting
one another due to a potential growing linearly with distance and
being overall proportional to the external magnetic field
 $h$.
The relative
motion of two particles in their center-of-mass frame is described by the Schr{\"o}dinger equation
\begin{equation}\label{Ai0}
\left(2 m_0 -\frac{1}{m}\frac{d^2}{d x^2}+ f |x|-E_n\right) \psi_n(x) =0 \,,
\end{equation}
where  $f\sim h$ is the``string tension''. If the kinks behave as Fermi particles, their wave
function must be anti-symmetric, $\psi_n(x)=-\psi_n(-x)$, and the energy
levels of the two-kink bound-states, determined by \eqref{Ai0}, are
given by
\begin{equation}
 E_n = 2m_0+ z_n\,{f^{2/3}}\,{m^{-1{\color{black}/3}}}, \quad n=1,2\ldots, \label{McWu}
\end{equation}
where the numbers $-z_n$ are the zeros of the Airy function,
$\mathrm{Ai}(-z_n)=0$. 
Energy spectra of this form, indicating kink
confinement, have been indeed observed \cite{Coldea10,Gr15,Wang15,Bera17,Faur17}
in quasi-1D quantum magnets close to the band minima at symmetry points of the Brillouin zone.

Several numerical techniques applied to microscopic model Hamiltonians
(e.g.\ the truncated conformal space approach \cite{Tak14,LT2015}, a
tangent-space method for matrix product states, the density matrix
renormalization group algorithm \cite{Bera17}) have been used to obtain the meson
(two-kink bound state) spectra in the whole Brillouin zone. Nowadays, the unprecedented 
increase of accuracy of such numerical techniques  allows one in some cases to directly compare
the experimental data with numerical results. Nevertheless, it is highly 
desirable to complement the direct numerical studies  with analytical calculations of the meson energy spectra in one-dimensional 
QFTs and spin-chain models for at least two reasons. 
First,   consistent first-principles analytic calculations  allow one to put the conclusions of the numerical
analysis on the firm ground. Second, analytic calculations are absolutely necessary for a deep and qualitative understanding of the 
underlying physics.

Although the confinement caused by the  mechanism outlined above  does not realize in exactly solvable models, it is quite common in non-integrable
deformations of integrable models induced by the discrete-symmetry breaking field $h$.   
Due to the absence of exact solutions, 
it is natural to restrict the analysis to the {\it{weak confinement regime}} corresponding to a small 
symmetry breaking field, and to employ some perturbation theory using $h$ as a small parameter. 
Two perturbative techniques nicely complementing one another have been used in the  literature.

The first more  rigorous  and consistent (but technically demanding) approach
is based on combining the Bethe-Salpeter equation \cite{FonZam2003,FZ06}
with a modified form factor expansion \cite{Del96,Rut09,Rut17P}. Up to now, 
this technique has been used for the calculation of  meson energy spectra only in two models of statistical mechanics
\begin{footnote}
{
In the high-energy physics, the Bethe-Salpeter equation was applied to the confinement problem 
by 't~Hooft \cite{Hooft74}, who considered a model for QCD
in one space and one time dimension in the limit of an infinite number of
colours.
}
\end{footnote}: in the IFT \cite{FonZam2003,FZ06,Rut05,Rut09}, and in the quantum Ising spin chain \cite{Rut08a}.
Both models have a very specific property, which was substantially exploited  in   derivation of 
the Bethe-Salpeter equations \cite{FonZam2003,Rut08a}:  the kink elementary excitations  in the deconfined phase of these models
 do not interact with each other, but behave as free Fermi-particles. This property  does not hold in other integrable models,
such as the  Potts and  sine-Gordon QFTs,   XXZ spin chain, etc. In these models  
particles strongly interact at small distances already in the deconfined phase at $h=0$. 
This short-range interaction is encoded  in the non-trivial factorizable scattering matrix, 
which is the key characteristic of the integrable model.  An extension of the systematic perturbative approach  exploiting
the Bethe-Salpeter  equation  to confinement 
in  such systems was identified by Fonseca and Zamolodchikov  \cite{FZ06} in 2006 as an important open problem. 

The second 
perturbative technique is not so rigorous, but, instead,  rather heuristic and intuitive. 
Its main advantages are the simplicity of calculations and  transparency of physical interpretation.
In this  approach, the two kinks forming a meson 
are treated as classical particles, which  move along the line and attract one another
with a constant force.  The kinetic energies of these particles are given by the dispersion 
relation of kinks. The energy spectrum  of their bound states is determined
in this technique by the semiclassical (or canonical) quantization of the
classical kink dynamics.  To leading order in $h$, the meson energy spectra 
 obtained by the  two aforementioned methods, coincide both for the IFT, and for the Ising spin chain. 
 
 One more important advantage of the second (heuristic) method is that it can be applied, after a proper modification,
 to models, in which kinks are not free, but interact with each other already  in the deconfined phase at $h=0$. 
This modification  was introduced in paper [\onlinecite{RutP09}], in which the meson mass spectrum in the Potts field
theory was  studied. It was shown there, that the strong short-range interaction between kinks in the deconfined phase of 
this model can be accounted for by the semiclassical Bohr-Sommerfeld quantization  condition by adding the two-kink 
scattering phase to its left-hand side. As a result, the semiclassical meson mass spectrum determined by the 
modified Bohr-Sommerfeld rule carries information about the non-trivial kink-kink scattering in the deconfined phase. 
 By means of this improved semiclassical technique, the  meson energy spectra were later  calculated in several models exhibiting confinement, including   the XXZ spin chain in a staggered magnetic field \cite{Rut18}, the XXZ spin ladder \cite{Lagnese_2022}, the transverse-field
 Ising ladder \cite{Ram20}, and the  thermally deformed  tricritical Ising model \cite{Mus22}.

In this work, we continue our study of the kink confinement in the  gapped antiferromagnetic XXZ spin chain in a weak staggered 
magnetic field,  initiated in  \cite{Rut18}. The Hamiltonian of the model is given by
\begin{align}\nonumber
\mathcal{H}(h)&=-\frac{J}{2}\sum_{j=-\infty}^\infty\!\!\left(\sigma_j^x\sigma_{j+1}^x+\sigma_j^y\sigma_{j+1}^y+
\Delta\,
\sigma_j^z\sigma_{j+1}^z
\right)\\
&-h\sum_{j=-\infty}^\infty (-1)^j\sigma_j^z.\label{XXZH}
\end{align}
Here the index $j$ enumerates the spin-chain sites, $\sigma_j^{\mathfrak{a}}$ are
the Pauli matrices, $\mathfrak{a} = x, y, z$, $J>0$  is the 
coupling constant, $\Delta<-1$ is the anisotropy parameter,  
$h$ is the strength of the staggered magnetic field.
Model \eqref{XXZH} has been used by Bera {\it et al.} \cite{Bera17} for the interpretation of their
 neutron scattering investigations of the  magnetic excitations  
in the quasi-1D antiferromagnetic 
compound  ${\rm {SrCo_2V_2O_8}}$. The effective staggered field accounts in the mean-field approximation for 
the weak inter-chain interaction in a three-dimensional (3D) array of parallel
spin chains in the 3D-ordered phase of such compounds, as it was suggested by Shiba \cite{Shiba80}.

Exploiting the integrability of model \eqref{XXZH}  in the deconfined phase at $h=0$, we perform two alternative perturbative calculations of  the meson energy spectra  in the weak confinement regime $h\to +0$ to the first order in the small parameter $h$. 
First, we present the details of the calculation announced previously \cite{Rut18}, which employs the non-rigorous  heuristic procedure outlined above.
Then, we derive the Bethe-Salpeter equation for model \eqref{XXZH} in the two-kink approximation. From the perturbative solution of this equation, we 
calculate in a systematic  fashion the meson energy spectra, and justify previously obtained results. Furthermore, we derive from the perturbative
solutions of the Bethe-Salpeter equation  the explicit formulas for the  two-kink contribution to the Dynamical Structure Factors (DSF) of the local spin operators for model
\eqref{XXZH}  at zero temperature in the weak confinement regime. 

The paper is structured in the following way. In   Sections \ref{Sec2} and \ref{Sec:DNS0},  we  recall some well-known properties of the XXZ spin-1/2  infinite chain in the gapped antiferromagnetic phase  at zero magnetic field. Section \ref{Sec2} contains information about 
some basic properties of  the  low-energy excitations in this model and the structure of their Hilbert space.
Section \ref{Sec:DNS0} is devoted to the  DSF of local spin 
operators in model  \eqref{XXZH} at $h=0$. By means of a straightforward unified calculation procedure, we derive 
new explicit formulas for the transverse and longitudinal DSF in the deconfined phase   in the two-kink approximation. Starting from
Sections \ref{BSE1}, we proceed to the analysis of the confinement in model \eqref{XXZH} induced by a weak staggered magnetic field $h>0$.
In Section \ref{BSE1} we  classify the meson bound states in the weak confinement regime, and
describe the heuristic calculation of their energy spectra. The  Bethe-Salpeter equation for  model \eqref{XXZH} is derived in Section \ref{SecBSE}. The perturbative solution of this equation in several  asymptotical regimes  is given in Section \ref{WK}. Using the results of this asymptotic analysis, we derive  the initial terms of the small-$h$ expansion for the
meson energy spectra, and justify the results of the previous non-rigorous heuristic calculations of these spectra.
Section \ref{DSFconf} contains the calculation of the two-kink contribution to the transverse and longitudinal DSF in the  confinement regime in the 
leading order in the staggered  field $h$. Concluding remarks are presented in Section \ref{Conc}. Finally,   some technical details are relegated
to four Appendixes. 
\section{Infinite XXZ spin chain at zero magnetic field \label{Sec2}}
In this section, we remind some well-known properties of the XXZ spin-1/2 chain \eqref{XXZH} at zero  staggered magnetic field. 
At $h=0$, the Hamiltonian \eqref{XXZH}  reduces to the form $H=\mathcal{H}(0)$:
\begin{equation}\label{XXZM}
{H}=-\frac{J}{2}\sum_{j=-\infty}^\infty\!\!\left(\sigma_j^x\sigma_{j+1}^x+\sigma_j^y\sigma_{j+1}^y+
{\Delta}\,\sigma_j^z\sigma_{j+1}^z
\right).
\end{equation}
Three phases are realized in the infinite 
spin chain \eqref{XXZM} at zero temperature in different regions of the anisotropy parameter $\Delta$: 
the ferromagnetic phase at $\Delta>1$, the critical  phase (spin-fluid, Luttinger liquid)
at $-1<\Delta<1$, and the gapped (massive) antiferromagnetic phase at $\Delta<-1$. 
Only the gapped antiferromagnetic phase will be considered in this paper. We shall use the standard parametrisation 
for the anisotropy parameter $\Delta<-1$:
\begin{align}
{\Delta}=({q}+{q}^{-1})/2=-\cosh \eta, \\
q=-\exp(- \eta)\in(-1,0),\quad \eta>0.
\end{align}

The Hamiltonian \eqref{XXZM} commutes with the $z$-projection of the total spin
\begin{equation}
S^z=\frac{1}{2}\sum_{j=-\infty}^\infty \sigma_j^z.
\end{equation}
For short,  the operator $S^z$ will be called the "total spin" in the sequel.
The Hamiltonian \eqref{XXZM} commutes as well with the unitary operator 
$
U=\otimes_{j\in\mathbb{Z}} \,\sigma_j^x,
$
 and with the 
 translation operator  by one chain site ${T}_1$, that acts on the Pauli matrices as 
\begin{equation}\label{T1s}
{T}_1^{-1}  \sigma_j^\mathfrak{a} {T}_1= \sigma_{j+1}^\mathfrak{a}.
\end{equation} 
Note, that 
\begin{equation}\label{Us}
 U\sigma_j^{y,z}  U^{-1}= -\sigma_{j}^{y,z},  \quad U  \sigma_j^{x}  U^{-1}= \sigma_{j}^x.
\end{equation} 
It is also useful to introduce the modified translation operator $\widetilde{T}_1=T_1 U$, which, of course, commutes with
the Hamiltonian  \eqref{XXZM}  as well. Its action on the Pauli matrices can be read from equations \eqref{T1s},
\eqref{Us}:
\begin{equation}\label{tilT1}
\widetilde{T}_1^{-1}   \sigma_j^{y,z} \widetilde{T}_1= -\sigma_{j+1}^{y,z},  \quad \widetilde{T}_1^{-1} \sigma_j^{x} \widetilde{T}_1= \sigma_{j+1}^x.
\end{equation}

The structure of the ground-states and low-energy excitations of the infinite chain  \eqref{XXZM} 
in the gapped antiferromagnetic phase is well known  \cite{Jimbo94}. Since this structure is qualitatively the same
for all $\Delta<-1$, it can be well understood by considering the Ising limit case $\Delta\to-\infty$, 
where the Hamiltonian simplifies drastically. 
In this limit, it is convenient to rescale the Hamiltonian \eqref{XXZM} and to add to it a suitable 
(infinite in the thermodynamic limit)  constant:
\begin{equation}\label{HIs}
{H}_I(\varepsilon)\equiv\frac{{H}}{J|\Delta|}+Const={H}_I^{(0)}+\varepsilon V,
\end{equation}
where $\varepsilon=|\Delta|^{-1}$ is a small parameter, and
\begin{eqnarray}\label{HI}
&&{H}_I^{(0)}=\frac{1}{2}\sum_{j=-\infty}^\infty (\sigma_j^z\sigma_{j+1}^z+1), \\\nonumber
&&V=\sum_{j=-\infty}^\infty(\sigma_j^+\sigma_{j+1}^-+\sigma_j^-\sigma_{j+1}^+),
\end{eqnarray}
with $\sigma_j^\pm= \frac{1}{2}(\sigma_j^x\pm i\sigma_j^y)$.

The model  \eqref{XXZM}  considered on a finite chain 
is solvable by the Bethe Ansatz method \cite{Orb58}, see also \cite{Tak09,Zv10,Dug15} for further references.
In the thermodynamic limit, the Hilbert space $\mathcal{L}$ of low-energy states of model  \eqref{XXZM} 
can be represented as the direct sum 
of four subspaces 
\begin{equation}
\mathcal{L} =\mathcal{L}_{00}\oplus\mathcal{L}_{11}\oplus\mathcal{L}_{01}\oplus\mathcal{L}_{10}.
\end{equation} 
The subspaces $\mathcal{L}_{\mu\nu}$ will be called the  topological sectors. The subspaces $\mathcal{L}_{00}\oplus\mathcal{L}_{11}$ and 
$\mathcal{L}_{01}\oplus\mathcal{L}_{10}$ represent the topologically neutral and topologically charged sectors, respectively. 
Each subspace $\mathcal{L}_{\mu\nu}$, in turn, can be decomposed into the sum 
of $n$-particle subspaces, with $n$ even for the neutral topological sectors, and $n$ odd for the charged topological sectors:
\begin{align}
&\mathcal{L}_{00}=\oplus_{m=0}^\infty\, \mathcal{L}_{00}^{(2m)}, \quad 
\mathcal{L}_{11}=\oplus_{m=0}^\infty\, \mathcal{L}_{11}^{(2m)},\\
&\mathcal{L}_{01}=\oplus_{m=0}^\infty\, \mathcal{L}_{01}^{(2m+1)}, \quad 
\mathcal{L}_{10}=\oplus_{m=0}^\infty \,\mathcal{L}_{10}^{(2m+1)}.
\end{align} 
Two vacuum subspaces  $\mathcal{L}_{00}^{(0)}$, $\mathcal{L}_{11}^{(0)}$ are one-dimensional, while 
all other subspaces $\mathcal{L}_{\mu\nu}^{(n)}$, $n>0$ have infinite dimensions.
\subsection{Vacuum sector}
There are two   
degenerate ground states $|vac\rangle^{(\mu)}$, $\mu=0,1$,  showing a N\'eel-type order, 
\begin{align}\label{vac1}
&\phantom{.}^{(1)} \langle vac| \sigma_j^z|vac\rangle^{(1)} =(-1)^j\bar{\sigma},\\
&\phantom{.}^{(0)} \langle vac| \sigma_j^z|vac\rangle^{(0)}=-(-1)^j\bar{\sigma}.
\end{align}
with  the staggered spontaneous magnetization  \cite{Baxter1973,Baxter1976,Iz99}
\begin{equation}\label{sig}
\bar{\sigma}(\eta)=\prod_{n=1}^\infty \left(
\frac{1-e^{-2 n \eta}}{1+e^{-2 n\eta}}
\right)^2.
\end{equation}
In the Ising limit $\eta\to\infty$ these ground states become the pure N\'eel states:
\begin{subequations}\label{vac0}
\begin{align}
\lim_{\eta\to\infty}|vac\rangle^{(1)}=|0\rangle^{(1)}:  \quad \ldots\downarrow{\color{blue}\
\underline{\stackrel{0}{\uparrow}}}\stackrel{1}{\downarrow}\stackrel{2}{\uparrow}\downarrow\ldots,\\
\lim_{\eta\to\infty}|vac\rangle^{(0)}=|0\rangle^{(0)}: \quad \ldots\uparrow{\color{blue}\underline{\stackrel{0}{\downarrow}}}\stackrel{1}{\uparrow}\stackrel{2}{\downarrow}\uparrow\ldots, 
\end{align}
\end{subequations}
and $\bar{\sigma}(\eta)\to1$.

The Hamiltonian symmetries corresponding to the operators $T_1$ and $U$ are spontaneously broken in the 
antiferromagnetic phase,
\begin{eqnarray}\label{T1vac}
T_1 |vac\rangle^{(1)}= |vac\rangle^{(0)},\quad T_1 |vac\rangle^{(0)}= |vac\rangle^{(1)},\\
U |vac\rangle^{(1)}= |vac\rangle^{(0)},\quad U |vac\rangle^{(0)}=  |vac\rangle^{(1)}.\label{Uvac}
\end{eqnarray}
On the other hand, the antiferromagnetic vacua  $  |vac\rangle^{(\mu)}$ are invariant with the respect to the modified translation operator
 $\widetilde{T}_1$,
\begin{equation}\label{T1vacA}
\widetilde{T}_1 |vac\rangle^{(\mu)}= |vac\rangle^{(\mu)},\;\;{\rm with}\;\; \mu=0,1.
\end{equation}

The ground-state energy $E_N(\Delta)$ of the periodic  chain having $N$ sites increases linearly with $N$ in the thermodynamic limit:
\begin{equation}\label{VE}
\lim_{N\to\infty} \frac{E_N(\Delta)}{N}= \frac{J}{2} C(\Delta).
\end{equation}
The ground-state energy per lattice site defined by the above equation 
is explicitly known due to Yang and Yang \cite{YY66_1,YY66_2}. The
ground state energy $E_N(\Delta)$ diverges in the thermodynamic limit $N\to\infty$.
In order to get rid of it, 
it is convenient to redefine the Hamiltonian \eqref{XXZM} by adding  an appropriate constant term
\begin{equation}\label{XXZ1}
{H}_1=-\frac{J}{2}\sum_{j=-\infty}^\infty\!\!\Big(\sigma_j^x\sigma_{j+1}^x+\sigma_j^y\sigma_{j+1}^y+
{\Delta}\,\sigma_j^z\sigma_{j+1}^z+ C(\Delta)
\Big),
\end{equation}
such that 
\begin{equation}\label{vacH0}
{H}_1|vac\rangle^{(\mu)}=0,
\end{equation}
with $\mu=0,1$.
\subsection{One-kink sector}
The elementary excitations are topologically charged,  represented \cite{Jimbo94} by  the
kinks $|K_{\mu\nu}({p})\rangle_s$ interpolating between the vacua $\mu$ and $\nu$,
and characterized by the quasimomentum $ {p}\in \mathbb{R}$, and by the 
$z$-projection of the spin $s=\pm1/2$,
\begin{subequations}
\begin{eqnarray} \label{omK}
&&{H}_1|K_{\mu\nu}({p})\rangle_s=\omega(p)\,|K_{\mu\nu}({p})\rangle_s,\\\label{T1K}
&&\widetilde{T}_1 |K_{\mu\nu}({p})\rangle_s=e^{i {p} }\,|K_{\mu\nu}({p})\rangle_{-s},\\\label{SK}
&&S^z |K_{\mu\nu}({p})\rangle_s=s\,|K_{\mu\nu}({p})\rangle_s~.
\end{eqnarray}
\end{subequations}
They also satisfy the symmetry properties:
\begin{subequations}\label{SymK}
\begin{eqnarray} 
&&{T}_1 |K_{\mu\nu}({p})\rangle_s=e^{i {p} }\,|K_{\nu\mu}({p})\rangle_{s},\\
&&U |K_{\mu\nu}({p})\rangle_s= |K_{\nu\mu}({p})\rangle_{-s},\\
 &&|K_{\mu\nu}({p+\pi})\rangle_s= \varkappa(\mu,s)\,|K_{\mu\nu}({p})\rangle_s,
\end{eqnarray}
\end{subequations}
where
\begin{align} \label{kap0}
\varkappa(0,1/2)= \varkappa(1,-1/2)=1, \\\nonumber
   \varkappa(1,1/2)= \varkappa(0,-1/2)=-1.
\end{align}
The quantum number $s=\pm1/2$ will be called ``the spin" for short.
Since the kinks carry spin $\pm1/2$, they are also often called ``spinons".  We shall use both terms as synonyms.
The spinon dispersion law  was found by Johnson, Krinsky, and McCoy \cite{McCoy73},
\begin{equation}\label{dl}
\omega({p},\eta)=I \,\sqrt{1-k^2 \cos^2 {p}},
\end{equation}
where
\begin{equation}\label{Iet}
I=\frac{2 J K}{\pi}\sinh \eta,
\end{equation}
and $K$ [$K'$] is the complete elliptic integral of modulus $k$ [$k'=\sqrt{1-k^2}$] such that 
\begin{equation}\label{KpK}
\frac{K'}{K }= \frac{\eta}{\pi}.
\end{equation}
Note, that the spinon dispersion law \eqref{dl} coincides up to a numerical factor and re-parametrization with the 
kink dispersion law \cite{Rut08a}
\begin{equation}\label{dIs}
\omega_{{I\!s}}({p},h_x)=2(1+h_x) \,\sqrt{1-\frac{4 h_x}{(1+h_x)^2} \cos^2 (p/2)},
\end{equation}
in the ferromagnetic Ising spin chain in the transverse magnetic field $h_x$. The latter model is defined by the Hamiltonian
\begin{equation}\label{HIsa}
{H}_{{I\!s}}=-\sum_{j=-\infty}^\infty(\sigma_j^z\sigma_{j+1}^z+h_x \sigma_j^x).
\end{equation}
The ferromagnetic phase in this well-studied integrable model is realized at $|h_x|<1$.

The dispersion law \eqref{dl} can be parametrized in terms of the Jacobi elliptic functions of modulus $k$:
\begin{align}\label{pe}
{p}(\alpha)=-\frac{\pi}{2}+\mathrm{am}\,\left(\frac{2 K \alpha}{\pi},k\right),\\
\omega(\alpha)=I \, \mathrm{dn}\,\left(\frac{2 K \alpha}{\pi},k\right)=J \sinh\eta\,\, \frac{d p(\alpha)}{d\alpha},\label{le}
\end{align}
where $\alpha$ is the rapidity variable \begin{footnote}
{The rapidity variable $\alpha$ is simply related with the rapidity variable $\lambda$  used previously in \cite{Rut18}: 
$\alpha=\pi-\lambda$. The definition of the rapidity $\alpha$ adopted here has been  changed in order to harmonize
 it  with notations in the monograph  
\cite{Jimbo94} by Miwa and Jimbo, see equation (7.18) there.}
\end{footnote}.

The space $\mathcal{L}^{(1)}$ of one-kink states is the sum of two subspaces  $\mathcal{L}^{(1)}=\mathcal{L}_{01}^{(1)}\oplus \mathcal{L}_{10}^{(1)}$, which are  spanned by
basis vectors $|K_{01}({p})\rangle_s$ and $|K_{10}({p})\rangle_s$, respectively. These basis vectors are normalised by the condition
\begin{equation}\label{Knorm}
\phantom{.}_{s} \langle K_{\nu\mu}(p) |K_{\mu'\nu'}(p')\rangle_{s'}=\pi\delta_{\mu\mu'}\delta_{\nu\nu'}\delta_{ss'}\delta(p-p'),
\end{equation}
for $p,p'\in[0,\pi)$.

Commonly used 
 are also the kink states $ |\mathcal{K}_{\mu\nu}(\xi)\rangle_{s}$ parametrised by the complex spectral parameter 
$\xi(\alpha)=-i e^{i\alpha}$. These states differ from $|K_{\mu\nu}(p)\rangle_{s}$ by the numerical factor
$\sqrt{p'(\alpha})$:
\begin{equation}
 |\mathcal{K}_{\mu\nu}(\xi)\rangle_{s}=\sqrt{\frac{\omega(p)}{J\sinh \eta}}\,\,|K_{\mu\nu}(p)\rangle_{s}.
 \end{equation}
Note, that a different notation $|\xi\rangle_{\epsilon;(i)}$ has been widely used \cite{Jimbo94}
for the one-kink states $ |\mathcal{K}_{\mu\nu}(\xi)\rangle_{s}$, with $\epsilon=\mathrm{sign}\,s$, and $i=\nu$.

The completeness relations for the projection operators on the subspaces $\mathcal{L}_{01}^{(1)}$ and $ \mathcal{L}_{10}^{(1)}$ read:
\begin{subequations}\label{PP}
\begin{align} 
\mathcal{P}_{01}^{(1)}=\sum_{s=\pm1/2}\int_0^\pi \frac{dp}{\pi}|K_{01}(p)\rangle_{s}\phantom{.}_{s} \langle K_{10}(p) |=\\\nonumber
\sum_{s=\pm1/2}\int_0^\pi \frac{d\alpha}{\pi}|\mathcal{K}_{01}[\xi(\alpha)]\rangle_{s}\phantom{.}_{s} \langle \mathcal{K}_{10}[\xi(\alpha)] |,\\
\mathcal{P}_{10}^{(1)}=\sum_{s=\pm1/2}\int_0^\pi \frac{dp}{\pi}|K_{10}(p)\rangle_{s}\phantom{.}_{s} \langle K_{01}(p) |=\\\nonumber
\sum_{s=\pm1/2}\int_0^\pi \frac{d\alpha}{\pi}|\mathcal{K}_{10}[\xi(\alpha)]\rangle_{s}\phantom{.}_{s} \langle \mathcal{K}_{01}[\xi(\alpha)]|.
 \end{align} 
\end{subequations}

It is instructive to describe the kink states $|K_{\mu\nu}({p})\rangle_s$ explicitly in the 
Ising limit $\eta \gg1$ by means of the Rayleigh-Schr\"odinger 
perturbation theory in the small parameter $\varepsilon=1/\cosh \eta$ for the Hamiltonian \eqref{HIs}. 
To this end, one can first consider the localized kink states $|\mathbf{K}_{\mu \nu}(j)\rangle$, which interpolate
between  vacua $|0\rangle^{(\mu)}$ to the left, and $|0\rangle^{(\nu)}$ to the right of the bond $( j,j+1)$:
    \begin{eqnarray*}
&&|\mathbf{K}_{10}(j)\rangle:\quad \ldots\downarrow{\color{blue}\
\underline{\stackrel{0}{\uparrow}}}\stackrel{1}{\downarrow}\uparrow\downarrow\ldots \downarrow 
{\color{dgreen} {\stackrel{j}{\uparrow}}}\,\,{\color{red}\mid}{\color{dgreen}{\stackrel{j+1}{\uparrow}}
}\downarrow\uparrow\ldots, \quad \text{at even } j,\\
&&|\mathbf{K}_{10}(j)\rangle:\quad \ldots\downarrow{\color{blue}\
\underline{\stackrel{0}{\uparrow}}}\stackrel{1}{\downarrow}\uparrow\downarrow\ldots \uparrow 
{\color{dgreen} {\stackrel{j}{\downarrow}}}\,\,{\color{red}\mid}{\color{dgreen} {\stackrel{j+1}{\downarrow}}}
\uparrow\downarrow\ldots, \quad \text{at odd } j,\\
&&|\mathbf{K}_{01}(j)\rangle:\quad \ldots\uparrow{\color{blue}\underline{\stackrel{0}{\downarrow}}}\stackrel{1}{\uparrow}\downarrow\uparrow\ldots\downarrow 
{\color{dgreen} {\stackrel{j}{\uparrow}}}\,\,{\color{red}\mid}{\color{dgreen}{\stackrel{j+1}{\uparrow}}
}\downarrow\uparrow\ldots, \quad \text{at odd } j,\\
&&|\mathbf{K}_{01}(j)\rangle:\quad \ldots\uparrow{\color{blue}\underline{\stackrel{0}{\downarrow}}}\stackrel{1}{\uparrow}\downarrow\uparrow\ldots\uparrow 
{\color{dgreen} {\stackrel{j}{\downarrow}}}\,\,{\color{red}\mid}{\color{dgreen}{\stackrel{j+1}{\downarrow}}
}\uparrow\downarrow\ldots, \quad \text{at even } j.
\end{eqnarray*}
These  states  are the eigenvectors of the zero-order Hamiltonian ${H}_I^{(0)}$, which are characterized by the same (unit) eigenvalue:
\begin{equation}
{H}_I^{(0)}|\mathbf{K}_{\mu\nu}(j)\rangle=|\mathbf{K}_{\mu\nu}(j)\rangle.
\end{equation}
The localized kink states  $|\mathbf{K}_{\mu \nu}(j)\rangle$ are normalized by the 
condition
\begin{equation}\label{nc}
\langle \mathbf{K}_{\nu\mu}(j) |\mathbf{K}_{\mu'\nu'}(j')\rangle=\delta_{\mu\mu'}\delta_{\nu\nu'}\delta_{jj'}.
\end{equation}
Their transformation properties under the action of the symmetry operators  read:
\begin{eqnarray*}
&&{T}_1 |\mathbf{K}_{\mu\nu}(j)\rangle=|\mathbf{K}_{\nu\mu}(j-1)\rangle,\\
&&U  |\mathbf{K}_{\mu\nu}(j)\rangle=|\mathbf{K}_{\nu\mu}(j)\rangle,\\
&&\widetilde{T}_1 |\mathbf{K}_{\mu\nu}(j)\rangle=|\mathbf{K}_{\mu\nu}(j-1)\rangle.
\end{eqnarray*}

The degeneracy in the excitation energy is removed in the first oder in $\varepsilon$: 
 \begin{equation}\label{omI}
{H}_I(\varepsilon)|K_{\mu\nu}^I(p)\rangle_s=[1-2 \varepsilon \cos(2p)]|K_{\mu\nu}^I(p)\rangle_s+O(\varepsilon^2).
\end{equation} 
The first-oder perturbative result for the kink energy  in this equation
recovers 
 two initial terms in the Taylor expansion  in $\varepsilon$ of the exact kink energy \eqref{dl}:
\begin{equation}\label{omT}
\frac{\omega(p)}{J|\Delta|}=1-2 \varepsilon \cos(2 p)+\varepsilon^2\,\left[
\frac{3}{2}-\cos(4{p})
\right]+O(\varepsilon^3).
\end{equation}

The first-order eigenstates  $|K_{\mu\nu}^I(p)\rangle_s$ in \eqref{omI} denote the kink Bloch states in the Ising limit:
\begin{subequations}\label{Kis}
\begin{eqnarray}
&& |K_{10}^I(p)\rangle_{1/2}=\sum_{m=-\infty}^\infty e^{i\, (2m+1)p} 
 |\mathbf{K}_{10}(2m)\rangle,
 \\
 && |{K}_{01}^I(p)\rangle_{1/2}=\sum_{m=-\infty}^\infty e^{2 i\, mp} 
 |\mathbf{K}_{01}(2m-1)\rangle,\\
 && |K_{10}^I(p)\rangle_{-1/2}=\sum_{m=-\infty}^\infty e^{2 i\, mp} 
 |\mathbf{K}_{10}(2m-1)\rangle,\\
  && |K_{01}^I(p)\rangle_{-1/2}=\sum_{m=-\infty}^\infty e^{i\, (2m+1)p} 
 |\mathbf{K}_{01}(2m)\rangle.
 \end{eqnarray}
 \end{subequations}
Due to  \eqref{nc}, the   kink  Bloch states \eqref{Kis}  satisfy at $p,p'\in [0,\pi)$ the normalisation condition
 \begin{equation} 
\phantom{.}_{s} \langle K_{\nu\mu}^I(p) |K_{\mu'\nu'}^I(p')\rangle_{s'}= \pi\delta_{\mu\mu'}\delta_{ss'}\delta(p-p').
\end{equation}
These states  satisfy also equations \eqref{T1K}, \eqref{SK},
\eqref{SymK},  and \eqref{PP}. All these properties indicate that the states $|K_{\mu\nu}^{I}(p)\rangle_s$ indeed represent the Ising limit of the one-kink
topological excitations $|K_{\mu\nu}(p)\rangle_s$ in the infinite antiferromagnetic XXZ spin chain \eqref{XXZM}:
\begin{equation}\label{limK1}
\lim_{\varepsilon\to 0}|K_{\mu\nu}(p)\rangle_s=|K_{\mu\nu}^{I}(p)\rangle_s.
\end{equation}

Note, that  equation \eqref{pe} describing the relation between the momentum and rapidity variables 
reduces in the Ising limit $\eta\to\infty$ to the simple linear dependence:
\begin{equation}\label{prap}
p(\alpha)=-\frac{\pi}{2}+\alpha +O(e^{-\eta}).
\end{equation}

The opposite limit $\eta\ll 1$ corresponds to the scaling regime.  In this limit, the kink dispersion law \eqref{dl} at small momenta $p$ 
takes the relativistic form:
\begin{equation}
\omega(p,\eta)= I k\sqrt{m^2+p^2} \left[
1+O\left(\frac{p^4}{m^2+p^2}\right)
\right],
\end{equation}
where 
\begin{equation}\label{kmass}
m=\frac{k'}{k}=4 \exp[-\pi^2/(2\eta)]\,\left(1+O(\exp[-\pi^2/\eta])\right).
\end{equation}
is the kink mass. Equations \eqref{pe}, \eqref{le} reduce in the scaling limit to the form
\begin{equation}
p(\beta)=m \sinh \beta, \quad \omega(\beta )= I\, k\, m\cosh \beta,
\end{equation}
where $\beta=(\alpha-\frac{\pi}{2})\frac{\pi}{\eta}$ is the rescaled rapidity variable.
\subsection{Two-kink sector \label{2kink}}
The subspace $\mathcal{L}^{(2)}$ of two-kink excitation of model \eqref{XXZM} in the antiferromagnetic phase $\Delta<-1$ is the 
direct sum of two subspaces $\mathcal{L}^{(2)}=\mathcal{L}_{00}^{(2)}\oplus \mathcal{L}_{11}^{(2)}$. The space 
$\mathcal{L}_{00}^{(2)}$ is spanned by the two-kink states $|K_{01}(p_1)K_{10}(p_2)\rangle_{s_1s_2}$, while the basis of the
space $\mathcal{L}_{11}^{(2)}$ is formed by the states $|K_{10}(p_1)K_{01}(p_2)\rangle_{s_1s_2}$. The projector operators
onto these two subspaces read:
\begin{subequations}\label{PP2}
\begin{align} 
\mathcal{P}_{00}^{(2)}=\sum_{s_1=\pm1/2\atop{s_2=\pm1/2}}\iint_{\Gamma} \frac{dp_1dp_2}{\pi^2}
|K_{01}(p_1)K_{10}(p_2)\rangle_{s_1s_2}\cdot\\\nonumber
\phantom{.}_{s_2 s_1} \langle K_{01}(p_2) K_{10}(p_1) |,\\\label{PP11}
\mathcal{P}_{11}^{(2)}=\sum_{s_1=\pm1/2\atop{s_2=\pm1/2}}\iint_{\Gamma} \frac{dp_1dp_2}{\pi^2}
|K_{10}(p_1)K_{01}(p_2)\rangle_{s_1s_2}\cdot\\\nonumber
\phantom{.}_{s_2 s_1} \langle K_{10}(p_2) K_{01}(p_1) |,
 \end{align} 
 \end{subequations}
where $\Gamma=\{p_1,p_2\in \Gamma | 0\le p_2<p_1<\pi \}$ is the fundamental triangular region in the plane $\langle p_1,p_2\rangle$.

The two-kink states $|K_{\mu\nu}(p_1)K_{\nu\mu}(p_2)\rangle_{s_1s_2}$
 are characterized by momenta $p_1,p_2\in\mathbb{R} $ and spins $s_1,s_2\in \{1/2,-1/2\}$ of particular kinks. 
 The energy of such a state is the sum of energies of particular kinks:
 \begin{align}\label{H1S}
 &{H}_1|K_{\mu\nu}(p_1)K_{\nu\mu}(p_2)\rangle_{s_1s_2}=[\omega(p_1)+\omega(p_2)]\cdot\\\nonumber
&	
|K_{\mu\nu}(p_1)K_{\nu\mu}(p_2)\rangle_{s_1s_2}.
 \end{align}
Besides, these states have the following properties:
\begin{subequations}\label{prop}
\begin{align}\label{T2K}
&{T}_1^2 |K_{\mu\nu}(p_1)K_{\nu\mu}(p_2)\rangle_{s_1s_2}=e^{2i(p_1+p_2)}\cdot\\\nonumber
&|K_{\mu\nu}(p_1)K_{\nu\mu}(p_2)\rangle_{s_1s_2},\\
&{T}_1 |K_{\mu\nu}(p_1)K_{\nu\mu}(p_2)\rangle_{s_1s_2}=e^{i(p_1+p_2)}\cdot\\\nonumber
&|K_{\nu\mu}(p_1)K_{\mu\nu}(p_2)\rangle_{s_1s_2},\\
&U |K_{\mu\nu}(p_1)K_{\nu\mu}(p_2)\rangle_{s_1s_2}= |K_{\nu\mu}(p_1)K_{\mu\nu}(p_2)\rangle_{-s_1-s_2},\\\label{tilT}
&\widetilde{T}_1 |K_{\mu\nu}(p_1)K_{\nu\mu}(p_2)\rangle_{s_1s_2}=e^{i(p_1+p_2)}\cdot\\\nonumber
&|K_{\mu\nu}(p_1)K_{\nu\mu}(p_2)\rangle_{-s_1-s_2},\\
&S^z |K_{\mu\nu}(p_1)K_{\nu\mu}(p_2)\rangle_{s_1s_2}=(s_1+s_2)\cdot\\\nonumber
&|K_{\mu\nu}(p_1)K_{\nu\mu}(p_2)\rangle_{s_1s_2},\\\label{pishift}
&|K_{\mu\nu}(p_1)K_{\nu\mu}(p_2)\rangle_{s_1s_2}=\\\nonumber
&\varkappa(\mu,s_1)|K_{\mu\nu}(p_1+\pi)K_{\nu\mu}(p_2)\rangle_{s_1s_2}\\
&=\varkappa(\nu,s_2)|K_{\mu\nu}(p_1)K_{\nu\mu}(p_2+\pi)\rangle_{s_1s_2},\nonumber
\end{align}
\end{subequations}
where $\varkappa(\mu,s)$ is given by \eqref{kap0}.

It is useful to define an alternative  basis in the  subspace  of two-kink states with zero total spin $S^z = 0$:
\begin{align}\nonumber
&|K_{\mu\nu}(p_1)K_{\nu\mu}(p_2)\rangle_\pm\equiv
\frac{1}{\sqrt{2}}\Big(
|K_{\mu\nu}(p_1)K_{\nu\mu}(p_2)\rangle_{1/2,-1/2}\\
&\pm
|K_{\mu\nu}(p_1)K_{\nu\mu}(p_2)\rangle_{-1/2,1/2}
\Big).\label{baspm1}
\end{align}
The  modified translation operator  $ \widetilde{T}_1$ becomes diagonal in this basis:
\begin{equation}\label{trmod}
 \widetilde{T}_1|K_{\alpha\beta}(p_1)K_{\beta\alpha}(p_2)\rangle_\pm=
\pm e^{i(p_1+p_2)}|K_{\alpha\beta}(p_1)K_{\beta\alpha}(p_2)\rangle_\pm.
\end{equation}
Due to \eqref{pishift}, these states transform in the following way under the shift of the kink momenta  by $\pi$:
\begin{align}\nonumber
&|K_{\mu\nu}(p_1)K_{\nu\mu}(p_2)\rangle_\pm=(-1)^{\mu}|K_{\mu\nu}(p_1+\pi)K_{\nu\mu}(p_2)\rangle_\mp=\\\label{ShM}
&(-1)^{\mu}|K_{\mu\nu}(p_1)K_{\nu\mu}(p_2+\pi)\rangle_\mp.
\end{align}
The two-kink scattering can be described by the Faddeev-Zamolodchikov commutation relations:
\begin{subequations}\label{FZC}
\begin{align}\label{FZ2}
|K_{\mu\nu}(p_1)K_{\nu\mu}(p_2)\rangle_{ss}=w_0(p_1,p_2)
 |K_{\mu\nu}(p_2)K_{\nu\mu}(p_1)\rangle_{ss},\\
|K_{\mu\nu}(p_1)K_{\nu\mu}(p_2)\rangle_{\pm}=w_\pm(p_1,p_2)
|K_{\mu\nu}(p_2)K_{\nu\mu}(p_1)\rangle_\pm.\label{FZpm}
\end{align}
\end{subequations}
\begin{figure}
\centering
\subfloat[ 
]
{\label{et1}
\includegraphics[width=.9\linewidth]{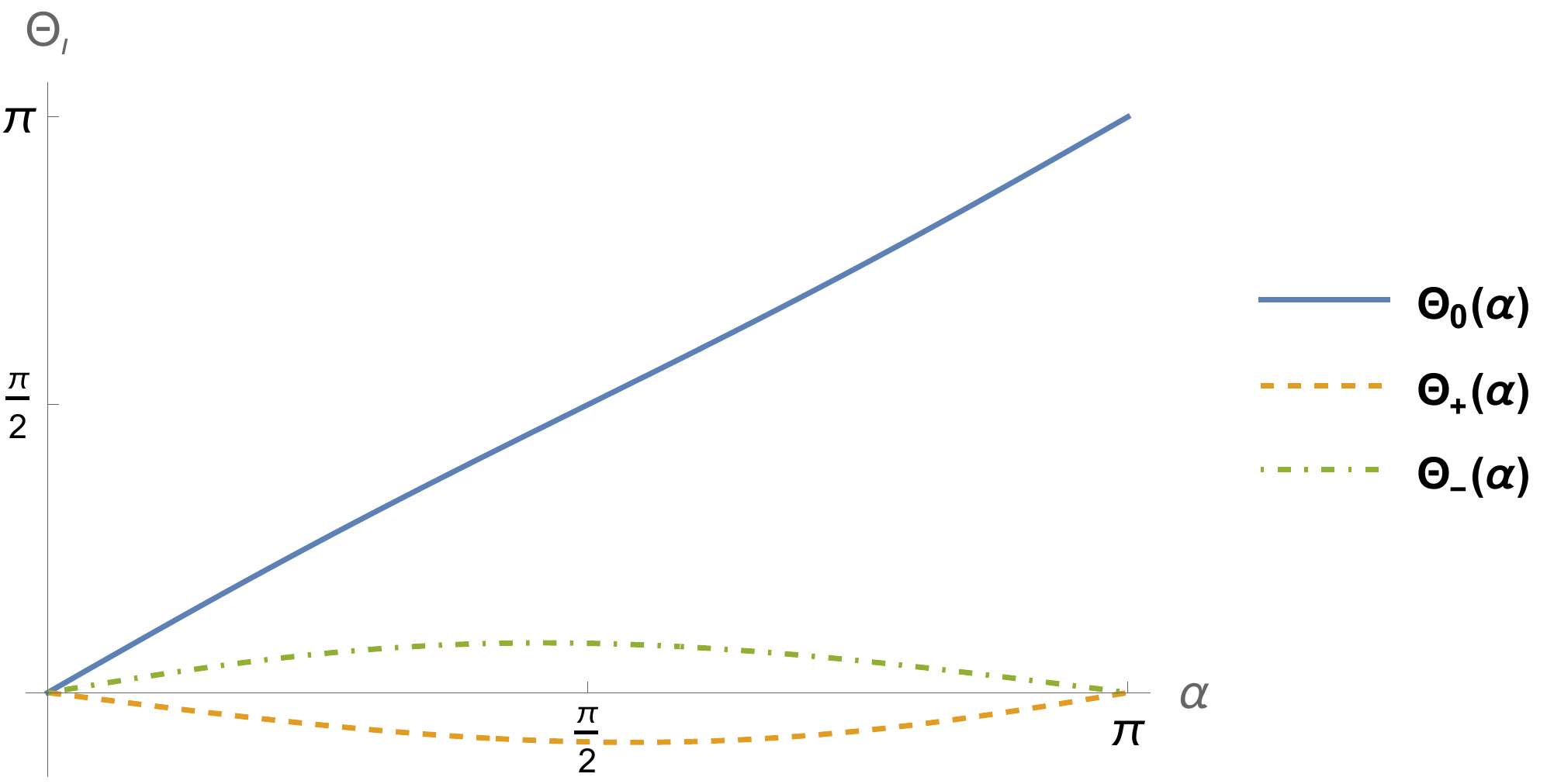}}

\subfloat[
]
	{\label{et2}
\includegraphics[width=.9\linewidth]{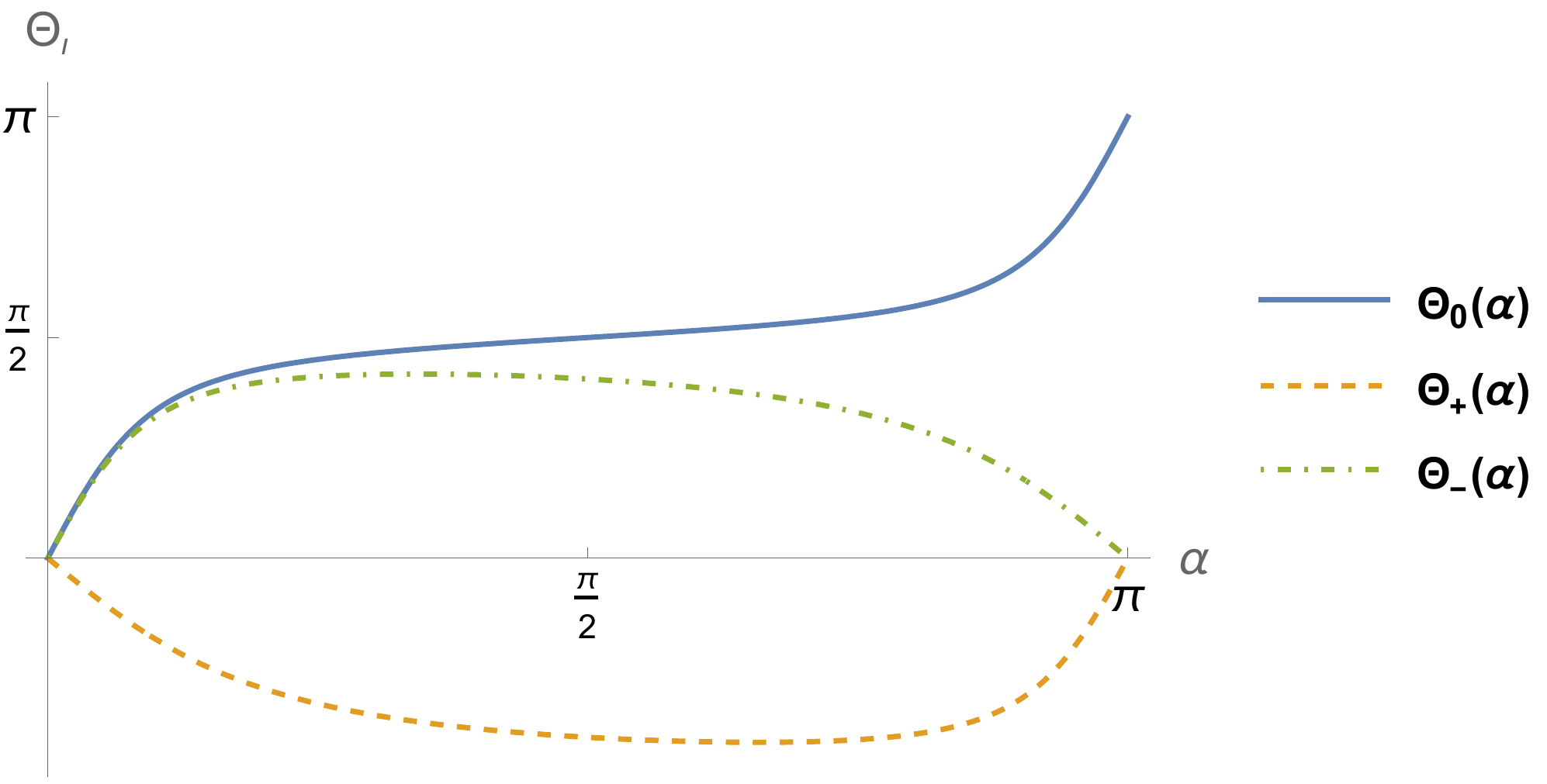}}

\caption{Scattering phases $\Theta_\iota(\alpha)$, with $\iota=0,\pm$,  
defined by \eqref{Th0}, \eqref{Thpm}, versus the rapidity $\alpha$
 at (a) $\eta=2$, and (b)  $\eta=0.3$.\label{fig:SPh}}

\end{figure}

The three scattering amplitudes $w_\iota(p_1,p_2)$, with $\iota=0,\pm$, can be  parametrized by the 
rapidity variable $\alpha$, 
\begin{subequations}\label{scph}
\begin{align}\label{wio}
&w_\iota(p_1,p_2)=\exp[-i\pi +i \theta_\iota(p_1,p_2)],\\
&\theta_\iota(p_1,p_2)=\Theta_\iota(\alpha_1-\alpha_2),\label{thet}\\
\label{Th0}
&\Theta_0(\alpha)=
 \alpha+\sum_{n=1}^\infty \frac{e^{-n\eta}\sin(2\alpha n)}{n \cosh(n \eta)},\\\label{Thpm}
&\Theta_\pm(\alpha)=\Theta_0(\alpha)+\chi_\pm (\alpha),\\
&\chi_+(\alpha)=-i \ln\left(- \frac{\sin[(\alpha+i \eta)/2]}{\sin[(\alpha-i \eta)/2]} \right),\\
&\chi_-(\alpha)=-i \ln\left( \frac{\cos[(\alpha+i \eta)/2]}{\cos[(\alpha-i \eta)/2]} \right),
\end{align}
\end{subequations}
where $p_j=p(\alpha_j)$, $j=1,2$,  and $\Theta_\iota(\alpha)$ are the scattering phases.
Figures \ref{et1} and \ref{et2} illustrate the rapidity dependencies of the scattering phases $\Theta_\iota(\alpha)$ at
$\eta=2$, and $\eta=0.3$, respectively. 
The  scattering amplitude $w_0(p_1,p_2)$ was found by 
Zabrodin \cite{Zabr92},  and the whole two-kink scattering matrix was determined by 
Davies {\it et al.} \cite{Miwa93}. 

The two-kink states $|\mathcal{K}_{\mu\nu}(\xi_1)\mathcal{K}_{\nu\mu}(\xi_2)\rangle_{s_1s_2}$ parametrized by the complex spectral parameters $\xi_{1,2}=-i e^{i\alpha_{1,2}}$ are simply related with $|K_{\mu\nu}(p_1)K_{\nu\mu}(p_2)\rangle_{s_1s_2}$:
\begin{align}\label{Kxi}
&|\mathcal{K}_{\mu\nu}(\xi_1)\mathcal{K}_{\nu\mu}(\xi_2)\rangle_{s_1s_2}=\\\nonumber
&\frac{\sqrt{\omega(p_1)\omega(p_2)}}{J\sinh\eta}|K_{\mu\nu}(p_1)K_{\nu\mu}(p_2)\rangle_{s_1s_2}.
 \end{align}

The different notation $|\xi_2,\xi_1\rangle_{\epsilon_2,\epsilon_1;(i)}$ has been commonly used  \cite{Jimbo94} 
for the two-kink states $| \mathcal{K}_{\mu\nu}(\xi_1)\mathcal{K}_{ \nu\mu}(\xi_2)\rangle_{s_1 s_2}$, 
with  $i=\mu$, and $\epsilon_{1,2}=\mathrm{sign}\, s_{1,2}$.

The commutation relation  \eqref{FZC} can be rewritten for the two-kink states \eqref{Kxi} in the matrix form:
 \begin{align}\label{Sma}
 &  |\mathcal{K}_{\mu\nu}(\xi_1)\mathcal{K}_{\nu\mu}(\xi_2)\rangle_{s_1 s_2}=\sum_{s_1',s_2'=\pm1/2 }\mathcal{S}_{s_1s_2}^{s_1's_2'}(\alpha_1-\alpha_2)\cdot \\\nonumber
&      |\mathcal{K}_{\mu\nu}(\xi_2)\mathcal{K}_{\nu\mu}(\xi_1)\rangle_{s_2' s_1'}.
  \end{align}
Another equivalent representation of the same commutation relation is given 
in the Appendix A.1 of the monograph \cite{Jimbo94} by Jimbo and Miwa:
  \begin{align}\label{CR}
   |\mathcal{K}_{\mu\nu}(\xi_2)\mathcal{K}_{\nu\mu}(\xi_1)\rangle_{s_2 s_1}=\\\nonumber
    -\sum_{s_1',s_2'=\pm1/2 }R_{s_1's_2'}^{s_1s_2} (\xi_1/\xi_2)  |\mathcal{K}_{\mu\nu}(\xi_1)\mathcal{K}_{\nu\mu}(\xi_2)\rangle_{s_1' s_2'},
  \end{align}
  \begin{equation}\label{RR}
     R(\xi)=\frac{1}{\kappa(\xi)}\begin{pmatrix}
     1 &&&\\
     & \frac{(1-\xi^2) q}{1-q^{2}\xi^2}&\frac{(1-q^{2}) \xi}{1-q^{2}\xi^2}&\\
    & \frac{(1-q^{2}) \xi}{1-q^{2}\xi^2}& \frac{(1-\xi^2) q}{1-q^{2}\xi^2}&\\
    &&&1
     \end{pmatrix},
  \end{equation}
  where $\xi_j =-ie^{i \alpha_j}$, 
  \begin{eqnarray}\label{kap}
&&\kappa(\xi)=\xi\frac{(q^{4}\xi^2;q^{4})\,(q^{2}\xi^{-2};q^{4})}
  {(q^{4}\xi^{-2};q^{4})\,(q^{2}\xi^{2};q^{4})},\\
 && \kappa(\xi_1/\xi_2)=\exp[i \,\Theta_0(\alpha_1-\alpha_2)],
  \end{eqnarray}
  and 
  \begin{equation}
  (z;p)=\prod_{n=0}^\infty(1-z \,p^n).
  \end{equation}
 \begin{figure}[htb]
\centering
\includegraphics[width=\linewidth, angle=00]{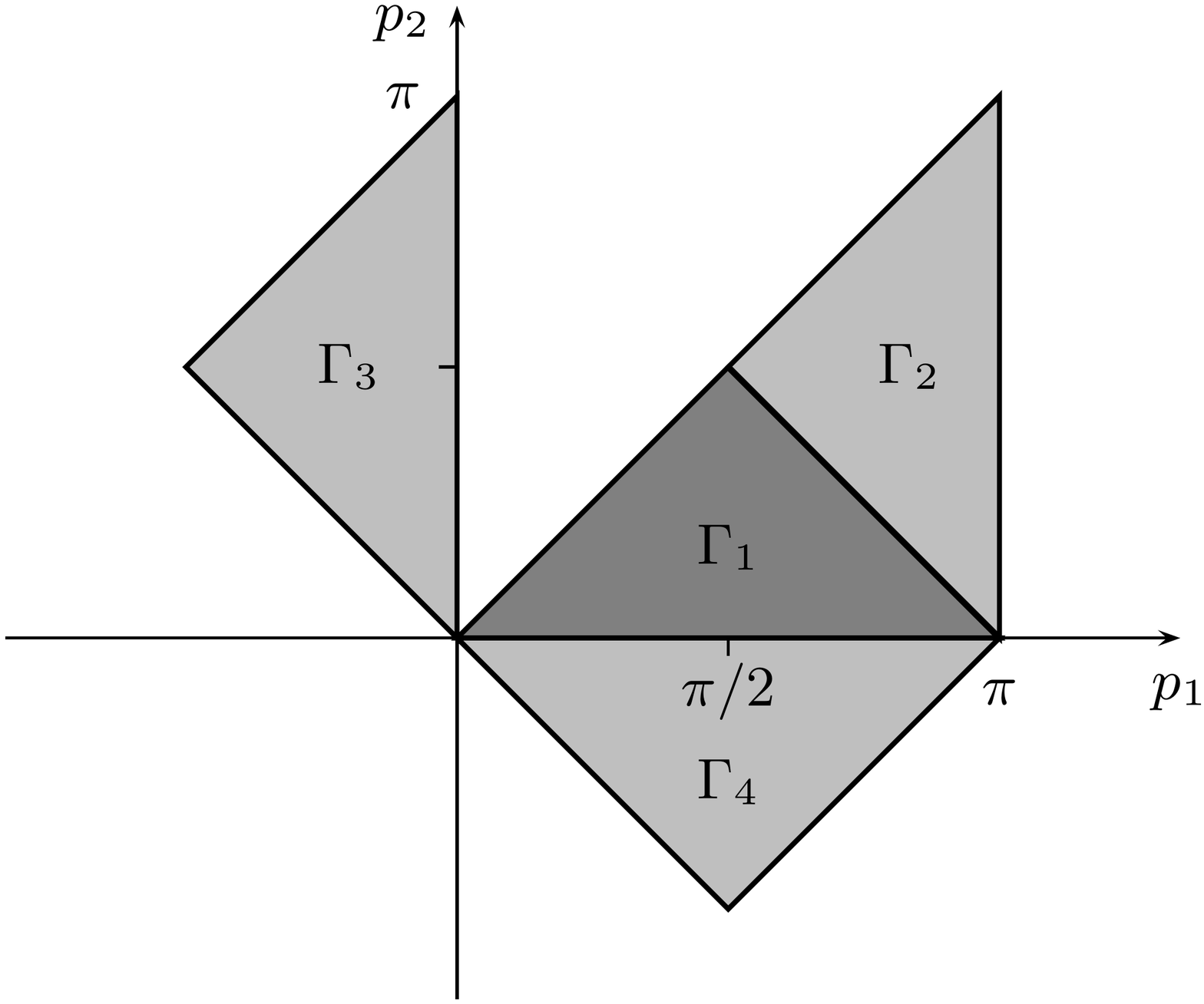}
\caption{\label{Gam}  Integration regions $\Gamma=\Gamma_1\cup\Gamma_2$ and  $\widetilde{\Gamma}=\Gamma_1\cup\Gamma_4$
in equations \eqref{PP2} and \eqref{PP11A}, respectively.
Triangular regions  $\Gamma_2$, $\Gamma_3$, and $\Gamma_4$ are equivalent in the sense \eqref{equi}.} 
\end{figure}

The $\mathcal{S}$-matrix in equation \eqref{Sma} is simply related with the $R$-matrix in equation \eqref{CR}
\begin{equation}
\mathcal{S}_{s_1s_2}^{s_1's_2'}(\alpha_1-\alpha_2)=-R_{s_2's_1'}^{s_2s_1}(\xi_2/\xi_1).
\end{equation}

The fundamental triangular region $\Gamma=\Gamma_1\cup \Gamma_2$ shown  
 in Figure \ref{Gam} represents an elementary cell in the two-dimensional momentum space $\mathbb{R}_{\mathbf{p}}^2$, 
$\langle p_1,p_2\rangle\in\mathbb{R}_{\mathbf{p}}^2$. To be more specific, let us introduce the following equivalence relations
in the momentum plane $\mathbb{R}_{\mathbf{p}}^2$:
\begin{equation}\label{equi}
\langle p_1,p_2\rangle\sim \langle p_1+n_1 \pi,p_2\rangle\sim \langle p_1 ,p_2+n_2\pi\rangle\sim \langle p_2 ,p_1\rangle,
\end{equation}
with integer $n_1,n_2\in \mathbb{Z}$. Denote by $\langle\langle p_1,p_2\rangle\rangle$ the equivalence class of the point
$\langle p_1,p_2\rangle$. A region $\mathfrak{E}\subset \mathbb{R}_{\mathbf{p}}^2$ will be called the elementary cell 
 in the momentum plane, if 
each equivalence class $\langle\langle p_1,p_2\rangle\rangle\in\langle\langle\mathbb{R}_{\mathbf{p}}^2 \rangle\rangle $ has 
just one representative point in  $\mathfrak{E}$.

 The triangular region $\Gamma$ in Figure \ref{Gam} gives an example of the elementary cell. One can easily show using equations
\eqref{pishift} and unitarity of the two-kink scattering matrix \eqref{RR}, that the integration region $\Gamma$ in equations \eqref{PP2}
can be replaced by any other elementary cell  $\mathfrak{E}$. In particular, the triangular region $\Gamma$ can be replaced in \eqref{PP2}
by the square elementary cell $\tilde{\Gamma}={\Gamma_1}\cup\Gamma_4$ shown in Figure \ref{Gam}. This leads to the following 
representation for the projection operator $\mathcal{P}_{11}^{(2)}$,
\begin{align}
\label{PP11A}
\mathcal{P}_{11}^{(2)}=\sum_{s_1=\pm1/2\atop{s_2=\pm1/2}}\iint_{\tilde{\Gamma}} \frac{dp_1dp_2}{\pi^2}
|K_{10}(p_1)K_{01}(p_2)\rangle_{s_1s_2}\cdot\\\nonumber
\phantom{.}_{s_2 s_1} \langle K_{10}(p_2) K_{01}(p_1) |,
 \end{align} 
which  will be used later.

In the Ising limit $\eta\to\infty$, the basis in the two-particle sector is formed by the  localised two-kink states 
$|\mathbf{K}_{\mu\nu}(j_1)\mathbf{K}_{\nu\mu}(j_2)\rangle$, with 
$-\infty<j_1<j_2<\infty$.  Four examples of such localized two-kink states are shown below:
\begin{align*}
&|\mathbf{K}_{10}(2)\mathbf{K}_{01}(5)\rangle:\quad \ldots\uparrow\downarrow{\color{blue}\
\underline{\stackrel{0}{\uparrow}}}\stackrel{1}{\downarrow}
{\color{dgreen}\stackrel{2}{\uparrow}}{\color{red}\mid}{\color{dgreen}\stackrel{3}{\uparrow}}\stackrel{4}{\downarrow}
{\color{dgreen}\stackrel{5}{\uparrow}}{\color{red}\mid}{\color{dgreen}\stackrel{6}{\uparrow}}\stackrel{7}{\downarrow}\stackrel{8}{\uparrow}
\ldots\quad s=1,\\
&|\mathbf{K}_{10}(3)\mathbf{K}_{01}(6)\rangle:\quad \ldots\uparrow\downarrow{\color{blue}\
\underline{\stackrel{0}{\uparrow}}}\stackrel{1}{\downarrow}
\stackrel{2}{\uparrow}{\color{dgreen}\stackrel{3}{\downarrow}}{\color{red}\mid}{{\color{dgreen}\stackrel{4}{\downarrow}}
\stackrel{5}{\uparrow}}{\color{dgreen}\stackrel{6}{\downarrow}}{\color{red}\mid}{\color{dgreen}\stackrel{7}{\downarrow}}\stackrel{8}{\uparrow}
\ldots\quad s=-1,\\
&|\mathbf{K}_{10}(3)\mathbf{K}_{01}(7)\rangle:\quad \ldots\uparrow\downarrow{\color{blue}\
\underline{\stackrel{0}{\uparrow}}}\stackrel{1}{\downarrow}
\stackrel{2}{\uparrow}{\color{dgreen}\stackrel{3}{\downarrow}}{\color{red}\mid}{{\color{dgreen}\stackrel{4}{\downarrow}}
\stackrel{5}{\uparrow}}{\color{black}\stackrel{6}{\downarrow}}{\color{dgreen}\stackrel{7}{\uparrow}}{\color{red}\mid}
{\color{dgreen}\stackrel{8}{\uparrow}}
\ldots\quad \;\;s=0,\\
&|\mathbf{K}_{10}(2)\mathbf{K}_{01}(6)\rangle:\quad \ldots\uparrow\downarrow{\color{blue}\
\underline{\stackrel{0}{\uparrow}}}\stackrel{1}{\downarrow}
{\color{dgreen}\stackrel{2}{\uparrow}}{\color{red}\mid}{\color{dgreen}\stackrel{3}{\uparrow}}{{\color{black}\stackrel{4}{\downarrow}}
\stackrel{5}{\uparrow}}{\color{dgreen}\stackrel{6}{\downarrow}}{\color{red}\mid}{\color{dgreen}\stackrel{7}{\downarrow}}
{\color{black}\stackrel{8}{\uparrow}}
\ldots\quad \;\;s=0.\\
\end{align*}
In the right column, $s$ stands for the ($z$-projection of the) total spin of the two-kink state.
The elementary properties of these states are:
\begin{eqnarray*}
&&\langle \mathbf{K}_{\mu\nu}(j_2) \mathbf{K}_{\nu\mu}(j_1) |\mathbf{K}_{\mu'\nu'}(j_1')\mathbf{K}_{\nu'\mu'}(j_2')\rangle=\delta_{\mu\mu'}\delta_{j_1j_1'}
\delta_{j_2j_2'}, \\
&&T_1|\mathbf{K}_{\mu\nu}(j_1)\mathbf{K}_{\nu\mu}(j_2)\rangle=|\mathbf{K}_{\nu\mu}(j_1-1)\mathbf{K}_{\mu\nu}(j_2-1)\rangle,\\
&&U|\mathbf{K}_{\mu\nu}(j_1)\mathbf{K}_{\nu\mu}(j_2)\rangle=|\mathbf{K}_{\nu\mu}(j_1)\mathbf{K}_{\mu\nu}(j_2)\rangle,\\
&&\widetilde{T}_1|\mathbf{K}_{\mu\nu}(j_1)\mathbf{K}_{\nu\mu}(j_2)\rangle=|\mathbf{K}_{\mu\nu}(j_1-1)\mathbf{K}_{\nu\mu}(j_2-1)\rangle,\\
&&{H}_I^{(0)}|\mathbf{K}_{\mu\nu}(j_1)\mathbf{K}_{\nu\mu}(j_2)\rangle=4|\mathbf{K}_{\mu\nu}(j_1)\mathbf{K}_{\nu\mu}(j_2)\rangle.
\end{eqnarray*}

The two-particle Bloch states $|K_{\mu\nu}^I(p_1)K_{\nu\mu}^I(p_2)\rangle_{s_1 s_2}$  
characterised by the quasimomenta 
$ {p_1,p_2}\in \mathbb{R}$ and the 
kink spins $s_{1,2}=\pm 1/2$ can be defined at $\varepsilon=0$ as follows:
\vspace{.3cm}
\begin{widetext}
\begin{subequations}\label{K2Is}
\begin{align}
&|K_{10}^I(p_1)K_{01}^I(p_2)\rangle_{-1/2,-1/2}=\!\!\!\!\sum_{m_1=-\infty}^\infty\sum_{m_2=m_1}^\infty\!\!
\Big[e^{2  i p_1 m_1+i p_2 (2m_2+1)}-
 e^{2i p_2 \,m_1+i p_1 (2m_2+1)}e^{i(p_1-p_2)}
\Big]|\mathbf{K}_{10}(2m_1-1)\mathbf{K}_{01}(2m_2)\rangle,\\
&|K_{10}^I(p_1)K_{01}^I(p_2)\rangle_{1/2,1/2}=\!\!\sum_{m_1=-\infty}^\infty\sum_{m_2=m_1+1}^\infty\!\!
\Big[e^{i p_1 (2m_1+1)+2 i p_2 m_2}-
e^{i p_2 (2m_1+1)+2i p_1 m_2}e^{i(p_1-p_2)}
\Big]|\mathbf{K}_{10}(2m_1)\mathbf{K}_{01}(2m_2-1)\rangle,\\
&|K_{10}^I(p_1)K_{01}^I(p_2)\rangle_{1/2,-1/2}= e^{i(p_1+p_2)}
 \sum_{m_1=-\infty}^\infty\sum_{m_2=m_1+1}^\infty
\Big[e^{i p_1 2m_1+i p_2 2m_2}-
e^{i p_2 2m_1+i p_1 2m_2}
\Big]|\mathbf{K}_{10}(2m_1)\mathbf{K}_{01}(2m_2)\rangle,\\
&|K_{10}^I(p_1)K_{01}^I(p_2)\rangle_{-1/2,1/2}=
\!\!\!\sum_{m_1=-\infty}^\infty\sum_{m_2=m_1+1}^\infty
\Big[e^{i p_1  2m_1+i p_2 2m_2}-
e^{i p_2 2m_1+i p_1 2m_2}
\Big]
\,|\mathbf{K}_{10}(2m_1-1)\mathbf{K}_{01}(2m_2-1)\rangle,\\&
|K_{01}^I(p_1)K_{10}^I(p_2)\rangle_{s_1,s_2} =U\, |K_{10}^I(p_1)K_{01}^I(p_2)\rangle_{-s_1,-s_2}.
\end{align}
\end{subequations}
\end{widetext}
Since these states have properties \eqref{PP2}, \eqref{prop}, and also 
satisfy equations \eqref{H1S}, and \eqref{CR}
to the first order in $\varepsilon$, they can be identified with  the Ising limit of the two-kink eigenstates of the 
Hamiltonian \eqref{XXZ1}:
\begin{equation}\label{lim2K}
\lim_{\varepsilon\to 0}|K_{\mu\nu}(p_1)K_{\nu\mu}(p_2)\rangle_{s_1s_2}=|K_{\mu\nu}^I(p_1)K_{\nu\mu}^I(p_2)\rangle_{s_1s_2}.
\end{equation}
\subsection{$n$-kink sector}
The basis in the $n$-kink subspace $\mathcal{L}^{(n)}$ is formed by the states 
\begin{equation}\label{Kbas}
|K_{\mu_1\mu_2}(p_1)K_{\mu_2\mu_3}(p_2)\ldots)K_{\mu_n\mu_{n+1}}(p_n)\rangle_{s_1s_2\ldots s_n}, 
\end{equation}
with  $\mu_i=0,1$, $\mu_i\ne \mu_{i+1}$, $s_i=\pm1/2$, and 
\[
0\le p_n<p_{n-1}<\ldots<p_1<\pi.
\]
Two notes are in order.
\begin{itemize}
\item
It is not difficult to generalise equations \eqref{K2Is}  and to write  the basis $n$-kink state \eqref{Kbas} explicitly in the Ising limit $\eta\to\infty$  for any $n$ in the form of the ``Bethe-Ansatz wave function".
\item 
Equations \eqref{vac0}, \eqref{Kis}, and \eqref{K2Is} represent the zero-order terms in the Taylor $\varepsilon$-expansions  of the
vacuum, one-kink and two-kink eigenstates of the Hamiltonian  \eqref{HIs}. Few subsequent terms in these Taylor expansions can be  straightforwardly calculated by  means of the Rayleigh-Schr\"odinger 
perturbation theory in $\varepsilon\to0$ applied to  Hamiltonian \eqref{HIs}. In particular,  two
initial terms in the  Taylor expansion in $\varepsilon$ of the  vacuum $|vac\rangle^{(1)}$
read \cite{Jimbo94}:
\begin{equation}\label{vacC1}
|vac\rangle^{(1)}=|0\rangle^{(1)}+\frac{\varepsilon}{2}\sum_{j=-\infty}^\infty|K_{10}(j)K_{01}(j+2)\rangle+O(\epsilon^2).
\end{equation}
\end{itemize}
\subsection{Two-kink form factors of local spin operators}
The matrix elements of local operators between the vacuum and $n$-particle basis states are commonly called ``form factors".
We collect below the well-known explicit formulas  for  the two-kink form factors of  the local spin operators $\sigma_0^\pm$ and $\sigma_0^z$. 

All non-vanishing two-particle form factors of the spin operators $\sigma_0^\pm, \sigma_0^z$ can be expressed in terms of four 
functions $X^{1}(\xi_1,\xi_2)$, $X^{0}(\xi_1,\xi_2)$, and $X_\pm^z(\xi_1,\xi_2)$:
\begin{subequations}\label{XXF}
\begin{align}\nonumber
&X^1(\xi_1,\xi_2)=\phantom{.}^{(1)}\langle vac|\sigma_0^+|\mathcal{K}_{10}(\xi_1)\mathcal{K}_{01}(\xi_2)\rangle_{-1/2,-1/2}=\\
&\phantom{.}^{(0)}\langle vac|\sigma_0^-|\mathcal{K}_{01}(\xi_1)\mathcal{K}_{10}(\xi_2)\rangle_{1/2,1/2},\label{xX}\\\nonumber
&X^0(\xi_1,\xi_2)=\phantom{.}^{(1)}\langle vac|\sigma_0^-|\mathcal{K}_{10}(\xi_1)\mathcal{K}_{01}(\xi_2)\rangle_{1/2,1/2}=\\\label{XX0}
&\phantom{.}^{(0)}\langle vac|\sigma_0^+|\mathcal{K}_{01}(\xi_1)\mathcal{K}_{10}(\xi_2)\rangle_{-1/2,-1/2},\\\nonumber
&X_+^z(\xi_1,\xi_2)=\phantom{.}^{(1)}\langle vac|\sigma_0^z|\mathcal{K}_{10}(\xi_1)\mathcal{K}_{01}(\xi_2)\rangle_+=\\
&-\phantom{.}^{(0)}\langle vac|\sigma_0^z|\mathcal{K}_{01}(\xi_1)\mathcal{K}_{10}(\xi_2)\rangle_+,\\\nonumber
&X_-^z(\xi_1,\xi_2)=\phantom{.}^{(1)}\langle vac|\sigma_0^z|\mathcal{K}_{10}(\xi_1)\mathcal{K}_{01}(\xi_2)\rangle_-=\\
&\phantom{.}^{(0)}\langle vac|\sigma_0^z|\mathcal{K}_{01}(\xi_1)\mathcal{K}_{10}(\xi_2)\rangle_-,
\end{align}
\end{subequations}
where  $\xi_{1,2}=-i \,e^{i\alpha_{1,2}}$, and 
\begin{align}\nonumber
&|\mathcal{K}_{\mu\nu}(\xi_1)\mathcal{K}_{\nu\mu}(\xi_2)\rangle_\pm\equiv
\frac{1}{\sqrt{2}}\Big(
|\mathcal{K}_{\mu\nu}(\xi_1)\mathcal{K}_{\nu\mu}(\xi_2)\rangle_{1/2,-1/2}\\
&\pm
|\mathcal{K}_{\mu\nu}(\xi_1)\mathcal{K}_{\nu\mu}(\xi_2)\rangle_{-1/2,1/2}
\Big).\label{baspm1A}
\end{align}
The functions $X^{j}(\xi_1,\xi_2)$ and $X_\pm^{z}(\xi_1,\xi_2)$ admit the following explicit representations:
\begin{align}\label{XX}
&X^{j}(\xi_1,\xi_2)=\rho^2\frac{(q^4;q^4)^2}{(q^2;q^2)^3}\cdot\\\nonumber
&\frac{(-q\xi_1\xi_2)^{1-j}\xi_{\color{black}2}\,\, \gamma(\xi_{\color{black}2}^2/\xi_{\color{black}1}^2)\,
\theta_{q^8}(-\xi_1^{-2}\xi_2^{-2}q^{4j})}{\theta_{q^4}(\xi_1^{-2}q^3)\,\theta_{q^4}(\xi_2^{-2}q^3)},\\\label{Xpl}
&X_+^{z}(\xi_1,\xi_2)=\frac{\sqrt{2}\,e^{-\eta/4}g(\alpha_1+\alpha_2,\eta)}{\sin[(\alpha_1-\alpha_2-i\eta)/2]}\, X^{0}(\xi_1,\xi_2),\\
&X_-^{z}(\xi_1,\xi_2)=-X_+^{z}(-\xi_1,\xi_2),\label{Xmn}
\end{align}
where
\begin{align}
&\gamma(\xi)\equiv\frac{(q^4 \xi;q^4;q^4)(\xi^{-1};q^4;q^4)}{(q^6 \xi;q^4;q^4)(q^2\xi^{-1};q^4;q^4)},\\
&\rho\equiv (q^2;q^2)^2\frac{(q^4;q^4;q^4)}{(q^6;q^4;q^4)},\\
&(x;y;z)\equiv \prod_{m,n=0}^\infty(1-x\, y^n z^m),\\
&\theta_x(y)=(x;x)(y;x)(x y^{-1};x),\\
&g(\alpha,\eta)=\frac{\vartheta_1(\frac{ \alpha}{2i\eta}|e^{-\pi^2/\eta})}{\vartheta_4(\frac{\alpha}{4i\eta}|e^{-\pi^2/(4\eta)})}.
\end{align}
Here $\vartheta_i(u|p)$  denote the elliptic theta-functions:
\begin{align}\label{theta}
&\vartheta_1(u|p)=2 p^{1/4}\sin(\pi u)\cdot \\\nonumber
&\prod_{n=1}^\infty(1-p^{2n})\left(1-2p^{2n}\cos(2\pi u)+p^{4n}\right),\\\nonumber
&\vartheta_4(u|p)=
\prod_{n=1}^\infty(1-p^{2n})\left(1-2p^{2n-1}\cos(2\pi u)+p^{2(2n-1)}\right),\\\nonumber
&\vartheta_2(u|p)=\vartheta_1(u+1/2|p), \quad \vartheta_3(u|p)=\vartheta_4(u+1/2|p).
\end{align}

The two-kink form factors of the $\sigma_0^\pm $ operators  were determined by means of the vertex-operator formalism
by Jimbo and Miwa \cite{Jimbo94}.  In equations \eqref{XX}, we essentially follow the notations   of  \cite{Kar98,Caux_2008}.
The explicit formulas for the form factors of the $\sigma_0^z$ operator in the XYZ spin-1/2 chain were obtained by Lashkevich \cite{Lash02}.
The XXZ limit  of these  formulas used in \eqref{Xpl} and   \eqref{Xmn} can be found in \cite{Dug15}. The explicit expressions for the form factors of all three
spin operators 
$\sigma_0^\mathfrak{a}$, $\mathfrak{a}=x,y,z$ in the XYZ  spin chain
were presented by Lukyanov and Terras in \cite{LukTer03}.

Form factors \eqref{XXF} satisfy a number of symmetry relations. We shall mention some of them. 
The first one reads:
\begin{equation}
X_\pm^{z}(-q\xi_1,\xi_2)=\mp e^{i[p(\alpha_2)-p(\alpha_1)]}X_\pm^{z}(-q\xi_2,\xi_1).
\end{equation}
Two other equalities
\begin{align}
X^{1}(\xi_1,\xi_2)=-X^{1}(\xi_1,-\xi_2)=X^{1}(-\xi_1,\xi_2),\\
X^{0}(\xi_1,\xi_2)=-X^{0}(-\xi_1,\xi_2)=X^{0}(\xi_1,-\xi_2),\nonumber
\end{align}
are consistent with \eqref{pishift}.

The equality  
\begin{align}
\phantom{.}^{(\mu)}\langle vac|\sigma_0^\mathfrak{a}|\mathcal{K}_{\mu\nu}(-q  \xi_1)\mathcal{K}_{\nu\mu}(\xi_2)\rangle_{s_1,s_2}
=\\
\phantom{.}^{(\nu)}\langle vac|\sigma_0^\mathfrak{a}|\mathcal{K}_{\nu\mu}( \xi_2)\mathcal{K}_{\mu\nu}(-q^{-1}\xi_1)
\rangle_{s_2,s_1}\nonumber
\end{align}
is the particular case of the form factor ``Riemann-Hilbert axiom" \eqref{RH}, which is discussed in Appendix  \ref{fA}.

Note, that Jimbo and Miwa  originally used in  \cite{Jimbo94} a  different notation for the two-particle Bloch states and form factors \eqref{XXF}, namely:
\begin{equation}
\phantom{.}^{(j)}\langle vac|\sigma_1^+|\xi_2,\xi_1\rangle_{--;(j)}=X^{j}(\xi_1,\xi_2),
\end{equation}

In the Ising limit $\varepsilon\to0$, the form factors \eqref{XX}, \eqref{Xpl},  \eqref{Xmn} have the following leading asymptotic behavior: 
\begin{align*}
&X^1(\xi_1,\xi_2)\cong-2 e^{i \alpha_1}\sin(\alpha_1-\alpha_2),\\
&X^0(\xi_1,\xi_2)\cong 2\varepsilon e^{i \alpha_1}\sin(\alpha_1-\alpha_2)\cos(\alpha_1+\alpha_2),\\
&X_+^z(\xi_1,\xi_2)\cong2 \varepsilon\sqrt{2}\, e^{i (\alpha_1+\alpha_2)/2}\sin(\alpha_1-\alpha_2)\sin\frac{\alpha_1+\alpha_2}{2},\\
&X_-^z(\xi_1,\xi_2)\cong2 i\varepsilon\sqrt{2}\, e^{i (\alpha_1+\alpha_2)/2}\sin(\alpha_1-\alpha_2)\cos\frac{\alpha_1+\alpha_2}{2}.
\end{align*}
These asymptotical formulas can be obtained directly by  means of the Rayleigh-Schr\"odinger 
perturbation theory in $\varepsilon\to0$ applied to  Hamiltonian \eqref{HIs}, and exploiting 
equations \eqref{K2Is}, \eqref{lim2K}, and \eqref{vacC1}.
\section{Dynamical structure factors at $h=0$ \label{Sec:DNS0}}
Theoretical study of the dynamical spin-structure factors for the XXZ model \eqref{XXZM} has a long history, see \cite{Caux_2008} for references.
In the Ising limit $\Delta\to-\infty$, both transverse and longitudinal  DSFs in the two-kink approximation were calculated by 
Ishimura  and Shiba \cite{Shiba_80}.  For  $\Delta<-1$, the calculation of the two-kink contribution to the transverse DSF  
 by means of the vertex operator approach was initiated by    Bougourzi, Karbach, and M\"uller  \cite{Kar98}, 
and completed by  Caux, Mossel, and Castillo \cite{Caux_2008}. The results of calculation of the two-kink contribution to the longitudinal DSF 
at $\Delta<-1$ by the same method were reported recently   by Castillo \cite{Cas20}. Later, the 
explicit representation for the longitudinal  DSF in the antiferromagnetically ordered phase $\Delta<-1$ was obtained 
by Babenko {\it et al. }\cite{Bab21} by means of the thermal form factor expansion method.

In this Section, we describe the calculation of the two-kink contributions to the  DSFs 
for model \eqref{XXZM} in the gapped antiferromagnetic phase $\Delta<-1$. As in papers \cite{Kar98,Caux_2008,Cas20}, we perform calculations
in the thermodynamic limit using the vertex operator approach. However, we shall apply  a slightly different, rather transparent derivation  procedure, 
which is equally suitable  for the  calculation of both transverse and longitudinal DSFs at $h=0$. The same procedure will be used in Section \ref{DSFconf} 
for the calculation of the DFSs for  model \eqref{XXZH} in the weak confinement regime at a small $h>0$,

Let us start from the finite-size version of the XXZ spin chain model defined by the Hamiltonian
\begin{equation}\label{XXZN}
{H}_N=-\frac{J}{2}\sum_{j=1}^N \left(\sigma_j^x\sigma_{j+1}^x+\sigma_j^y\sigma_{j+1}^y+
{\Delta}\,\sigma_j^z\sigma_{j+1}^z
\right).
\end{equation}
The periodic boundary conditions are implied,
and the number of sites is a multiple of four, $N\mod 4=0$. It is well known from the Bethe Ansatz solution,  that  model \eqref{XXZN} has at $\Delta<-1$,  aside from the true ground state $|0\rangle$, also the pseudo-vacuum state  $|1\rangle$. The one-site translation 
operator $T_1$ acts on these states in the following way:
\begin{equation}
T_1 |0\rangle=|0\rangle,\quad T_1|1\rangle=-|1\rangle, 
\end{equation} 
and their energies become degenerate in the thermodynamic limit:
\[
\lim_{N\to\infty}(E_1-E_0)=0.
\] 
The states $|\Phi_\mu\rangle$, $\mu=0,1$, defined by equations
\[
|\Phi_1\rangle=\frac{|0\rangle+|1\rangle}{\sqrt{2}}, \quad |\Phi_0\rangle=\frac{|0\rangle-|1\rangle}{\sqrt{2}}
\] 
transform under the one-site translation in accordance with relations
\begin{equation}\label{T1 Fi}
T_1 |\Phi_1\rangle=|\Phi_0\rangle,\quad T_1 |\Phi_0\rangle=|\Phi_1\rangle,
\end{equation} 
which are similar to equations \eqref{T1vac}. In the thermodynamic  limit, the states $|\Phi_\mu\rangle$
reduce to the  N\'eel-type ordered vacua of the infinite chain:
\begin{equation}\label{lFi}
\lim_{N\to\infty}|\Phi_\mu\rangle= |vac\rangle^{(\mu)}.
\end{equation} 
The dynamical structure factor in the state  $|vac\rangle^{(\mu)}$  can be defined  as follows \cite{Caux_2008}:
\begin{align}\nonumber
&{S}_\mu^{\mathfrak{a}\mathfrak{b}}(\mathrm{k},\omega)=\lim_{N\to\infty}\frac{1}{4 N}\sum_{j_1,j_2=1}^N e^{-i \mathrm{k} (j_1-j_2)}\int_{-\infty}^\infty dt\, e^{i\omega t}\cdot\\
&\langle \Phi_\mu|\sigma_{j_1}^\mathfrak{a}(t) \sigma_{j_2}^\mathfrak{b} (0)|\Phi_\mu\rangle,\label{StFN}
\end{align}
where $\sigma_j^\mathfrak{a}(t)=e^{i H_N t }\sigma_j^\mathfrak{a} e^{-i H_N t} $.

Taking into account \eqref{T1s}, \eqref{T1 Fi}, and \eqref{lFi}, on can easily proceed in \eqref{StFN} to the thermodynamic limit:
\begin{align}\label{Smuab}
&S_\mu^{\mathfrak{a}\mathfrak{b}}(\mathrm{k},\omega)=\frac{1}{8 }\sum_{j=-\infty}^\infty e^{-i\mathrm{k} j}\int_{-\infty}^\infty dt \,e^{i\omega t}\cdot\\\nonumber
&\Big[\phantom{.}^{(\mu)} \langle vac|\sigma_j^\mathfrak{a}(t) \sigma_{0}^\mathfrak{b} (0)|vac\rangle^{(\mu)}+\\
&\phantom{.}^{(\mu)} \langle vac|\sigma_{j+1}^\mathfrak{a}(t) \sigma_{1}^\mathfrak{b} (0)|vac\rangle^{(\mu)}\Big] ,\nonumber
\end{align}
where 
\begin{equation}\label{sigt}
\sigma_j^\mathfrak{a}(t)=e^{i H t }\sigma_j^\mathfrak{a} e^{-i H t}, 
\end{equation}
and the Hamiltonian $H$ is given by \eqref{XXZM}.

It follows immediately from equations \eqref{T1vacA}, and \eqref{T1s}, that the right-hand side of \eqref{Smuab} does not depend on $\mu$.
So, one can drop the index $\mu$ in $S_\mu^{\mathfrak{a}\mathfrak{b}}(\mathrm{k},\omega)$, and define the dynamic structure factor $S^{\mathfrak{a}\mathfrak{b}}(\mathrm{k},\omega)$
in the infinite chain as follows:
\begin{equation}
S^{\mathfrak{a}\mathfrak{b}}(\mathrm{k},\omega)=S_1^{\mathfrak{a}\mathfrak{b}}(\mathrm{k},\omega).
\end{equation}
The two-spinon contribution to the structure factor then takes the form
\begin{align}\label{DSF20}
&S_{(2)}^{\mathfrak{a}\mathfrak{b}}(\mathrm{k},\omega)=\frac{1}{8 }\sum_{j=-\infty}^\infty e^{-i\mathrm{k} j}\int_{-\infty}^\infty dt \,e^{i\omega t}\cdot\\\nonumber
&\Big[\phantom{.}^{(1)} \langle vac|\sigma_j^\mathfrak{a}(t)\mathcal{P}_{11}^{(2)} \sigma_{0}^\mathfrak{b} (0)|vac\rangle^{(1)}+\\
&\phantom{.}^{(1)} \langle vac|\sigma_{j+1}^\mathfrak{a}(t)\mathcal{P}_{11}^{(2)}  \sigma_{1}^\mathfrak{b} (0)|vac\rangle^{(1)}\Big] ,\nonumber 
\end{align}
where $\mathcal{P}_{11}^{(2)}$ is the projection operator \eqref{PP11} onto the two-spinon subspace $\mathcal{L}_{11}^{(2)}$.

Two dynamical structure factors are of particular importance:
the transverse DSF $S^{+-}(\mathrm{k},\omega)$, and the longitudinal DSF $S^{zz}(\mathrm{k},\omega)$. Subsequent calculations of these two functions are
slightly different and will be described separately. 
\subsubsection{Transverse DSF \label{trDNS}}
After substitution of \eqref{PP11}  into \eqref{DSF20} and straightforward manipulations exploiting \eqref{sigt} and \eqref{H1S} one obtains:
\begin{widetext}
\begin{align}\nonumber
&S_{(2)}^{+-}(\mathrm{k},\omega)=\frac{1}{8}\sum_{j=-\infty}^\infty e^{-i\mathrm{k} j}\int_{-\infty}^\infty dt 
\iint_{\tilde{\Gamma}}\frac{dp_1\,dp_2}{\pi^2}\,e^{i[\omega-\omega(p_1)-\omega(p_2)] t} \Big[\phantom{.}^{(1)} \langle vac|\sigma_j^+|K_{10}(p_1)K_{01}(p_2)\rangle_{-1/2,-1/2}\cdot\\
&\phantom{.}_{-1/2,-1/2} \langle K_{10}(p_2)K_{01}(p_1)|\sigma_0^-|vac\rangle^{(1)}+
\phantom{.}^{(1)} \langle vac|\sigma_{j+1}^+|K_{10}(p_1)K_{01}(p_2)\rangle_{-1/2,-1/2}\cdot
\phantom{.}_{-1/2,-1/2} \langle K_{10}(p_2)K_{01}(p_1)|\sigma_1^-|vac\rangle^{(1)}\Big]=\nonumber\\\label{Spm}
& \frac{\pi}{4}\,\sum_{j=-\infty}^\infty e^{-i\mathrm{k} j}
\iint_{\tilde{\Gamma}}\frac{dp_1\,dp_2}{\pi^2} \delta[\omega-\omega(p_1)-\omega(p_2)] \Big[\phantom{.}^{(1)} \langle vac|\sigma_j^+|K_{10}(p_1)K_{01}(p_2)\rangle_{-1/2,-1/2}\cdot\\
&\phantom{.}_{-1/2,-1/2} \langle K_{10}(p_2)K_{01}(p_1)|\sigma_0^-|vac\rangle^{(1)}+
\phantom{.}^{(1)} \langle vac|\sigma_{j+1}^+|K_{10}(p_1)K_{01}(p_2)\rangle_{-1/2,-1/2}\cdot
\phantom{.}_{-1/2,-1/2} \langle K_{10}(p_2)K_{01}(p_1)|\sigma_1^-|vac\rangle^{(1)}\Big].\nonumber
\end{align}
The summation over  $j\in\mathbb{Z}$ in \eqref{Spm} can be split into two sums over 
even $j=2m$ and odd $j=2m+1$, with $m\in\mathbb{Z}$.  Exploiting  equalities 
\begin{equation}
\widetilde{T}_1|vac\rangle^{(1)}=|vac\rangle^{(1)},\quad  \sigma_{j+1}^\pm =\widetilde{T}_1^{-1} \sigma_j^{\mp} \widetilde{T}_1, \quad  \widetilde{T}_1 |K_{10}(p_1)K_{01}(p_2)\rangle_{s_1,s_2}=
 e^{i(p_1+p_2)}|K_{10}(p_1)K_{01}(p_2)\rangle_{-s_1,-s_2},
\end{equation} 
that follow from \eqref{T1vacA}, \eqref{tilT1}, and \eqref{tilT},  one obtains  from  \eqref{Spm}:
\begin{align}\nonumber
&S_{(2)}^{+-}(\mathrm{k},\omega)=\frac{\pi}{4}\sum_{m=-\infty}^\infty 
\iint_{\tilde{\Gamma}}\frac{dp_1\,dp_2}{\pi^2} \,e^{2i\, m(p_1+p_2-\mathrm{k}) }\,\delta[\omega-\omega(p_1)-\omega(p_2)] \Big\{\,\Big[\phantom{.}^{(1)} \langle vac|\sigma_0^+|K_{10}(p_1)K_{01}(p_2)\rangle_{-1/2,-1/2}\cdot\\\nonumber
&\phantom{.}_{-1/2,-1/2} \langle K_{10}(p_2)K_{01}(p_1)|\sigma_0^-|vac\rangle^{(1)}+
\phantom{.}^{(1)} \langle vac|\sigma_{0}^-|K_{10}(p_1)K_{01}(p_2)\rangle_{1/2,1/2}\cdot
\phantom{.}_{1/2,1/2} \langle K_{10}(p_2)K_{01}(p_1)|\sigma_0^+|vac\rangle^{(1)}\Big]+\\\nonumber
&e^{i(p_1+p_2-\mathrm{k})}\,\Big[\phantom{.}^{(1)} \langle vac|\sigma_0^-|K_{10}(p_1)K_{01}(p_2)\rangle_{1/2,1/2}\cdot
\phantom{.}_{-1/2,-1/2} \langle K_{10}(p_2)K_{01}(p_1)|\sigma_0^-|vac\rangle^{(1)}+\\
&\phantom{.}^{(1)} \langle vac|\sigma_{0}^+|K_{10}(p_1)K_{01}(p_2)\rangle_{-1/2,-1/2}\cdot
\phantom{.}_{1/2,1/2} \langle K_{10}(p_2)K_{01}(p_1)|\sigma_0^+|vac\rangle^{(1)}\Big]\Big\}.\label{Str}
\end{align}

Using the Poisson summation formula
\begin{equation}\label{PoS}
\sum_{m=-\infty}^\infty e^{2i\, m Q }=\pi \sum_{l=-\infty}^\infty\delta(Q-\pi l),
\end{equation}
the integral representation \eqref{Str} of the two-spinon transverse DSF $S_{(2)}^{+-}(\mathrm{k},\omega)$ can be simplified to the form:
\begin{align}
&S_{(2)}^{+-}(\mathrm{k},\omega)=\frac{1}{4} \iint_{\tilde{\Gamma}}  dp_1 dp_2\sum_{l=-\infty}^\infty\,\delta(p_1+p_2-\mathrm{k}-\pi l)\,
\delta[\omega-\omega(p_1)-\omega(p_2)] \, \mathcal{G}^{+-}(p_1,p_2|\mathrm{k} )=\nonumber\\
&\frac{1}{4} \int_0^\pi dP   \int_0^{\pi/2} dp  \sum_{l=-\infty}^\infty\,
\delta(P-\mathrm{k} -\pi l)
\,\delta[\omega-\epsilon(p|P)] \, \mathcal{G}^{+-}(P/2+p,P/2-p|\mathrm{k} ),\label{Str3}
\end{align}
where
\begin{align}\label{Gpm}
 &\mathcal{G}^{+-}(p_1,p_2|\mathrm{k} )= \frac{d\alpha(p_1)}{dp_1} \frac{d\alpha(p_2)}{dp_2}\,\Big\{|X^1(\xi_1,\xi_2)|^2+|X^0(\xi_1,\xi_2)|^2+\\
& e^{i (p_1+p_2-\mathrm{k} )} \Big[
X^0(\xi_1,\xi_2)\left(X^1(\xi_1,\xi_2)\right)^*+\left( X^0(\xi_1,\xi_2)\right)^* X^1(\xi_1,\xi_2)\nonumber
\Big]\Big\}
.
\end{align}
\end{widetext}

\begin{figure}[ht]
\centering
\includegraphics[width=\linewidth, angle=00]{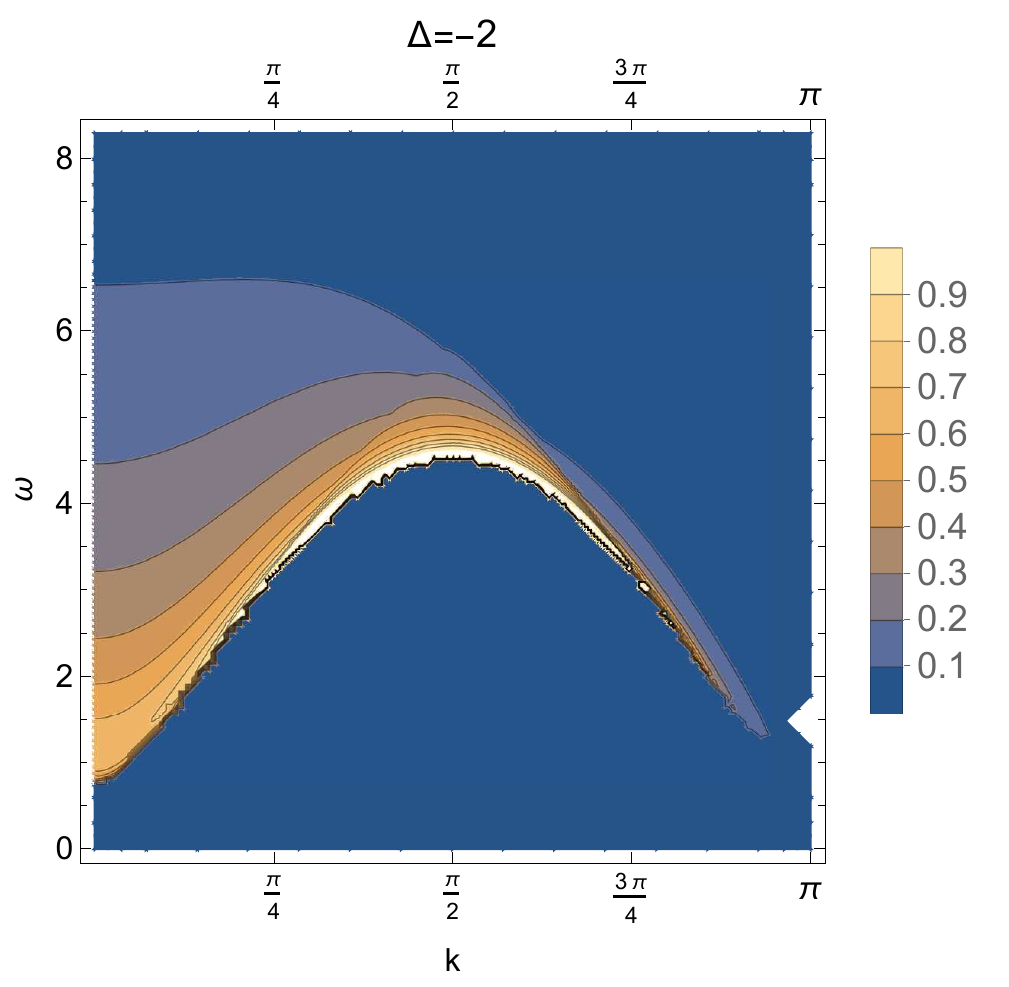}
\caption{\label{tDNS}  Two-spinon transverse DSF \eqref{SFtr} at $\Delta=-2$, $J=1$. } 
\end{figure}

Here notations \eqref{xX}, \eqref{XX0} have been used,  $\xi_i=-i e^{i \alpha(p_i)}$, $i=1,2$, and the rapidity $\alpha(p)$ corresponding to the momentum $p$ is determined by the inversion of equation \eqref{pe}.
In equation \eqref{Str3}, we have changed the integration variables to $P=p_1+p_2$ and $p=(p_1-p_2)/2$, and  used the notation
\begin{equation}\label{Energ}
\epsilon(p|P)=\omega(P/2+p)+\omega(P/2-p)
\end{equation}
 for the total energy of  two spinons.
Properties of this function, which plays the key role in the subsequent analysis, are described in detail in Appendix \ref{2E}.

It is clear from \eqref{Str3}, that only one term  survives  in the infinite sum in $l$ at any real   $\mathrm{k}\notin \pi \mathbb{Z}$.
This is the $l=0$ term, if  $\mathrm{k}\in (0,\pi)$. In the latter case, one obtains after integration in $P$: 
\begin{align}\label{Str4}
S_{(2)}^{+-}(\mathrm{k},\omega)= \int_0^{\pi/2} dp  \,
\delta[\omega-\epsilon(p|P)] \cdot\\
 \mathcal{G}_0^{+-}(P/2+p,P/2-p)\Big|_{P=\mathrm{k} },\label{Str4}\nonumber
\end{align}
where 
\begin{equation}
 \mathcal{G}_0^{+-}(p_1,p_2)=\frac{1}{4} 
 \frac{d\alpha(p_1)}{dp_1} \frac{d\alpha(p_2)}{dp_2}\, |X^1(\xi_1,\xi_2)+X^0(\xi_1,\xi_2)|^2.
\end{equation}
The remaining integration in \eqref{Str4} is performed by making use of the $\delta$-function in the integrand:
\begin{align}\label{SFtr}
S_{(2)}^{+-}(\mathrm{k},\omega)=  \sum_{i=1}^{\mathfrak{N}(P,\omega)}\frac{ \mathcal{G}_0^{+-}(P/2+p,P/2-p)}{|\partial_p \epsilon(p|P)|}
\Bigg|_{\substack{\,p=p^{(i)}\\\!\!\!P=\mathrm{k} }},
\end{align}
for kinematically allowed energies 
\begin{equation}\label{kinE}
\omega\in\left(\min_p\epsilon(p|P),\max_p\epsilon(p|P)\right).
\end{equation}
Here $p^{(i)}=p^{(i)}(P,\omega)$ are the solutions of the equation, 
\begin{equation}\label{epOm}
\epsilon(p|P)=\omega,
\end{equation}
such that $0<p=p^{(i)}<\pi/2$, and the number $\mathfrak{N}(P,\omega)$ of such solutions   takes values 1 or 2, depending on the values of $P$ and $\omega$.  

It is convenient to slightly change notations for the solutions of equation \eqref{epOm}. Namely, we  shall denote the solution 
of \eqref{epOm} by $p_a(P,\omega)$, if $\partial_p \epsilon(p|P)\Big|_{p=p_a}>0$, and use the notation  $p_b(P,\omega)$ for the 
 solution, such that $\partial_p \epsilon(p|P)\Big|_{p=p_b}<0$.
The solutions $p_{a,b}\in(0,\pi/2)$ of equation \eqref{epOm} are shown in Figure \ref{Fig:Energ}  in  Appendix \ref{2E}, and their explicit expressions  are given in equation  \eqref{p12} therein.

Figure \ref{tDNS} illustrates the frequency and momentum dependence of the two-spinon transverse DSF at $\Delta=-2$  calculated from equation \eqref{SFtr}. An alternative explicit representation for the same transverse DSF was obtained by Caux, Mossel, and Castillo  \cite{Caux_2008}. There is a strong numerical evidence that both representations are equivalent: we compared numerically predictions for the transverse DSF calculated from \eqref{SFtr} with results presented in Figure 5 of paper \cite{Caux_2008} and found an excellent agreement at all $\Delta$, $\mathrm{k}$, and $\omega$.
\subsubsection{Longitudinal DSF \label{lDNS}}
Calculations of the longitudinal DSF are very similar to those described in Section \ref{trDNS}. The main difference is that only the 
two-kink configurations with zero total spin $s_1+s_2=0$ contribute  to the form factor expansion of the longitudinal DSF. Proceeding to the 
basis \eqref{baspm1} in the subspace of such two-kink states, we obtain from \eqref{DSF20}:
\begin{widetext}
\begin{align}\nonumber
&S_{(2)}^{zz}(\mathrm{k},\omega)=\frac{1}{8}\sum_{j=-\infty}^\infty e^{-i\mathrm{k} j}\int_{-\infty}^\infty dt 
\iint_{\tilde{\Gamma}}\frac{dp_1\,dp_2}{\pi^2}\,e^{i[\omega-\omega(p_1)-\omega(p_2)] t}
\sum_{\iota=\pm}
 \Big[\phantom{.}^{(1)} \langle vac|\sigma_j^z|K_{10}(p_1)K_{01}(p_2)\rangle_{\iota}\cdot\\
&\phantom{.}_{\iota} \langle K_{10}(p_2)K_{01}(p_1)|\sigma_0^z|vac\rangle^{(1)}+
\phantom{.}^{(1)} \langle vac|\sigma_{j+1}^z|K_{10}(p_1)K_{01}(p_2)\rangle_{\iota}\cdot
\phantom{.}_{\iota} \langle K_{10}(p_2)K_{01}(p_1)|\sigma_1^z|vac\rangle^{(1)}\Big]=\nonumber\\\label{Sz}
& \frac{\pi}{4}\,\sum_{j=-\infty}^\infty e^{-i\mathrm{k} j}
\iint_{\tilde{\Gamma}}\frac{dp_1\,dp_2}{\pi^2} \delta[\omega-\omega(p_1)-\omega(p_2)] \sum_{\iota=\pm}\Big[\phantom{.}^{(1)} \langle vac|\sigma_j^z|K_{10}(p_1)K_{01}(p_2)\rangle_{\iota}\cdot\\
&\phantom{.}_{\iota} \langle K_{10}(p_2)K_{01}(p_1)|\sigma_0^z|vac\rangle^{(1)}+
\phantom{.}^{(1)} \langle vac|\sigma_{j+1}^z|K_{10}(p_1)K_{01}(p_2)\rangle_{\iota}\cdot
\phantom{.}_{\iota} \langle K_{10}(p_2)K_{01}(p_1)|\sigma_1^z|vac\rangle^{(1)}\Big].\nonumber
\end{align}
After splitting the summation over  $j\in\mathbb{Z}$   into two sums over 
even $j=2m$ and odd $j=2m+1$,   and exploiting  equalities 
\begin{equation}
\widetilde{T}_1|vac\rangle^{(1)}=|vac\rangle^{(1)},\quad  \sigma_{j+1}^z =-\widetilde{T}_1^{-1} \sigma_j^{z} \widetilde{T}_1, \quad  \widetilde{T}_1 |K_{10}(p_1)K_{01}(p_2)\rangle_{\iota}=\iota\,
 e^{i(p_1+p_2)}|K_{10}(p_1)K_{01}(p_2)\rangle_{\iota},
\end{equation} 
that follow from \eqref{T1vacA}, \eqref{tilT1}, and \eqref{trmod},  one finds from  \eqref{Sz}:
\begin{align}\label{Sz1}
&S_{(2)}^{zz}(\mathrm{k},\omega)=\frac{\pi}{2}\sum_{m=-\infty}^\infty 
\iint_{\tilde{\Gamma}}\frac{dp_1\,dp_2}{\pi^2} \,e^{2i\, m(p_1+p_2-\mathrm{k} ) }\,\delta[\omega-\omega(p_1)-\omega(p_2)] \cdot\\
&\sum_{\iota=\pm}\Big[\big|\phantom{.}^{(1)} \langle vac|\sigma_0^z|K_{10}(p_1)K_{01}(p_2)\rangle_{\iota}\big|^2
(1-\iota \,e^{i(p_1+p_2-\mathrm{k} )})\Big].\nonumber
\end{align}
Application of the  Poisson summation formula \eqref{PoS}
to \eqref{Sz1} yields:
\begin{equation}
S_{(2)}^{zz}(\mathrm{k},\omega)=\frac{1}{2} \iint_{\tilde{\Gamma}}  dp_1 dp_2\sum_{l=-\infty}^\infty\,\delta(p_1+p_2-\mathrm{k} -\pi l)\,
\delta[\omega-\omega(p_1)-\omega(p_2)] \, \mathcal{G}^{zz}(p_1,p_2|\mathrm{k} ),\label{Sz1X}
\end{equation}
where
\begin{align}\label{Gzm}
 &\mathcal{G}^{zz}(p_1,p_2|\mathrm{k} )= \frac{d\alpha(p_1)}{dp_1} \frac{d\alpha(p_2)}{dp_2}\,\Big\{|X_+^z(\xi_1,\xi_2)|^2[1-e^{i(p_1+p_2-\mathrm{k} )}]+
 |X_-^z(\xi_1,\xi_2)|^2[1+e^{i(p_1+p_2-\mathrm{k} )}]
\Big]\Big\}.
\end{align}
\end{widetext}

\begin{figure}[ht]
\centering
\includegraphics[width=\linewidth, angle=00]{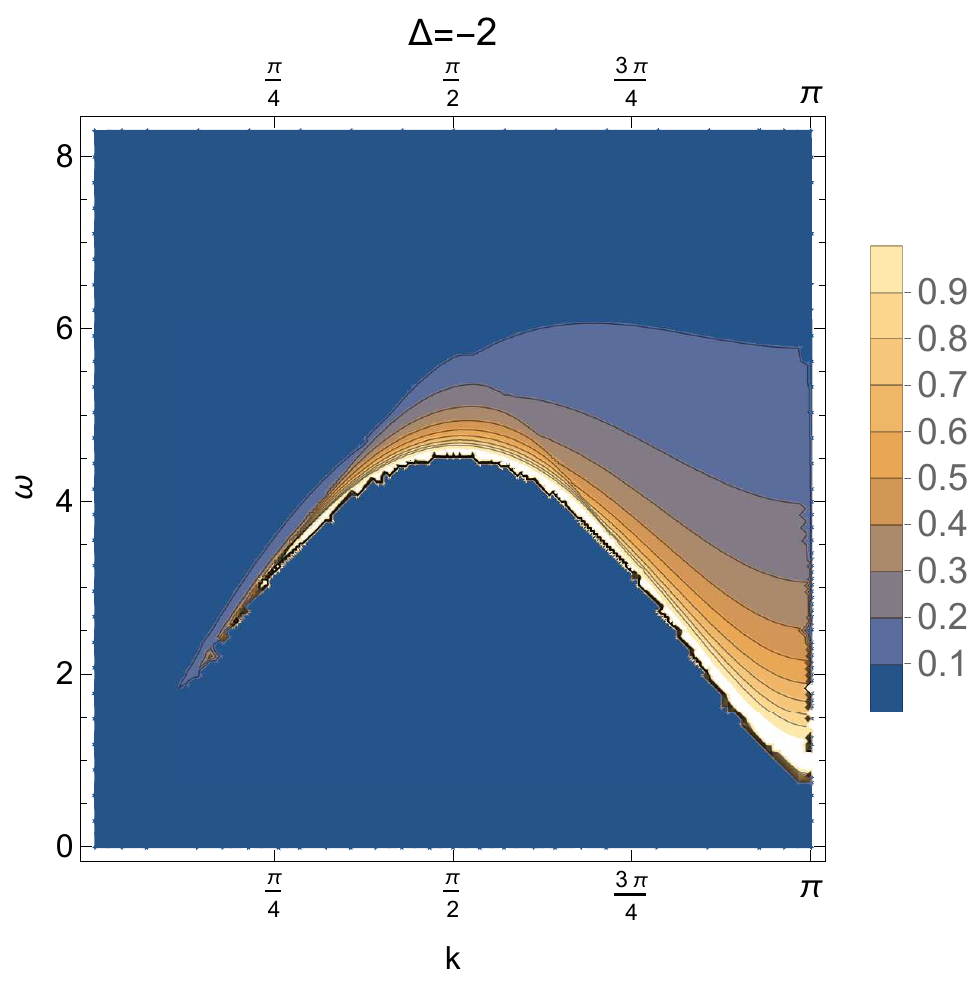}
\caption{\label{LDNS}  Two-spinon longitudinal DSF \eqref{lDN} at $\Delta=-2$, $J=1$. } 
\end{figure}

Only the $l=0$ term survives in the infinite sum in $l$ in the right-hand side of \eqref{Sz1} at $0\le \mathrm{k}<\pi$.
Following the steps used previously in 
Section \ref{trDNS} for the calculation of the transverse DSF, one obtains
\begin{align}\label{Str4a}
S_{(2)}^{zz}(\mathrm{k},\omega)= \int_0^{\pi/2} dp  \,
\delta[\omega-\epsilon(p|P)] \cdot\\
 \mathcal{G}_0^{zz}(P/2+p,P/2-p)\Big|_{P=\mathrm{k} },\label{Str4}\nonumber
\end{align}
where 
\begin{equation}\label{G0zz}
 \mathcal{G}_0^{zz}(p_1,p_2)=
 \frac{d\alpha(p_1)}{dp_1} \frac{d\alpha(p_2)}{dp_2}\, |X_-^z(\xi_1,\xi_2)|^2.
\end{equation}
Note, that the form factor $X_+^z(\xi_1,\xi_2)$ does not contribute to the longitudinal DSF $S_{(2)}^{zz}(\mathrm{k},\omega)$
at $0\le \mathrm{k}<\pi$.
The final result for the two-spinon longitudinal DSF for the kinematically allowed energies \eqref{kinE} reads
\begin{equation}\label{lDN}
S_{(2)}^{zz}(\mathrm{k},\omega)=  \sum_{i=1}^{\mathfrak{N}(P,\omega)}\frac{ \mathcal{G}_0^{zz}(P/2+p,P/2-p)}{|\partial_p \epsilon(p|P)|}
\Bigg|_{\substack{\,p=p^{(i)}\\\!\!\!P=\mathrm{k} }},
\end{equation}
where the  notations $p^{(i)}$ and $\mathfrak{N}(P,\omega)$, that were introduced in Section  \ref{trDNS}   after equation \eqref{SFtr}, have been used.

We did not try to compare this explicit expression for the two-spinon longitudinal DSF with rather cumbersome formulas for this quantity
reported by Castillo \cite{Cas20},  and by Babenko {\it et al. }\cite{Bab21}. Instead, we have checked that  our
formula \eqref{lDN} perfectly reproduces the $\omega$ dependencies of the function 
$S_{(2)}^{zz}(\mathrm{k},\omega)$ at several fixed values of the momentum $\mathrm{k}$, 
which are plotted in Figure 4 in paper \cite{Bab21}.

The frequency and momentum dependence  of the two-spinon   longitudinal  DSF  given by \eqref{lDN} 
at $\Delta=-2$ is shown in Figure \ref{LDNS}.
\section{XXZ spin chain in a weak staggered magnetic field \label{BSE1}}
Application of the staggered magnetic field $h>0$ breaks integrability of the XXZ spin-chain model. It also explicitly 
breaks the symmetry of the model Hamiltonian \eqref{XXZH} with respect to the inversion of all spins and to the one-site translation.
However, the Hamiltonian \eqref{XXZH} still commutes with operators $S^z$, $\widetilde{T}_1$, and 
$T_2={T}_1^2=\widetilde{T}_1^2$:
\begin{equation}\label{com}
[\mathcal{H}(h),S^z]=[\mathcal{H}(h),\widetilde{T}_1]=[\mathcal{H}(h),{T}_2]=0.
\end{equation} 
Note also, that 
\begin{equation}\label{T1S}
\widetilde{T}_1 S^z+S^z\widetilde{T}_1=0.
\end{equation}

Let us first consider the ground state eigenvalue problem
\begin{equation}
\mathcal{H}_N(h)\, |vac(h,N)\rangle=E_{vac}(h,N)|vac(h,N)\rangle
\end{equation}
for the finite-$N$ version of model model  \eqref{XXZH}  defined by the Hamiltonian 
\begin{align}\nonumber
\mathcal{H}_N(h)&=-\frac{J}{2}\sum_{j=-N/2}^{N/2-1}\!\!\left[\sigma_j^x\sigma_{j+1}^x+\sigma_j^y\sigma_{j+1}^y+
\Delta\left(
\sigma_j^z\sigma_{j+1}^z+1
\right)
\right]\\
&-h\sum_{j=-N/2}^{N/2-1} (-1)^j\sigma_j^z,\label{XXZH1}
\end{align}
with even $N$, and supplemented with periodic boundary conditions.

In the thermodynamic limit, one finds by means of the straightforward perturbative analysis at small $h>0$:
\begin{align}\label{vach}
&|vac(h)\rangle\equiv \lim_{N\to\infty}|vac(h,N)\rangle=|vac\rangle^{(1)}+O(h),\\
&e_{vac}(h)\equiv \lim_{N\to\infty}\frac{E_{vac}(h,N)}{N}=
\frac{J}{2} C(\Delta) - h\,\bar{\sigma}(\eta)+O(h^2),
\end{align}
where $|vac\rangle^{(1)}$ is the first ground state of the infinite chain at $h=0$ determined by equations \eqref{vacH0} and \eqref{vac1},
$e_{vac}(h)$ is the ground-state energy per lattice site, the constant $C(\Delta)$ was defined in \eqref{VE}, and $\bar{\sigma}(\eta)$ is the zero-field  spontaneous staggered magnetization \eqref{sig}.
\subsection{Classification of meson states}
As in equation \eqref{XXZ1}, we redefine Hamiltonian \eqref{XXZH} by adding a constant term in order to get rid of the 
vacuum energy:
\begin{align}\nonumber
&\mathcal{H}_1(h)=\sum_{j=-\infty}^\infty\!\!
\bigg\{-\frac{J}{2}\left[\sigma_j^x\sigma_{j+1}^x+\sigma_j^y\sigma_{j+1}^y+
\Delta\,
\sigma_j^z\sigma_{j+1}^z
\right]\\
&-(-1)^j h\, \sigma_j^z-e_{vac}(h)\bigg\},\label{XXZH1A}\\
&\mathcal{H}_1(h)|vac(h)\rangle=0.
\end{align}
The meson states $|\pi_{s,\iota,n}(P)\rangle$ can be classified by the 
quasimomentum  $P\in[0,\pi)$, the spin $s=0,\pm1$ and two further quantum numbers  $\iota=0,\pm$, and $n=1,2,\ldots$
The quantum numbers  $\iota$  and $s$ are not independent: we assign 
$\iota=0$ for $s=\pm1$, and $\iota=\pm $, if $s=0$.
Operators $\mathcal{H}_1(h)$, $T_1^2$, and $S^z$ act on the meson states as follows:
\begin{subequations}\label{mesA}
\begin{align}\label{mes}
&\mathcal{H}_1(h)|\pi_{s,\iota,n}(P)\rangle=E_{\iota,n}(P)|\pi_{s,\iota,n}(P)\rangle,\\\label{pT1}
&T_1^2|\pi_{s,\iota,n}(P)\rangle=e^{2 i P}|\pi_{s,\iota,n}(P)\rangle,\\
&S^z|\pi_{s,\iota,n}(P)\rangle=s|\pi_{s,\iota,n}(P)\rangle.
\end{align}
\end{subequations}
At $h=0$, the meson states  $|\pi_{s,\iota,n}(P)\rangle$ decouple into some  linear combinations of two-kink states described
in Section \ref{2kink}: 
\begin{align*}
&|\pi_{1,0,n}(P)\rangle\to |K_{10}(p_1)K_{01}(p_2)\rangle_{1/2,1/2},\\
&|\pi_{-1,0,n}(P)\rangle\to |K_{10}(p_1)K_{01}(p_2)\rangle_{-1/2,-1/2},\\
&|\pi_{0,+,n}(P)\rangle\to |K_{10}(p_1)K_{01}(p_2)\rangle_+,\\
&|\pi_{0,-,n}(P)\rangle\to |K_{10}(p_1)K_{01}(p_2)\rangle_-,
\end{align*}
with $e^{i(p_1+p_2)}=e^{iP}$.

The action of the modified translation operator $\widetilde{T}_1$ on the meson states  can be found by combination of \eqref{mesA} 
with \eqref{com}, and \eqref{T1S}.

In the case $s=\pm1$, with a proper choice of the overall phases of  states
$|\pi_{s,\iota=0,n}(P)\rangle$, one may always set up the condition
\begin{equation}\label{piT1}
\widetilde{T}_1 |\pi_{s,\iota=0,n}(P)\rangle=e^{i P}  |\pi_{-s,\iota=0,n}(P)\rangle.
\end{equation}
It follows immediately from \eqref{piT1} and \eqref{com}, that the meson states $|\pi_{1,0,n}(P)\rangle$ and 
$|\pi_{-1,0,n}(P)\rangle$ indeed have  the same energy $E_{0,n}(P)$, as it was already anticipated in \eqref{mes}.

If $s=0$, the index $\iota$ can take two values $\iota=\pm$, and one should put in analogy with \eqref{trmod}:
\begin{equation}\label{0T1}
\widetilde{T}_1 |\pi_{0,\iota,n}(P)\rangle=\iota e^{i P} |\pi_{0,\iota,n}(P)\rangle.
\end{equation}
By analytical continuation of equations \eqref{mesA},  \eqref{piT1}, and \eqref{0T1} to all $P\in\mathbb{R}$,
one finds:
\begin{equation}
|\pi_{s,\iota=0,n}(P+\pi)\rangle=e^{i \chi_s}|\pi_{s,\iota=0,n}(P)\rangle,
\end{equation}
with $s=\pm1$, and
\begin{equation}
|\pi_{s=0,\iota,n}(P+\pi)\rangle=e^{i \phi_\iota}|\pi_{s=0,-\iota,n}(P)\rangle,
\end{equation}
with $\iota=\pm$, and some real functions $\chi_s(P)$,  $\phi_\iota(P)$.

The meson energy spectra $E_{\iota,n}(P)$ must obey the following symmetry relations:
\begin{align}\label{parit}
&E_{\iota,n}(-P)=E_{\iota,n}(P),\\\label{trP}
&E_{\iota,n}(P+\pi)=E_{-\iota,n}(P),
\end{align}
with $\iota=0,\pm$, $n=1,2,\ldots$, and $P\in \mathbb{R}$.
One more equality 
\begin{equation}\label{refP}
E_{\iota,n}(\pi-P)=E_{-\iota,n}(P)
\end{equation}
is the direct consequence of  \eqref{parit}, \eqref{trP}. 

In what follows, we shall concentrate on the calculation of the meson energy spectra $E_{\iota,n}(P)$
in the interval $0<P<\pi/2$. Due to 
equalities \eqref{parit}-\eqref{refP}, this is sufficient to determine the dispersion laws $E_{\iota,n}(P)$ at all $P\in \mathbb{R}$ .
\subsection{Heuristic calculation of the meson energy spectra\label{heur}}
In paper \cite{Rut18}, a  heuristic  procedure of the calculation  of the   meson energy spectra in  model \eqref{XXZH} was briefly announced. Now we proceed to the detailed description of this heuristic calculation, which is
based on techniques  developed previously in  papers
\cite{Rut08a,RutP09}.

Let us treat the two kinks as classical particles  moving along the line, and attracting one another with a linear potential. Their Hamiltonian will be taken in the form
 \begin{equation}\label{Hk}
 H(x_1,x_2,p_1,p_2)=\omega(p_1)+\omega(p_2)+f\cdot(x_2-x_1),
 \end{equation}
 where $\omega(p)$ is the kink dispersion law \eqref{dl}.
 The kink spatial coordinates $x_1,x_2\in \mathbb{R}$ are subjected to the constraint 
 \begin{equation}\label{x1x2}
 -\infty<x_1<x_2<\infty,
  \end{equation}
  that
results from the  local "hard-sphere interaction"  of two particles \begin{footnote}{
Another equivalent possibility  \cite{Rut05,Rut08a,FZ06} is to remove  constraint (\ref{XX}) and to replace the linear potential $f\,(x_2-x_1)$ in (\ref{Hk}) by
$f\,|x_2-x_1|$.
}
\end{footnote}at $x_1=x_2$.

After the canonical transformation
\begin{subequations}\label{Canon}
\begin{align}
X=\frac{x_1+x_2}{2}, \quad x=x_1-x_2,\\
P=p_1+p_2, \quad p=\frac{p_1-p_2}{2},
\end{align}
\end{subequations}
the Hamiltonian \eqref{Hk} takes the form
\begin{equation}\label{Hk1}
H(p,x|P)=\epsilon(p|P)-f\, x,
\end{equation}
where $\epsilon(p|P)$  is given by \eqref{Energ}, and  $x<0$.

The total energy-momentum conservation laws read:
\begin{align}
&\epsilon(p(t)|P)-f\, x(t)=E=Const,\\
&P(t)=Const.
\end{align}

The classical evolution in the ``center of mass frame" is determined by the canonical equations of motion:
\begin{subequations}\label{can}
\begin{align}
&\dot{X}(t)=\frac{\partial \epsilon(p|P)}{\partial P},\\
&\dot{P}(t)=0,\\\label{xt}
&\dot{x}(t)=\frac{\partial \epsilon(p|P)}{\partial p},\\
&\dot{p}(t)=f.\label{pt}
\end{align}
\end{subequations}
This classical evolution  strongly depends on the values of the 
conserved total momentum   $P$, and energy $E$.

Let us start from  the case of a small enough total momentum $P$ of two particles:
\begin{equation}\label{Psm}
0\le P \le P_c(\eta),
\end{equation}
where the critical momentum $P_c(\eta)$ is given by \eqref{Pc}. In this case,  the kinetic energy $\epsilon(p|P)$ monotonically increases in $p$ in the interval $(0,\pi/2)$, see Figure \ref{fig:Lb}.
The dynamics of the system is qualitatively different in two 
regimes: 
\begin{equation}\label{1R}
\epsilon(0|P)<E<\epsilon(\pi/2|P), 
\end{equation}
and 
\begin{equation}\label{2R}
E>\epsilon(\pi/2|P).
\end{equation}
\begin{enumerate}
\item
In the first regime \eqref{Psm}, \eqref{1R}, equation $\epsilon(p|P)=E$ has two solutions $p=\pm p_a$ 
bounding the kinematically allowed region 
\begin{equation}
-p_a<p<p_a
\end{equation}
in the interval  
$p\in (-\pi/2,\pi/2)$ of the momentum $p$ variable, see Figure \ref{fig:Lb}.
Let us choose the initial conditions for equations \eqref{xt}, \eqref{pt} as follows: $x(0)=0$ and $p(0)=-p_a<0$, see Figure \ref{fig:L}. 
Due to \eqref{pt}, the momentum $p$   linearly increases in time
$p(t)=-p_{a} +f\, t$ until the moment 
\begin{equation}\label{t1}
 t_1 =\frac{ 2 p_a }{f}, 
 \end{equation}
 when $p (t)$ reaches the value $p(t_1-0)=p_a>0$.
The spatial coordinate $x$, in turn,  decreases from the value $x(0)=0$ to its minimal value 
\begin{equation}\label{xmin}
x_{min}=\frac{\epsilon(0|P)-E}{f}<0
\end{equation}
at $t=t_1/2$, and then increases up to the initial  zero value at the time moment $t_1$, $x(t_1)=0$. After the subsequent 
elastic reflection 
from the infinite potential well at $x=0$, the momentum changes its sign: $p(t_1+0)=-p(t_1-0)=-p_a$. 
Then the whole cycle described above repeats. So,  in this first regime,  
the coordinate $x(t)$ and the momentum $p(t)$ of the relative motion of two particles are periodic functions of time with  period $t_1$, 
and 
\begin{equation}\label{p1t}
p(t)=-p_a+ \{{t}/{t_1}\}\,t_1\,f, \quad \text{at   }t>0,
\end{equation}
where
$\{ z\}$ denotes the fractional part of $z$.
\begin{figure}
\centering
\subfloat[ 
Evolution in the spacial coordinate $x$.
]
{\label{fig:La}
\includegraphics[width=.75\linewidth]{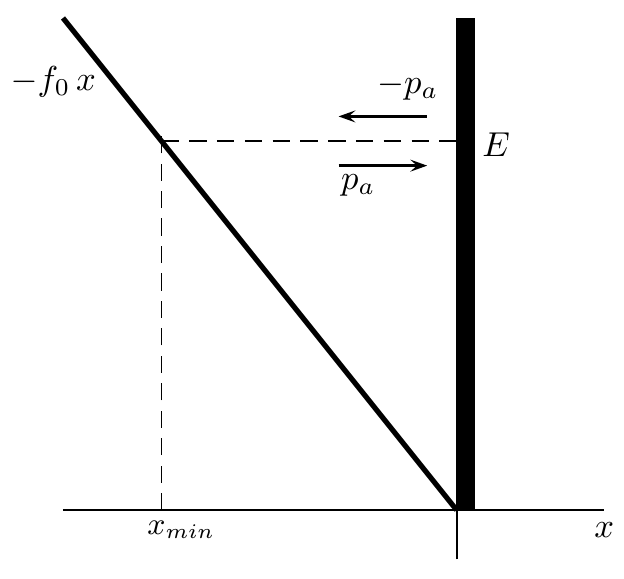}}

\subfloat[
Evolution in the momentum  $p$.
]
	{\label{fig:Lb}
\includegraphics[width=.95\linewidth]{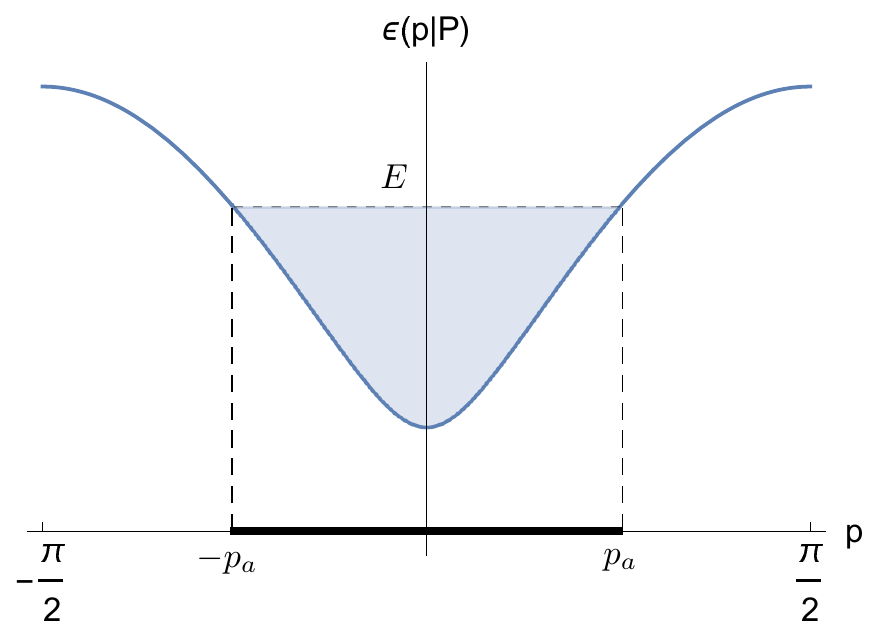}}

\caption{\label{fig:L} Evolution of the classical Hamiltonian system \eqref{xt}, \eqref{pt} 
in spatial coordinate $x$ (a), and in momentum $p$ (b), at $P=0$ in the first dynamical regime \eqref{1R}. 
}
\end{figure}

Figure \ref{fig:dn1} illustrates the ``lentils-pod-like" world paths $x_1(t)<x_2(t)$ of two particles in the first dynamical regime.
One can easily see from the canonical equations of motion \eqref{can}, that both particles drift together in this regime 
with the average velocity
\begin{equation}\label{Xt}
\langle \dot{X}\rangle:=\frac{1}{t_1}\int_0^{t_1} dt\, \dot{X}(t)=\frac{\omega(p_{1a})-\omega(p_{2a})}{p_{1a}-p_{2a}},
\end{equation}
where 
\begin{equation}\label{p12a}
p_{1a}=P/2+p_a,
\quad p_{2a}=P/2-p_a.
\end{equation}
\begin{figure}[ht]
\centering
\includegraphics[width=.95\linewidth, angle=00]{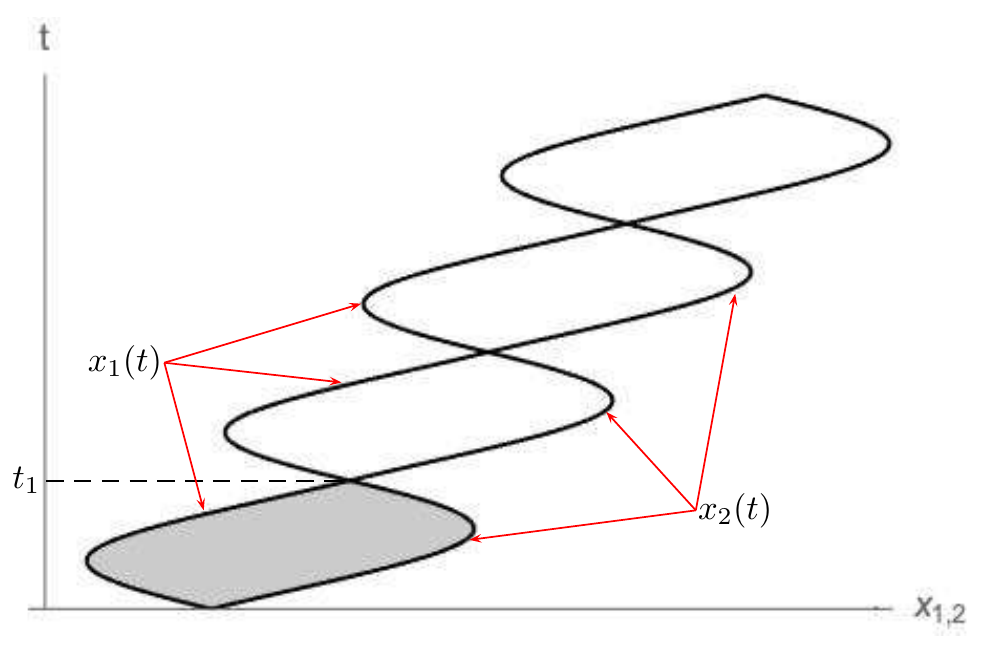}
\caption{\label{fig:dn1}  Lentils-pod-like world paths $x_1(t)$ and $x_2(t)$ of two particles 
in the first  regime. The dynamics of the particles is determined by the Hamiltonian \eqref{Hk}, the period $t_1$ is 
given by  \eqref{t1}. } 
\end{figure}

\item
At higher energies \eqref{2R}, the kinematically allowed regions in the $p$-variable extend to the whole real axis. 
The momentum $p(t)$ linearly increases with time 
\begin{equation}\label{ptA}
p(t)=p(0)+f\, t, \quad t\in\mathbb{R},
\end{equation}
 while the coordinate $x(t)$ oscillates in the interval 
  $x_{min}\le x(t)\le x_{max}$, where $x_{min}$ is given by \eqref{xmin}, and
\begin{equation}
 x_{max}=\frac{\epsilon(\pi/2|P)-E}{f}.
\end{equation}
Since $x_{max}< 0$, the two kinks never meet and display periodic Bloch oscillations \cite{Bloch29} 
along the spin chain with time  period
$t_2={\pi}/{f}
$.
Figure~\ref{fig:dn2} shows such Bloch oscillations of two particles in real space. 
The time-dependencies of their spatial coordinates
can  be easily found explicitly:
\begin{align}\label{x12t}
&x_1(t)=f^{-1}\,\omega(ft+p_{10})+C_1,\\\nonumber
&x_2(t)=-f^{-1}\,\omega(-ft+p_{20})+C_2,
\end{align}
where 
\begin{align}
p_{10}+p_{20}=P,\\
C_2-C_1=E/f.\label{C12E}
\end{align}
The kinks do not drift along the chain in the this  regime:
\begin{equation}\label{dX0}
\langle \dot{X}\rangle:=\int_t^{t+t_2} dt'\, \dot{X}(t')=0.
\end{equation}
\begin{figure}[ht]
\centering
\includegraphics[width=.95\linewidth, angle=00]{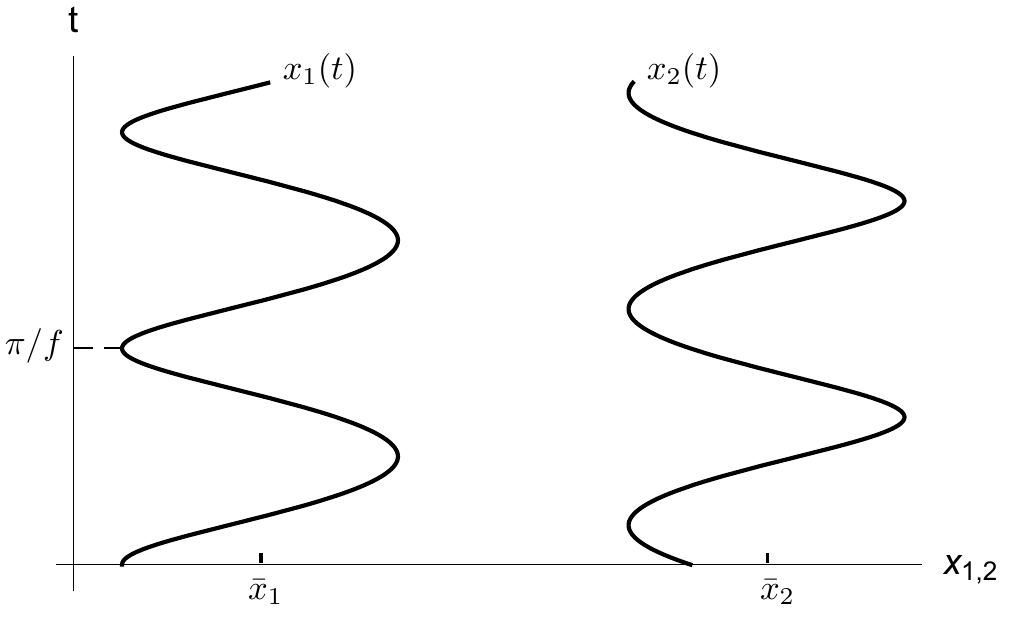}
\caption{\label{fig:dn2} Bloch oscillations of two particles 
in the second dynamical regime. The  period of oscillation is $\pi/f$, $x_1(t)$ and $x_1(t)$ are given by \eqref{x12t}.} 
\end{figure}
\end{enumerate}

At 
\begin{equation}\label{Pp}
P_c(\eta)<P<\pi/2,
 \end{equation}
 the profile of the function $\epsilon(p|P)$ changes, as it is described in Appendix \ref{2E} and  shown in Figure \ref{Ea}.  
In this case, the function $\epsilon(p|P)$ 
has a local maximum at $p = 0$, and takes its minimum value  $\epsilon_m(P,\eta)$ at
$p = \pm p_m(P,\eta)$ where $p_m(P,\eta)$ and $\epsilon_m(P,\eta)$ are  given by equations \eqref{pmin}, and \eqref{epm}, 
respectively. 
As the result, the classical evolution of the system in the third regime under condition \eqref{Pp} and
\begin{equation}\label{Ee}
\epsilon_m(P)<E<\epsilon(0|P)
 \end{equation}
 becomes more complicated. 

\begin{figure}[ht]
\centering
\includegraphics[width=.95\linewidth, angle=00]{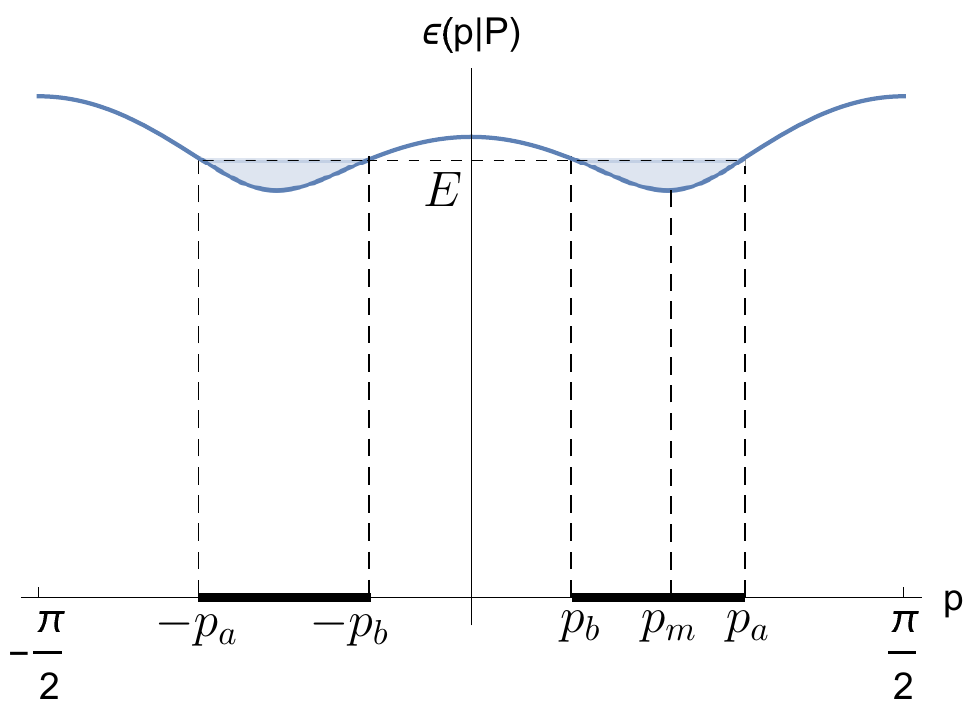}
\caption{\label{fig:2Lac}  Two kinematically allowed regions $(-p_a,-p_b)$ and $(p_b,p_a)$ in the third dynamical regime
\eqref{Pp}, 
\eqref{Ee}.} 
\end{figure}

Let $p=p_b$, and 
$x=0$ at $t=0$. Then the momentum $p$ linearly increases in time 
\begin{align}
&p(t)=p_b+f\,t,\quad \text{at  }0<t<t_3,\\\label{t3}
&t_3=\frac{p_a-p_b}{f},
\end{align}
in the right lacuna in Figure \ref{fig:2Lac}, and at 
$t=t_3$ reaches the value $p_a$. During the time interval $(0,t_3)$, the spatial coordinate 
decreases to the value $x_{min}=[\epsilon_{m}(P)-E]/f<0$, and then returns to the initial zero value: $x(t_3)=0$.
After the elastic reflection from the infinite potential wall at $x=0$, the sign of the momentum $p$ changes: $p(t_3+0)=-p_a$.
During the subsequent time interval $t_3<t<2t_3$, the momentum $p$ linearly increases in time in the left lacuna,
\begin{equation}
p(t)=-p_a+f\,t,
\end{equation}
By the end of this time interval $p(2t_3-0)=-p_b$, and $x(2t_3)=0$. After the second scattering from the infinite potential wall 
at $x=0$, the momentum changes the sign and returns to its initial value: $p(2t_3+0)=p_b$.
So, the momentum $p(t)$ and the spatial coordinate $x(t)$ are periodic functions of time with period $2t_3$.

\begin{figure}[ht]
\centering
\includegraphics[width=.95\linewidth, angle=00]{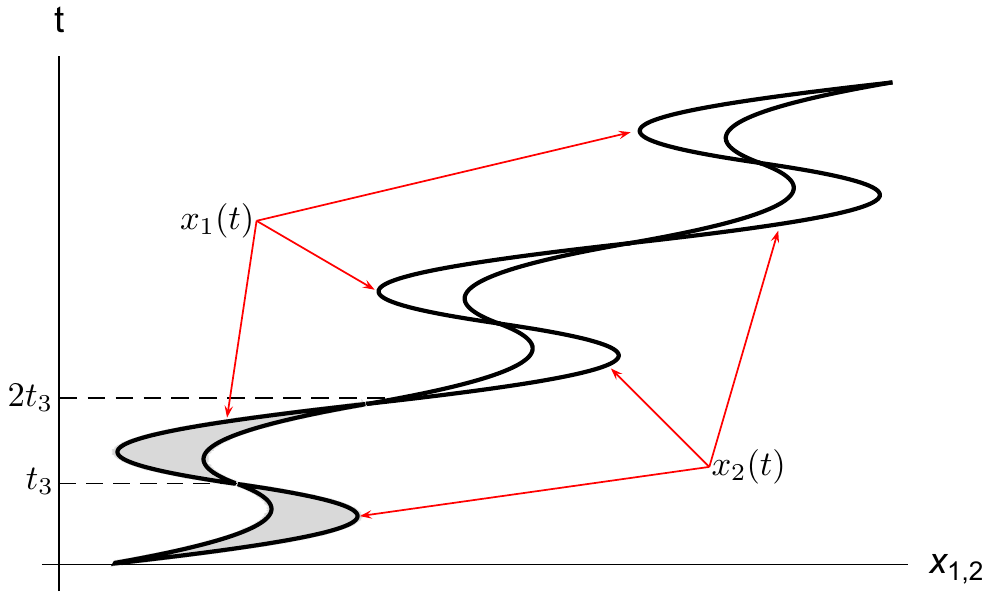}
\caption{\label{fig:dn3}  World paths $x_1(t)$ and $x_2(t)$ of two particles 
in the third dynamical  regime. The half-period $t_3$ is 
given by  \eqref{t3}. } 
\end{figure}

Evolution of the spatial coordinates of two particles in this third regime is shown in Figure \ref{fig:dn3}.
The two particles drift together  with the average velocity
\begin{equation}
\langle \dot{X}\rangle=\frac{\omega(p_{1a})-\omega(p_{1b})-\omega(p_{2a})+\omega(p_{2b)}}{2(p_{1a}-p_{1b})},
\end{equation}
where 
\begin{equation}\label{p12b}
p_{1b}=P/2+p_b, \quad  p_{2b}=P/2-p_b.
\end{equation}

With increasing energy, the points $-p_b$ and $p_b$ in Figure~\ref{fig:2Lac}  approach one another, and finally  merge in 
the origin, when the energy exceeds the
value $E=\epsilon(0|P)$. At higher energies in the interval 
\begin{equation}
\epsilon(0|P)<E<\epsilon(\pi/2|P),
\end{equation}
the kinematically allowed region fills the interval $(-p_a,p_a)$, and the classical evolution of the system is described by equations 
\eqref{t1}, \eqref{p1t},  and  \eqref{Xt}, corresponding to the regime (I).
Upon further increase of the energy into the interval \eqref{2R}, the system falls into the Bloch oscillatory regime (II) 
characterized by equations \eqref{x12t} and \eqref{dX0}.

\begin{figure}
\begin{center}
\includegraphics[width=.48\textwidth]{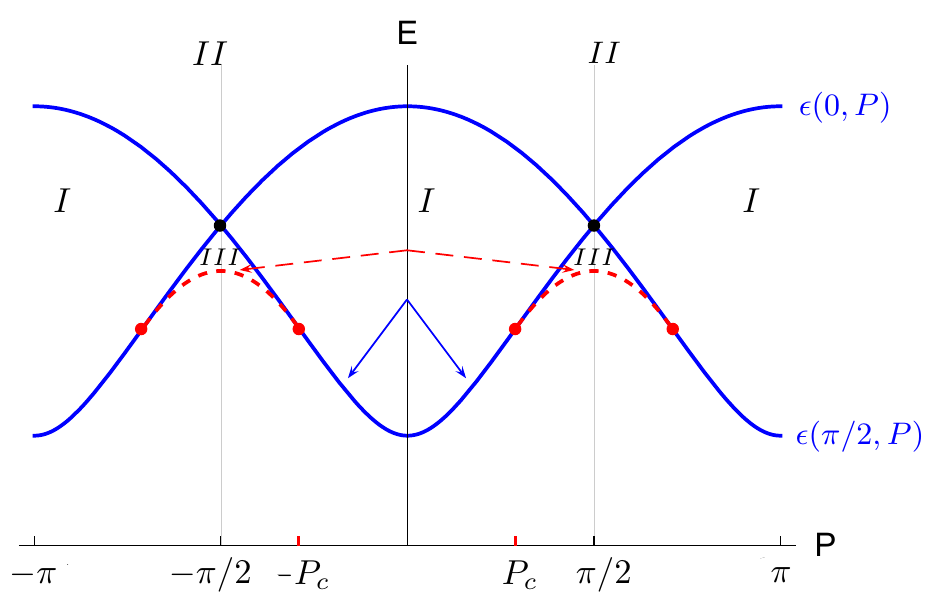}
\caption{\label{fig:RegE} Structure of the meson energy spectra in the energy-momentum plane.  Detailed explanations are in the text.
}
\end{center}
\end{figure}

It is straightforward to extend the analysis described above  from the interval $(0,\pi/2)$ of the 
total momentum to  all $P\in \mathbb{R}$. The result is illustrated  in Figure~\ref{fig:RegE}, which shows
 the regions in the $PE$-plane, where the dynamical regimes (I), (II), and  (III) are realized. 
The whole diagram is symmetric with respect to the reflection 
$P\to-P$ and $\pi$-periodic in the momentum $P$.
Two solid  blue  curves in Figure~\ref{fig:RegE}
plot the functions $\epsilon(0,P)$, and $\epsilon(\pi/2,P)$. The right dashed red  curve displays the function 
$\epsilon_m(P)$ in the interval $P_c<P<\pi-P_c$, where $\epsilon_m(P)$ and $P_c$  are given by \eqref{epm} and \eqref{Pc},
respectively. The left  dashes red curve is the mirror reflection of the right one with respect to the ordinate axis. 
The whole diagram in  Figure~\ref{fig:RegE} corresponds to a generic fixed value of the parameter $\eta>0$. 

Let us  now turn to the quantization of the  Hamiltonian dynamics described above. 
Two approximate schemes can be used at small $f>0$. 
\subsubsection{Semiclassical quantization}
In order to quantize the states well inside the regions (I), (II), or (III) in Figure~\ref{fig:RegE} far enough from their boundaries, 
it is natural to use the semiclassical Bohr-Sommerfeld quantization rule. This rule states, that the increase  of the phase   $\Delta \Phi$  of the semiclassical wave-function corresponding to one cycle of the periodical phase trajectory $x(t), p(t)$ 
in some potential well profile must be  a multiple of $2\pi$: 
 \begin{equation}\label{Den}
 \Delta \Phi=2\pi n.
 \end{equation}

In the first dynamical regime (I), the phase shift $ \Delta \Phi$ consists of three terms,
  \begin{equation}\label{Den3}
 \Delta \Phi= \Delta_1 \Phi+\Delta_2 \Phi+\Delta_3 \Phi,
 \end{equation}
 where the first one 
   \begin{equation}\label{dPhi1}
    \Delta_1 \Phi=\oint dx\, p(x)
   \end{equation}  
   is associated with the one cycle of the classical movement  over the closed periodical phase trajectory,
   and the second one 
\begin{equation}
    \Delta_2 \Phi=\frac{\pi}{2}
\end{equation}
represents the familiar phase shift of the wave function at the left turning point $x_{min}$, see Figure  \ref{fig:La} and ref.
[\onlinecite{LL3}].
The third phase
shift  $\Delta_3 \Phi$ is associated with the right turning point $x=0$.
It results from the mutual elastic scattering of two kinks, that meet together at some point $x_1=x_2$. Just before their collision,  the left and the right kinks had momenta  $p_{1a}$ and $p_{2a}$  given by  equation \eqref{p12a}, $p_{1a}> p_{2a}$. After the collision, the left and the right kinks get momenta  $p_{2a}$, and  $p_{1a}$, respectively. Accordingly, the phase shift $\Delta_3 \Phi$
must be identified (to the zero order in $f$) with the  scattering phase \eqref{scph} of the $\iota$-th meson mode:
\begin{equation}
\Delta_3 \Phi=\theta_\iota(p_{1a},p_{2a}),
   \end{equation}  
   where $\iota=0,\pm$.
So, the WKB energy levels $E_{\iota, n}(P)$ (with $n\gg1$) of the $\iota$-th meson mode 
are determined in the first dynamical regime by the quantization condition:
\begin{equation} \label{WKB}
\Delta_1 \Phi+\frac{\pi}{2}+\theta_\iota\left(P/2+p_a,P/2-p_a\right)=2 \pi n,
\end{equation}
where $n$ is the number of the energy level,  $n=1,2,\ldots$.  

The phase  shift $\Delta_1 \Phi$  defined by \eqref{dPhi1} can be rewritten as 
\begin{equation}\label{dPh1}
\Delta_1 \Phi=2\int_{x_{min}}^0 dx \, p(x),
\end{equation}
where the momentum $p(x)$  monotonically increases with $x$ from zero 
at the left turning point $p(x_{min})=0$, to the positive value $p(0)=p_a>0$ at the right turning point. 
The integral in the right-hand side of \eqref{dPh1}  can be further transformed as follows:
\begin{align}\label{calc}
&\Delta_1 \Phi=2\int_0^{p_a}dp \,\, \frac{d x(p)}{dp}\, \,p=\frac{2}{f}\int_0^{p_a}dp \,\, \dot{x}\,p =\\\nonumber
&\frac{2}{f}\int_0^{p_a}dp \,\, \frac{\partial \epsilon(p|P)}{\partial p}\,p =
\frac{1}{f}\left[ 2 E_{\iota, n}(P)\,p_a  -\int_{-p_a}^{p_a}dp \, \epsilon(p|P)\right].
\end{align}
The canonical equations of motion \eqref{pt} and \eqref{xt} have been used in the second  and third equalities, respectively. 
 In the last equality, we have  integrated by parts, and then used equation
 \begin{equation}
 E_{\iota, n}(P)=\epsilon(\pm p_a|P).
 \end{equation}
Thus,  the semiclassical quantization rule \eqref{WKB} predicts the following energy spectrum of the two-kink bound states
in the first regime:
\begin{align}\label{EnSem1}
&  2E_{\iota, n}(P)\,p_a  -\int_{-p_a}^{p_a}dp \, \epsilon(p|P)=\\\nonumber
&  f\left[2\pi  \,(n-1/4)-\theta_\iota\left(P/2+p_a,P/2-p_a\right)\right].
 \end{align}

The quantity  $f \,\Delta_1 \Phi$ standing in the left-hand side of this equation equals at $E=E_{\iota, n}(P)$ to the area of the dashed region
in Figure \ref{fig:Lb}.
On the other hand, the ratio $\Delta_1 \Phi/f$ admits 
 an alternative geometrical interpretation: 
\begin{align*}
 &\frac{\Delta_1 \Phi}{f}= \frac{1}{f}\oint p\, dx=-\frac{1}{f}\oint x\, dp=-\frac{1}{f}\int_{-p_a}^{p_a} x\, dp=\\
& -\int_{0}^{t_1} x(t)\, dt=\int_{0}^{t_1} [x_2(t)-x_1(t)]\, dt=S,
\end{align*} 
where $S$ is the area of the dashed  "lentil seed" in Figure~\ref{fig:dn1}.

 In the third (III) dynamical regime, the functions $p(t)$ and $x(t)$ are periodic with the period $2t_3$, 
 and one can still use the Bohr-Sommerfeld rule  \eqref{Den}, and equations
 \eqref{Den3},  \eqref{dPhi1} for the semiclassical quantization. However, now the three terms in the right-hand side of 
 \eqref{Den3} split into two contributions corresponding to the left and right lacunas in Figure \ref{fig:2Lac}:
 \begin{equation*}
 \Delta_i \Phi= \Delta_{i}^{(l)}\Phi+\Delta_{i}^{(r)} \Phi,
 \end{equation*}
where $i=1,2,3$, and 
 \begin{align*}
& \Delta_1^{(l)} \Phi = \Delta_1^{(r)} \Phi =\\
&\frac{1}{f}\left[  E_{\iota, n}(P)\,(p_a-p_b)  -\int_{p_b}^{p_a}dp \, \epsilon(p|P)\right],\\
 &\Delta_2^{(l)} \Phi =\Delta_2^{(r)} \Phi=\frac{\pi}{2},\\
 &\Delta_3^{(l)} \Phi =\theta_\iota(p_{2b},p_{1b}), \quad 
 \Delta_3^{(r)} \Phi =\theta_\iota(p_{1a},p_{2a}).
\end{align*} 
Thus,  the the Bohr-Sommerfeld semiclassical quantization rule \eqref{Den} leads 
in the third regime to the following  energy spectrum $E_{\iota, n}(P)$ of the $\iota$-th  two-kink bound-states mode:
\begin{align}\label{EnSem3}
 & 2E_{\iota, n}(P)\,(p_a-p_b)  -2\int_{p_b}^{p_a}dp \, \epsilon(p|P)=
 f\bigg[2\pi  \,(n-1/2)-\\\nonumber
& {\theta_\iota\left(P/2+p_a,P/2-p_a\right)+\theta_\iota\left(P/2+p_b,P/2-p_b\right)}\bigg],\\\nonumber
&   E_{\iota, n}(P)=\epsilon(p_a|P)= \epsilon(p_b|P).
 \end{align}
Note, that quantities   $f\, \Delta_1 \Phi$ and $ \Delta_1 \Phi/f$ are now equal to the areas of the dashed regions in 
Figure \ref{fig:2Lac}, and Figure \ref{fig:dn3}, respectively.

The Bohr-Sommerfeld quantization rule \eqref{Den}  cannot be directly applied in the second dynamical regime, 
since the momentum $p(t)$ monotonically increases with time (see \eqref{pt}), and, therefore,  the phase trajectories 
in the $(x,p)$-plane  are not closed. Nevertheless, the semiclassical energy spectra $E_{\iota, n}(P)$ can be  partly 
recovered by means of the following simple arguments \cite{Rut08a}. 

In the second (II) dynamical regime, the two kinks do not meet together, and oscillate around certain positions 
$\bar{x}_1$ and $\bar{x}_2$ in the spin chain, see  Figure \ref{fig:dn2}. These two kinks cannot drift along the chain, and, therefore, 
the velocity $v_{\iota, n}(P)$ of their bound state is zero. Since
$v_{\iota, n}(P)=\partial E_{\iota, n}(P)/\partial P=0$, the energy $E_{\iota, n}(P)$ does not depend on $P$. On the other hand, it is clear 
from equations \eqref{x12t}, \eqref{C12E}, that the shift of the second kink to the right by $\Delta x$:
\begin{equation}\label{dx}
\bar{x}_1\to \bar{x}_1, \quad \bar{x}_2\to \bar{x}_2+\Delta x,
\end{equation}
leads to the proportional increase of the two-kink energy: 
\begin{equation}\label{dE}
E\to E+ f \Delta x.
\end{equation}
Recalling that the spin chain is  discrete with  unit lattice spacing, and the antiferromagnetic ground state is invariant with 
respect to the translation by two lattice sites, we can argue, that the translation parameter $\Delta x$ in equations \eqref{dx} and \eqref{dE}
must take integer even values. As a result, the energy spectrum of the two-kink bound states in the second semiclassical regime
must form the  equidistant Zeeman ladders
\begin{equation}\label{EnSem2}
E_{\iota, n}(P) = 2 n f+ A_{\iota},
\end{equation}
with some constants $A_{\iota}$. These constants will be determined later  (see equation \eqref{Sem3A}) in   Section \ref{SemReg}. 
Equation \eqref{Sem3A} will be derived in Appendix \ref{SR3} in   the more rigorous  approach based on the 
Bethe-Salpeter equation.

Equations \eqref{EnSem1}, \eqref{EnSem2}, and \eqref{EnSem3},   represent the leading terms of the semiclassical expansions in 
integer powers of $f\to0$ of the  meson energy spectra $E_{\iota, n}(P)$, that hold well inside the regions (I), (II), and (III) in 
Figure~\ref{fig:RegE}, respectively.  However, these semiclassical expansions cannot be applied close to boundaries of these regions. 

In the vicinity of the curves separating the region (I) from the neighboring regions (II) and (III) in 
Figure~\ref{fig:RegE}, one should use instead the 
{\it crossover expansions}, which will be described later in  Section \ref{crosr}.
On the other hand, close to the bottom boundaries of the regions (I) and (III), the small-$f$ asymptotics of the meson spectra
are determined by the {\it {low energy expansions}}  
in fractional powers of $f$. Few initial terms
in these expansions can be obtained by means of the canonical quantization of the Hamiltonian dynamics of the model \eqref{Hk}. 
\subsubsection{Canonical quantization \label{CanQuan}}
There are three low-energy expansions for the  meson energy spectra, which hold at small $f$ in different regions 
of the $(E,P)$-plane shown in Figure~\ref{fig:RegE}. 
\begin{enumerate}
\item The first low-energy expansion holds at $P \in (-P_c(\eta), P_c(\eta))$ and energies slightly above the minimal value $\epsilon(0|P)$.
In this region, which is  indicated by the solid blue arrows in Figure~\ref{fig:RegE}, the 
momentum $|p|<p_{a}$ is small, and the effective kinetic energy $\epsilon(p|P)$  can be expanded  in $p$ to the second order:
\begin{equation}\label{epsE}
\epsilon(p|P)=
\epsilon(0|P)+\frac{\epsilon''(0|P)}{2} \,p^2+\ldots.
\end{equation}
\item The second low-energy expansion describes the meson energy spectra in the regions lying 
slightly above the red dashed curves in Figure~\ref{fig:RegE}, and indicated there by 
dashed red arrows.   In this regime, the effective kinetic energy $\epsilon(p|P)$ has the profile of the kind shown in Figure \ref{fig:2Lac}, 
and the meson energy $E$  is slightly above the minimum value $\epsilon_m(P)=\epsilon(p_m|P)$ given by \eqref{epm}.
\item The third low-energy expansion holds close to the points $P=\pm P_c+ \pi n$ with $n\in \mathbb{Z}$ at the meson energies $E$ slightly above its lower bound $\epsilon(0|P_c)$. Corresponding profiles of the effective kinetic energy are shown by blue dashed lines in Figures \ref{Ea}, \ref{Ec}.
\end{enumerate}

In this Section we shall restrict our attention to the first low-energy expansion and derive its first three terms using the canonical
quantization of the classical dynamics determined by the Hamiltonian 
\begin{equation}\label{Hk1A}
H(p,x|P)=\epsilon(0|P)+\frac{\epsilon''(0|P)}{2} \,p^2-f x.
\end{equation}
All three low-energy expansions will be derived later in Section \ref{WK} in the more rigorous approach based on the Bethe-Salpeter equation.

So, in this Section the analysis will be restricted to the case $P \in (-P_c(\eta), P_c(\eta))$, with a small positive $E- \epsilon(0|P)$.
After the replacement $p\to-i \partial_x$, the classical Hamiltonian \eqref{Hk1A} describing the relative motion 
of two kinks, transforms into its quantum counterpart:
\begin{equation}\label{Hq}
\hat{H}=\epsilon(0|P)-\frac{\epsilon''(0|P)}{2} \partial_x^2 - f \, x.
\end{equation}
This second-order differential operator acts on the  wave functions $\psi(x)$ that vary in the half-line $x<0$ and vanish at 
$x\to-\infty$. The eigenvalues of the Hamiltonian \eqref{Hq} determine the meson energy spectrum $E_{n}(P)$ in this
approximate 
quantization scheme. In order to complete the eigenvalue problem  for  $E_{n}(P)$, 
one has to  supplement the differential equation 
\begin{equation}\label{Hq2}
\left[\epsilon(0|P)-\frac{\epsilon''(0|P)}{2} \partial_x^2 - f \, x\right]\psi_{ n}(x)=E_{ n}(P) \,\psi_{ n}(x),
\end{equation}
defined in the negative half-axis $x<0$, with the appropriate boundary condition for the eigenfunctions $\psi_{ n}(x)$ at the origin $x=0$. 

For particles, which  are free at $f=0$, the boundary condition at $x=0$ is determined by their statistics. 
 In the best studied free-fermionic case, which is realized in the  IFT \cite{McCoy78,FonZam2003, FZ06} and in the Ising spin chain \cite{Rut08a}, the relevant one 
 is the Dirichlet boundary condition 
 \begin{equation}\label{DirBC}
 \psi_{ n}(0)=0.
\end{equation}
  The resulting energy spectrum reads in this case:
\begin{equation}\label{LE2E}
E_n(P)=\epsilon(0|P)+f^{2/3} [\epsilon''(0|P)/2]^{1/3} z_n,
\end{equation}
where $-z_n$,  $n=1,2,\ldots$, are the zeros of the Airy function. For the IFT, the meson mass spectrum 
of the this structure was predicted by McCoy and Wu \cite{McCoy78}  in 1978.

The right-hand side of \eqref{LE2E} represents two 
initial terms of the  low-energy expansion in integer powers of the small parameter $f^{1/3}$ for the meson energy spectrum.
In the canonical quantization scheme, it is not difficult 
to calculate a few subsequent terms of this expansion following the procedure developed by Fonseca and Zamolodchikov  for the IFT, see Appendix B of  reference [\onlinecite{FonZam2003}]. One can show this way, that the next non-vanishing term in the low energy expansion for $E_n(P)$ in the free-fermionic case is of order $f^{4/3}$:
\begin{equation}\label{LE2A}
E_n(P)=\epsilon(0|P)+f^{2/3} [\epsilon''(0|P)/2]^{1/3} z_n+O(f^{4/3}),
\end{equation}
cf. equation (B.19) in \cite{FonZam2003}.

In the case of  free bosons, one should choose the Neumann boundary condition $\psi_n'(0)=0$, instead of \eqref{DirBC}. 
As the result, the 
meson energy spectrum is given by equation \eqref{LE2A}, in which $z_n$ are replaced by the numbers $z_n'$, such that $(-z_n')$ are the zeros of the derivative of the  Airy function. 

In the case of the XXZ spin-chain  \eqref{XXZH}, the choice of the boundary condition for equation \eqref{Hq2} is not so evident, since the kinks are not free  at $h=0$. Their strong short-range interaction is completely characterized  at $h=0$  by the scattering amplitudes $w_\iota(p_1,p_2)$ given by equations \eqref{scph}. In a certain sense, these scattering amplitudes determine 
also the statistics of kinks due to the Faddeev-Zamolodchikov commutation relations \eqref{FZC}.  Since 
\begin{equation}
\lim_{p_1\to p_2}w_\iota(p_1,p_2)=-1,
\end{equation}
for all $\iota=0,\pm$, the kinks behave almost like free fermions in  mutual scatterings with  a small momentum transfer.
And since only small momenta $p=(p_1-p_2)/2$ are relevant in the considered low-energy dynamical regime (see equation 
\eqref{epsE}),
it is tempting to assume, that  the differential equation \eqref{Hq2} should be supplemented with the Dirichlet boundary condition.
However, this is not correct. 
We will show below, that, instead,  the correct choice of the boundary condition for  
equation  \eqref{Hq2} is the Robin boundary condition
\begin{equation}\label{Rob}
\psi(0)-a_{\iota}(P) \psi'(0)=0,
\end{equation}
where 
\begin{align}\label{scL}
&a_\iota(P)=-\frac{1}{2}\,\partial_p \theta_\iota(P/2+p,P/2-p)\Big|_{p=0}=\\\nonumber
&-  \frac{J \sinh \eta}{\omega(P/2)}\,\,\, \frac{d \Theta_\iota(\alpha)}{d\alpha}\bigg|_{\alpha=0}.
\end{align}
denotes the scattering length in the $\iota$-th two-kink scattering channel. 
With this new boundary condition, the spectrum of the Sturm-Liouville  problem \eqref{Hq2}
modifies to the form:
\begin{align}\label{les}
&E_{\iota, n}(P)=\epsilon(0|P)+f^{2/3} [\epsilon''(0|P)/2]^{1/3} z_n+ \\
&
f\, a_\iota (P) \nonumber
+O(f^{4/3}).
\end{align}
\renewcommand{\theenumi}{\arabic{enumi}}

The  justification  of the Robin boundary condition \eqref{Rob} for the 
differential equation \eqref{Hq2} is the following.

The first and  most important reason, is that the resulting  low-energy spectrum \eqref{les} 
will be confirmed later in Appendix \ref{LE} in the more consistent calculations based on the perturbative solution
of the Bethe-Salpeter equation.
 
Second, the  low-energy expansion \eqref{les} is consistent with the semiclassical expansion \eqref{EnSem1} 
 in the following sense. Both expansions describe 
the small-$f$ asymptotical behavior of the meson energy spectrum $E_{\iota, n}(P)$ in the first (I) region shown in Figure \ref{fig:RegE}. The semiclassical expansion \eqref{EnSem1} can be used at $n\gg 1$, while the low-energy expansion
\eqref{les} holds in the narrow strip above the bottom boundary of the region (I),
 i.e. at small enough $E_{\iota, n}(P)-\epsilon(0|P)>0$, and $P\in (-P_c,P_c)$. In the crossover region, 
 the two asymptotical expansions
 \eqref{EnSem1} and  \eqref{les}  must be equivalent. Indeed, the semiclassical asymptotical formula \eqref{EnSem1} can be reduced to the form
 \eqref{les}  in the crossover region, using  the large-$n$ asymptotics \cite{AbrSt} for the zeros of the Airy function
 \begin{equation}
 z_n=\left[\frac{3 \pi}{8}(4n-1)\right]^{2/3}[1+O(n^{-2})],
 \end{equation}
 together with formulas
 \begin{align*}
&\theta_\iota(P/2+p_a,P/2-p_a)=-2p_a a_\iota(P)+O(p_a^2),\\
& E_{\iota, n}(P)=\epsilon(0|P)+\frac{\epsilon''(0|P)}{2}p_a^2+O(p_a^4).
 \end{align*}

Third, equation \eqref{Rob} can be interpreted as the {\it {effective}} boundary condition arising in a certain  modification
of the Sturm-Liouville problem \eqref{Hq2}, \eqref{DirBC}. In order to introduce the latter, let us first modify our 
original phenomenological classical model of two particles by adding to its Hamiltonian \eqref{Hk}  some 
interaction potential $u(x_1-x_2)$, that mimics the short-range interaction between kinks in the XXZ spin chain \eqref{XXZH} at $h=0$. 
Accordingly, the potential $u(x)$ should vanish at  distances much larger than the correlation length $\xi$.
For simplicity, we shall assume, that $u(x)=0$ at $|x|>r$, with some $r\sim \xi$. After the canonical quantization of this 
modified classical model under the assumptions that the two particles are fermions, we obtain the modified
Sturm-Liouville problem in the half line $x<0$, consisting of the second-order differential equation
\begin{equation}\label{Hq2a}
\left[\epsilon(0|P)-\frac{\epsilon''(0|P)}{2} \partial_x^2 - f \, x+u(x)\right]\psi_{ n}(x)=E_{ n}(P) \,\psi_{ n}(x),
\end{equation}
and the Dirichlet boundary condition \eqref{DirBC}. 

At $f=0$,  the  energy spectrum is continuous, 
\[
E_p(P)=\epsilon(0|P)+\frac{\epsilon''(0|P)}{2}p^2,
\]
and the corresponding eigenfunction can be written  at $x<-r$ as
\begin{equation}\label{wf}
\psi_{p}(x)=\sin[p x-\varphi(p)].
\end{equation}
The scattering phase  $\varphi(p)$  corresponding to the short-range potential $u(x)$ must be  identified (up to the 
factor $1/2$) with the scattering phase \eqref{thet}:
\begin{equation}
\varphi(p)=\frac{1}{2}\,\theta_\iota(P/2+p,P/2-p).
\end{equation}
At small $p$, this scattering phase becomes proportional to the scattering length \eqref{scL},
\[
\varphi(p)=-p\, a_\iota(P)+O(p^2),
\]
and the wave function \eqref{wf} reduces to the form
\begin{equation}\label{wf1}
\psi_{p}(x)=\sin\left\{p[ x+a_\iota(P)]+O(p^2)\right\}.
\end{equation}
So, the wave function \eqref{wf} satisfies at small $p$  the  {\it effective Dirichlet boundary condition} at $x=-a_\iota(P)$,
\begin{equation}\label{BCa}
\psi_{p}[-a_\iota(P)]=O(p^2),
\end{equation}
or equivalently, the Robin boundary condition at $x=0$:
\begin{equation}\label{Rob2}
\psi_p(0)-a_{\iota}(P) \psi_p'(0)=O(p^2).
\end{equation}

At $f>0$ the solution of equation \eqref{Hq2a} vanishing at $x\to-\infty$ is given  at $x<-r$  by the Airy function:
\begin{equation}\label{Ai}
\psi_{ n}(x)=\mathrm{Ai}\,[-z(x)],
\end{equation}
where
\begin{equation}
z(x)=\left[\frac{2}{\epsilon''(0|P)}\right]^{1/3}f^{-2/3}\,[f x+E_{\iota, n}(P)-\epsilon(0|P)].
\end{equation}
Clearly, the function \eqref{Ai} must satisfy at small $f\to +0$ and finite fixed $n=1,2,\ldots$, the same effective Robin boundary condition \eqref{Rob}, as the  function \eqref{wf1}.  

As in the case $f=0$, one can
replace with sufficient accuracy the effective Robin boundary condition \eqref{Rob} for the Airy function \eqref{Ai} by the effective Dirichlet boundary condition in the shifted point: $\psi_{n}[-a_\iota(P)]=0$. The resulting characteristic equation $z[-a_\iota(P)]=z_n$, leads to 
the meson energy spectrum \eqref{les}.
 \section{Bethe-Salpeter equation \label{SecBSE}}
 In this Section we derive the Bethe-Salpeter equation for the XXZ spin chain model \eqref{XXZH1A} and describe  its essential properties.
\subsection{Two-kink approximation}
The energy spectrum of mesons at $h>0$ is determined by the eigenvalue problem \eqref{mesA}. This problem is extremely  difficult, 
because the interaction term $\sim h$ in the Hamiltonian \eqref{XXZH1A} does not conserve the number of kinks. 
Accordingly, the meson state solving equations  \eqref{mesA} must contain contributions of $2n$-kink states with all $n=1,2,\ldots$:
\[
|\pi_{s,\iota,n}(P)\rangle=|\pi_{s,\iota,n}^{(2)}(P)\rangle+{{|\pi_{s,\iota,n}^{(4)}(P)\rangle+\ldots}},
\]
where $|\pi_{s,\iota,n}^{(2n)}(P)\rangle$ is a linear combinations of the $2n$-kink states, 
\[
|K_{\mu_1\mu_2}(p_1)K_{\mu_2\mu_3}(p_2)\ldots)K_{\mu_{2n}\mu_{2n+1}}(p_{2n})\rangle_{s_1s_2\ldots s_{2n}}, 
\]
with $\mu_1=\mu_{2n+1}=1$, $s_1+\dots+s_{2n}=s$, and $\exp{[2 i (p_1+\dots+p_{2n})]}=\exp{(2 iP)}$.
As in the cases of the IFT \cite{FZ06} and Ising spin-chain model \cite{Rut08a}, the key simplification is provided by the two-kink approximation. 
It implies that one replaces the exact Hamiltonian eigenvalue problem \eqref{mesA} by its projection onto the two-kink subspace 
$\mathcal{L}_{11}^{(2)}$:
\begin{subequations}\label{mesB1}
\begin{align}\label{mes2a}
&\mathcal{H}_1^{(2)}(h)|\tilde{\pi}_{s,\iota,n}(P)\rangle=\tilde{E}_{\iota,n}(P)|\tilde{\pi}_{s,\iota,n}(P)\rangle,\\\label{T1til}
&T_1^2|\tilde{\pi}_{s,\iota,n}(P)\rangle=e^{2 i P}|\tilde{\pi}_{s,\iota,n}(P)\rangle,\\
&S^z|\tilde{\pi}_{s,\iota,n}(P)\rangle=s|\tilde{\pi}_{s,\iota,n}(P)\rangle,
\end{align}
\end{subequations}
where
\begin{align}\nonumber
&\mathcal{H}_1^{(2)}(h)=\mathcal{P}_{11}^{(2)}\sum_{j=-\infty}^\infty\!\!
\bigg\{-\frac{J}{2}\big[\sigma_j^x\sigma_{j+1}^x+\sigma_j^y\sigma_{j+1}^y+\\
&\Delta\,
\sigma_j^z\sigma_{j+1}^z
\big]
-h\, [(-1)^j \, \sigma_j^z-\bar{\sigma}  ]\bigg\}\mathcal{P}_{11}^{(2)}.\label{XXZ2}
\end{align}
Here $\mathcal{P}_{11}^{(2)}$ is the projection operator \eqref{PP11} onto the two-kink subspace $\mathcal{L}_{11}^{(2)}$, and 
$\bar{\sigma}$ is the zero-field spontaneous magnetization   \eqref{sig}.  
Tildes distinguish solutions of equations \eqref{mesB1} from those of the exact eigenvalue problems \eqref{mesA}.

Action of the modified translation operator $\widetilde{T}_1$ on the two-kink meson states  $|\tilde{\pi}_{s,\iota,n}(P)\rangle$
is determined by relations 
\begin{subequations}\label{mesB}
\begin{align}\label{mes2}
&\widetilde{T}_1 |\tilde{\pi}_{s,0,n}(P)\rangle=e^{i P}  |\tilde{\pi}_{-s,0,n}(P)\rangle, \; \text{for } s=\pm1,\\\label{tilTiot}
&\widetilde{T}_1 |\tilde{\pi}_{0,\iota,n}(P)\rangle=\iota e^{i P} |\tilde{\pi}_{0,\iota,n}(P)\rangle, \; \text{for } \iota=\pm.
\end{align}
\end{subequations}
 For $P,P'\in [0,\pi)$, we shall normalise the  meson states by the condition:
\begin{equation}\label{normPi}
\langle\tilde{\pi}_{s,\iota,n}(P)|\tilde{\pi}_{s',\iota',n'}(P')\rangle=\pi \delta_{s,s'}\delta_{\iota,\iota'}\delta_{n,n'}\delta(P-P').
\end{equation}

In the momentum representation, equation \eqref{mes2a} takes the form:
\begin{align}\label{BS0}
&[\omega(p_1)+\omega(p_2)-\tilde{E}(P)]\Phi_{s_1,s_2}(p_1,p_2|P)=\\\nonumber
&h\sum_{m=-\infty}^\infty\sum_{s_1',s_2'=\pm1/2}\iint_{\tilde{\Gamma}}\frac{dp_1'dp_2'}{\pi^2}\, \Phi_{s_1',s_2'}(p_1',p_2'|P)\cdot\\\nonumber
&\exp [2 i m(p_1+p_2-p_1'-p_2')] \cdot\\
&\phantom{.}_{s_2s_1} \langle K_{10}(p_2) K_{01}(p_1) |Q
| K_{10}(p_1') K_{01}(p_2') \rangle_{s_1's_2'},\nonumber
\end{align}
 where
\begin{equation}\label{Q}
Q=\sigma_{0}^z-\sigma_{1}^z-2 \bar{\sigma},
\end{equation}
the integration region $\tilde{\Gamma}$ is shown in Figure \ref{Gam},
and $\Phi_{s_1,s_2}(p_1,p_2|P)$ denotes the wave function 
\begin{align}\label{Phiphi}
\Phi_{s_1,s_2}(p_1,p_2|P)=\phantom{.}_{s_2 s_1} \langle K_{10}(p_2) K_{01}(p_1) |\tilde{\pi}(P)\rangle,
\end{align}
corresponding to the meson state $ |\tilde{\pi}(P)\rangle$.
It follows immediately from \eqref{T2K} and \eqref{T1til} that
\begin{equation}
\Phi_{s_1,s_2}(p_1,p_2|P)\left[e^{2i(P-p_1-p_2)}-1\right]=0.
\end{equation}

Let $p_1+p_2\in[0,\pi)$. Then equation \eqref{BS0} after summation over $m$ takes the form
\begin{align}\label{BS1}
&[\omega(p_1)+\omega(p_2)-\tilde{E}(P)]\Phi_{s_1,s_2}(p_1,p_2|P)=\\\nonumber
&h\sum_{s_1',s_2'=\pm1/2}\iint_{\tilde{\Gamma}}\frac{dp_1'dp_2'}{\pi}\,
\delta (p_1+p_2-p_1'-p_2')\cdot\\\nonumber
&\phantom{.}_{s_2s_1} \langle K_{10}(p_2) K_{01}(p_1) |Q
| K_{10}(p_1') K_{01}(p_2') \rangle_{s_1's_2'} \cdot\\
&\Phi_{s_1',s_2'}(p_1',p_2'|P).\nonumber
\end{align}
The subsequent analysis will be performed separately for the cases of the meson spin $s=1$ and $s=0$.
\subsection{s=1}
The wave function \eqref{Phiphi} of a meson with spin $s=1$ has only one component with $s_1=s_2=1/2$. 
We shall use the notation $\Phi_{\iota}(p_1,p_2|P)$ with $\iota=0$ for this wave function:
\begin{equation}\label{PhiphiA}
\Phi_{\iota=0}(p_1,p_2|P)=\phantom{.}_{1/2 ,1/2} \langle K_{10}(p_2) K_{01}(p_1) |\tilde{\pi}_{s=1, \iota=0}(P)\rangle.
\end{equation} 
Due to \eqref{FZ2}, it satisfies the following symmetry relation:
\begin{equation}
\Phi_0(p_1,p_2|P)=w_0(p_2,p_1)\Phi_0(p_2,p_1|P).
\end{equation}
For $P,(p_1+p_2)\in[0,\pi)$, we define the reduced meson wave function  $\phi_0(p|P)$
by the relation:
\begin{equation}\label{phi0P}
\Phi_0(p_1,p_2|P)=\pi\,e^{-i p_1}\,\delta(p_1+p_2-P)\,\phi_0\left(\frac{p_1-p_2}{2}|P\right).
\end{equation} 
The reduced wave function $\phi_0(p|P)$ can be analytically continued to the entire real axis $p\in \mathbb{R}$, where it 
satisfies  the following symmetry relations:
\begin{align}\label{phpi}
&\phi_0(p+\pi|P)=\phi_0(p|P),\\\label{phS}
&\phi_0(-p|P)=\,\tilde{w}_0\!\left(P/2+p,P/2-p\right)\,\phi_0(p|P),
\end{align}
where $\widetilde{w}_0(p_1,p_2)=e^{i(p_2-p_1)}w_0(p_1,p_2)$.

Substitution of \eqref{phi0P} into \eqref{BS1} leads to the following integral equation for the function $\phi_0(p|P)$:
\begin{align} \label{BSQ2}
&[\epsilon(p|P)-\tilde{E}_{\iota=0}(P)]\phi_0(p|P)=\\
&\frac{4 h\bar{\sigma}}{\pi}\int_{0}^{\pi/2} dp' G_0(p,p'|P) \phi_0(p'|P),\nonumber
\end{align} 
where $\epsilon(p|P)$ is given by \eqref{Energ}, 
and 
\begin{align}\label{G0}
 &G_0(p,p'|P) =\frac{e^{i(p_1-p_1')}}{4\bar{\sigma}}\cdot\\\nonumber
& {\phantom{.}_{1/2 ,1/2}\langle K_{10}(p_2) K_{01}(p_1) |Q
| K_{10}(p_1') K_{01}(p_2') \rangle_{1/2,1/2}},
\end{align}
with 
\begin{equation}\label{ppsh}
p_{1,2}=P/2\pm p, \quad p_{1,2}'=P/2\pm p'.
\end{equation} 
The integral kernel \eqref{G0} has the following symmetry properties:
\begin{align}
&G_0(p,p'|P)=G_0^*(p',p|P),\\\label{perG0}
&G_0(p,p'|P)=G_0(p+\pi,p'|P)=G_0(p,p'+\pi|P),\\
&G_0(-p,p'|P)=\widetilde{w}_0(P/2+p,P/2-p)G_0(p,p'|P),\\
&G_0(p,-p'|P)=\widetilde{w}_0(P/2-p',P/2+p')G_0(p,p'|P).\label{G0S}
\end{align}
It follows from equations \eqref{phS} and \eqref{G0S} that the integrand in the right-hand side of \eqref{BSQ2} is an even function
of the integration variable $p'$. Therefore, integration in this variable in \eqref{phS} can be extended to the interval $(-\pi/2,\pi/2)$:
\begin{align} \label{BSQ3}
[\epsilon(p|P)-\tilde{E}_{0}(P)]\phi_0(p|P)=\\
{f}\int_{-\pi/2}^{\pi/2} \frac{dp'}{\pi} \,G_0(p,p'|P) \phi_0(p'|P),\nonumber
\end{align} 
where the $f=2 h\bar{\sigma}$ is the string tension, cf. \cite{FZ06,Rut09}.
\subsection{s=0 \label{sec:s=0}}
The wave function \eqref{Phiphi} of a meson with zero spin $s=0$ has two components,
$\Phi_{1/2,-1/2}(p_1,p_2|P)$ and $\Phi_{-1/2,1/2}(p_1,p_2|P)$, which must satisfy the system of two
coupled linear integral equations \eqref{BS1}. The sum over spins $s_1',s_2'$ in the right-hand sides of these equations
reduces  to  two terms due to the  restriction $s_1'+s_2'=0$. In order to decouple these two equations, we 
proceed 
to the basis \eqref{baspm1} in equations \eqref{Phiphi} and \eqref{BS1}. 
To this end, let us first consider the scalar product 
\begin{equation}\label{scpr}
\phantom{.}_{\iota} \langle K_{10}(p_2) K_{01}(p_1) |\tilde{\pi}_{s=0,\iota'}(P)\rangle.
\end{equation}
Here and throughout this Section \ref {sec:s=0},  the indices  $\iota,\iota'$ take two values $\iota,\iota'=\pm$.
Exploiting \eqref{0T1}, one finds:
\begin{align}
&\phantom{.}_{\iota} \langle K_{10}(p_2) K_{01}(p_1) |\tilde{\pi}_{s=0,\iota'}(P)\rangle=\\\nonumber
&\phantom{.}_{\iota} \langle K_{10}(p_2) K_{01}(p_1) |\widetilde{T}_1^{-1}\widetilde{T}_1|\tilde{\pi}_{s=0,\iota'}(P)\rangle=\\\nonumber
&\iota\,\iota' e^{i(P-p_1-p_2)}\phantom{.}_{\iota} \langle K_{10}(p_2) K_{01}(p_1) |\tilde{\pi}_{s=0,\iota'}(P)\rangle.
\end{align}
Therefore, the following equality holds
\[
[1-\iota\,\iota' e^{i(P-p_1-p_2)}]\phantom{.}_{\iota} \langle K_{10}(p_2) K_{01}(p_1) |\tilde{\pi}_{s=0,\iota'}(P)\rangle=0,
\]
for any  $p_1,p_2,P\in\mathbb{R}$ and $\iota,\iota'=\pm$. It is easy to understand from this equality, 
that, if 
\begin{equation}\label{mom}
(p_1+p_2),P\in [0,\pi), 
\end{equation}
then (i) the scalar product \eqref{scpr} vanishes at $\iota\ne\iota'$, and (ii)
the scalar product \eqref{scpr}  also vanishes at $\iota=\iota'$, if $P\ne p_1+p_2$. This allows us to define in the region \eqref{mom}
the wave functions $\Phi_{\iota}(p_1,p_2|P)$, $\phi_\iota(p|P)$  for the $s=0$  meson states with definite parity $\iota=\pm$ 
as follows:
\begin{align}
&\phantom{.}_{\iota} \langle K_{10}(p_2) K_{01}(p_1) |\tilde{\pi}_{s=0,\iota'}(P)\rangle=\delta_{\iota\iota'}\,
\Phi_{\iota}(p_1,p_2|P),\\
&\Phi_{\iota}(p_1,p_2|P)=\pi\,\delta(p_1+p_2-P)\,\phi_\iota\!\left(\frac{p_1-p_2}{2}|P\right).\label{iot}
\end{align}
Due to \eqref{FZpm}, \eqref{ShM}, these  wave functions satisfy the following symmetry relation:
\begin{align}
&\Phi_\pm(p_1,p_2|P)=w_\pm(p_2,p_1)\Phi_\pm(p_2,p_1|P),\\\label{phisym}
&\phi_\pm(-p|P)=w_\pm\!\left(P/2+p,P/2-p\right)\,\phi_\pm(p|P),\\
&\phi_\pm(p+\pi|P)=\phi_\pm(p|P).
\end{align}

The  integral equations  \eqref{BS1} decouple in new notations and transform to the form:
\begin{align} \label{BSQC}
&[\epsilon(p|P)-\tilde{E}_{\iota}(P)]\phi_\iota(p|P)=\\
&\frac{4 h\bar{\sigma}}{\pi}\int_{0}^{\pi/2} dp' G_\iota(p,p'|P) \phi_\iota(p'|P),\nonumber
\end{align} 
where 
\begin{align}\label{Gio}
 G_\iota(p,p'|P) =\frac{ {\phantom{.}_{\iota}\langle K_{10}(p_2) K_{01}(p_1) |Q
| K_{10}(p_1') K_{01}(p_2') \rangle_\iota}}{4\bar{\sigma}},
\end{align}
with $p_{1,2}=P/2\pm p$, and $p_{1,2}'=P/2\pm p'$.

The symmetry properties of the kernels \eqref{Gio}
read:
\begin{align}\label{G0So}
&G_\iota(p,p'|P)=G_\iota^*(p',p|P),\\\nonumber
&G_\iota(p,p'|P)=G_\iota(p+\pi,p'|P)=G_\iota(p,p'+\pi|P),\\\nonumber
&G_\iota(-p,p'|P)={w}_\iota(P/2+p,P/2-p)G_\iota(p,p'|P),\\\nonumber
&G_\iota(p,-p'|P)={w}_\iota(P/2-p',P/2+p')G_\iota(p,p'|P).
\end{align}

The integrand in the integral in \eqref{BSQC} is even in  variable $p'$
due to \eqref{phisym} and \eqref{G0So}. Extending the integration interval in this integral to $(-\pi/2,\pi/2)$, 
we represent equation \eqref{BSQC} in the form similar to  \eqref{BSQ3}:
\begin{align} \label{BSQd}
&[\epsilon(p|P)-\tilde{E}_{\iota}(P)]\phi_\iota(p|P)=\\
&f\int_{{-\pi/2}}^{\pi/2} \frac{dp'}{\pi}\, G_\iota(p,p'|P) \phi_\iota(p'|P).\nonumber
\end{align}
\subsection{s=0,1}
From now on,   we will permit the index $\iota$ in the Bethe-Salpeter equation 
\eqref{BSQd}  to take three values 
$\iota=0,\pm1$.
This allows us to   combine the integral equations  \eqref{BSQd}  with \eqref{BSQ3} and to describe  in the unified manner  the meson states with different spins $s$: with $s=1$ at $\iota=0$, and with $s=0$ at $\iota=\pm$. 

The Bethe-Salpeter integral equations \eqref{BSQd} constitute three eigenvalue problems that determine  in the two-kink approximation three sets of the meson dispersion laws $\{\tilde{E}_{\iota, n}(P)\}_{n=1}^\infty$, with 
$\iota=0,\pm1$, and $0\le P<\pi$.
These equations  are to some extent similar to the Bethe-Salpeter equation derived in 
 \cite{Rut08a} for the ferromagnetic Ising spin-chain model in the 
confinement regime, see equation (37) there.

The normalisation condition  following from \eqref{normPi}, \eqref{PhiphiA}, \eqref{phi0P}, \eqref{iot} for the solutions of these equations reads:
\begin{equation}\label{normphi}
\int_0^{\pi/2}\frac{dp}{\pi} \,|\phi_{\iota,n}(p|P)|^2=1,
\end{equation}
with $\iota=0,\pm$, and $n=1,2,\ldots$.

Let us summarize some symmetry properties of functions that stand in  equations \eqref{BSQd}.
\begin{itemize}
\item{Periodicity.}
\begin{align}
&\epsilon(p|P)=\epsilon(p+\pi|P),\\\label{PerG}
&G_\iota(p,p'|P)=G_\iota(p+\pi,p'|P)=G_\iota(p,p'+\pi|P).\\
&\phi_\iota(p|P)=\phi_\iota(p+\pi|P).
\end{align}
\item{Complex conjugation.}
\begin{equation}\label{Gcon}
[G_\iota(p,p'|P)]^*=G_\iota(-p,-p'|P).
\end{equation}
\item{Reflection symmetries.}
\begin{align}
&\epsilon(-p|P)=\epsilon(p|-P)=\epsilon(p|P),\\\label{R1G}
&G_\iota(-p,p'|P)=W_\iota(p|P) G_\iota(p,p'|P),\\\label{R2G}
&G_\iota(p,-p'|P)=W_\iota(-p'|P) G_\iota(p,p'|P),\\
&\phi_\iota(-p|P)=W_\iota(p|P) \phi_\iota(p|P),\label{refl}
\end{align}
where 
\begin{equation}\label{Wio}
W_\iota(p|P)=
\exp[-2 i p\, \delta_{\iota,0}]\,w_\iota(P/2+p,P/2-p). 
\end{equation}
\end{itemize}
Note, that 
\begin{subequations}
\begin{align}
&W_\iota(p+\pi|P)=W_\iota(p|P), \text{  for }  \iota=0,\pm,\\
&W_0(p+\pi/2|\pi-P)=W_0(p|P),\\
&W_\pm(p+\pi/2|\pi-P)=W_\mp(p|P).
\end{align}
\end{subequations}

By complex conjugating equation \eqref{BSQd} and taking into account \eqref{Gcon}, one can see that, if $\phi_\iota(p|P)$ solves 
the uniform integral equation \eqref{BSQd}, the  function $[\phi_\iota(-p|P)]^*$ must solve the same equation as well. Since the solution
of equation \eqref{BSQd} is unique up to a numerical factor, we conclude that $[\phi_\iota(-p|P)]^*=C\, \phi_\iota(p|P)$, with some 
constant $C$, such that $|C|=1$. Without loss of generality, we shall put $C=-1$, yielding
\begin{equation}
[\phi_\iota(-p|P)]^*=-\, \phi_\iota(p|P).
\end{equation}

It is well known \cite{Jimbo94}, that the two-kink matrix elements of the  $\sigma_0^z$ operator
\begin{equation}\label{MEs}
 {\phantom{.}_{s_2,s_1}\langle K_{10}(p_2) K_{01}(p_1) |\sigma_0^z
| K_{10}(p_1') K_{01}(p_2') \rangle_{s_1',s_2'}}
\end{equation}
have the so-called {\it kinematic singularities} - simple poles at coinciding in- and out-momenta. 
The kinematic simple poles of the matrix element \eqref{MEs} with $s_1=s_2=s_1'=s_2'$ are located 
at {\color{black}{four hyperplanes determined by any of  the equalities}}
\begin{equation}\label{kp}
p_1'=p_1, \quad p_2'=p_2, \quad p_1'=p_2,  \quad  p_2'=p_1,
\end{equation}
while the kinematic simple poles of  \eqref{MEs} at $s_1\ne s_2$, $s_1=s_1'$, $s_2= s_2'$ lie at  {\color{black}{two hyperplanes}}
 $p_1'=p_1,$ and  $p_2'=p_2$. 
 Accordingly, the matrix element of the operator $Q=\sigma_0^z-\sigma_1^z-2 \bar{\sigma}$ in the right-hand side of \eqref{Gio} also has simple poles located at the hyperplanes \eqref{kp}.
Two such simple poles merge, if $p_1+p_2=p_1'+p_2'$. This leads to the second order poles
at $p'=\pm p$ in the integral kernels  $G_\iota(p,p'|P)$ determined by \eqref{G0}, \eqref{Gio}.
These kernels  can be represented as  sums of two terms
\begin{equation}\label{resi}
G_\iota(p,p'|P)=G_\iota^{(sing)}(p,p'|P)+G_\iota^{(reg)}(p,p'|P),
\end{equation}
where (i) the first term has  second order poles at 
$p'=\pm p$, while the second term  is regular at real $p,p',P$; and (ii) both
 functions $G_\iota^{(sing)}(p,p'|P)$ and $G_\iota^{(reg)}(p,p'|P)$ satisfy the symmetry relations
\eqref{PerG}, \eqref{R1G},  \eqref{R2G}. 

The explicit form of the singular part $G_\iota^{(sing)}(p,p'|P)$ of the kernel is obtained in Appendix \ref{MEL}.
In order to present the  final result in a compact form, we proceed to the complex variables 
\begin{equation}\label{zzv}
z=e^{2 i p},\quad  z'=e^{2 i p'},\quad v=e^{iP},
\end{equation}
and introduce the notations
\begin{align}\label{Giz}
\mathcal{G}_\iota(z,z'|v)=G_\iota(p,p'|P),\\
\mathcal{W}_{\iota}(z|v)=W_\iota(p|P),\label{Wz}
\end{align}
for $\iota=0,\pm$. For any $v$ such that $|v|=1$, the functions $\mathcal{W}_{\iota}(z|v)$ are analytical and single-valued in $z$ in 
some open vicinity
of the unit circle $S_1= \{z \big|\,|z|=1$\}, and the kernels $\mathcal{G}_\iota(z,z'|v)$ are single valued in the vicinity of $S_1\times S_1$.

Equation \eqref{resi} in new notations takes the form
\begin{equation}\label{rsG}
\mathcal{G}_\iota(z,z'|v)=\mathcal{G}_\iota^{(sing)}(z,z'|v)+\mathcal{G}_\iota^{(reg)}(z,z'|v),
\end{equation}
where $\mathcal{G}_\iota^{(reg)}(z,z'|v)$ is regular in $z,z'$ in some open vicinity of $S_1\times S_1$. For the singular term 
$\mathcal{G}_\iota^{(sing)}(z,z'|v)$, we obtained in Appendix \ref{MEL} the following explicit representation:
\begin{subequations}\label{singG}
\begin{align}\label{sG}
&\mathcal{G}_\iota^{(sing)}(z,z'|v)=\mathcal{G}_\iota^{(s)}(z,z'|v)+\mathcal{W}_\iota(z'|v)\mathcal{G}_\iota^{(s)}(z,z'^{-1}|v),\\
&\mathcal{G}_\iota^{(s)}(z,z'|v)=-\frac{ z z'}{(z'-z e^{-\delta})^2}+\frac{z\, \delta_{\iota,0}}{2(z'-z e^{-\delta})}+\\\nonumber
&\frac{  \mathcal{W}_\iota(z'|v)}{ \mathcal{W}_\iota(z|v)}\left[-\frac{ z z'}{(z'-z e^{\delta} )^2}-\frac{  z'\,\delta_{\iota,0}}{2(z'-z e^{\delta})}\right].\end{align}
\end{subequations}
where $\delta\to+0$.
\subsection{Singular integral equations in the unit circle \label{SIE}}
It is convenient to rewrite the  Bethe-Salpeter equations \eqref{BSQd}  in complex variables \eqref{zzv}:
\begin{align} \label{BSQ4}
[\mathcal{E}(z|v)-\Lambda_\iota(v)]\psi_\iota(z|v)=\\
{f}\oint_{S_1} \frac{dz'}{2\pi i z'} \,\mathcal{G}_\iota(z,z'|v) \psi_\iota(z'|v),\nonumber
\end{align} 
where the unit circle $S_1$ in the complex variable  $z'$ is passed in the counter-clockwise direction, and 
\begin{equation}\label{psph}
\mathcal{E}(z|v)=\epsilon(p|P), \quad \Lambda_\iota(v)=\tilde{E}_{\iota}(P),\quad 
\psi_\iota(z|v)=\phi_\iota(p|P).
\end{equation}

The function $\mathcal{E}(z|v)$ is algebraic in $z$. Its explicit expression is given in equation 
\eqref{OmepA} in Appendix \ref{2E}, where its analytic properties are also described in details.
Here we notice only the symmetry property $\mathcal{E}(z|v)=\mathcal{E}(z^{-1}|v)$.

The wave functions $\psi_\iota(z|v)$ with $\iota=0,\pm$ are single-valued and analytical 
in some open vicinity of the unit circle $S_1$, 
and satisfy there the symmetry relations
\begin{align}\label{psiR}
&\psi_\iota(z^{-1}|v)=\mathcal{W}_{\iota}(z|v)\psi_\iota(z|v),\\ 
&\psi_\iota(z^{-1}|v)=-\psi_\iota^*(z|v).\label{ConPs}
\end{align}
Taking \eqref{rsG}, \eqref{sG}, and \eqref{psiR} into account,  one can replace the kernel in the 
integrand in \eqref{BSQ4}, as
\begin{equation}
\mathcal{G}_\iota(z,z'|v) \to  \widetilde{\mathcal{G}}_\iota(z,z'|v) =2\mathcal{G}_\iota^{(s)}(z,z'|v)+\mathcal{G}_\iota^{(reg)}(z,z'|v).
\end{equation}
Then, the Bethe-Salpeter equation \eqref{BSQ4} takes the final form
\begin{align} \label{BSQ5}
[\mathcal{E}(z|v)-\Lambda_\iota(v)]\psi_\iota(z|v)=\\\nonumber
2{f}\oint_{S_1} \frac{dz'}{2\pi i z'} \,\mathcal{G}_\iota^{(s)}(z,z'|v) \psi_\iota(z'|v)+\\
{f}\oint_{S_1} \frac{dz'}{2\pi i z'} \,\mathcal{G}_\iota^{(reg)}(z,z'|v) \psi_\iota(z'|v),\nonumber
\end{align} 
with additional constraint \eqref{psiR}. In terms of the original momentum variables $p,p',P$, this equation reads
\begin{align} \label{BSQg}
&[\epsilon(p|P)-\tilde{E}_{\iota}(P)]\phi_\iota(p|P)=\\
&f\int_{{-\pi/2}}^{\pi/2} \frac{dp'}{\pi}\, [2G_\iota^{(s)}(p,p'|P)+ G_\iota^{(reg)}(p,p'|P)]\phi_\iota(p'|P),\nonumber
\end{align}
where $G_\iota^{(s)}(p,p'|P)=\mathcal{G}_\iota^{(s)}(z,z'|v)$.

Equation \eqref{BSQ5} belongs to the class of  uniform linear singular integral equations. For the general theory of singular integral equations see the monograph  \cite{Mus} by Muskhelishvili. The properties of the Bethe-Salpeter equation \eqref{BSQ5} are to much 
extent similar to the properties of its analogs in the IFT \cite{FZ06} and in the Ising spin chain \cite{Rut08a}. The main difference from 
the  latter model, which is  free-fermionic in the de-confined phase,  is the transformation of the solution of the Bethe-Salpeter equation under the reflection $z\to z^{-1}$.  The solution of equation \eqref{BSQ4} transforms according to formula \eqref{psiR}  under this reflection, whereas the solution 
of the analogous Bethe-Salpeter equation corresponding to the Ising spin chain only changes its sign \cite{Rut08a}. 

The wave function $\psi_\iota(z|v)$ can be viewed as a vector in the Hilbert space with the scalar product 
\begin{equation}
(\varphi,\psi )=\frac{1}{2\pi i} \oint_{S_1} \frac{dz}{ z}\,\varphi^*(z)\, \psi(z).
\end{equation}
For each   $\iota=0,\pm$, and $v\in S_1$, the integral equation \eqref{BSQ5} constitutes the eigenvalue problem 
\[
\mathbf{H}_{\iota}(v) \psi_\iota=\Lambda_\iota(v)\, \psi_\iota
\]
for the Hermitian operator $\mathbf{H}_{\iota}(v)$, defined by
\[
\mathbf{H}_{\iota}(v) \psi(z)=\mathcal{E}(z|v)\psi(z)-{f}\oint_{S_1} \frac{dz'}{2\pi i z'} \,\mathcal{G}_\iota(z,z'|v) \psi(z').
\]
The operator $\mathbf{H}_{\iota}(v)$ acts in the subspace of functions satisfying the symmetry relation 
 \eqref{psiR}. The spectrum of the operator $\mathbf{H}_{\iota}(v)$ is real, positive, and discrete.  
For its eigenvalues, we shall use notations 
 $\{\Lambda_{\iota, n}(v)\}_{n=1}^\infty$: $0<\Lambda_{\iota 1}(v)<\Lambda_{\iota 2}(v)\ldots$.  
 Corresponding eigenvectors will be denoted as $\psi_{\iota,n}(z|v)$. For given $\iota$ and $v$, the 
 eigenvectors with different $n$ are mutually orthogonal. They will be normalised by the condition
 \begin{equation}\label{psinorm}
 \frac{1}{2\pi i} \oint_{S_1} \frac{dz}{ z}\,\psi_{\iota,n}^*(z|v)\, \psi_{\iota,n}(z|v)=2,
 \end{equation}
which is just equation \eqref{normphi} rewritten in the variable \newline $z=e^{2ip}$.

Although the eigenvalue problem \eqref{BSQ5},  \eqref{psiR}  cannot be solved exactly, it admits perturbative solutions in the 
weak coupling limit $f\to +0$ in different asymptotical regimes, which will be described in Section \ref{WK}. In the rest of this Section, we shall introduce
three auxiliary function  $g_{\iota+}(z)$, $g_{\iota-}(z)$, and $U_\iota(z)$, which will be used later in the 
small-$f$ perturbative calculations of the eigenvalues $\{\Lambda_{\iota, n}(v)\}_{n=1}^\infty$.

First, we denote by  $g_{\iota\pm}(z)$ two functions:
\begin{equation} \label{gpm}
g_{\iota\pm} (z)= \oint_{S_1}\frac{dz'}{2\pi i}\, \frac{\psi_\iota(z')}{(z'-z)},
\end{equation}
where  $g_{\iota+}(z)$ is defined at $|z|<1$, and  $g_{\iota-}(z)$ is defined in the region
$|z|>1$. 
The evident properties of these function are:
\begin{enumerate}
\item  $g_{\iota+}(z)$ and $g_{\iota-}(z)$  are  analytical at $|z|< 1$ and at $|z|> 1$,  respectively. \label{anal}
\item  $g_{\iota+}(z)$ and $g_{\iota-}(z)$ can be continued to the unit circle $S_1$, where they are continuous together
with their derivatives.  
\item  \label{i5}
Relation with the function $\psi_\iota(z)$ at $|z|=1$:
\begin{equation}\label{pdig}
\psi_\iota(z)=\lim_{\delta\to+0}[g_{\iota+}(z e^{-\delta})-g_{\iota-}(z e^{\delta})].
\end{equation} 
\item 
It follows from  \eqref{ConPs} and \eqref{pdig}, that:
\begin{align}\label{gmin}
&g_{ \iota\pm}^*(z)=-
g_{\iota\mp}(z^*),& \text{for } |z|<1,\\\label{gcong}
&g_{ \iota\pm}^*(z)=-
g_{\iota\mp}(z^{-1}),& \text{for } |z|=1.
\end{align}
\item The following equality holds at $|z|>1$:
\begin{align}g_{\iota+}(z^{-1})=-\mathcal{W}_\iota(z)\, g_{\iota-}(z)+g_{\iota+}(0)-\\\nonumber
\oint_{S_1}\frac{dz'}{2\pi i}\, 
\frac{\mathcal{W}_\iota(z')-\mathcal{W}_\iota(z)}{(z'-z)}\psi_\iota(z').
\end{align}
\end{enumerate}
We denote by $g_{\iota,n\pm}(z|v)$ the auxiliary functions, associated according to 
definition \eqref{gpm} with the eigenfunctions  $\psi_{\iota,n}(z|v)$.
Exploiting equalities \eqref{ConPs}, \eqref{pdig},  and \eqref{gcong}, the normalisation condition \eqref{psinorm} can be rewritten in terms of  
$g_{\iota,n+}(z)$:
\begin{equation}\label{normg}
\int_{-\pi/2}^{\pi/2}\frac{dp}{\pi}\,g_{\iota,n+}(z)\, g_{\iota,n+}(z^{-1})\Big|_{z=\exp(2 i p)}=-1.
\end{equation}

Let us  also define  one more auxiliary function $U_\iota(z)$ inside the unit circle $|z|<1$:
\begin{equation}\label{U}
U_\iota(z)=[\mathcal{E}(z)-\Lambda_\iota-\delta_{\iota,0}\,f ]\,g_{\iota+}(z)+2f\,z\,g_{\iota+}'(z).
\end{equation}
Its analytic properties are similar to those of the function $\mathcal{E}(z)$, since $g_{\iota+}(z)$
is analytical at $|z|<1$. The function $U_\iota(z)$ can be analytically continued into the region 
$|z|>1$, where it admits the 
 following representation in terms of the 
function $g_{\iota-}(z)$:
\begin{align}\nonumber
&U_\iota(z)=[\mathcal{E}(z)-\Lambda_\iota-\delta_{\iota,0}\,f ]\,g_{\iota-}(z)-2f\,z\,g_{\iota-}'(z)-\\
&2f \,z\,\frac{\mathcal{W}_\iota'(z)}{\mathcal{W}_\iota(z)}\,g_{\iota-}(z)+f X_\iota(z),\label{U1}
\end{align}
where 
\begin{align}\label{XXa}
&X_\iota(z)=\oint_{S_1} \frac{dz'}{2\pi i z'} \,\mathcal{G}_\iota^{(reg)}(z,z') \psi_\iota(z')-\\\nonumber
&\oint_{S_1} \frac{dz'}{2\pi i } \,\frac{[\mathcal{W}_\iota(z')-\mathcal{W}_\iota(z)-(z'-z)\mathcal{W}_\iota'(z)]}{\mathcal{W}_\iota(z)}\cdot\\
&\left[
\frac{2z}{(z'-z)^2}+\frac{\delta_{\iota,0}}{z'-z}
\right] \psi_\iota(z')\,+\nonumber\\
&\delta_{\iota,0}\left\{-g_{\iota+}(0)+\frac{\mathcal{W}_\iota'(z)}{\mathcal{W}_\iota(z)}\lim_{z\to \infty} [z \,g_{\iota-}(z)]\right\}.\nonumber
\end{align}
Note, that the integrands in the integrals in the right-hand side of \eqref{XXa} are regular at $z'\in S_1$.
To prove \eqref{U1}, it is sufficient to subtract \eqref{U1} from \eqref{U}, and to check using \eqref{pdig}, that the resulting
equation is equivalent to \eqref{BSQ5}.

Equation \eqref{U} can be viewed as the first-order differential equation for the function $g_{\iota+}(z)$. The appropriate 
partial solution of this equation reads
\begin{align}\label{g+}
g_{\iota+}(z)=\frac{1}{2f}\int_{0}^z\frac{dz'}{ z'}
\left(
\frac{z}{z'}
 \right)^{\delta_{\iota,0}/2}
U_\iota(z')\cdot \\
\exp\Bigg\{\frac{i}{2f}\,[{\mathcal F}(z',\Lambda_\iota)-{\mathcal F}(z,\Lambda_\iota)]\Bigg\},\nonumber
\end{align}
where 
\begin{equation}
{\mathcal F}(z,\Lambda)=\int_{z_1}^z \frac{dt}{i t}\,[\mathcal{E}(t)-\Lambda]. \label{Fdef}
\end{equation}
As in reference \cite{Rut08a}, we have to put  to the origin  the initial integration point in the integral in \eqref{g+} 
in order to provide 
analyticity of the function  $g_{\iota+}(z)$ at $z=0$.  Any other choice of the initial integration point would lead to
an essential singularity of the right-hand side of \eqref{g+} at $z=0$. It follows from \eqref{OmepA}, \eqref{Omas}, that the function $\mathcal F(z,\Lambda)$, determined by \eqref{Fdef} is singular at
 $z \to 0$: ${\mathcal F}(z,\Lambda)\sim z^{-1/2}$. Nevertheless, the integral in $z'$ in equation \eqref{g+} converges, if the 
 integration path lies in the physical sheet $\mathfrak{L}_{++}$  described in Appendix \ref{2E}, and approaches the origin along the real axis either from the right, 
 or from the left side.
 
 The choice of the initial point $z_1$ in the integral in \eqref{Fdef} is the subject of convenience, since it has no effect on the difference
 $[{\mathcal F}(z',\Lambda_\iota)-{\mathcal F}(z,\Lambda_\iota)]$ in the right-hand side of \eqref{g+}. We shall put $z_1=1$ for 
 $0\le P\le\pi/2$, and $z_1=-1$  for $\pi/2<P< \pi$.

\begin{figure}[htb]
\centering
\includegraphics[width=\linewidth, angle=00]{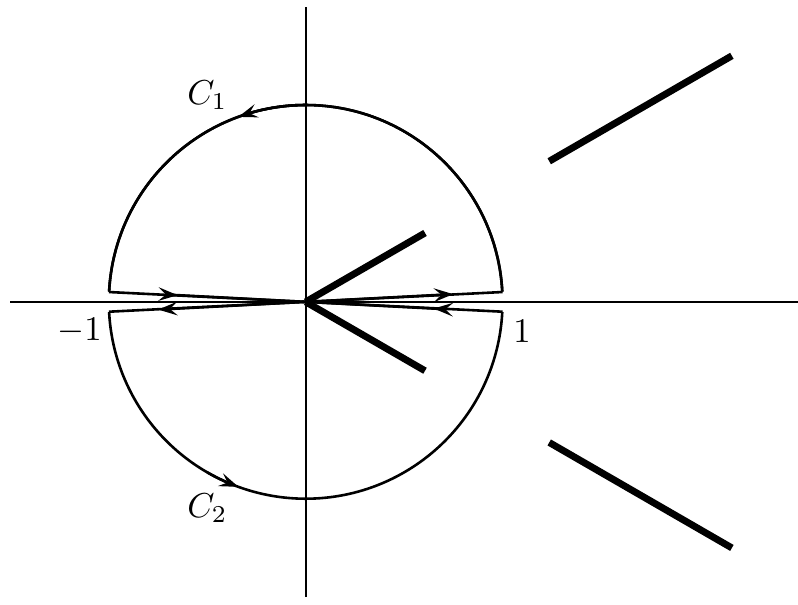}
\caption{\label{figC11}  Closed integration contours $C_1$ and $C_2$ in equation \eqref{C12} in the complex $z$-plane.
Solid straight lines display the branching cuts of the function $\mathcal{E}(z)$ defined by \eqref{OmepA}.} 
\end{figure}

The requirement of analyticity of the auxiliary function $g_{\iota+}(z)$ in the circle $|z|<1$ leads to two constraints:
\begin{equation}\label{C12A}
J_\beta(\Lambda_\iota)=0,
\end{equation}
{with }
\begin{equation}\label{C12}
J_\beta(\Lambda_\iota)=\oint_{C_\beta}\frac{dz}{z}\,{ z^{-\delta_{\iota,0}/2}}\,
U_\iota(z)
\exp\Big[\frac{i}{2f}\,{\mathcal F}(z,\Lambda_\iota)\Big],
\end{equation}
where $\beta=1,2$ and the integration contours $C_{1,2}$ are shown in Figure~\ref{figC11}.
Equalities \eqref{C12A} guarantee that the right-hand side in equation \eqref{g+} is a single-valued function of $z$ at $|z|<1$.

In what follows, we shall use also the notation $F(p,E)$ for the integral \eqref{Fdef}  expressed in terms of the momentum variable. In particular, 
at   $0\le P\le\pi/2$, we have:
\begin{align}\label{FpE}
F(p,E)={\mathcal F}(z,\Lambda)
=2\int_0^p dp' \, [\epsilon(p')-E],
\end{align}
with $z=\exp(2 ip)$ and  $\Lambda=E$.
\section{Weak coupling asymptotics \label{WK}}
In this Section we outline 
 the perturbative calculations of the  spectra $\{\Lambda_{\iota, n}\}_{n=1}^\infty$ of the eigenvalue problem \eqref{BSQ5},  \eqref{psiR}  in the limit of a small string tension $f\to +0$, and present the obtained results.  The details of these calculations, which are essentially based   on the asymptotical analysis of equations  \eqref{C12A}, are relegated to Appendix \ref{PSIE}.
 
In the limit $f\to +0$, the 
integrals  \eqref{C12} are determined 
due to the factor $\exp\Big[\frac{i}{2f}\,{\mathcal F}(z,\Lambda_\iota)\Big]$ in the integrand
by the contributions of the saddle points of the function ${\mathcal F}(z,\Lambda_\iota)$. 
These saddle points are located at the solutions of the equation 
\begin{equation}\label{ELam}
\mathcal{E}(z)= \Lambda_\iota.
\end{equation}
It is shown in Appendix \ref{2E}, that this saddle-point equation has four solutions $z_a,z_a^{-1},z_b,z_b^{-1}$,
which are determined by  \eqref{zab}. It turns out, however, that only the saddle points lying in the unit circle
$S_1$ contribute to the weak-coupling asymptotics of the eigenvalues $\Lambda_{\iota, n}$.
At different values of the parameters $v=e^{i P}$ and $\eta$, there are  zero, two, or four  such saddle points. 
Though for generic values of the parameters $P,\eta$, these points are well separated from  each other,
they merge in $S_1$ at certain particular values of $P$ and $\eta$. As the result, depending on the 
values of parameters $P,\eta$,  one has to distinguish nine  regimes, in which the eigenvalues $\Lambda_{\iota, n}$ have
different asymptotic expansions in the weak-coupling limit $f\to+0$. In what follows we first describe three semiclassical 
regimes: in the first one there are two well separated saddle points in the unit circle $S_1$, in the second regime there are no saddle points  in  $S_1$, and in the third regime there are four such saddle points.  Then, we proceed to  three low-energy expansions, which describe the meson energy spectra close to 
their low-energy edge at different values of the meson momentum. Finally, three
crossover asymptotical expansions are presented, which hold  close 
to the boundaries between the regions (I),  (II) and (III)   shown in Figure \ref{fig:RegE}. 
Due to  symmetry relations \eqref{parit}-\eqref{refP}, 
the calculation of the meson energy spectra $\tilde{E}_{\iota, n}(P)$ will be restricted without loss of generality 
to the momenta in the interval $P\in (0,\pi/2)$. 
\subsection{Semiclassical regimes \label{SemReg}}
\underline{First semiclassical regime.}

\noindent
The first semiclassical regime is realized, if the energy $E$ and momentum $P$ of the meson fall well inside
the region (I) shown in Figure  \ref{fig:RegE}. Location of the saddle points solving equation  \eqref{ELam} 
in this regime at $0<P<\pi/2$ is shown in Figure \ref{fig:S1}.  Two of them  $z_a$ and $z_a^{-1}$ lie in the 
unit circle $S_1$ in this case.  It is shown in 
 Appendix \ref{Sem1}, that the  meson energy spectrum at $f\to+0$ is 
 determined in the first semiclassical regime by contributions of these saddle points into the integrals \eqref{C12}.
To the leading order in $f\to+0$, the final result for the  meson energy spectrum in the first semiclassical regime reads:
\begin{align}\label{Mdl}
&2 \tilde{E}_{\iota, n}(P)\, p_a-\int_{-p_a}^{p_a}dp \,\epsilon(p|P)=\\\nonumber
&f\left[2\pi \left(n-\frac{1}{4}\right)-\theta_\iota(P/2+p_a,P/2-p_a)
\right]+O(f^2),
\end{align}
with $\tilde{E}_{\iota, n}(P)=\epsilon(p_a|P)$, and integer $n\gg 1$, in agreement with previously obtained result 
\eqref{EnSem1}.

Note, that due to \eqref{Mdl},   two sequential meson energies at given $P$ are separated 
in the first semiclassical regime by the small interval 
\begin{equation}\label{dEn}
\Delta \tilde{E}_{\iota, n}^{(\mathrm{I})}(P)\equiv \tilde{E}_{\iota, n+1}(P)-\tilde{E}_{\iota, n}(P)=\frac{\pi f}{p_a}+O(f^2).
\end{equation}
With increasing $n$, both $p_a$ and $\tilde{E}_{\iota, n}(P)$  increase as well, until they approach the 
values $\pi/2$ and $\epsilon(\pi/2|P)$, respectively, at a certain $n=\mathcal{N}(P|h)$. Further increase of 
$n$ leads to the crossover into the second semiclassical regime, which will be discussed later. 
The number $\mathcal{N}(P|h)$ of meson states with fixed $\iota=0,\pm$, and $P\in (0,P_c(\eta))$
in the first semiclassical regime can be found from \eqref{Mdl}:
\begin{equation}\label{Ntot}
 \mathcal{N}(P|h)=\frac{1}{2\pi f}\left[
 \pi\, \epsilon(\pi/2|P)-\int_{-\pi/2}^{\pi/2}{dp} \, \epsilon(p|P)\right]
+O(1 ).
\end{equation}
It diverges as $h^{-1}$ at $h\to0$.

In the scaling regime, i.e.  at small $\eta\ll1$ and $P\ll1$, the meson dispersion law \eqref{Mdl}
takes the relativistic form:
\begin{equation}
\tilde{E}_{\iota, n}(P)=I k\sqrt{M_{\iota, n}^2+P^2},
\end{equation}
where 
\begin{equation}
M_{\iota, n}=2 m \cosh(\beta_{\iota, n})
\end{equation}
is the meson mass, $m$ is the kink mass \eqref{kmass}, and the  rescaled rapidities $\beta_{\iota, n}$ solve the equation
\begin{align}\label{eqbe}
&\sinh(2\beta_{\iota, n})-2\beta_{\iota, n}=\\\nonumber
&\lambda\left[2\pi \left(n-\frac{1}{4}\right)-\Theta_\iota (2\alpha,\eta)\Big|_{\alpha=\beta_{\iota, n}\eta/\pi}
\right]+O(\lambda^2),
\end{align}
with $\lambda=\frac{f}{I k\, m^2}$.

Note, that in the scaling limit $\eta\to 0$ the kink scattering phases 
 $\Theta_\iota (\alpha,\eta)|_{\alpha=\beta\eta/\pi}$
 reduce to the soliton-soliton scattering phases  $\Theta_\iota^{(SG)} (\beta)$ of the sine-Gordon field theory 
in the asymptotically free regime \cite{Sm92}:
\begin{align}
&\lim_{\eta\to 0} \Theta_\iota (\alpha,\eta)\Big|_{\alpha=\beta\eta/\pi}= \Theta_{\iota}^{(SG)}(\beta),\\\nonumber
&\exp\left[ i \Theta_{0}^{(SG)}(\beta)\right]=\exp\left[ i \int_0^\infty \frac{dy}{y} \frac{\sin(2 \beta y)}{\cosh(\pi y)}e^{-\pi y}\right]=\\
&-\frac{\Gamma\left(\frac{1}{2}+\frac{\beta}{2\pi i}\right)\Gamma\left(-\frac{\beta}{2\pi i}\right)}{\Gamma\left(\frac{1}{2}-
\frac{\beta}{2\pi i}\right)\Gamma\left(\frac{\beta}{2\pi i}\right)},\\
& \Theta_{+}^{(SG)}(\beta)=\Theta_{0}^{(SG)}(\beta)-i \ln \frac{\pi -i \beta}{\pi +i \beta},\\
&\Theta_{-}^{(SG)}(\beta)=\Theta_{0}^{(SG)}(\beta).
\end{align}

\underline{Second semiclassical regime.}

\noindent
In the second semiclassical regime, the energy $E$ and momentum $P$ of a meson state are 
located well above the lower bound of the region (II) in Figure \ref{fig:RegE},  and all four solutions of 
equation \eqref{ELam} are real.
It is shown in 
Appendix \ref{SR3}, that the Bethe-Salpeter equation leads to the following small-$f$
asymptotics for the meson energies,
\begin{equation}\label{Sem3A}
\tilde{E}_{\iota, n}=2n f+\frac{2}{\pi}\int_{-\pi/2}^{\pi/2}{dp}  \, \omega(p)-f \delta_{\iota,0}, 
\end{equation}
in agreement with our previous result \eqref{EnSem2}.

\underline{Third semiclassical regime.}

\noindent

The third semiclassical regime is realized for the meson states with the energy and momentum  well inside the region (III) in Figure \ref{fig:RegE}. 
All four saddle points $z_a,z_a^{-1},z_b,z_b^{-1}$ are located in the unit circle $S_1$ in this case, being well separated
one from another. It is shown in Appendix \ref{Sem2}, that the small-$f$ asymptotics of the meson energy 
spectra in this regime is determined by contributions of these saddle points into the integrals \eqref{C12}. 
This leads to the following  meson energy spectrum $E_{\iota, n}(P)$ in the third semiclassical regime at $P_c(\eta)<P<\pi-P_c(\eta)$:
\begin{align}
\label{SE2}
&E_{\iota, n}(P)\, (p_a-p_b)-\int_{p_b}^{p_a}dp \,\epsilon(p|P)=f\bigg[\pi \left(n-\frac{1}{2}\right)+\\\nonumber
&\frac{\theta_\iota(P/2+p_b,P/2-p_b)\!-\!\theta_\iota(P/2+p_a,P/2-p_a)}{2}
\bigg]\!\!+O(f^2), \\
\nonumber
&E_{\iota, n}(P)=\epsilon(p_a|P)=\epsilon(p_b|P), \quad n\gg1,
\end{align}
with $0<p_b<p_a<\pi/2$, in agreement with \eqref{EnSem3}.

It follows from \eqref{SE2}, that  two sequential meson energies at momentum $P$ are separated in  this 
third semiclassical regime by the  interval 
\begin{equation}\label{dEn2}
\Delta {E}_{ n}^{(\mathrm{III})}(P)\equiv \tilde{E}_{\iota, n+1}(P)-\tilde{E}_{\iota, n}(P)=\frac{\pi f}{p_a-p_b}+O(f^2).
\end{equation}

\begin{figure}
\centering
\subfloat[ 
Semiclassical energy spectra of the double-degenerate  meson modes with  spin $s=\pm 1$.
]
{\label{fig:iot0}
\includegraphics[width=.9\linewidth]{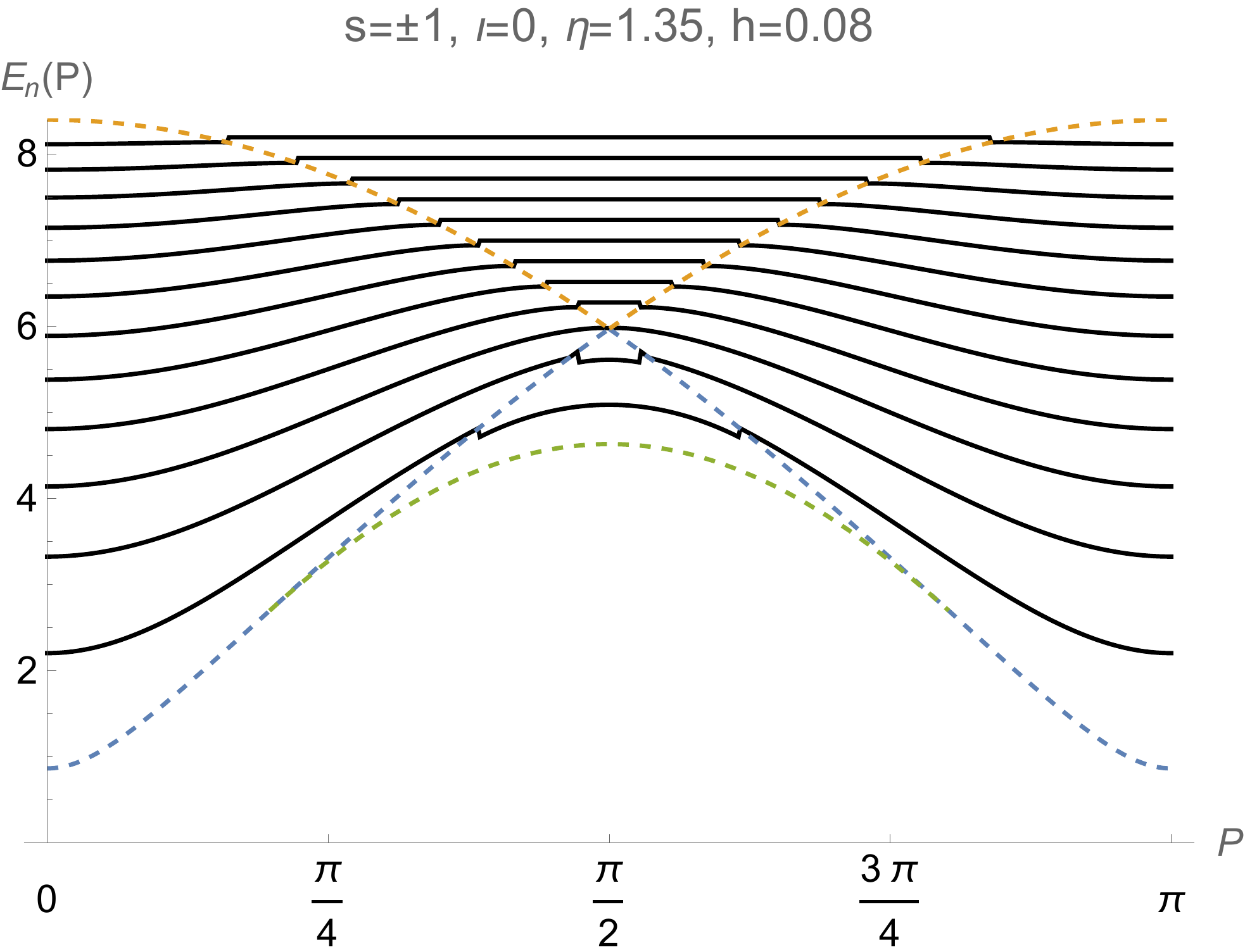}}

\subfloat[
Semiclassical energy spectra of the  meson modes with  spin $s=0$ and $\iota =-$.
]
	{\label{fig:iot2}
\includegraphics[width=.9\linewidth]{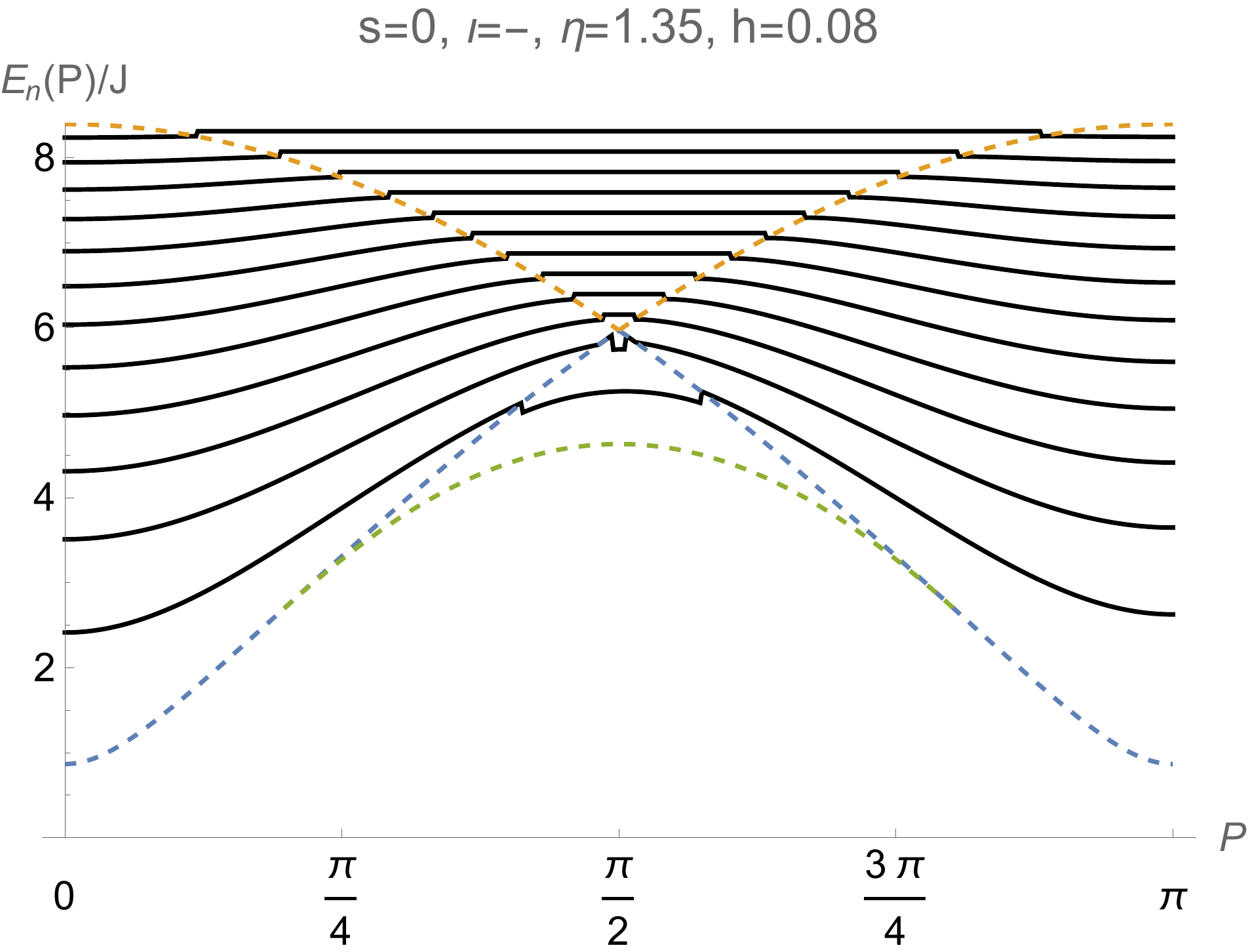}}

\caption{\label{figSemES} Semiclassical energy spectra of the meson modes 
for the model \eqref{XXZH} at $J=1$, $\Delta=-\cosh (1.35)$, and $h=0.08$,
calculated from \eqref{Mdl},  \eqref{Sem3A}, and \eqref{SE2}.
 }
\end{figure}

Figure \ref{figSemES} displays the semiclassical energy spectra of the two-spinon (meson) bound 
states calculated from \eqref{Mdl}, \eqref{Sem3A},  and  \eqref{SE2} at $J=1$, $\eta=1.35$, and $h=0.08$.
The energy spectra of mesons with spin $s=\pm 1$ are shown in Figure \ref{fig:iot0}.
They are symmetric with respect to the reflection $P\to \pi-P$. 
In contrast, the spectra of the $s=0$ meson modes with $\iota=-$ shown in Figure \ref{fig:iot2} are slightly
asymmetric, and transform after the  reflection  $P\to \pi-P$ into the spectra of the $s=0$  modes with $\iota=+$,
in accordance with equation \eqref{refP}.

As one can see in Figures \ref{figSemES}a,b, the semiclassical meson spectra have small discontinuities at the dashed lines separating  the 
regions (I), (II), and (III) in the $PE$-plane, which are shown in  Figure \ref{fig:RegE}. This indicates, that the semiclassical approximation
fails in crossover regions close to the dashed separatices. The meson energy spectra in these narrow  crossover regions will be presented in
 Section \ref{crosr}.  The resulting meson energy spectra, in which the semiclassical formulas  \eqref{Mdl}, \eqref{Sem3A},  \eqref{SE2} 
are modified in the crossover regions according to equations \eqref{cross1}-\eqref{cross2}, are continuous in the whole Brillouin zone.

\subsection{Low-energy regimes}
Formulas \eqref{Mdl}, and \eqref{SE2} represent the initial terms of the semiclassical asymptotic expansions 
for the meson energy spectra $\tilde{E}_{\iota, n}(P)$
in integer powers of the  string tension $f\to+0$. These semiclassical 
asymptotic expansions are supposed to work well for the meson states with large
quantum numbers $n\gg 1$. For the energy 
spectra  of mesons  with small $n=1,2,\dots$,
one should use instead the low-energy asymptotic expansions in fractional powers of $f$.
Three such low-energy expansions were introduced in \cite{Rut18} and discussed 
in Section \ref{CanQuan} in the frame of the heuristic approach exploiting the canonical
quantization of the Hamiltonian dynamics of the model \eqref{Hk}. 
 Now we shall describe briefly, how these low-energy expansions can be obtained in the 
 more rigorous approach based of the perturbative solution of the Bethe-Salpeter equation 
 \eqref{BSQ4}.

As  in the case of the semiclassical expansion, we start from equalities \eqref{C12A}, and replace the integrals 
$J_\beta(\Lambda_\iota)$ in the left-hand side  by their 
 saddle-point asymptotics at $f\to+0$. In contrast to the semiclassical regimes, however,
 the relevant saddle points of the function ${\mathcal F}(z,\Lambda_\iota)$ in the low-energy regimes are degenerate. 
 The kinds of the saddle-point degeneracy are  different in the three low-energy regimes.
 
 \underline{First low-energy regime.}

\noindent
The first low-energy regime  is realised at $P\in (-P_c(\eta), P_c(\eta))$ and energies $E$ slightly above the two-kink edge $\epsilon(0|P)$.
In this regime, the two Morse saddle points $z_a$ and $z_a^{-1}$ shown in Figure \ref{fig:S1}  approach with decreasing $E$
 the point $z=1$, and  finally merge there at $E=\epsilon(0|P)$. It is shown in
Appendix \ref{LE}, that the contribution of the resulting degenerate saddle point $z=1$ into the integral $J_1(\Lambda_\iota)+J_2(\Lambda_\iota)$
determines the first low-energy expansion of the meson dispersion laws. The final result coincides with formula 
 \eqref{les}. This asymptotical formula holds at $f\to +0$ and $-P_c(\eta)<P<P_c(\eta)$, for not very large $n=1,2,\ldots$. 

 \underline{Second low-energy regime.}

\noindent
In the second low-energy regime, the meson momentum lies in the interval $P\in (P_c(\eta),\pi-P_c(\eta))$, 
and the energy is slightly above the the two-kink edge, which is given by the function  $\epsilon(P,\eta)$ determined by 
equation \eqref{pmin} and plotted by the right red dashed line in Figure \ref{fig:RegE}. 
The Morse saddle points $z_a$ and $z_b$ shown in Figure \ref{fig:S2} merge at $E=\epsilon(P,\eta)$, as well as the saddle points 
$z_a^{-1}$ and $z_b^{-1}$. As the result, the small-$f$ asymptotics of the integrals  $J_{1}(\Lambda_\iota)$ and $J_{2}(\Lambda_\iota)$
are determined 
in second low-energy regime by the contributions of the degenerate saddle points $\exp\left(2i \,p_m(P,\eta)\right)$,  and $\exp\left(-2i \,p_m(P,\eta)\right)$, respectively,
with $p_m(P,\eta)$ given by \eqref{pmin}.

The calculation of these saddle-point  asymptotics is described in Appendix \ref{LE2}. The final result for the three initial terms in  the second low-energy
expansion of the meson energy spectrum reads:
\begin{align}\nonumber
&\tilde{E}_{\iota, n}(P)=\epsilon(0|P)+f^{2/3}\left[\frac{\partial_p^2\epsilon(p|P)|_{p=p_m}}{2}\right]^{1/3} x_n-\\
&\frac{f}{2}\partial_p \theta_\iota(P/2+p,P/2-p)\Big|_{p=p_m}+O(f^{4/3}),\label{slee}
\end{align}
where $x_n=z_{(n+1)/2}'$ at odd $n$,   $x_n=z_{n/2}$ at even $n$, and $n=1,2,\ldots$.
Formula \eqref{slee} was presented without derivation in paper \cite{Rut18}, see equation (41) there.

\underline{Third low-energy regime.}

\noindent
The third low-energy expansion describes the meson dispersion law close to the points 
$P=\pm P_c+\pi n$,  $E=\epsilon(0|P_c)$, with $P_c$ given by equation \eqref{Pc}, and $n\in \mathbb{Z}$. 
These points are shown in red in Figure \ref{fig:RegE}. Since $\epsilon''(0|P_c)=0$, the Taylor expansion
of the effective energy $\epsilon(p|P_c)$ takes the form:
\begin{equation}
\epsilon(p|P_c)=\epsilon(0|P_c)+\frac{\partial_p^4\epsilon(p|P_c)|_{p=0}}{24} p^4+O(p^6).
\end{equation}
Accordingly, all four saddle points of the function $\mathcal{F}(z,\Lambda)$ 
given  by \eqref{Fdef}  merge at $P=P_c$ and $\Lambda=\epsilon(0|P_c)$:
\[
z_a=z_a^{-1}=z_b=z_b^{-1}=1.
\]
Derivation of the third law-energy expansion is to much extent similar to the procedure described in Appendix \ref{LE2}.
The main difference, however, is that the momentum and energy variables must be rescaled, instead of equations \eqref{rEp},  in the following way:
\begin{subequations}\label{rEp1}
\begin{align}\label{tE1}
&p=t\, \mathfrak{p},\quad p'=t\, \mathfrak{p}',\\
&\tilde{E}_\iota=\epsilon(0|P_c)+t^4 \mathfrak{e}_\iota,
\end{align}
\end{subequations}
with $t=f^{1/5}$.

The final result for the third law-energy expansion reads:
\begin{align}\nonumber
&\tilde{E}_{\iota, n}(P_c)=\epsilon(0|P_c)+f^{4/5}\left[\frac{\partial_p^4\epsilon(p|P_c)|_{p=0}}{6}\right]^{1/5} c_n+\\\label{3le}
&f\,a_\iota(P_c)+O(f^{6/5}),
\end{align}
where $n=1,2,\ldots$, $a_\iota(P)$ is the scattering length \eqref{scL}, and $c_n$ are the  consecutive solutions of  equation (93) in 
reference \cite{Rut08a}: 
\begin{eqnarray}\label{eqcn}
&&\int_0^\infty dy \,  \bigg[\sin\bigg( \frac{y^5}{20}-y\, c_n\bigg)-
\exp\bigg(- \frac{y^5}{20}+y \,c_n\bigg)\bigg]\cdot \\&&\nonumber
\int_0^\infty dx \, x^2 \cos\bigg( \frac{x^5}{20}-x \,c_n\bigg)=
\int_0^\infty dx \,  \cos\bigg( \frac{x^5}{20}-x \,c_n\bigg)\cdot \\
&&\int_0^\infty dy \, y^2 \bigg[\sin\bigg( \frac{y^5}{20}-y\, c_n\bigg)+
\exp\bigg(- \frac{y^5}{20}+y\ c_n\bigg)\bigg], \nonumber
\end{eqnarray}
$c_1=1.787$, $c_2=3.544$, $c_3=5.086$, $c_4=6.518$. Note that  equation (93) in paper \cite{Rut08a} contains a misprint, which is 
corrected in \eqref{eqcn}.

The third law-energy expansions \eqref{3le}  were announced in reference  \cite{Rut18}, see equation (44) there.
\subsection{Crossover regimes \label{crosr}} 
There are three crossover regimes, which are realized in the vicinity of the
boundary curves separating the regions (I), (II) and (III) 
 in Figure \ref{fig:RegE}. In  these regimes, the meson energy $E_{\iota,n}(P)$ is close 
 to some local maximum value of the effective two-kink kinetic energy $\epsilon(p|P)$, see Figures \ref{Fig:Energ} (a), (b), (c). Perturbative calculation 
 of the meson energy spectra in these crossover regimes is based on the 
 asymptotic saddle point analysis of the integrals $J_\beta(\Lambda_\iota)$ 
 defined in equation \eqref{C12}. Since these calculations are to much extent
 similar to those outlined  above and described in Appendix \ref{PSIE}, we skip their details and present only the final results. 

\underline{First crossover regime.}

\noindent
The first crossover regime is realised close to the boundary separating the regions  (I) and (III) in Figure \ref{fig:RegE}.  At $P\in (P_c(\eta),\pi/2)$, it takes place at 
the meson energies $\tilde{E}_{\iota, n}(P)$ close to $\epsilon(0|P)$. 
At small $h\to0$, the meson energy spectrum 
$\tilde{E}_{\iota, n}(P)$ is determined in this case by equations:
\begin{align}\label{cross1}
&\frac{\mathrm{Ai}\,(\lambda_{\iota,n})}{\mathrm{Bi}\,(\lambda_{\iota,n})}=\\\nonumber
&\cot\left[-\frac{F(p_a,\tilde{E}_{\iota,n}(P))}{2f}
-\frac{\pi}{4}+\frac{\theta_\iota(P/2+p_a,P/2-p_a)}{2}
\right],
\end{align}
with $\tilde{E}_{\iota, n}(P)=\epsilon(p_a|P)$, and
\begin{align}
\lambda_{\iota,n}=\left(
\frac{2}{f^2\, \left|\partial_p^2\epsilon(p,P)\right|_{p=0}}
\right)^{1/3}\cdot\\\nonumber
[\tilde{E}_{\iota,n}(P)-\epsilon(0|P)-f a_{\iota}(P)].
\end{align}
Here and below the function 
$F(p,E)$ is determined by \eqref{FpE}, and $\mathrm{Bi}\,(\lambda)$
denotes the second solution
\begin{equation}
\mathrm{Bi}\,(\lambda)=\int_0^\infty \frac{dt}{\pi} \left[\sin\left(\frac{t^3}{3}+t \lambda\right)+\exp\left(-\frac{t^3}{3}+t \lambda\right)\right]
\end{equation}
 of the Airy differential equation.
 
\underline{Second crossover regime.}

\noindent
The second crossover regime is realised close to the boundary separating the regions  (I) and (II) in Figure~\ref{fig:RegE}.  At $P\in (P_c(\eta),\pi/2)$, it takes place at  the meson energy  $\tilde{E}_{\iota, n}(P)$  close to $\epsilon(\pi/2|P)$. 
In this case, the  small-$f$ asymptotics of the meson energies 
$\tilde{E}_{\iota, n}(P)$ is determined by solutions of two transcendent 
equations:
\begin{align}
&\frac{\mathrm{Ai}\,(\tilde{\lambda}_{\iota,n})}{\mathrm{Bi}\,(\tilde{\lambda}_{\iota,n})}=
\tan\left[\frac{F(\pi/2,\tilde{E}_{\iota,n}(P))}{2f}
-\frac{\pi}{2} \, \delta_{\iota,0}
\right],\\
\label{cross2}
&\tilde{\lambda}_{\iota,n}=\left(
\frac{2}{f^2\, \left|\partial_p^2\epsilon(p,P)\right|_{p=\pi/2}}
\right)^{1/3}\cdot\\\nonumber
&\left[\tilde{E}_{\iota,n}(P)-\epsilon(\pi/2|P)+\frac{f}{2}\partial_p \theta_\iota(P/2+p,P/2-p)\Big|_{p=\pi/2}\right].
\end{align}

\underline{Third crossover regime.}

\noindent Finally, the third crossover regime takes place at $P=\pm \pi/2+\pi n$, 
with $n\in \mathbb{Z}$, at energies $\tilde{E}_{\iota, n}(\pi/2)$  close 
$\epsilon(0|\pi/2)=\epsilon(\pi/2|\pi/2)$. These points separate the regions (II) and (III) in Figure~\ref{fig:RegE}, see also 
Figure \ref{Eb}. The energy spectra  $\tilde{E}_{\iota, n}(\pi/2)$ in this regime
are determined by solutions of two equations:
\begin{align}\nonumber
&\frac{[\mathrm{Bi}(\breve{\lambda}_{\iota,\iota,n})+i\, \mathrm{Ai}(\breve{\lambda}_{\iota,\iota,n})][\mathrm{Bi}(\breve{\lambda}_{\iota,-\iota,n})+i\, \mathrm{Ai}(\breve{\lambda}_{\iota,-\iota,n})]}
{[\mathrm{Bi}(\breve{\lambda}_{\iota,\iota,n})-i\, \mathrm{Ai}(\breve{\lambda}_{\iota,\iota,n})][\mathrm{Bi}(\breve{\lambda}_{\iota,-\iota,n})-i\, \mathrm{Ai}(\breve{\lambda}_{\iota,-\iota,n})]}=\\\label{cross3}
&\exp\left[i
\frac{F(\pi/2,\tilde{E}_{\iota,n}(\pi/2))}{f}- i \pi \delta_{\iota,0}
\right],\\
&\breve{\lambda}_{\iota,\iota',n}=\left(
\frac{2}{f^2\, \left|\partial_p^2\epsilon(p,\pi/2)\right|_{p=0}}
\right)^{1/3}\cdot\\\nonumber
&\left[\tilde{E}_{\iota,n}(\pi/2)-\epsilon(0|\pi/2)-f a_{\iota'}(\pi/2)\right],
\end{align}
where $a_{\iota}(P)$ is the scattering length \eqref{scL}. 

Note, that 
due to the symmetry \eqref{refP},
the energy spectra $\tilde{E}_{\iota,n}(P)$ of mesons with 
 opposite parities 
$\iota=\pm$ coincide  at $P=\pi/2$.
\section{Dynamic structure factors in the confinement regime \label{DSFconf}}
In this Section we describe the effect of a weak longitudinal staggered magnetic field on the 
structure factors of the spin operators. In the thermodynamic limit, these structure factors are defined as follows:
\begin{align}\label{SFh}
&S^{\mathfrak{a}\mathfrak{b}}(\mathrm{k} ,\omega|h)=\frac{1}{8 }\sum_{j=-\infty}^\infty e^{-i\mathrm{k} j}\int_{-\infty}^\infty dt \,e^{i\omega t}\cdot\\\nonumber
&\Big[\langle vac(h)|\sigma_j^\mathfrak{a}(t) \sigma_{0}^\mathfrak{b} (0)|vac(h)\rangle+\\
&\langle vac(h)|\sigma_{j+1}^\mathfrak{a}(t) \sigma_{1}^\mathfrak{b} (0)|vac(h)\rangle\Big],\nonumber
\end{align}
where $|vac(h)\rangle$ is the ground state of the infinite spin chain with the Hamiltonian \eqref{XXZH}. As in Section \ref{Sec:DNS0}, we shall limit 
our attention to the case $\mathrm{k}\in(0,\pi)$ without loss of generality.

Two approximations will be used in calculation of the structure factors \eqref{SFh}.
First, the analysis will be limited to the leading order in the weak staggered magnetic field $h$.
In accordance with equation  \eqref{vach},  this allows one to replace the vacuum state $|vac(h)\rangle$ in equation
\eqref{SFh} by its zero-field counterpart $|vac\rangle^{(1)}$. Second, as in Section \ref{Sec:DNS0}, the analysis will be restricted 
solely to the two-spinon contribution to the structure factor. In the leading order in $h$, the latter is given by equation \eqref{DSF20}.
The operator $\mathcal{P}_{11}^{(2)}$ in this equation denotes the projector onto the two-kink subspace $\mathcal{L}_{11}^{(2)}$. This
projector operator admits representation \eqref{PP11} in terms of the two-kink basis states  $|K_{10}(p_1)K_{01}(p_2)\rangle_{s_1s_2}$.
However, since such two-kink states do not diagonalize the Hamiltonian \eqref{XXZH} at any non-zero staggered 
magnetic field $h$, representation
 \eqref{PP11}  of the operator $\mathcal{P}_{11}^{(2)}$ cannot be used for calculation of the DSF \eqref{DSF20} at $h>0$. 
 Instead, we shall use the following expansion of this projector operator in the basis of meson states  $|\tilde\pi_{s,\iota}(P)\rangle$ determined by equations 
 \eqref{mesB1}, \eqref{mesB}, and \eqref{normPi}:
 
 \begin{equation}\label{P112}
 \mathcal{P}_{11}^{(2)}=\sum_{s=0,\pm1} \mathcal{P}_{s}^{(2)}.
  \end{equation}
 Here 
 \begin{equation} \label{P110}
\mathcal{P}_{s}^{(2)}=\int_{0}^\pi \frac{dP}{\pi}\sum_{n=1}^\infty
|\tilde\pi_{s,\iota=0,n}(P)\rangle
\langle \tilde\pi_{s,\iota=0,n}(P) |
  \end{equation}
  for $s=\pm1$, and 
  \begin{align}\label{Prs0}
  &\mathcal{P}_{s=0}^{(2)}= \mathcal{P}_{s=0,\iota=+}^{(2)}+ \mathcal{P}_{s=0,\iota=-}^{(2)},\\\nonumber
&\mathcal{P}_{s=0,\iota}^{(2)}=\int_{0}^\pi \frac{dP}{\pi}\sum_{n=1}^\infty
|\tilde\pi_{s=0,\iota,n}(P)\rangle
\langle \tilde\pi_{s=0,\iota,n}(P) |.
   \end{align}
\subsection{Transverse DSF \label{Sec_Tr}} 
Formulas \eqref{DSF20}, \eqref{P112}, \eqref{P110}  lead  in the adopted approximation to the following 
representation for the transverse DSF:
\begin{align}\label{DSFtr}
&S_{(2)}^{+-}(\mathrm{k} ,\omega|h)=\frac{1}{8 }\sum_{j=-\infty}^\infty e^{-i\mathrm{k} j}\int_{-\infty}^\infty dt \,e^{i\omega t}\cdot\\\nonumber
&\Big[\phantom{.}^{(1)} \langle vac|\sigma_j^+(t)\mathcal{P}_{s=-1}^{(2)} \sigma_{0}^{-} (0)|vac\rangle^{(1)}+\\
&\phantom{.}^{(1)} \langle vac|\sigma_{j+1}^+(t)\mathcal{P}_{s=-1}^{(2)}  \sigma_{1}^- (0)|vac\rangle^{(1)}\Big] .\nonumber 
\end{align}
Integration in $t$ and summation in $j$ in the right-hand side can be performed 
using equations \eqref{mes}, \eqref{piT1}, and    following the procedure described in Section \ref{Sec:DNS0}. The result reads:
 \begin{equation} \label{Spmconf}
 S_{(2)}^{+-}(\mathrm{k},\omega|h)=\sum_{n=1}^\infty \delta[\omega-\tilde{E}_{\iota=0,n}(P)]\,I_n^{+-}(P|h)\Big|_{P=\mathrm{k}},
 \end{equation}
where $\tilde{E}_{\iota=0,n}(P)$ is the dispersion law of two degenerate meson modes with $\iota=0$ and $s=\pm 1$
in the two-kink approximation, and $ I_n^{+-}(P|h)$ are the intensities corresponding to these modes:
\begin{align}\label{Inpm}
 I_n^{+-}(P|h)=\frac{\pi}{4}\,\big|\phantom{.}^{(1)} \langle vac| \sigma_0^+|\tilde\pi_{s=-1,\iota=0,n}(P)\rangle+\\
  \phantom{.}^{(1)} \langle vac| \sigma_0^-|\tilde\pi_{s=1,\iota=0,n}(P)\rangle\big|^2.\nonumber
 \end{align}
The matrix elements of the $\sigma_0^\pm$ operators in the right-hand side  can be 
expressed in terms of the wave function $\phi_{\iota=0,n}(p|P)$ solving the 
Bethe-Salpeter equation \eqref{BSQ3} and normalized by the condition \eqref{normphi}:
\begin{widetext}
\begin{equation}\label{mel}
\phantom{.}^{(1)}\! \langle vac| \sigma_0^\pm|\tilde\pi_{s=\mp1,\iota=0,n}(P)\rangle=
\int_0^{\pi/2}\frac{dp}{\pi}\,\phi_{\iota=0,n}(p|P)e^{-i(p+P/2)}
\phantom{.}^{(1)}\!\langle vac| \sigma_0^\pm
| K_{10}(P/2+p)K_{01}(P/2-p)\rangle_{\mp1/2,\mp1/2}.
 \end{equation} 
 \end{widetext}
The matrix elements of the $\sigma_0^\pm$ operators between the vacuum and two-kink states are, in turn, 
simply related due to \eqref{Kxi}, \eqref{xX}, \eqref{XX0} with the form factors $X^1(\xi_1,\xi_2)$ and $X^0(\xi_1,\xi_2)$ given by equation \eqref{XX}.

Thus, equations \eqref{Spmconf} -  \eqref{mel} describe the transverse DSF in the antiferromagnetic 
XXZ spin chain in the confinement  regime in the two-kink approximation. Substitution of obtained in Section \ref{WK}
perturbative solutions of the 
Bethe-Salpeter equation \eqref{BSQ3} instead of the wave function $\phi_{\iota=0,n}(p|P)$ in the right-hand side of 
\eqref{mel} yields the explicit asymptotical formulas for the transverse DSF in the limit $h\to  +0$. In what follows, we shall restrict our analysis to the 
semiclassical approximations for the meson wave function, which were described
in Section \ref{SemReg}.

In the semiclassical regimes, the wave function  $\phi_{\iota=0,n}(p|P)$ becomes highly oscillating in $p$ at $h\to 0$, while 
the matrix element in the integrand in the right-hand side  of \eqref{mel} remains regular and smooth in this limit. As the result, the semiclassical
asymptotics of the integral in the right-hand side of \eqref{mel} is determined by contributions of the saddle points, which are located at the solutions of the  equation 
\begin{equation}\label{enP}
\epsilon(p|P)=\tilde{E}_n(P). 
\end{equation}
\subsubsection{First semiclassical regime}
There is only one such saddle point $p=p_a$ in the interval $(0,\pi/2)$ in the first semiclassical regime
at $0<P<P_c(\eta)$. Accordingly, in the leading order in $h\to 0$, we can apply the asymptotical formula \eqref{phi0}
for the function  $\phi_{\iota=0,n}(p|P)$ and then perform integration in $p$ in the right-hand side of \eqref{mel}  using the 
Dirac delta-function $\delta(p-p_a)$. As the result, we obtain in this case:
\begin{equation}\label{InP}
 I_n^{+-}(P|h)=\frac{\pi f}{p_a}\,  S_{(2)}^{+-}(P,\omega|0)\Big|_{\omega=\tilde{E}_n(P)},
\end{equation}
where  $S_{(2)}^{+-}(P,\omega|0)$ is the 
two-kink contribution to the transverse DSF at zero staggered magnetic field 
given by equation \eqref{SFtr}. 

The result \eqref{InP} for the intensity $I_n^{+-}(P|h)$ holds in the first semiclassical regime at $0<P<P_c(\eta)$.
On the other hand, at   $\pi-P_c(\eta)<P<\pi$ and $\epsilon(\pi/2|P)<\tilde{E}_n(P)<\epsilon(0|P)$,
 the unique solution of  equation \eqref{enP} in the interval  $(0,\pi/2)$
  is $p=p_b$. In this case, we obtain instead of \eqref{InP}:
\begin{equation}\label{InPA}
 I_n^{+-}(P|h)=\frac{\pi f}{\pi/2-p_b}\,  S_{(2)}^{+-}(P,\omega|0)\Big|_{\omega=\tilde{E}_n(P)}.
\end{equation}

It is instructive to rewrite \eqref{InP}, \eqref{InPA}  in the equivalent form:
\begin{equation}\label{InP1}
 I_n^{+-}(P|h)=\Delta{E}_{n}^{(\mathrm{I})}(P)\,\cdot  S_{(2)}^{+-}(P,\omega|0)\Big|_{\omega=\tilde{E}_n(P)},
\end{equation}
where 
\begin{align}
&\Delta{E}_{n}^{(\mathrm{I})}(P	)\equiv \tilde{E}_{\iota, n+1}(P)-\tilde{E}_{\iota, n}(P)=\\
&\begin{cases}
\frac{\pi f}{p_a}+O(f^2), &  0<P<P_c(\eta),\\\nonumber
\frac{\pi f}{\pi/2-p_b}+O(f^2)  , & \pi-P_c(\eta)<P<\pi
\end{cases}
\end{align}
is the small interval between two sequential energies  of the $\iota$-meson  mode at given  $P$  in the first semiclassical regime. 

Let us sum both sides of equality \eqref{InP1}   in $n$ at some fixed $P\in(0,P_c(\eta))$:
\begin{equation}\label{sumR}
\sum_{n=1}^{\mathcal{N}(P|h)} I_n^{+-}(P|h)=\sum_{n=1}^{\mathcal{N}(P|h)}\Delta{E}_{n}^{(\mathrm{I})}(P) \,S_{(2)}^{+-}(P,E|0)),
\end{equation}
where the upper  limit of summation $\mathcal{N}(P|h)$ is given by equation \eqref{Ntot}. 
Due to \eqref{Spmconf}, we have in the left-hand side the integral 
\begin{equation}\label{Inth}
\int_{\epsilon(0|P)}^{\epsilon(\pi/2|P)}dE \,S_{(2)}^{+-}(P,E|h),
\end{equation}
while  the right-hand side of \eqref{sumR} represents
the Riemann sum approximating at small  $f$ the integral 
\begin{equation}\label{Int0}
\int_{\epsilon(0|P)}^{\epsilon(\pi/2|P)}dE \,S_{(2)}^{+-}(P,E|0).
\end{equation}
The function $S_{(2)}^{+-}(P,E|0)$ vanishes  at energies $E$ outside the interval $(\epsilon(0|P),\epsilon(\pi/2|P))$, 
as well as the function $S_{(2)}^{+-}(P,E|h)$ at $h\to0$. 
Thus, we obtain from \eqref{sumR}:
\begin{equation}
\lim_{h\to +0}\int_{0}^\infty dE \,S_{(2)}^{+-}(P,E|h)=\int_{0}^\infty dE \,S_{(2)}^{+-}(P,E|0).
\end{equation}
Despite the confinement of kinks induced by the arbitrary weak staggered magnetic field $h>0$,
the two-kink transverse DSF integrated over the  energy is continuous in $h$ at $h\to+0$.  
 \subsubsection{Third semiclassical regime}
 There are two saddle points $p_b< p_a$ in the interval $(0,\pi/2)$ in the third semiclassical regime 
 at $P\in(P_c(\eta),\pi-P_c(\eta))$.  To the leading order in $h$,
 the reduced meson wave function $\phi_{0,n}(p|P)$ is determined by equations  \eqref{ps0}, \eqref{Cab}, and \eqref{psph}.
The delta-functions in the right-hand side of \eqref{ps0} give rise  to two terms after integration in $p$ 
in the right-hand side of \eqref{mel}. Then, substitution of the result in \eqref{Inpm} and subsequent straightforward calculations yield:
\begin{equation}
 I_n^{+-}(P|h)=\Delta {E}_{ n}^{(\mathrm{III})}\Big[
 S_{(2)}^{+-}(P,\omega|0)+Z_n^{+-}(P,\omega)
 \Big]\Big|_{\omega=\tilde{E}_n(P)},
\end{equation}
where $\Delta {E}_{ n}^{(\mathrm{III})}$ is given by \eqref{dEn2}, $S_{(2)}^{+-}(P,\omega|0)$ is the transverse DSF at zero magnetic field 
given by \eqref{SFtr}, and the oscillating in $n$ term
\begin{align}\nonumber
&Z_n^{+-}(P,\omega)\!=\!\frac{(-1)^{n-1}\,J^2 \sinh^2 \eta}{2
[|\epsilon'(p_a)\,\epsilon'(p_b)|\,\omega(p_{1a})\omega(p_{2a})\omega(p_{1b})\omega(p_{2b})]^{1/2}}\cdot\\\label{Zpm}
&\mathrm{Re}\bigg\{\left[
X^1(\xi_{1a},\xi_{2a})+X^0(\xi_{1a},\xi_{2a})\right]\cdot\\\nonumber
&\left[X^1(\xi_{1b},\xi_{2b})+X^0(\xi_{1b},\xi_{2b})\right]^*\cdot\\\nonumber
&\exp \left[\frac{i}{2}[\theta_0(p_{1b},p_{2b})-\theta_0(p_{1a},p_{2a})]\right]
\bigg\},
\end{align}
results from the interference of  contributions of two saddle points $p_a$ and $p_b$.  In equation \eqref{Zpm}, we have used 
notations   \eqref{p12a}, and \eqref{p12b}. 
\subsection{Longitudinal  DSF} 
The longitudinal DSF $S_{(2)}^{zz}(\mathrm{k},\omega|h)$ in the weak confinement regime in the adopted approximation is given by equation 
\eqref{DSF20}, in which the projection operator $\mathcal{P}_{11}^{(2)}$ is replaced by  the operator  $\mathcal{P}_{s=0}^{(2)}$ 
given by \eqref{Prs0}, and   $\mathfrak{a}= \mathfrak{b}=z$. Exploiting equalities \eqref{tilT1}, \eqref{T1vacA},  and \eqref{0T1}, equation 
\eqref{DSF20} can be then simplified to the form:
\begin{align}\label{DSFlg}
&S_{(2)}^{zz}(\mathrm{k},\omega|h)=\sum_{j=-\infty}^\infty \frac{e^{-i\mathrm{k} j}}{4 }\,\int_{-\infty}^\infty dt \,e^{i\omega t}\cdot\\\nonumber
&\phantom{.}^{(1)} \langle vac|\sigma_j^z(t)\mathcal{P}_{s=0}^{(2)} \sigma_{0}^{z} (0)|vac\rangle^{(1)}.
\end{align}

The summation in the right-hand side should be performed separately for even, and for odd $j$.
Denoting corresponding partial sums as $\Sigma_{even}$ and $\Sigma_{odd}$, respectively, we obtain 
after straightforward calculations exploiting equalities \eqref{tilT1}, \eqref{T1vac}, \eqref{mes2}, and \eqref{tilTiot}:
\begin{align}\label{SiEv}
&\Sigma_{even}=
\frac{1}{2}\int_0^\pi dP\!\!\sum_{m=-\infty}^\infty e^{2mi (P-\mathrm{k})}\cdot\\\nonumber
&\sum_{\iota=\pm}\sum_{n=1}^\infty
\delta\left[
\omega-\tilde{E}_{\iota,n}(P)
\right]
\left|
\phantom{.}^{(1)} \langle vac|\sigma_0^z|\tilde{\pi}_{s=0,\iota,n}(P)\rangle
\right|^2,\\\label{SiOdd}
&\Sigma_{odd}=
\frac{1}{2}\int_0^\pi dP\sum_{m=-\infty}^\infty e^{2mi (P-\mathrm{k} )}\sum_{\iota=\pm}(-\iota )\cdot\\\nonumber
&e^{i (P-\mathrm{k} )}\sum_{n=1}^\infty
\delta\left[
\omega-\tilde{E}_{\iota,n}(P)
\right]
\left|
\phantom{.}^{(1)} \langle vac|\sigma_0^z|\tilde{\pi}_{s=0,\iota,n}(P)\rangle
\right|^2,
\end{align}
with 
\begin{equation}
S_{(2)}^{zz}(\mathrm{k},\omega|h)=\Sigma_{even}+\Sigma_{odd}.
\end{equation}
 Using the Poisson summation formula \eqref{PoS}, and assuming that $0<\mathrm{k}<\pi$,  formulas \eqref{SiEv}, \eqref{SiOdd}
 can be simplified to the form:
 \begin{align}\label{siev}
 &\Sigma_{even}=\frac{1}{2}\sum_{\iota=\pm}\sum_{n=1}^\infty
 \delta\left[
\omega-\tilde{E}_{\iota,n}(P)
\right]\,I_{\iota,n}^{zz}(P|h)\Big|_{P=\mathrm{k} },\\\label{siod}
 &\Sigma_{odd}=\frac{1}{2}\sum_{\iota=\pm}(-\iota)\sum_{n=1}^\infty
 \delta\left[
\omega-\tilde{E}_{\iota,n}(P)
\right]\,I_{\iota,n}^{zz}(P|h)\Big|_{P=\mathrm{k} },
 \end{align}
 where
  \begin{equation}\label{Inzz}
 I_{\iota,n}^{zz}(P|h)=\pi\,\left|\phantom{.}^{(1)} \langle vac| \sigma_0^z|\tilde\pi_{s=0,\iota,n}(P)\rangle\right|^2.
 \end{equation}
 Adding \eqref{siev} and \eqref{siod}, we get finally:
 \begin{equation} \label{Szzconf}
 S_{(2)}^{zz}(\mathrm{k} ,\omega|h)=\sum_{n=1}^\infty \delta[\omega-\tilde{E}_{\iota,n}(P)]\,I_{\iota,n}^{zz}(P|h)\Big|_{
 \begin{smallmatrix}P=\mathrm{k} \\
 \iota=-
 \end{smallmatrix}}.
 \end{equation}
Note, that the meson  states with $s=0$, $\iota=+$ do not contribute to the longitudinal structure factor $S_{(2)}^{zz}(\mathrm{k} ,\omega|h)$: 
the quantum interference leads to the mutual cancellation of   contributions of two magnetic sub-lattices of the antiferromagnetic ground state 
$|vac\rangle^{(1)}$.
\begin{figure}
\centering
\subfloat[Transverse modes with $s=\pm1$. ]{
	\label{Eba}
\includegraphics[width=.95\linewidth]{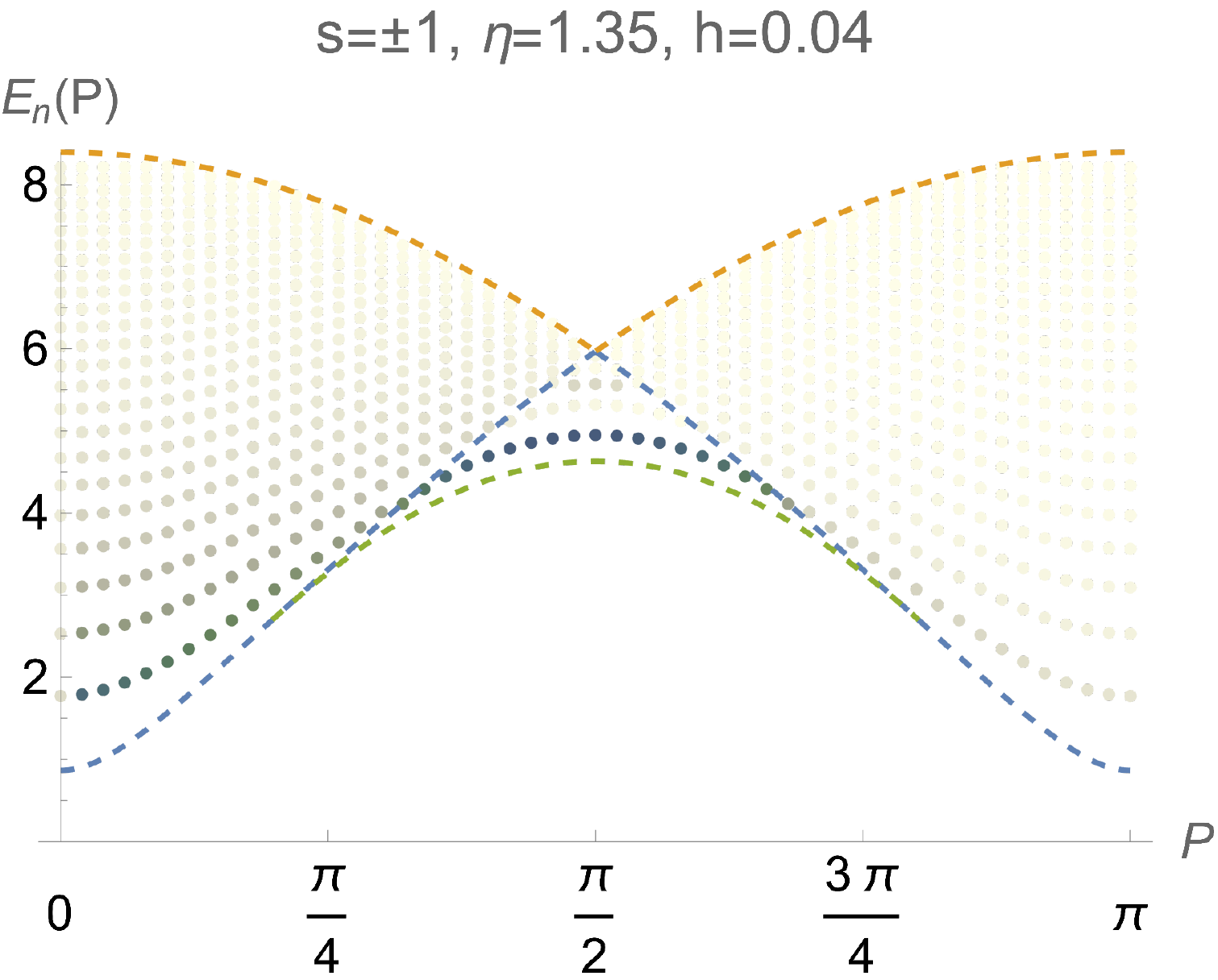}}

\subfloat[Longitudinal modes modes with $s=0$, $\iota=-$. ]{
	\label{Eca}
\includegraphics[width=.95\linewidth]{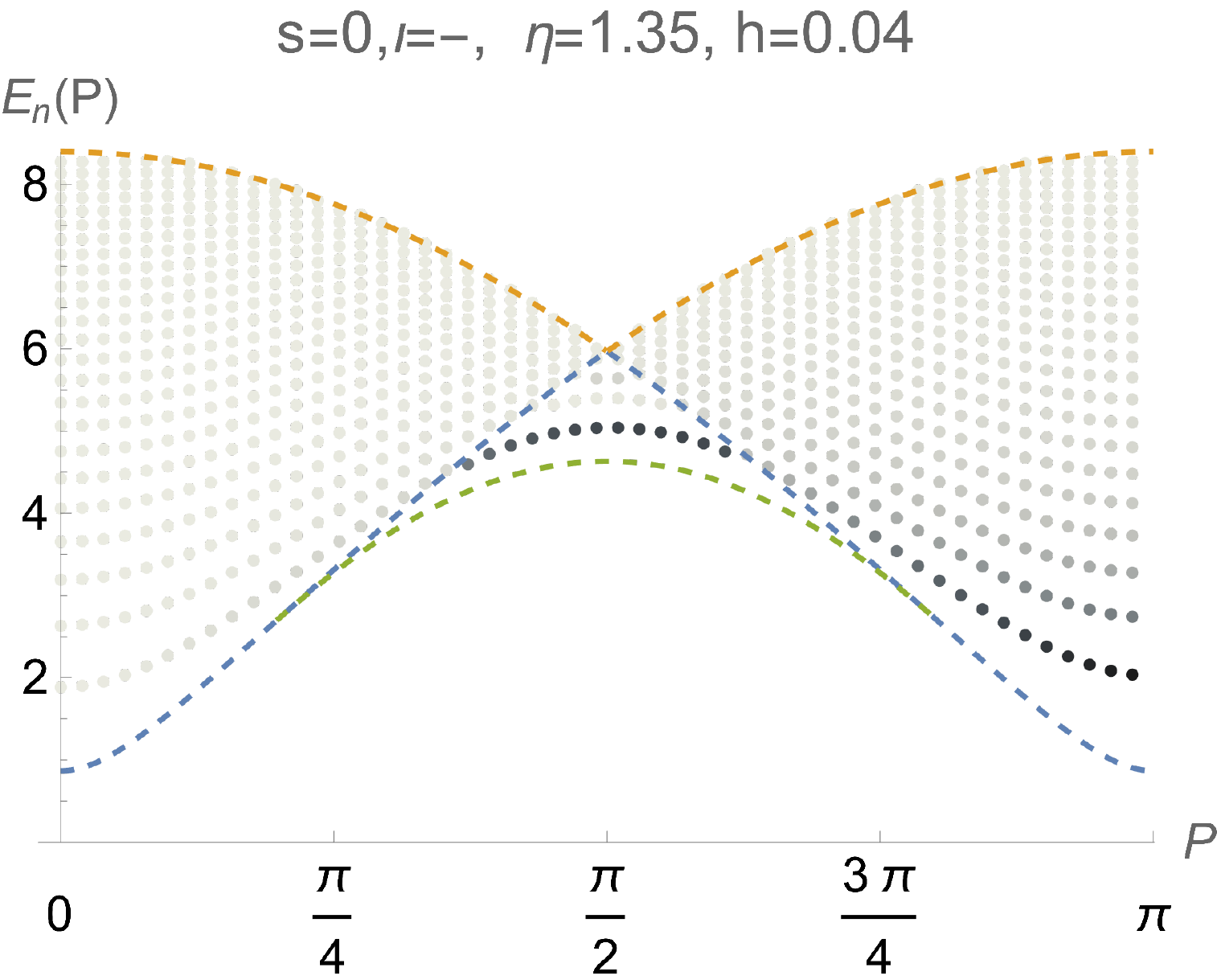}}

\caption{Semiclassical energy spectra of the  transverse (a) and  longitudinal (b) meson modes  at $\eta=1.35$, $h=0.04$. 
Darkness of the dotts
characterizes the intensities  of the meson modes. Dashed lines are the same as in Figure \ref{figSemES}. \label{Fig:Intens}}
\end{figure}

The matrix element of the $\sigma_0^z$ operator in the right-hand side of \eqref{Inzz} can be 
expressed in terms of the  wave function $\phi_{\iota=-,n}(p|P)$ solving equation \eqref{BSQ3} and satisfying the 
normalisation condition \eqref{normphi}:
\begin{align}\nonumber
&\phantom{.}^{(1)}\! \langle vac| \sigma_0^z|\tilde\pi_{s=0,\iota=-,n}(P)\rangle=
\int_0^{\pi/2}\frac{dp}{\pi}\,\phi_{\iota=-,n}(p|P)\cdot\\
&\phantom{.}^{(1)}\!\langle vac| \sigma_0^z
| K_{10}(P/2+p)K_{01}(P/2-p)\rangle_{-}.\label{melM}
 \end{align} 

As in the previous Section \ref{Sec_Tr}, we restrict our analysis to the calculation of the semiclassical small-$h$ asymptotics of the
DSF. These calculations for the longitudinal DSF are very similar to those described in Section \ref{Sec_Tr}, but now 
they are based on equations \eqref{Inzz}-\eqref{melM}, instead of \eqref{Spmconf}-\eqref{mel}. Skipping the details of these calculations, 
we present below only the final results.

In the first semiclassical regime,  the function $ I_{\iota=-,n}^{zz}(P|h)$ reads:
\begin{equation}\label{InzzA}
 I_{\iota=-,n}^{zz}(P|h)=\Delta{E}_{n}^{(\mathrm{I})}(P)\cdot  S_{(2)}^{zz}(P,\omega|0)\Big|_{\omega=\tilde{E}_n(P)},
\end{equation}
and $S_{(2)}^{zz}(P,\omega|0)$ is the longitudinal DSF at $h=0$  given by equation \eqref{lDN}. The latter
simplifies in the first semiclassical region to the form:
\begin{equation}
S_{(2)}^{zz}(P,\omega|0)= \frac{ \mathcal{G}_0^{zz}(P/2+p,P/2-p)}{|\partial_p \epsilon(p|P)|}
\Bigg|_{p=p_0},
\end{equation}
where $p_0=p_a$ at $P\in (0,P_c(\eta))$, $p_0=p_b$ at $P\in (\pi/2,\pi-P_c(\eta))$, and 
$\mathcal{G}_0^{zz}(p_1,p_2)$ is determined  by \eqref{G0zz}.

In the third semiclassical regime
\begin{equation}\label{secSem}
 I_{\iota=-,n}^{zz}(P|h)=\Delta {E}_{ n}^{(\mathrm{III})}\Big[
  S_{(2)}^{zz}(P,\omega|0)+Z_n^{zz}(P,\omega)
  \Big]\Big|_{\omega=\tilde{E}_n(P)},
\end{equation}
where $\Delta {E}_{ n}^{(\mathrm{III})}$ and $S_{(2)}^{+-}(P,\omega|0)$  are given by equations \eqref{dEn2}, and 
 \eqref{lDN}, respectively, and the oscillating in $n$ term reads:
\begin{align}\nonumber
&Z_n^{zz}(P,\omega)\!=\!\frac{(-1)^{n-1}\,2\,J^2 \sinh^2 \eta}{
[\left|\epsilon'(p_a)\,\epsilon'(p_b)\right|\,\omega(p_{1a})\omega(p_{2a})\omega(p_{1b})\omega(p_{2b})]^{1/2}}\cdot\\\label{Zmin}
&\mathrm{Re}\bigg\{
X_-^z(\xi_{1a},\xi_{2a})
\left[X_-^z(\xi_{1b},\xi_{2b})\right]^*\cdot\\\nonumber
&\exp \left[\frac{i}{2}[\theta_-(p_{1b},p_{2b})-\theta_-(p_{1a},p_{2a})]\right]
\bigg\}.
\end{align}

\begin{figure}
\centering
\subfloat[Intensities $I_n^{+-}(P|h)$ of the transverse modes with $s=\pm1$. ]{
	\label{TrI}
\includegraphics[width=.95\linewidth]{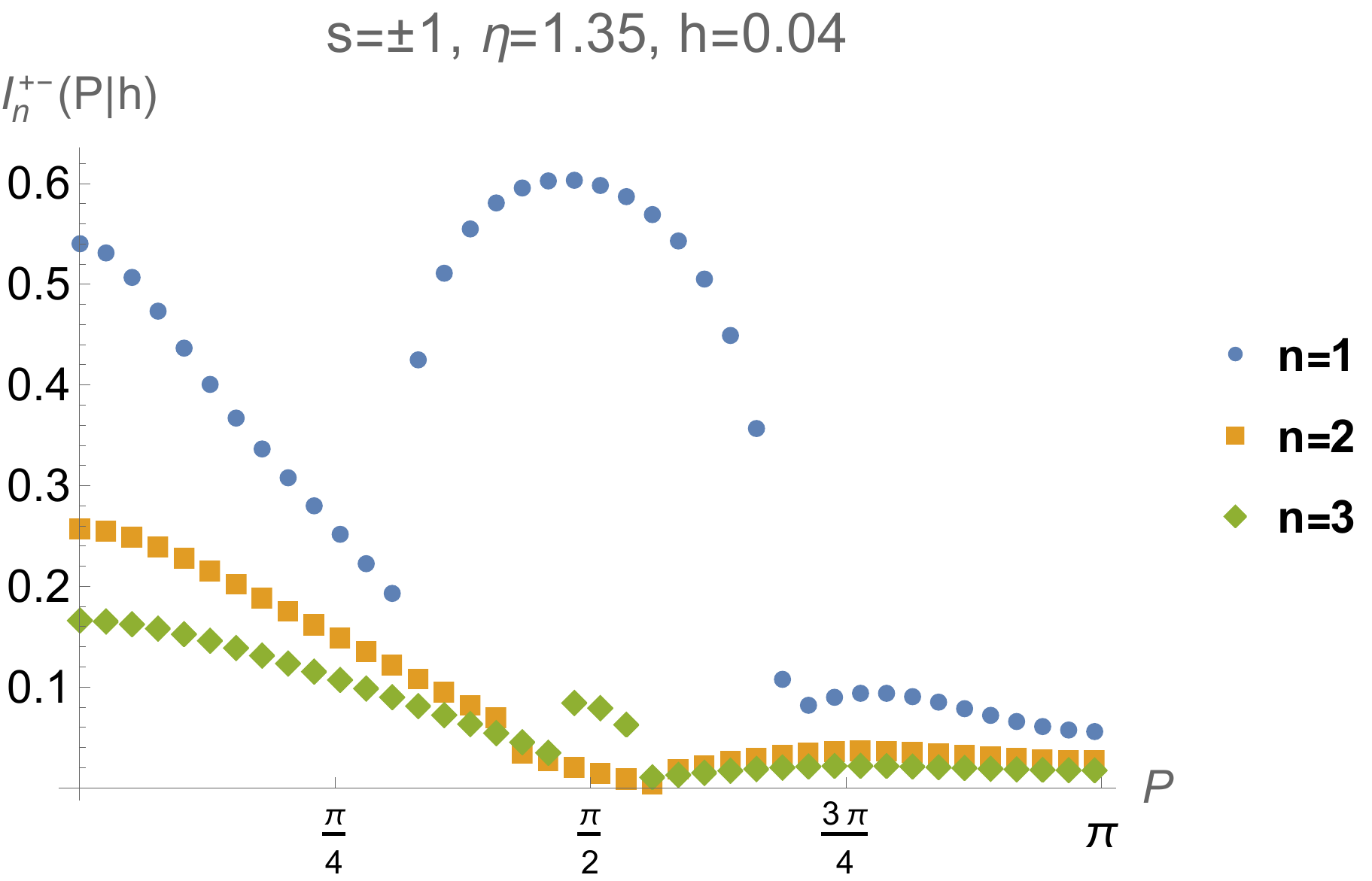}}

\subfloat[Intensities $I_{-,n}^{zz}(P|h)$ of the longitudinal  modes with $s=0$, $\iota=-$. ]{
	\label{Li}
\includegraphics[width=.95\linewidth]{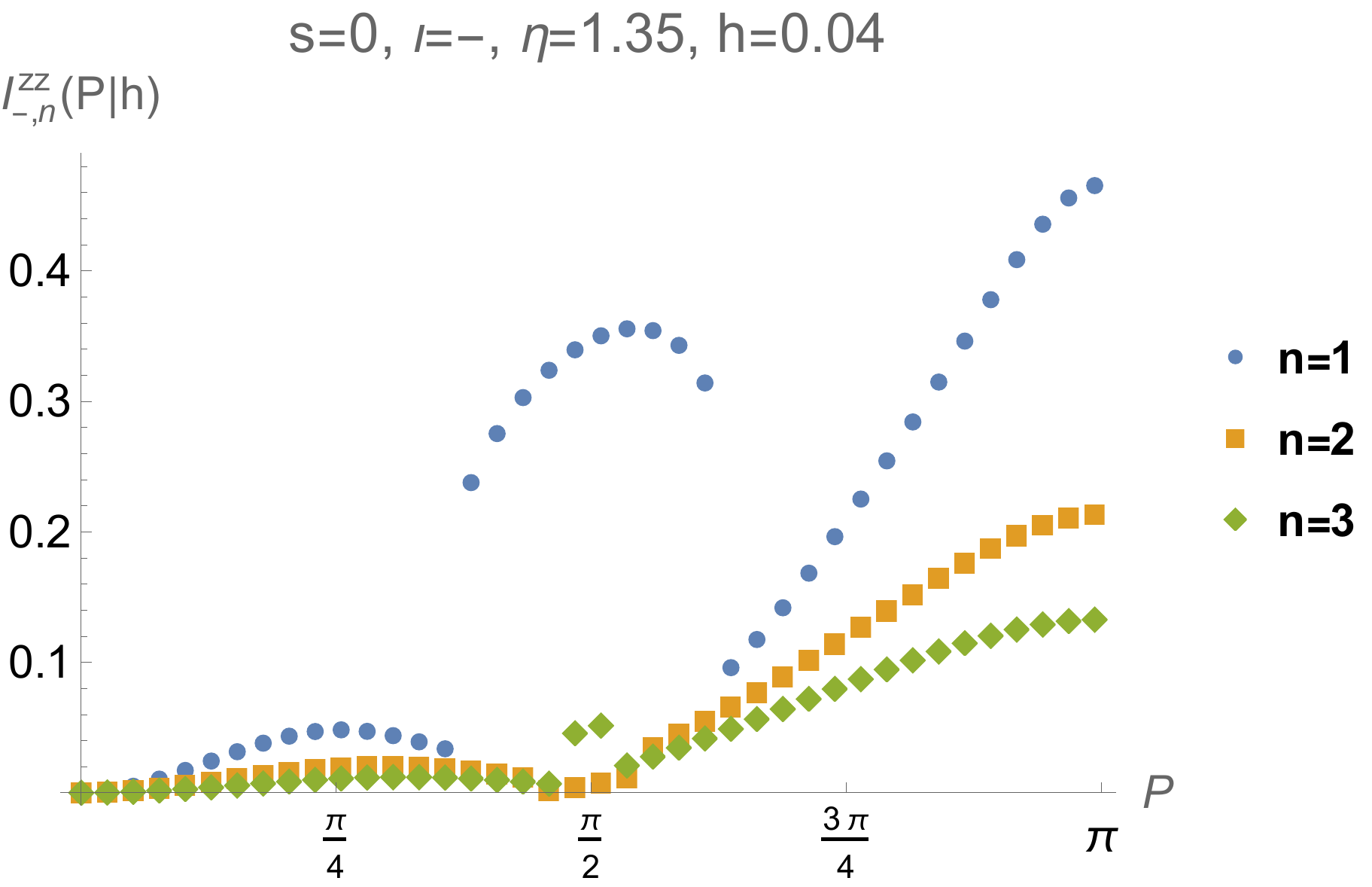}}

\caption{Momentum dependences of intensities of three lowest meson modes at  $\eta=1.35$, $h=0.04$. 
\label{Fig:Inte}}. 
\end{figure}

The obtained results are illustrated in Figure \ref{Fig:Intens}. It displays the similar  semiclassical 
meson energy spectra determined by equations \eqref{Mdl}, \eqref{SE2}, and \eqref{refP}
at $\eta=1.35$, as those in Figure \ref{figSemES}, but at the smaller value of the magnetic field $h=0.04$. 
The dispersion curves for the two degenerate meson modes with $s=\pm1$ plotted
in Figure \ref{Eba}, and Figure \ref{Eca} show the energy spectra of the  modes with $s=0$ and $\iota=-$.
The darkness of the points in Figure \ref{Fig:Intens}  characterizes the  intensities  of the corresponding modes.
For the transverse modes shown in Figure   \ref{Eba}, these intensities are determined by equations \eqref{InP1}, \eqref{Inzz}, while 
for the longitudinal modes with $s=0$ and $\iota=-$ shown in Figure \ref{Eca}, the intensities are given by  equations \eqref{InzzA}, \eqref{secSem}.
The DSF intensities of the $s=0$  meson modes  with $\iota=+$ are equal to zero.

The intensities of the meson dispersion curves  in Figure \ref{Fig:Intens} display very different qualitative 
behaviour in three regions
of the $(PE)$-plane shown in Figure \ref{fig:RegE}. In the first  regions (I), the intensities 
$I_n^{+-}(P|h)$ and $I_{-,n}^{zz}(P|h)$ monotonically decrease with
increasing $n$. In the third  region (III), the dependence of the intensities  is non-monotonic and alternating in $n$. In the second region
(II), the intensities vanish in the adopted two-spinon semiclassical approximation.

These features of the meson DSF are clearly seen in Figure \ref{Fig:Inte} showing  rather peculiar momentum dependencies 
of  intensities of three lowest transverse (Figure \ref{TrI}) and longitudinal (Figure \ref{Li}) meson modes.
\section{Conclusions\label{Conc}}
We have investigated the main properties of 
 the spinon bound-state (``meson'')
excitations in the infinite XXZ spin chain in the massive antiferromagnetic phase
in the weak confinement regime, which takes place in this model in the presence of a weak staggered longitudinal magnetic field $h$. We analytically calculated the small-$h$ asymptotics of the meson energy spectra in the whole Brillouin zone 
using two different perturbative schemes.  In the first, less rigorous, but more 
physically transparent approach, the  meson energy spectra were obtained by 
 quantization of the Hamiltonian dynamics of two  classical particles moving along the line and attracting one another with a linear potential. The results for the
 meson energy spectra obtained this way were confirmed and extended  by means of a 
 more rigorous and systematic  technique exploiting  the perturbative solution of the  Bethe-Salpeter  integral equation \eqref{BSQ5}, which was derived 
for the XXZ spin-chain  model  \eqref{XXZH} in Section \ref{SecBSE}.  Based on this perturbative analysis, we have described  nine asymptotic regimes,  which are realized in different regions of the meson energy-momentum plane.  A similar structure of the meson energy spectra in the weak confinement regime was previously found  \cite{Rut08a} in the Ising spin-chain model \eqref{HIsa} perturbed by a weak  uniform longitudinal magnetic field. This is not surprising, since the 
kink dispersion laws in models \eqref{XXZM} and \eqref{HIsa} are the same up to a re-parametrization.  
 Finally, using the perturbative solution of the Bethe-Salpeter equation, explicit formulas were  obtained  for the  two-kink contribution to the transverse and longitudinal dynamical structure factors of the local spin operators for the XXZ spin chain  model in the weak confinement regime. 

 Strictly speaking, nine asymptotic expansions, the  initial terms of which are 
presented in Section \ref{WK}, describe in model   \eqref{XXZH}  the small-$h$  asymptotic behavior  of the meson energy spectra  in the whole Brillouin zone.
We expect, however, that a rather accurate numerical description of all meson energy spectra at small $h$ can be provided solely by the three semiclassical expansions \eqref{Mdl}, 
\eqref{Sem3A}, and \eqref{SE2}. This suggests, in particular, that  the semiclassical
formulas can be used not only to describe the energy spectra  of mesons with large quantum number $n\gg 1$, but  may also  work well for lowest lying mesons with $n=1,2,\ldots$. The high efficiency of the semiclassical formulas for description of the 
energy  spectra of light mesons in different QFT and spin-chain models exhibiting confinement was confirmed in various works \cite{LT2015,Kor16,Lagnese_2020,Mus22}.

We believe that, even though the energy spectra of magnetic excitations can be 
determined by direct numerical methods, such as  the  DMRG,  the matrix product state (MPS) approach, and other techniques, 
 the analytically obtained in this paper formulas for the meson energy spectra and the magnetic DSFs will be helpful for understanding and interpretation 
of the results on the inelastic neutron-scattering and  terahertz  spectroscopy experiments in quasi-1D antiferromagnetic crystals
 in the confinement regime. 
 
 {\color{black}The main advantage of our analytic approach  is that, in contrast to the direct numerical methods that  typically require 
considerable computational   time and efforts  in order  to determine the magnetic excitation energy and DSF at  certain fixed values 
of the  model parameters and quasi-momentum, the analytic perturbative techniques developed in this paper provide  explicit formulas for
these quantities  in the whole Brillouin zone in a wide range  of   parameters $\Delta$ and $h$. This allows us, in particular, to
predict the non-monotonic $n$-dependence of the DSF at wave-vectors $\mathrm{k}$ close to $\pi/2$, see Figures \ref{Fig:Intens}, and \ref{Fig:Inte},
and to elucidate the non-trivial role of the elastic two-spinon scattering in forming the meson energy spectra in the semiclassical and low-energy 
regimes, see equations \eqref{EnSem1}, \eqref{les},  \eqref{SE2}, and \eqref{slee}.}

There are several directions for further study. 
The procedure developed in Section \ref{SecBSE} can be applied to a derivation of the
Bethe-Salpeter equation in other QFTs and spin-chain models exhibiting confinement, which are integrable, but not free in the deconfined phase. 

For experimental applications, it would be interesting to extend the analysis of the 
spinon confinement in the XXZ spin-chain model \eqref{XXZH} to the case of nonzero temperatures. It would be also interesting to study theoretically the effect of a uniform transverse  magnetic field on the spinon confinement in this model. 
\begin{acknowledgments} I am thankful to Frank G\"ohmann for many fruitful discussions, and to Fedor Smirnov for helpful correspondence.   
This work was supported by Deutsche Forschungsgemeinschaft (DFG) via Grant BO 3401/7-1.
\end{acknowledgments}
\appendix
\section{Properties of the function $\epsilon(p|P)$ \label{2E}}
In this Appendix we describe in details the properties of the "kinetic energy of two spinons in the  center momentum frame" 
\begin{equation}\label{Energ1}
\epsilon(p|P)=\omega(P/2+p)+\omega(P/2-p),
\end{equation}
where $\omega(p)$ is the spinon dispersion law \eqref{dl}.

The symmetry properties of this function
\begin{align}
&\epsilon(p|P)=\epsilon(p+\pi|P)=\epsilon(-p|P)=\epsilon(p|-P),\\\label{epsym2}
&\epsilon(p|P)=\epsilon(p+\pi/2|-P+\pi),\\
&\epsilon(p|\pi/2)=\epsilon(p+\pi/2|\pi/2)=\epsilon(-p+\pi/2|\pi/2).\label{epsym3}
\end{align}
follow immediately from its definition.

\begin{figure}
\centering
\subfloat[ $0\le P<\pi/2$]
{\label{Ea}
\includegraphics[width=.9\linewidth]{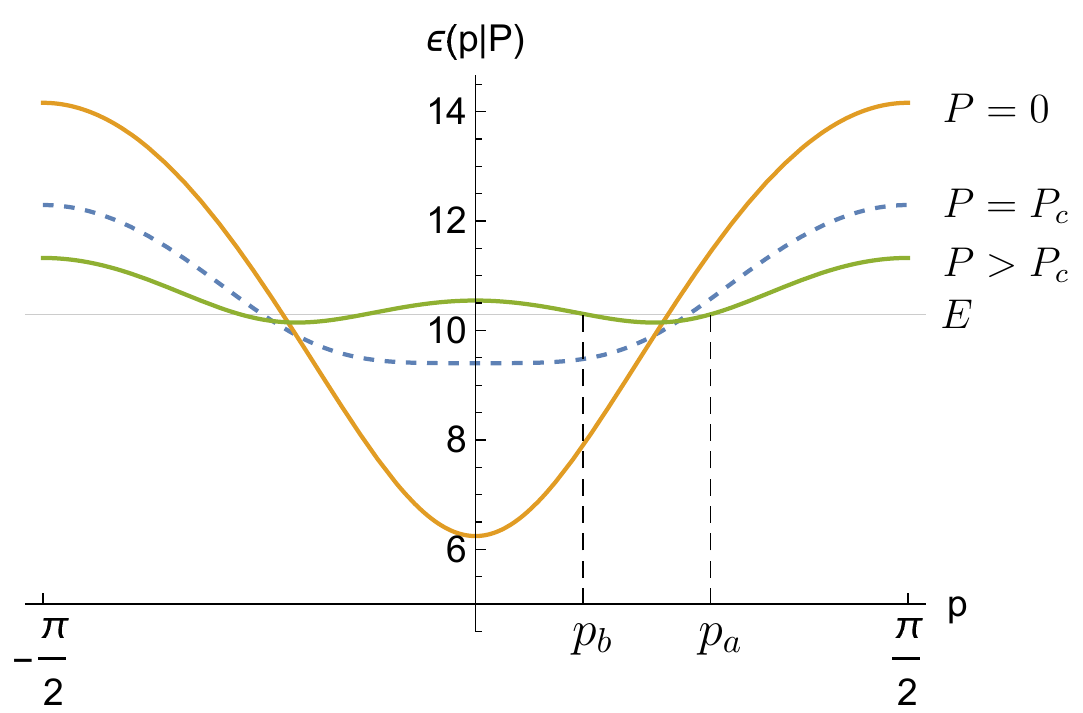}}

\subfloat[$P=\pi/2$]{
	\label{Eb}
\includegraphics[width=.85\linewidth]{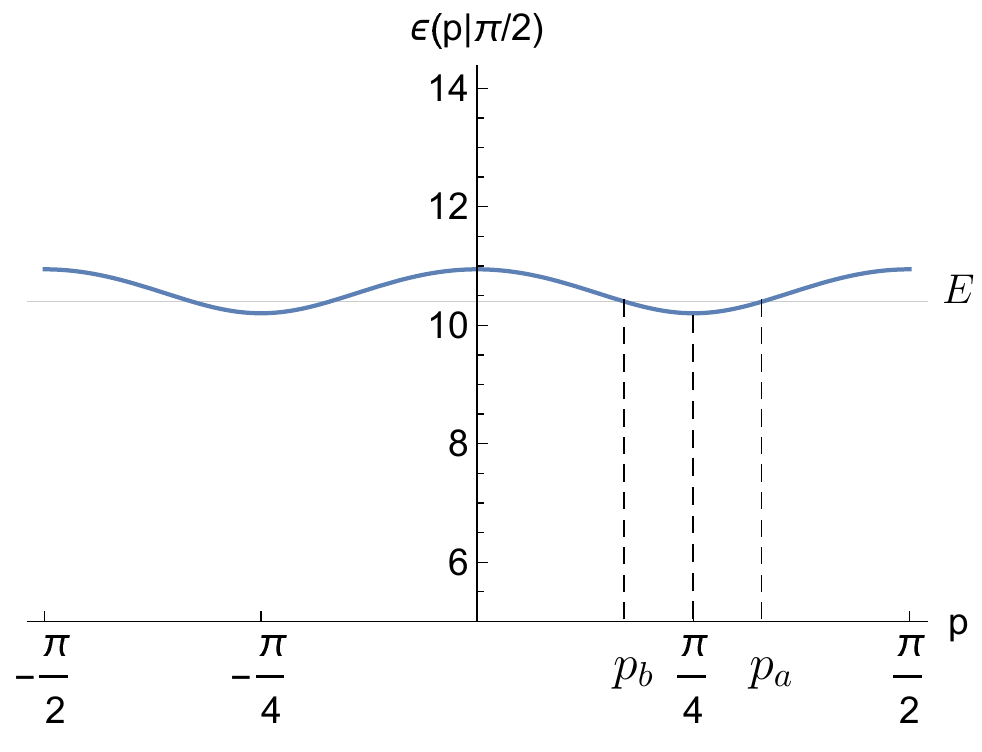}}

\subfloat[$\pi/2<P\le\pi$]{
	\label{Ec}
\includegraphics[width=1.\linewidth]{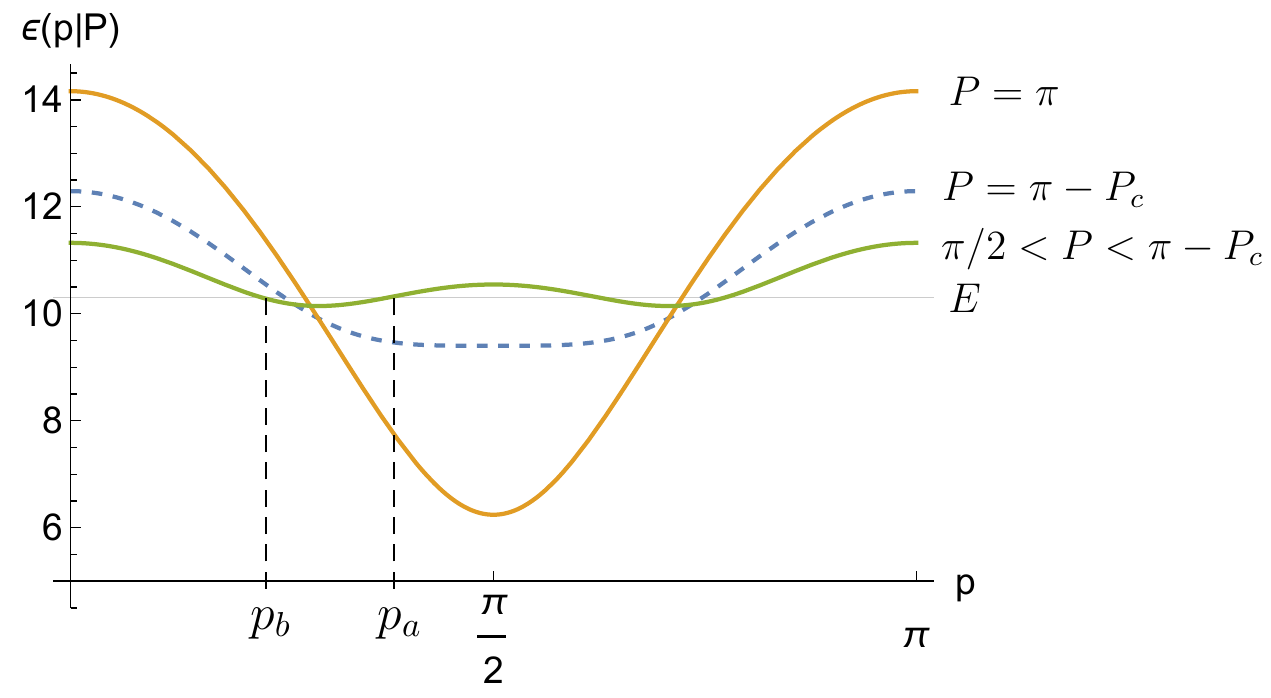}}

\caption{Energy of two spinons $\epsilon(p|P)$ defined by \eqref{Energ1} at different $P$ as the function of $p$ at 
$\eta=\mathrm{arccosh}\, 5$.  The critical momentum $P_c(\eta)$ is determined by \eqref{Pc}. \label{Fig:Energ}}. 

\end{figure}

Evolution of the $p$-dependence of the function $\epsilon(p|P)$ with increasing $P\in[0,\pi]$ is shown in Figure \ref{Fig:Energ}.
At $P=0$, the function $\epsilon(p|P)$ takes the minimum value at $p=0$, and monotonically increases with increasing $p$  
in the interval $0\le p\le\pi/2$, see Figure \ref{Ea}. This qualitative behavior of the $p$-dependence of the function $\epsilon(p|P)$
does not change with increasing $P$ until the latter reaches the  critical value  $P_c(\eta)$  determined by the conditions
\begin{equation}
\partial_p^2 \epsilon(p|P_c)\Big|_{p=0}=0, \quad 0<P_c<\pi/2.
\end{equation}
The critical total momentum $P_c(\eta)$ can be represented in terms of the complementary elliptic modulus $k'(\eta)$
\begin{equation}\label{Pc}
P_c(\eta)=\mathrm{arccos}\,\frac{1-k'(\eta)}{1+k'(\eta)},
\end{equation}
which in turn can be expressed as the squared  ratio of two elliptic theta-functions \eqref{theta}:
\begin{equation}\label{kprim}
k'(\eta)=\left(\frac{\vartheta_4(0,e^{-\eta})}{\vartheta_3(0,e^{-\eta})}\right)^2=\left(\frac{\vartheta_2(0,e^{-\pi^2/\eta})}{\vartheta_3(0,e^{-\pi^2/\eta})}\right)^2.
\end{equation}
At larger $P$ in the interval $P_c(\eta)<P<\pi/2$, the function  $\epsilon(p|P)$ becomes non-monotonic in $p$ at $0<p<\pi/2$. It has a local maximum at $p=0$, and takes the minimum value $\epsilon_m$ at $p=p_{m}\in (0,\pi/4)$, where
\begin{align}\label{epm}
&\epsilon_m(P,\eta)= I(\eta) \, [1+k'(\eta)]\,\sin P,\\\label{pmin}
&p_{m}(P,\eta)=\frac{1}{2}\,\mathrm{arccos}\left(\,\frac{\cos P}{\cos P_c(\eta)}\right).
\end{align}
Here the constant $I(\eta)$ is determined by  \eqref{Iet}.

The $p$-dependence of the function $\epsilon(p,P)$ at $P=\pi/2$ is shown in Figure \ref{Eb}.
The additional symmetries \eqref{epsym3} holding at $P=\pi/2$ lead to the following equalities:
\begin{equation}
\epsilon(0,\pi/2)=\epsilon(\pi/2,\pi/2), \quad p_{m}(\pi/2,\eta)=\pi/4.
\end{equation}

Finally, Figure \ref{Ec} displays the $p$-dependence of the function $\epsilon(p|P)$ at $\pi/2<P\le\pi$. Due to the symmetry 
\eqref{epsym2}, the curves on this Figure are shifted  to the right  by $\pi/2$  with respect to their counterparts in Figure  \ref{Ea}.

As one can see from Figure \ref{Fig:Energ}, the number $\mathfrak{N}(P,\omega)$ of real solutions $p^{(i)}\in(0,\pi/2)$ of the equation $\epsilon(p|P)=\omega$ for $P\in(0,\pi)$, and $\omega$ 
in the kinematically allowed interval \eqref{kinE}, is the following:
\begin{equation}
\mathfrak{N}(P,\omega)=\begin{cases}
2, &  P_c(\eta)< P \le \pi/2, \quad\quad \;\;\;  \omega<\epsilon(0,P), \\
2, & \pi/2<P< \pi-P_c(\eta), \quad \omega<\epsilon(\pi/2,P), \\
1, & \text{otherwise}.
\end{cases}
\end{equation}

\begin{figure}[htb]
\centering
\includegraphics[width=\linewidth, angle=00]{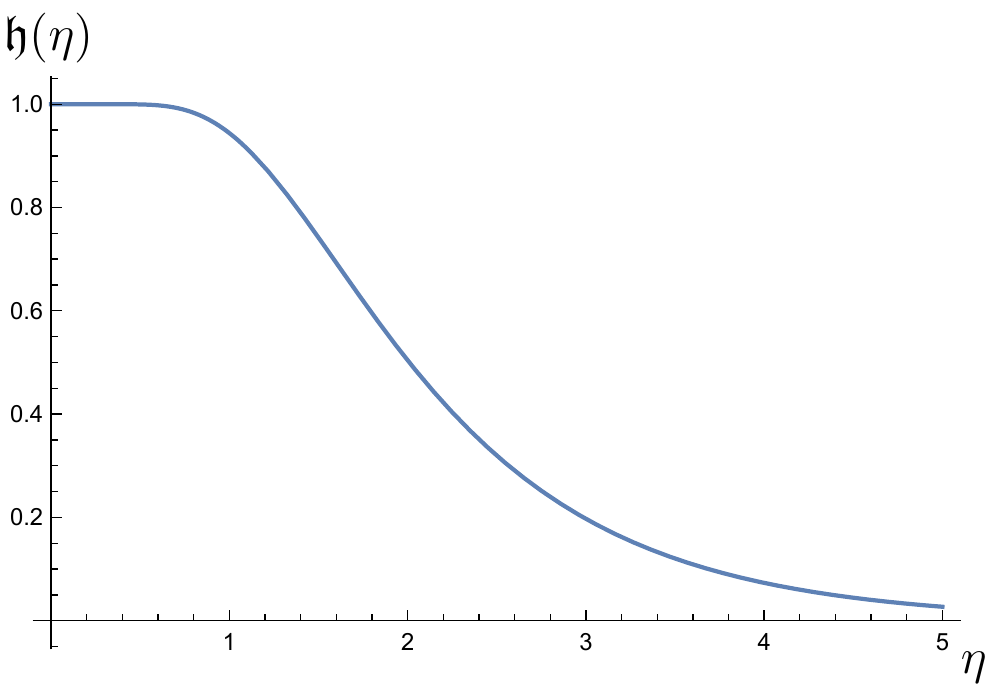}
\caption{\label{hf}  Variation of the parameter $\mathfrak{h}$ determined by \eqref{hfr} and \eqref{kprim} with $\eta$.} 
\end{figure}

To describe the properties of the function $\epsilon(p|P)$ analytically continued into the complex $p$-plane, we   proceed to the variables: 
\begin{equation}\label{hfr}
z=e^{2 i p},\quad v=e^{i P}, \quad \mathfrak{h}=\frac{1-k'}{1+k'}.
\end{equation}
Due to \eqref{kprim}, the parameter $\mathfrak{h}$ depends on $\eta$, and  varies in the interval $(0,1)$ at $\eta\in(0,\infty)$, see Figure \ref{hf}.
In new variables \eqref{hfr}, the function $\epsilon(p|P)$ can be written as:
\begin{equation}\label{OmepA}
\epsilon(p|P) =\mathcal{E}(z|v)=\frac{I k}{2}\, \Omega(z),
\end{equation}
where 
\begin{equation}\label{Ome}
\Omega(z)=\left(\mathfrak{h}+\frac{1}{\mathfrak{h}}-\frac{z}{v}-\frac{v}{z}\right)^{1/2}+
\left(\mathfrak{h}+\frac{1}{\mathfrak{h}}-{v}{z}-\frac{1}{z v}\right)^{1/2}.
\end{equation}
Note, that the function $\mathcal{E}(z|v)$ stands in the left-hand side of the Bethe-Salpeter equation 
\eqref{BSQ5}.

The algebraic function $\Omega(z)$  has   in the complex plane six square-root branching points:
\begin{equation}\label{BP}
0,\,\mathfrak{h}v,\,\mathfrak{h}v^{-1},\, \mathfrak{h}^{-1}v, \,\mathfrak{h}^{-1}v^{-1},\,\infty.
\end{equation}
Its Riemann surface $\mathfrak{L}$ has four sheets $\mathfrak{L}_{\alpha\beta}$ with $\alpha,\beta=\pm1$, which are distinguished by the signs of the
first and the second terms in the right-hand side of \eqref{Ome} at $z = 1$. 
\begin{figure}[htb]
\centering
\includegraphics[width=\linewidth, angle=00]{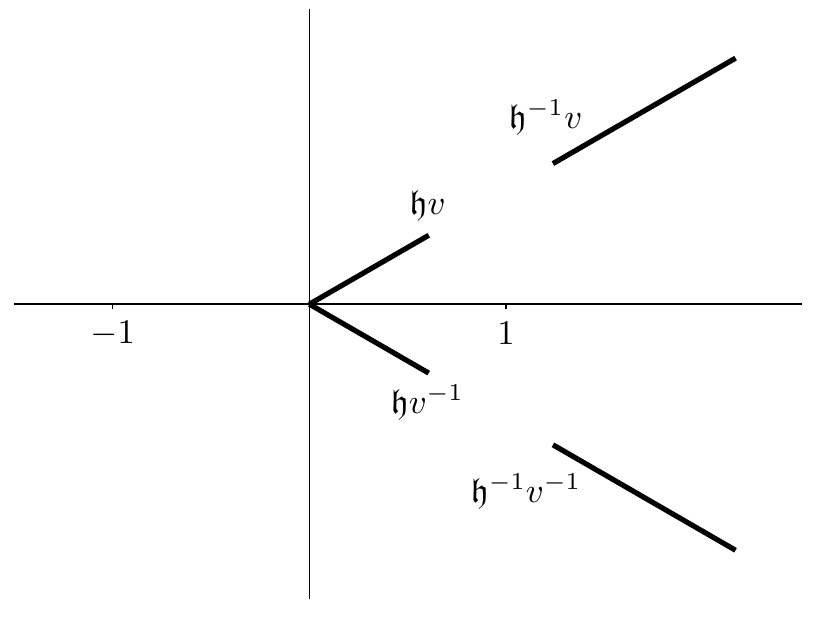}
\caption{\label{OmA}  Branching points of function $\Omega(z)$ determined by \eqref{Ome}, and cuts on $z$-plane for 
$|v|=1$, and $\mathfrak{h}\in(0,1)$.}
\end{figure}
As in \cite{Rut08a}, we draw four cuts on the $z$-plane shown in Figure~\ref{OmA}, in order to separate these sheets. The sheet 
$\mathfrak{L}_{++}$ will be called  "the physical sheet".

Note, that for $v=e^{i P}$ with $0<P<\pi$ and real $\mathfrak{h}$, the function $\Omega(z)$ has in the physical sheet the following 
asymptotical behavior
at real  $z\to 0$:
\begin{align}
\Omega(z)=\begin{cases}\label{Omas}
{2 {z^{-1/2}} \sin \frac{P}{2}}+O(z^{1/2}), & \text{if  } z >0,\\
{2 {|z|^{-1/2}}\cos \frac{P}{2}}+O(|z|^{1/2}), & \text{if  }  z <0.
\end{cases}
\end{align}

The equation 
\begin{equation}
\Omega(z) =\lambda
\end{equation}
has four solutions $z_a,z_a^{-1},z_b,z_b^{-1}$ in the Riemann surface $\mathfrak{L}$, which are given by 
\begin{align}\label{zab}
&z_{\alpha}+z_{\alpha}^{-1}=-\frac{v+v^{-1}}{(v-v^{-1})^2}\lambda^2\mp\\\nonumber
&2\left[
1+\frac{(\mathfrak{h}+\mathfrak{h}^{-1}) \lambda^2}{(v-v^{-1})^2}+\frac{\lambda^4}{(v-v^{-1})^4}
\right]^{1/2},
\end{align}
with $\alpha=a,b$. At certain values of parameters $v,\mathfrak{h}$ the point $z_a$, or $z_b$, or both of them, lie 
in the unit circle $|z|=1$. In this case, the corresponding solution of equation \eqref{epOm} can be recovered from 
\eqref{hfr},  \eqref{OmepA}  and \eqref{zab}:
\begin{align}\label{p12}
&p_{a,b}(P,\omega)=\frac{1}{2}\mathrm{arccos}\,\Bigg\{\left(\frac{\omega}{I\, k\sin P}\right)^2{\cos P}\,\mp\\\nonumber
&\sqrt{
\left[
1-\left(\frac{\omega}{I\,(1-k') \sin P}\right)^2
\right]
\left[
1-\left(\frac{\omega}{I\,(1+k') \sin P}\right)^2
\right]
}\Bigg\}.
\end{align}
\section{Form factors of the $\sigma_0^z$ operator  \label{fA}}
The form factors of the $\sigma_0^z$ operator are defined as the matrix element between the $\mu$-th vacuum and the 
$n$-kink state with even $n$:
\begin{align}\label{ff}
&f^{\mu}(\alpha_1,\ldots,\alpha_{n})_{s_1,\ldots,s_{n}}=\\
&\phantom{.}^{(\mu_{1})}\langle vac|\sigma_0^z|\mathcal{K}_{\mu_{n+1}\mu_{n}}(\xi_{n})\ldots\mathcal{K}_{\mu_2\mu_1}(\xi_1)
\rangle_{s_n,\ldots,s_1},\nonumber
\end{align}
where $\mu_1=\mu_{n+1}=\mu$, and $\xi_j=-i e^{i \alpha_j}$. For indices $\mu\in\{0,1\}$, we  shall use the following notations: 
$\overline0=1$, $\overline1=0$. Since $\mu_{j+1}=\overline{\mu_j}$, we have 
$\mu_j=\mu_1$ for odd $j$, and $\mu_j=\overline{\mu}_1$, for even $j$. The form factors 
\eqref{ff} of the $\sigma_0^z$ operator
are non-zero only if $s_1+\ldots+s_n=0$.
The form factors corresponding to the two different antiferromagnetic vacua are simply related with one another:
\begin{equation}
f^{0}(\alpha_1,\ldots,\alpha_{n})_{-s_1,\ldots,-s_{n}}=-f^{1}(\alpha_1,\ldots,\alpha_{n})_{s_1,\ldots,s_{n}}.
\end{equation}

Note, that a different notation  has been used in \cite{Jimbo94} for the form factors \eqref{ff}:
\begin{equation}
\phantom{.}_{(i)}\langle vac|\sigma_1^z|\xi_1,\ldots,\xi_{n}\rangle_{\nu_1,\ldots,\nu_{n};i}=
f^{\mu}(\alpha_1,\ldots,\alpha_{n})_{s_1,\ldots,s_{2n}},
\end{equation}
where $i=\mu$, and  $\nu_j=2 s_j.$

For the two-particle form factors, we have explicit expressions in terms of the functions
$X_\pm^z(\xi_1,\xi_2)$ defined by equations \eqref{Xpl}, \eqref{Xmn}:
\begin{align}\label{form2}
&f^{1}(\alpha_1,\alpha_{2})_{1/2,-1/2}=\frac{X_+^z(\xi_2,\xi_1)-X_-^z(\xi_2,\xi_1)}{\sqrt{2}},\\\nonumber
&f^{1}(\alpha_1,\alpha_{2})_{-1/2,1/2}=\frac{X_+^z(\xi_2,\xi_1)+X_-^z(\xi_2,\xi_1)}{\sqrt{2}}.
\end{align}
For the $n$-particles form factors, rather cumbersome integral representations were obtained by 
Jimbo and Miwa \cite{Jimbo94}. These form factors are meromorphic functions of the rapidities $\alpha_1,\ldots,\alpha_n$. 
They satisfy a set of equalities listed below, which are 
very much similar to the Smirnov's axioms  \cite{Sm92} for the form factors in integrable field theories.

1. Riemann-Hilbert axioms:
\begin{align}
&f^{\mu}(\alpha_1,\ldots,\alpha_l+\pi,\ldots,\alpha_{n})_{s_1,\ldots,s_l,\ldots,s_{n}}=\\\nonumber
&\varkappa(\overline{\mu_{l}},s_l)f^{\mu_{1}}(\alpha_1,\ldots,\alpha_l,\ldots,\alpha_{n})_{s_1,\ldots,s_l,\ldots,s_{n}},\\\label{RH}
&f^{\mu}(\alpha_1,\ldots,\alpha_{n-1},\alpha_{n}+2i\eta)_{s_1,\ldots,s_{n}}=\\
&f^{\overline{\mu}}(\alpha_{n},\alpha_1,\ldots,\alpha_{n-1})_{s_{n,}s_1,\ldots,s_{n-1}},\nonumber
\end{align}
where $\varkappa(\mu,s)$ is given by \eqref{kap0}.

2. Symmetry property:
\begin{align}
&
f^{\mu}(\alpha_1,\ldots,\alpha_l,\alpha_{l+1},\ldots,\alpha_{n})_{s_1,\ldots,s_l,s_{l+1},\ldots,s_n}\cdot \\\nonumber
& \mathcal{S}_{s_l' s_{l+1}'}^{s_l s_{l+1}}(\alpha_l-\alpha_{l+1})=
\\\nonumber
&f^{\mu}(\alpha_1,\ldots,\alpha_{l+1},\alpha_{l},\ldots,\alpha_{n})_{s_1,\ldots,s_{l+1}',s_{l}',\ldots,s_n}.
\end{align}

3. At fixed real $0\le\alpha_1,\ldots,\alpha_{n-1}<\pi$, the form factors \eqref{ff} as  functions of the complex variable 
$\alpha_n$ lying in the  rectangle $0\le\mathrm{Re}\, \alpha_n<\pi$,  $0\le\mathrm{Im}\, \alpha_n \le \eta$,  have the simple 
{ annihilation poles} at $\alpha_j+i \eta$, with $j=1,\ldots,n-1$. The residue
 at such a pole at $\alpha_n=\alpha_{n-1}+i \eta$ reads:
 \begin{widetext}
\begin{align}\label{resAn}
&-2 i \,\, \mathrm{res}_{\alpha_n=\alpha_{n-1}+i \eta}\, f^{\mu}(\alpha_1,\ldots,\alpha_{n-2},\alpha_{n-1},\alpha_{n})_{s_1, \ldots,s_{n-2},s_{n-1},s_n}=
\delta_{-s_n,s_{n-1}}\,  f^{\mu}(\alpha_1,\ldots,\alpha_{n-2})_{s_1, \ldots,s_{n-2}}-\\
&\delta_{-s_n,\tau_0}\,f^{\overline{\mu}}(\alpha_1,\ldots,\alpha_{n-2})_{s_1', \ldots,s_{n-2}'}\,
 \mathcal{S}_{\tau_1 s_1}^{\tau_0 s_1' }(\alpha_{n-1}-\alpha_1)\,
 \ldots\,
  \mathcal{S}_{\tau_{n-3} s_{n-3}}^{\tau_{n-4} s_{n-3}' }(\alpha_{n-1}-\alpha_{n-3})\,\mathcal{S}_{s_{n-1} s_{n-2}}^{\tau_{n-3} s_{n-2}' }(\alpha_{n-1}-\alpha_{n-2})
.\nonumber
\end{align}
 \end{widetext}
For natural $n$ and $m$, the matrix element of the $\sigma_0^z$ operator between the $n$- and $m$-kink states 
is non-zero only for even   $(n+m)$. In this case, it can be expressed in terms of the  form factors \eqref{ff} by means of the crossing relation \cite{Jimbo94}:
\begin{widetext}
\begin{align}\label{cross}
&\phantom{.}_{s_n,\ldots,s_1}\langle\mathcal{K}_{\mu_{n+1}\mu_n}(\xi_{n})\ldots\mathcal{K}_{\mu_2\mu_1}(\xi_1)|\sigma_0^z|\mathcal{K}_{\mu_1\mu_2}(\xi'_1)\ldots
\mathcal{K}_{\mu_m\mu_{m+1}}(\xi'_{m})\rangle_{s'_1,\ldots,s'_{m}}=\\\nonumber
&\phantom{.}^{(\mu_{n+1})}\langle vac|\sigma_0^z|\mathcal{K}_{\mu_{n+1}\mu_n}(-q\xi_{n})\ldots
\mathcal{K}_{\mu_2\mu_1}(-q\xi_1)
\mathcal{K}_{\mu_1\mu_2}(\xi'_1)\ldots
\mathcal{K}_{\mu_m\mu_{m+1}}(\xi'_{m})\rangle_{-s_n,\ldots,-s_1,s'_1,\ldots,s'_m}.
\end{align}
\end{widetext}
\section{Two-kink matrix elements of $\sigma_0^z$ \label{MEL}}
In this Appendix we derive  the explicit formulas \eqref{singG} for the singular part of the 
integral kernel $\mathcal{G}_\iota^{(sing)}(z,z'|v)$ of the Bethe-Salpeter equation \eqref{BSQ4}.
First, we check that these formulas hold in the Ising limit.
Then  we turn to the general case of arbitrary $\eta>0$, and consider the one- and two-kink matrix elements of the operator
$(\sigma_0^z-\bar\sigma)$:
\begin{subequations}\label{MES}
\begin{align}\label{Ymn1}
&\phantom{.}_{s}\langle K_{\nu\mu}(p)|(\sigma_0^z-\bar{\sigma})|K_{\mu\nu}(p')\rangle_{s'}=Y_{\mu\nu}(p|p')_{s}\,\delta_{s,s'},\\
&\phantom{.}_{s_2,s_1}\langle K_{\mu\nu}(p_2)K_{\nu\mu}(p_1)|(\sigma_0^z-\bar{\sigma})|K_{\mu\nu}(p_1')K_{\nu\mu}(p_2')\rangle_{s_1',s_2'}     =\nonumber\\
&Y_{\mu\nu}(p_2,p_1|p_1',p_2')_{s_2,s_1|s_1',s_2'},\label{Y2}
\end{align} 
\end{subequations}
and also the two-kink matrix elements of the operator $(\sigma_1^z+\bar{\sigma})$. 
Due to the translation symmetry relations \eqref{tilT1}, and \eqref{tilT}, the latter are simply related with analogous
matrix elements of the operator $(\sigma_0^z-\bar\sigma)$:
 \begin{align}
 &\phantom{.}_{s_2,s_1}\langle K_{\mu\nu}(p_2)K_{\nu\mu}(p_1)|(\sigma_1^z+\bar{\sigma})|K_{\mu\nu}(p_1')K_{\nu\mu}(p_2')\rangle_{s_1',s_2'}     =\nonumber\\
&- e^{i(p_1'+p_2'-p_1-p_2)}\,Y_{\mu\nu}(p_2,p_1|p_1',p_2')_{-s_2,-s_1|-s_1',-s_2'}.
\end{align}  
We shall use also two further notations:
\begin{subequations}\label{Yspm}
\begin{align}
&Y_{\mu\nu}(p_2,p_1|p_1',p_2')_s=Y_{\mu\nu}(p_2,p_1|p_1',p_2')_{s,s|s,s},\\\label{Y2spm}
&Y_{\mu\nu}(p_2,p_1|p_1',p_2')_\pm=\\
&\phantom{.}_{\pm}\langle K_{\mu\nu}(p_2)K_{\nu\mu}(p_1)|(\sigma_0^z-\bar{\sigma})|K_{\mu\nu}(p_1')K_{\nu\mu}(p_2')\rangle_{\pm}  =\nonumber\\
&\frac{1}{2}\sum_{s,s'=\pm1/2}(-1)^{s-s'}
Y_{\mu\nu}(p_2,p_1|p_1',p_2')_{-s,s|s',-s'}.\nonumber
\end{align} 
\end{subequations}
In the case of coinciding in- and out- total momenta of two kinks
\begin{equation}\label{consM}
p_1+p_2=p_1'+p_2'=P,
\end{equation}
we proceed in \eqref{Yspm} to the variables $p=(p_1-p_2)/2$, $p'=(p_1'-p_2')/2$, and define two further functions:
\begin{subequations}\label{YspmA}
\begin{align}\label{Y2spmA}
&\mathcal{Y}_{\mu\nu}(p,p'|P)_s=Y_{\mu\nu}(p_2,p_1|p_1',p_2')_s,\\\label{Y2pmX}
&\mathcal{Y}_{\mu\nu}(p,p'|P)_\pm=Y_{\mu\nu}(p_2,p_1|p_1',p_2')_\pm,
\end{align} 
\end{subequations}
where 
\begin{equation}\label{p12p}
p_{1,2}=\pm p+P/2, \quad  p_{1,2}'=\pm p'+P/2.
\end{equation}
In these notations, the integral kernels $G_\iota(p,p'|P)$ (with $\iota=0,\pm$) 
defined by equations \eqref{G0} and \eqref{Gio} take the form:
\begin{subequations}
\begin{align}\label{G0pm}
&G_0(p,p'|P)=\frac{e^{i(p-p')}}{4\bar\sigma}\sum_{s=\pm1/2}
\mathcal{Y}_{10}(p,p'|P)_s,\\\label{pmG}
&G_\pm(p,p'|P)=\frac{\mathcal{Y}_{10}(p,p'|P)_\pm}{2\bar\sigma}.
\end{align} 
\end{subequations}

We describe the structure of the kinematic singularities of the 
matrix elements \eqref{MES}
and derive equations \eqref{singG} for any $\eta>0$.

In the Ising limit $\eta\to\infty$, the staggered spontaneous magnetization \eqref{sig} and the scattering amplitudes \eqref{wio}, \eqref{Wio},  \eqref{Wz} reduce to:
\begin{subequations}\label{Islim}
\begin{align}
&\bar{\sigma}(\eta)=1,\\
&w_0(p_1,p_2)=-e^{i(p_1-p_2)}, 
\; w_\pm(p_1,p_2)=-1, \\
&W_\iota(p|P)=\mathcal{W}_{\iota}(z|v)=-1.
\end{align}
\end{subequations}
The spin operators $\sigma_j^z$ are diagonal in the basis of the localized $n$-kink states.
In particular,
\begin{align}\label{1K}
&(\sigma_0^z-1)|\mathbf{K}_{\mu\nu}(j)\rangle=\chi_{\mu\nu}^{(1)}(j)|\mathbf{K}_{\mu\nu}(j)\rangle,\\
&(\sigma_0^z-1)|\mathbf{K}_{\mu\nu}(j_1)\mathbf{K}_{\nu\mu}(j_2)\rangle=
\chi_{\mu\nu}^{(2)}(j_1,j_2)|\mathbf{K}_{\mu\nu}(j_1)\mathbf{K}_{\nu\mu}(j_2)\rangle,\label{2K}
\end{align}
where $j_1<j_2$, and 
\begin{align}\label{chi1}
&\chi_{01}^{(1)}(j)=\begin{cases}
-2 , & j\ge 0,\\
0, & \text{otherwise},
\end{cases},\\\nonumber
&\chi_{10}^{(1)}(j)=\begin{cases}
-2 , & j< 0,\\
0, & \text{otherwise},
\end{cases},\\
&\chi_{01}^{(2)}(j_1,j_2)=\begin{cases}
-2 , & j_1\ge 0,\\
-2 , & j_2< 0,\\
0, & \text{otherwise},
\end{cases},\\\nonumber
&\chi_{10}^{(2)}(j_1,j_2)=\begin{cases}
-2 , & j_1<0,\; j_2 \ge 0,\\
0, & \text{otherwise}.
\end{cases}
\end{align}
The matrix elements of the spin operator $(\sigma_0^z-1)$ between the one-kink  Bloch states \eqref{Kis}
states can be easily found from \eqref{1K}, \eqref{chi1}. Due to \eqref{limK1}, the result yields
the matrix elements \eqref{1K} in the Ising limit $\eta\to\infty$:
\begin{subequations}\label{Yis}
\begin{align}
&Y_{01}(p|p')_{1/2}=
\frac{2}{1-\exp[2 i (p-p'+i0)]},\\
&Y_{10}(p|p')_{1/2}=
\frac{2 \exp[ i (p'-p)]}{1-\exp[2 i (p'-p+i0)]},\\
&Y_{01}(p|p')_{-1/2}=
-\frac{2 \exp[ i (p'-p)]}{1-\exp[2 i (p'-p-i0)]},\\
&Y_{10}(p|p')_{-1/2}=
-\frac{2}{1-\exp[2 i (p-p'-i0)]}.
\end{align}
\end{subequations}
In agreement with the general theory \cite{Jimbo94}, these one-kink matrix elements have the  simple poles at $e^{ip}=e^{ip'}$. The mathematical origin of these kinematic poles in the Ising limit is transparent from the above calculation. 

Similarly, one can calculate using \eqref{2K} the matrix elements of the operator $(\sigma_0^z-1)$  between the two-kink Bloch states \eqref{K2Is}. Due to \eqref{lim2K}, the result gives us the matrix elements \eqref{Y2},  \eqref{Yspm} in the Ising limit. This way, 
one  obtains at $\eta\to\infty$:
\begin{widetext}
\begin{subequations}\label{Y2Is}
\begin{align}\label{Y2ss}
& Y_{10}(p_2,p_1|p_1',p_2')_s=-\frac{1}{2 \bar\sigma}\bigg[
Y_{10}(p_1|p_1')_{s}Y_{01}(p_2|p_2')_{s}+\frac{w_0(p_1',p_2')}{w_0(p_1,p_2)}
Y_{01}(p_1|p_1')_{s}Y_{10}(p_2|p_2')_{s}+\\\nonumber
&w_0(p_1',p_2')\,Y_{10}(p_1|p_2')_{s}Y_{01}(p_2|p_1')_{s}+\frac{1}{w_0(p_1,p_2)}\,Y_{10}(p_2|p_1')_{s}Y_{01}(p_1|p_2')_{s}\bigg],\\
&Y_{10}(p_2,p_1|p_1',p_2')_{-1/2, 1/2|1/2, -1/2}=-\frac{1}{2\sin(p_1-p_1'+i0)\sin(p_2-p_2'-i0)}-
\frac{1}{2\sin(p_1-p_1'-i0)\sin(p_2-p_2'+i0)}+\nonumber\\
&\frac{1}{2\sin(p_1-p_2'+i0)\sin(p_2-p_1'-i0)}+
\frac{1}{2\sin(p_1-p_2'-i0)\sin(p_2-p_1'+i0)},\\
&Y_{10}(p_2,p_1|p_1',p_2')_{1/2,-1/2|-1/2, 1/2}=\exp[i(p_1'+p_2'-p_1-p_2)]\,Y_{10}(p_2,p_1|p_1',p_2')_{-1/2, 1/2|1/2, -1/2},\\
&Y_{10}(p_2,p_1|p_1',p_2')_{-1/2,1/2|-1/2, 1/2}=Y_{10}(p_2,p_1|p_1',p_2')_{1/2,-1/2|1/2, -1/2}=0,\\\label{Y2pm}
&Y_{10}(p_2,p_1|p_1',p_2')_{\pm}=\frac{1}{2}\,\left[Y_{10}(p_2,p_1|p_1',p_2')_{1/2,-1/2|-1/2, 1/2}+
Y_{10}(p_2,p_1|p_1',p_2')_{-1/2,1/2|1/2, -1/2}\right].
\end{align}
\end{subequations}
\end{widetext}
Of course, since formulas \eqref{Y2Is} relate to the Ising limit $\eta\to\infty$, one should use substitutions \eqref{Islim}, 
\eqref{Yis} for the quantities in the right-hand side of equation \eqref{Y2ss}.

The two-kink matrix elements \eqref{Y2Is} have the kinematic  simple  poles at $e^{i p_j}=e^{i p_k'}$, with $j,k=1,2$. 
After substitution of \eqref{Y2ss}, \eqref{Y2pm}  into   \eqref{G0pm}, and exploiting equation 
 \eqref{Giz}, one can easily show that equalities \eqref{rsG}, \eqref{singG} indeed hold in the Ising limit $\eta\to\infty$, 
 and furthermore:
\begin{equation}
\lim_{\eta\to\infty}\mathcal{G}_\iota^{(reg)}(z,z'|v)=0.
\end{equation}

Returning to the general case of arbitrary $\eta>0$, let us first relate the one- and two-kink matrix elements
of the operator $(\sigma_0^z-\bar\sigma)$  with their counterparts depending on the complex spectral 
parameter $\xi$:
\begin{widetext}
\begin{align}\label{1kink}
&\phantom{.}_{s}\langle K_{\nu\mu}(p)|(\sigma_0^z-\bar{\sigma})|K_{\mu\nu}(p')\rangle_{s}=
\frac{J\sinh\eta}{[{\omega(p)\omega(p')}]^{1/2}}\phantom{.}_{s}\langle \mathcal{K}_{\nu\mu}(\xi)|(\sigma_0^z-\bar{\sigma})|\mathcal{K}_{\mu\nu}(\xi')\rangle_{s},
\\\label{2Kink}
&\phantom{.}_{s_2,s_1}\langle{K}_{\mu\nu}(p_2){K}_{\nu\mu}(p_1)|(\sigma_0^z-\bar{\sigma})|{K}_{\mu\nu}(p'_1)
{K}_{\nu\mu}(p'_2)\rangle_{s'_1,s'_2}=
\frac{J^2\sinh^2\eta}{[\omega(p_1)\omega(p_2)\omega(p'_1)\omega(p'_2)]^{1/2}}\cdot\\\nonumber
&\phantom{.}_{s_2,s_1}\langle\mathcal{K}_{\mu\nu}(\xi_2)\mathcal{K}_{\nu\mu}(\xi_1)|(\sigma_0^z-\bar{\sigma})|\mathcal{K}_{\mu\nu}(\xi'_1)
\mathcal{K}_{\nu\mu}(\xi'_2)\rangle_{s'_1,s'_2}
\end{align}
\end{widetext}
In the above formulas, the momentum $p$ and $\xi$ variables are related  due to their parametric dependence 
on the rapidity   $\alpha$: by equation
\eqref{pe} for the former, and by the equality $\xi=-ie^{i\alpha}$ for the latter.

Using the crossing relation  \eqref{cross}, one can express the one- and two-kink matrix elements in terms of the two- and four-kink
form factors, respectively:
\begin{align}\label{K1f}
&\phantom{.}_{s}\langle \mathcal{K}_{\nu\mu}(\xi)|\sigma_0^z|\mathcal{K}_{\mu\nu}(\xi')\rangle_{s}=
f^\nu(\alpha',\alpha+i\eta)_{s,-s},\\\label{K2f}
&\phantom{.}_{s_2,s_1}\langle\mathcal{K}_{\mu\nu}(\xi_2)\mathcal{K}_{\nu\mu}(\xi_1)|\sigma_0^z|\mathcal{K}_{\mu\nu}(\xi'_1)
\mathcal{K}_{\nu\mu}(\xi'_2)\rangle_{s'_1,s'_2}=\\\nonumber
&f^\mu(\alpha_2',\alpha_1',\alpha_1+i\eta,\alpha_2+i\eta)_{s_2',s_1',-s_1,-s_2}.
\end{align}
Annihilation poles in the form factors in the right-hand side of equations \eqref{K1f}, \eqref{K2f} transform to the kinematic poles in the matrix elements
in the left-hand side of these equations.

For the one-kink matrix elements \eqref{1kink}, two initial terms in the Laurent  expansion in $(p-p')$ can be found from \eqref{K1f}, \eqref{form2},
\eqref{Xpl}, and \eqref{Xmn}:
\begin{subequations}\label{1Kpp}
\begin{align}
&\phantom{.}_{s }\!\langle {K}_{10}(p)|(\sigma_0^z\pm \bar{\sigma})|
{K}_{01}(p')\rangle_{s}=\\\nonumber
&\frac{i \bar{\sigma}}{p-p'\pm i0}+\left(s+ \frac{ \bar{\sigma}}{2}\right)+O(p-p'),\\
&\phantom{.}_{s }\!\langle {K}_{01}(p)|(\sigma_0^z\pm \bar{\sigma})|
{K}_{10}(p')\rangle_{s}=\\\nonumber
&\frac{i \bar{\sigma}}{p'-p\pm i0}+\left(s- \frac{ \bar{\sigma}}{2}\right)+O(p-p'),
\end{align}
\end{subequations}
where $s=\pm1/2$. We have added the infinitesimal shifts $\pm i 0$ in the pole terms in the right-hand sides to provide agreement of 
these equations  with formulas \eqref{Yis} in  the Ising limit $\eta\to \infty$. 

The structure of the kinematic singularities in the two-kink matrix elements \eqref{2Kink} can be recovered from
equations \eqref{K2f} and \eqref{resAn}. It turns out, in particular, that the structure of all kinematic singularities of 
the two-kink matrix element $Y_{10}(p_2,p_1|p_1',p_2')_s$ defined by equations \eqref{Y2spm}, \eqref{Y2} is completely
characterized  by equation \eqref{Y2ss} in the following sense:  at any $\eta>0$, the difference of the right- and left-hand sides of this
equation  is a regular function of   kink momenta $p_1,p_2,p_1',p_2'\in \mathbb{R}$. In other words, all kinematic poles of the 
matrix element $Y_{10}(p_2,p_1|p_1',p_2')_s$ at $\eta>0$ are contained in the right-hand side of equation \eqref{Y2ss}.

Merging of two kinematic simple poles leads to the second order pole at $p\to p'$ in the function $\mathcal{Y}_{10}(p,p'|P)_s$ 
defined by equation \eqref{Y2spmA}.
Combining \eqref{Y2spmA}, \eqref{Y2ss}, with  \eqref{Ymn1}, \eqref{p12p}, and  \eqref{1Kpp}, one obtains:
\begin{align}\label{maYs}
&\mathcal{Y}_{10}(p,p'|P)_s=\frac{\bar{\sigma}}{2(p-p'-i0)^2}-\frac{i \,s}{p-p'-i0}+\\\nonumber
&\frac{w_0(p_1',p_2')}{w_0(p_1,p_2)}
\left[\frac{\bar{\sigma}}{2(p-p'+i0)^2}+\frac{i\,s}{p-p'+i0}\right]+O(1),
\end{align}
where the dropped terms are regular at $p\to p'$.
After summation over the spin $s$, the first-order pole terms cancel:
\begin{align}
&\sum_{s=\pm1/2 }\mathcal{Y}_{10}(p,p'|P)_s=\frac{\bar{\sigma}}{(p-p'-i0)^2}+\\\nonumber
&\frac{w_0(p_1',p_2')}{w_0(p_1,p_2)}
\frac{\bar{\sigma}}{(p-p'+i0)^2}+O(1).
\end{align}
This leads to the following structure of the second-order pole singularity of the integral kernel 
$G_0(p,p'|P)$:
 at $p\to p'$:
\begin{align}\label{G0si}
&G_0(p,p'|P)=
\frac{1}{4(p-p'-i0)^2}+\frac{i}{4(p-p'-i0)}+\\\nonumber
&\frac{W_0(p'|P)}{W_0(p|P)}\left[
\frac{1}{4(p-p'+i0)^2}-\frac{i}{4(p-p'+i0)}
\right]+O(1),
\end{align} 
where  the scattering amplitude $W_0(p|P)$ is given by \eqref{Wio}.

The function $G_0^{(sing)}(p,p'|P)$ introduced in equation \eqref{resi} must have the
following properties.
\begin{enumerate}
\item It satisfies the same symmetry relations \eqref{PerG}, \eqref{Gcon}, \eqref{R1G},  \eqref{R2G},
as the function $G_0(p,p'|P)$.
\item The function $G_0^{(sing)}(p,p'|P)$  is regular at $p,p'\in \mathbb{R}$ apart from the points 
$p=\pm p'+\pi l$, with  $l\in \mathbb{Z}$, where it has the second order poles.
\item Near the point $p=p'$, the structure of the singularity of this function must be described by equation 
\eqref{G0si}.
\end{enumerate}
These properties define the function  $G_0^{(sing)}(p,p'|P)$ uniquely up to addition of some  regular 
function of $p,p'$.  

One can easily see, that  the function  
\[
G_0^{(sing)}(p,p'|P)=\mathcal{G}_0^{(sing)}(z,z'|v)
\]
determined by equations \eqref{zzv}, and \eqref{singG} at $\iota=0$ indeed satisfies all the 
constraints  listed above. This completes derivation of formula \eqref{singG} in the case $\iota=0$ at arbitrary $\eta>0$.

 Derivation of formula \eqref{singG} in the cases of $\iota=\pm$ is quite similar, so we can be brief. The integral kernel
 $G_\pm(p,p'|P)$ given by equation \eqref{pmG} has the second order pole at 
 $p=p'$. For the singular  at $p\to p'$ part of this function, we obtained the  following formula:
\begin{align}
&G_\pm(p,p'|P)=\frac{1}{4(p-p'-i0)^2}+\\\nonumber
&\frac{W_\pm(p'|P)}{W_\pm(p|P)}
\frac{1}{4(p-p'+i0)^2}+O(1).
\end{align}
As in the analogous equation \eqref{G0si}, the dropped terms are regular at $p=p'$. Combining this result with the
symmetry relations \eqref{PerG}, \eqref{R1G},  \eqref{R2G} for the functions $G_\pm(p,p'|P)$ leads finally  to equations 
\eqref{singG} for $\iota=\pm$.
\section{Perturbative solutions of the integral equation \eqref{BSQ4}\label{PSIE}}
In this Appendix, we present some technical details of the perturbative solution of the  Bethe-Salpeter integral equation \eqref{BSQ4}, and calculate initial
terms of the small-$f$  expansions for the meson dispersion laws  $\tilde{E}_{\iota, n}(P)$ in different asymptotical regimes 
described in Section \ref{WK}.
\subsection{First semiclassical regime. \label{Sem1}}
In the first semiclassical regime, there are two well separated saddle points of the function ${\mathcal F}(z,\Lambda_\iota)$  in the unit circle $S_1$. One should distinguish the cases  $0\le P<\pi/2$, and $\pi/2< P<\pi$. We shall concentrate on  the first case
$0\le P<\pi/2$. 

\begin{figure}
\centering
\subfloat[ 
First semiclassical regime.
]
{\label{fig:S1}
\includegraphics[width=.9\linewidth]{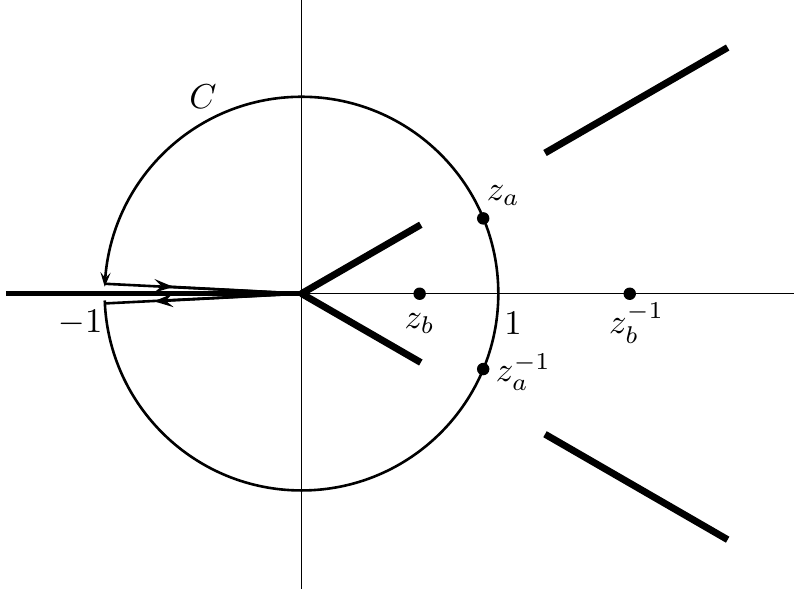}}

\subfloat[
Second semiclassical regime.
]
	{\label{fig:S3}
\includegraphics[width=.9\linewidth]{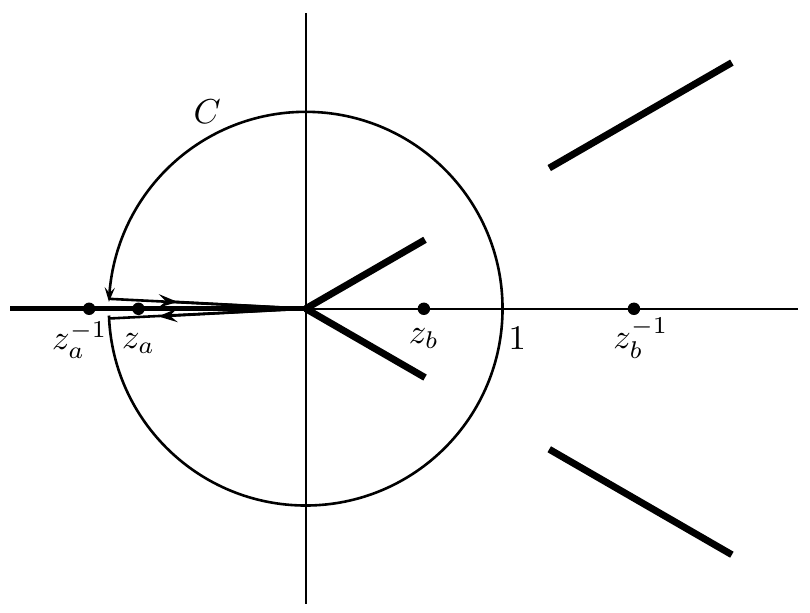}}

\subfloat[
Third semiclassical regime.
]
	{\label{fig:S2}
\includegraphics[width=.9\linewidth]{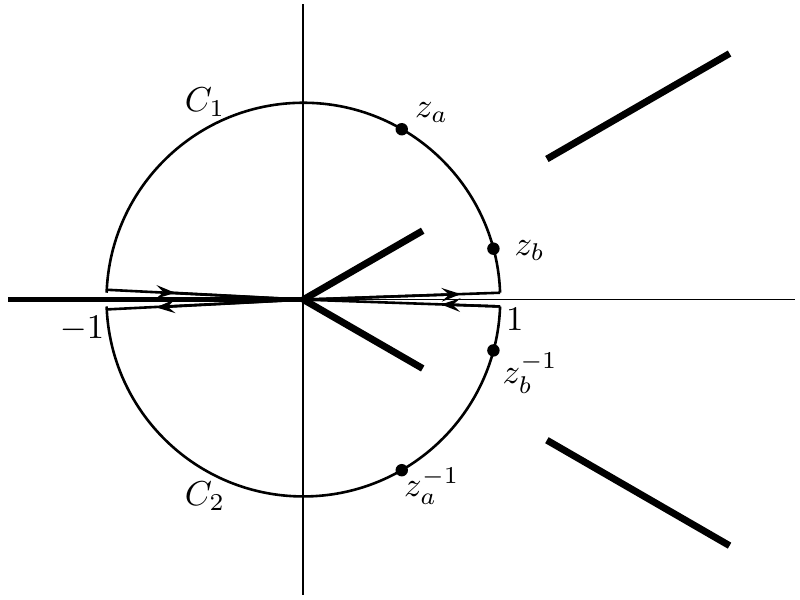}}

\caption{\label{figC12}  Integration contours  $C,C_1,C_2$  and saddle points 
$z_a,z_b,z_a^{-1},z_b^{-1}$ in equations \eqref{CA} and \eqref{C12A} at $0<P<\pi/2$.
Solid straight lines display the branching cuts of the function $\mathcal{F}(z,\Lambda_{\iota})$ defined by \eqref{Fdef}.}

\end{figure}

At $0\le P<\pi/2$, the first semiclassical regime is realized at $\mathcal{E}(z)|_{z=1}<\Lambda_\iota<\mathcal{E}(z)|_{z=-1}$. 
Configuration of four saddle points in the complex $z$-plane in this case is shown in Figure \ref{fig:S1}.
The saddle-points $z_b, z_b^{-1}$ are real, while  
$z_a, z_a^{-1}$ lie in the unit circle:  $|z_a|=1$. 

It follows from the definition \eqref{Fdef} of the function ${\mathcal F}(z,\Lambda_\iota)$, that it has, aside from
the six square-root branching points  \eqref{BP} common with those of the function $\mathcal{E}(z)$, also the 
logarithmic branching points at $z=0$, and at $z=\infty$. 

Accordingly, the function ${\mathcal F}(z,\Lambda_\iota)$ becomes single-valued in the physical sheet $\mathfrak{L}_{++}$, 
if we draw the extra branching cut  in it  along the negative real half-axis, as it is  
 shown in Figure \ref{figC12}. As it was mentioned in Section \ref{SIE}, we also put $z_1=1$ in equation  \eqref{Fdef}, so that 
 $\mathcal{F}(1,\Lambda_{\iota})=0$.
 
 Now let us turn to the equation
\begin{equation}\label{J12}
J(\Lambda_\iota)=0,
\end{equation}
with 
\begin{align}\nonumber
&J(\Lambda_\iota)=J_1(\Lambda_\iota)+J_2(\Lambda_\iota)=\\
\oint_{C}\frac{dz}{z}\,{ z^{-\delta_{\iota,0}/2}}\,
&U_\iota(z)
\exp\Big[\frac{i}{2f}\,{\mathcal F}(z,\Lambda_\iota)\Big],\label{CA}
\end{align}
where the integration contour $C=C_1+C_2$ is shown in Figure \ref{fig:S1}.
The small-$f$ asymptotic expansion of the integral \eqref{CA} is determined by contributions of two saddle points $z_a$ and $z_a^{-1}$. In order to calculate these contributions, one needs to find in the explicit form the small-$f$ expansion of the auxiliary function $U_\iota(z)$. 
The latter can be obtained following the procedure developed in references \cite{Rut08a,Rut09}, as it is described below.

Let us return to the Bethe-Salpeter equation \eqref{BSQ5} supplemented with the constraint \eqref{psiR}, and consider it now in the
 class of generalised functions in the unit circle $S_1$. 
 We denote by $\psi_\iota(z|f)$ the solution of this problem  to emphasize  its dependence on the parameter $f$, and expand it in the 
 latter  into the Neumann power series:
\begin{equation}\label{Neum}
\psi_\iota(z|f)=\mathcal{C}\left[\psi_\iota^{(0 )}(z)+\sum_{n=1}^\infty f^n \,\mathcal{A}_\iota^{(n)}(z)\right],
\end{equation}
where $\mathcal{C}$ is the normalisation constant, which will be determined later.
The leading term in this expansion can be easily found:
\begin{eqnarray}\label{psigen}
&\psi_\iota^{(0 )}(z)=\pi\Big[ \mathcal{W}_\iota(z_a)^{-1/2}\delta(p-p_a)+\\
&\mathcal{W}_\iota(z_a)^{1/2}\delta(p+p_a)\Big],\nonumber
\end{eqnarray}
where $p=\frac{\mathrm{arg}\,z}{2}$,
$p_a=\frac{\mathrm{arg}\,z_a}{2}$, $p\in(-\pi/2,\pi/2)$, $p_a\in(0,\pi/2)$. The  coefficients
$\mathcal{A}_\iota^{(n)}(z)$ can be, in principle, determined recursively from  \eqref{BSQ5}. After substitution of the 
expansion \eqref{Neum} into \eqref{gpm}, one obtains the Neumann expansions for the functions $g_{\iota \pm}(z|f)$:
\begin{equation}\label{Neumg}
g_{\iota \pm}(z|f)=\mathcal{C}\sum_{n=0}^\infty f^n\,\mathcal{B}_{\iota}^{(n)}(z),
\end{equation}
where
\begin{equation}\label{BC}
\mathcal{B}_{\iota}^{(0)}(z)=\frac{z_a\, \mathcal{W}_\iota(z_a)^{-1/2}}{z_a-z}+
\frac{z_a^{-1}\,  \mathcal{W}_\iota(z_a)^{1/2}}{z_a^{-1}-z},
\end{equation}
for $|z|\lessgtr 1$. In turn, substitution of \eqref{Neumg} either into \eqref{U}, or into \eqref{U1} yields the asymptotical expansion for the 
auxiliary function $U_\iota(z|f)$:
\begin{equation}\label{NeumU}
U_\iota(z|f)=\mathcal{C}\sum_{n=0}^\infty f^n\, \mathcal{U}_{\iota}^{(n)}(z),
\end{equation}
with
\begin{equation}\label{U0}
 \mathcal{U}_{\iota}^{(0)}(z)=  [\mathcal{E}(z)-\Lambda_\iota]\, \mathcal{B}_{\iota}^{(0)}(z).
\end{equation}
It is important to note, that the function $ \mathcal{U}_{\iota}^{(0)}(z)$ is regular at $z\in S_1$ together with all its higher derivatives,
in contrast with the functions  $\psi_\iota^{(0 )}(z)$, and $\mathcal{B}_{\iota}^{(0)}(z)$. It is possible to show, that the same is true
as well for the higher-order coefficients $ \mathcal{U}_{\iota}^{(n)}(z)$, with $n=1,2,\ldots$, in expansion \eqref{NeumU}. Note also the 
reflection relation
\begin{equation}\label{rU}
 \mathcal{U}_{\iota}^{(0)}(z_a^{-1})=-\mathcal{W}_\iota(z_a)\, \mathcal{U}_{\iota}^{(0)}(z_a),
\end{equation}
following from \eqref{U0}.

After substitution of \eqref{NeumU} into the integral  \eqref{CA}, one obtains in the straightforward fashion the asymptotic expansion 
 in $f\to+0$ for the left-hand side of \eqref{J12}:
\begin{equation}\label{J12A}
J(\Lambda_\iota|f)=2 i 
\sqrt{\frac{2 \pi f }{\epsilon'(p_a)}}\,
\sum_{n=0}^\infty f^n\,
\mathfrak{J}^{(n)}(\Lambda_\iota),
\end{equation}
with
\begin{align}\nonumber
&\frac{\mathfrak{J}^{(0)}(\Lambda_\iota)}{\mathcal{C}}=z_a^{-\delta_{\iota,0}/2}\,\mathcal{U}_{\iota}^{(0)}(z_a)\exp\left[
\frac{i \mathcal{F}(z_a,\Lambda_\iota)}{2 f}+\frac{i \pi}{4}
\right]+\\\label{J120}
&z_a^{\delta_{\iota,0}/2}\,\mathcal{U}_{\iota}^{(0)}(z_a^{-1})\exp\left[-
\frac{i \mathcal{F}(z_a,\Lambda_\iota)}{2 f}-\frac{i \pi}{4}
\right].
\end{align}
Equating \eqref{J12A} to zero and taking into account  \eqref{rU}, and  \eqref{J120},
one arrives to the  equation:
\begin{equation}
\exp\left[
\frac{i \mathcal{F}(z_a,\Lambda_\iota)}{ f}+\frac{i \pi}{2}
\right]=z_a^{\delta_{\iota,0}} \,\mathcal{W}_\iota(z_a)+O(f),
\end{equation}
which leads to the final expression  \eqref{Mdl}  for meson energy spectrum in the first semiclassical regime.

Let us now proceed to the calculation of  the normalisation constant 
$\mathcal{C}$, that stands in the right-hand sides of equations \eqref{Neum} and \eqref{Neumg}. 
At the first sight, one could  find $\mathcal{C}$ by substitution of the zero-order term \eqref{psigen} in the Neumann asymptotical expansion   
\eqref{Neum}  for the wave function $\psi_\iota(z|f)$ into the normalisation condition \eqref{psinorm}. However, this is not the case, since 
the product of the generalised function $\psi_\iota^{(0 )}(z)$ and its complex conjugate, that appears in the integrand in the normalisation
condition \eqref{psinorm}, is ill defined. Instead, we shall use the normalization condition in the form \eqref{normg}. We substitute in it
the integral representation \eqref{g+} for the auxiliary function $g_{\iota, n+}(z)$, in which the function  $U_\iota(z')$ is replaced 
by the zero-order term $\mathcal{C}\,\mathcal{U}_\iota^{(0)}(z')$ in its  expansion  \eqref{NeumU}: 
\begin{align}\label{g+0}
g_{\iota+}^{(0)}(z)=\frac{\mathcal{C}}{2f}\int_{\gamma_1(z)}\frac{dz'}{ z'}
\left(
\frac{z}{z'}
 \right)^{\delta_{\iota,0}/2}
\mathcal{U}_\iota^{(0)}(z')\cdot \\
\exp\Bigg\{\frac{i}{2f}\,[{\mathcal F}(z',\Lambda_\iota)-{\mathcal F}(z,\Lambda_\iota)]\Bigg\}.\nonumber
\end{align}

The integration in the $z'$-variable in equation \eqref{g+} runs  along the 
path $\gamma_1(z)$ lying in the physical sheet $\mathfrak{L}_{++}$ and connecting the points $0$ along the segments: 
\[
\gamma_1(z)=\left[0,e^{i (\pi-0)} \right]\cup\left[e^{i (\pi-0)} ,z]
\right].
\]
 Since we are interested in the case $\Lambda_\iota=
\Lambda_{\iota, n}$, we can replace due to \eqref{J12}, \eqref{CA}
the integration contour $\gamma_1(z)$ by the contour $\gamma_2(z)\subset \mathfrak{L}_{++}$: 
\[
\gamma_2(z)=C+\gamma_1(z)=\left[0,e^{i (\pi+0)} \right]\cup\left[e^{i (\pi+0)} ,z]
\right].
\]
At $f\to+0$, the  exponential factor in the integrand in the right-hand side of \eqref{g+0} highly oscillates in $z'\in S_1$, and the  asymptotical 
behaviour of the integral in \eqref{g+0} can be easily found by the steepest descent method. For leading asymptotics, we obtain this way
for  $z=e^{2 i p},$ $z_a=e^{2 i p_a}$:
\begin{align}\label{gio1}
&\frac{g_{\iota+}^{(0)}(z)}{\mathcal{C}}=\begin{cases}
\mathcal{B}_{\iota}^{(0)}(z),&\!\!  \text{if} \, p_a+\delta<|p|<\pi/2, \\
\sqrt{\frac{2\pi}{f \epsilon'(p_a)}}\,\mathcal{U}_\iota^{(0)}(z_a)\Xi(z),& \!\! \text{if} \, |p|<p_a-\delta,
\end{cases}\\\label{gio2}
&\frac{g_{\iota+}^{(0)}(z^{-1})}{\mathcal{C}}=\begin{cases}
\mathcal{B}_{\iota}^{(0)}(z^{-1}),&\!\!  \text{if} \, p_a+\delta<|p|<\pi/2, \\
\sqrt{\frac{2\pi}{f \epsilon'(p_a)}}\,\frac{\mathcal{U}_\iota^{(0)}(z_a^{-1})}{\Xi(z)},&\!\!  \text{if} \, |p|<p_a-\delta,
\end{cases}
\end{align}
where $\mathcal{B}_{\iota}^{(0)}(z)$ is given by \eqref{BC}, $\delta>0$ is some arbitrary small number independent of $f$,
and the phase factor
\begin{equation*}
\Xi(z)=\left(\frac{z}{z_a}\right)^{\delta_{\iota,0}/2}\exp\left\{\frac{i[\mathcal{F}(z_a,\Lambda_\iota)-\mathcal{F}(z,\Lambda_\iota)]}{2f}
-\frac{i\pi}{4}\right\}
\end{equation*}
highly oscillates in $z$.

Multiplying both sides of equations \eqref{gio1} and  \eqref{gio2}, and taking into account the formula
\begin{equation}
\mathcal{U}_\iota^{(0)}(z_a)\,\mathcal{U}_\iota^{(0)}(z_a^{-1})=\frac{[\epsilon'(p_a)]^2}{4}
\end{equation}
following from \eqref{U0} and \eqref{BC}, one obtains:
\begin{align}\label{g0pr}
&\frac{g_{\iota+}^{(0)}(z)g_{\iota+}^{(0)}(z^{-1})}{\mathcal{C}^2}=\\\nonumber
&\begin{cases}
\mathcal{B}_{\iota}^{(0)}(z)\mathcal{B}_{\iota}^{(0)}(z^{-1}),&  \text{if} \, p_a+\delta<|p|<\pi/2, \\
\frac{\pi\epsilon'(p_a)}{2f },& \text{if} \, |p|<p_a-\delta.
\end{cases}
\end{align}
At $f\to0$,  the right-hand side is large $\sim f^{-1}$ in the absolute value at $|p|<p_a-\delta$,
while in two intervals $p_a+\delta<p<\pi/2$ and $-\pi/2<p<-p_a-\delta$, this function is much smaller $\sim 1$. It is clear also, that 
the left-hand side of \eqref{g0pr} is of order $\sim f^{-1}$  in two narrow crossover regions $-p_a+\delta<|p|<p_a+\delta$.
 Therefore,  the main contribution 
 \[
 2(p_a-\delta)\, \frac{\pi\epsilon'(p_a)}{2f }\,\mathcal{C}^2
 \]
 to the normalisation integral in the left-hand side of \eqref{normg} arises from the interval $|p|<p_a-\delta$, where 
 the integrand almost does not depend on $p$. After sending $\delta\to0$, we get finally from \eqref{normg}:
\begin{equation}\label{normC}
\mathcal{C}^2=-\frac{f}{p_a \,\epsilon'(p_a)}.
\end{equation}
The obtained normalisation constant $\mathcal{C}$ being purely imaginary is determined by \eqref{normC} up to the sign. For the reduced 
wave function $\phi_\iota(p)$ defined by equations \eqref{phi0P}, \eqref{iot}    we get:
\begin{align}\nonumber
&\phi_{\iota}(p)=\pm i \pi \sqrt{\frac{f}{p_a \,\epsilon'(p_a)}}
\Big[{W}_\iota(p_a)^{-1/2}\delta(p-p_a)+\\
&{W}_\iota(p_a)^{1/2}\delta(p+p_a)\Big]+O(f^{3/2}).\label{phi0}
\end{align}
This result has been used in Section \ref{DSFconf} in calculations of the DSF in the confinement regime.
\subsection{Second  semiclassical regime \label{SR3}}
The second semiclassical regime is realized, if the energy $E$ and momentum $P$ of the meson  
fall well inside the region (II) shown in Figure \ref{fig:RegE}.  In this case, all four saddle points $z_a, z_a^{-1}, z_b, z_b^{-1}$ are real.
Figure \ref{fig:S3} displays their locations  at $0<P<\pi/2$, together with positions of the branching cuts 
of the function $\mathcal{F}(z,\Lambda_{\iota})$, and the integration contour $C$ in equation \eqref{CA}.
The small-$f$ asymptotics  of the integral  \eqref{CA} arises from the vicinity the saddle point $z_a\in (-1,0)$.
Two contributions of this saddle point cancel one another  in \eqref{CA}, if 
\[
\frac{{\mathcal F}(z,\Lambda_\iota)}{2f}\bigg|_{z=-1+i0}- \frac{{\mathcal F}(z,\Lambda_\iota)}{2f}\bigg|_{z=-1-i0}- \pi \,\delta_{\iota,0}=-2 \pi n,
\] 
with integer $n$. This requirement leads to the dispersion law
\begin{equation}
\tilde{E}_{\iota, n}=2n f+\frac{1}{\pi}\int_{-\pi/2}^{\pi/2}{dp} \, \epsilon(p|P)-f \delta_{\iota,0}, 
\end{equation}
or equivalently,  to formula \eqref{Sem3A}.
\subsection{Third semiclassical regime. \label{Sem2}}
In the third semiclassical regime, there are four well separated saddle points 
of the function ${\mathcal F}(z,\Lambda_\iota)$  in the unit circle $S_1$, see Figure \ref{fig:S2}. At $0\le P<\pi/2$ this regime  is realized at
$P_c(\eta)<P<\pi/2$ and $\epsilon_m(P,\eta)<\Lambda_\iota<\mathcal{E}(z)|_{z=1}$, where $P_c(\eta)$ and $\epsilon_m(P,\eta)$
are given by \eqref{Pc} and \eqref{epm}, respectively.  As in the first semiclassical regime at $0\le P<\pi/2$, we put $\mathcal{F}(1,\Lambda_\iota)=0$ and draw the extra branch cut 
in the physical sheet $\mathfrak{L}_{++}$ along the negative real half-axis.

The perturbation procedure described in Appendix \ref{Sem1}  should be slightly modified in the third semiclassical regime. 
Instead of equation \eqref{CA}, we have to use both constraints \eqref{C12A} with $\beta=1,2$. Equations \eqref{Neum}, \eqref{Neumg},
\eqref{NeumU}, and \eqref{U0} are still valid, but now we get:
\begin{align}\label{ps0}
&\psi_\iota^{(0 )}(z|f)=\pi\, \Big[\mathcal{C}_a \delta(p-p_a)+\mathcal{C}_b \delta(p-p_a)+\\
&\mathcal{C}_a\mathcal{W}_\iota(z_a)\,\delta(p+p_a)+\mathcal{C}_b\mathcal{W}_\iota(z_b)\,\delta(p+p_b)\Big],\nonumber\\
&\mathcal{B}_{\iota}^{(0)}(z)=\mathcal{C}_a 
\left[\frac{z_a}{z_a-z}+\frac{z_a^{-1}\, \mathcal{W}_\iota(z_a)}{z_a^{-1}-z}\right]+\\\nonumber
&\mathcal{C}_b 
\left[\frac{z_b}{z_b-z}+\frac{z_b^{-1}\, \mathcal{W}_\iota(z_b)}{z_b^{-1}-z}\right].
\end{align}
The  constants $\mathcal{C}_{a}$ and  $\mathcal{C}_{b}$ must satisfy the system of two  uniform  linear equations:
\begin{align}\label{LSys}
&-z_a^{-\delta_{\iota,0}/2}\sqrt{\epsilon'(p_a)}\exp\left[
\frac{i \mathcal{F}(z_a,\Lambda_\iota)}{2 f}+\frac{i \pi}{4}
\right] \,\mathcal{C}_a+\\\nonumber
&z_b^{-\delta_{\iota,0}/2}\sqrt{-\epsilon'(p_b)}\exp\left[
\frac{i \mathcal{F}(z_b,\Lambda_\iota)}{2 f}-\frac{i \pi}{4}
\right]\,\mathcal{C}_b=0,\\\nonumber
&z_a^{\delta_{\iota,0}/2}\sqrt{\epsilon'(p_a)}\exp\left[
\frac{-i \mathcal{F}(z_a,\Lambda_\iota)}{2 f}-\frac{i \pi}{4}
\right]\mathcal{W}_\iota(z_a)\,\mathcal{C}_a-\\\nonumber
&z_b^{\delta_{\iota,0}/2}\sqrt{-\epsilon'(p_b)}\exp\left[-
\frac{i \mathcal{F}(z_b,\Lambda_\iota)}{2 f}+\frac{i \pi}{4}
\right] \mathcal{W}_\iota(z_b) \,\mathcal{C}_b=0, 
\end{align}
that follows from \eqref{C12A}, \eqref{C12}  in the leading order in $f$.
After setting its determinant to zero, one obtains:
\begin{align}\label{DetC}
\exp\left(
\frac{i[\mathcal{F}(z_a,\Lambda_\iota)-\mathcal{F}(z_b,\Lambda_\iota)]}{ f}
\right)=-\frac{\mathcal{W}_\iota(z_a)\,z_a^{\delta_{\iota,0}} }{\mathcal{W}_\iota(z_b) z_b^{\delta_{\iota,0}} }.
\end{align}
This leads to the semiclassical meson energy spectrum $E_{\iota, n}(P)$ determined by equations \eqref{SE2}.

The ratio of the 
the coefficients $\mathcal{C}_{a}$ and $\mathcal{C}_{b}$ can be found from equations
\eqref{LSys}, \eqref{DetC},  and \eqref{SE2}:
\begin{equation}
\frac{\mathcal{C}_{a,n}}{\mathcal{C}_{b,n}}=(-1)^{n-1}\sqrt{\frac{-\epsilon'(p_b)}{\epsilon'(p_a)}}\left[\frac{W_\iota(p_b)}{W_\iota(p_a)}\right]^{1/2}.
\end{equation}
In order to  complete calculation of these coefficients, we  use the procedure described in Appendix  \ref{Sem1}, which exploits
 the normalisation condition \eqref{normg}. The result reads:
\begin{align}\label{Cab}
&\mathcal{C}_{a,n}=\varkappa_n \sqrt{\frac{f}{\epsilon'(p_a)(p_a-p_b)}}\,[W_\iota(p_a)]^{-1/2},\\\nonumber
&\mathcal{C}_{b,n}=(-1)^{n-1}\varkappa_n\sqrt{\frac{f}{-\epsilon'(p_b)(p_a-p_b)}}\,[W_\iota(p_b)]^{-1/2},
\end{align}
where $\varkappa_n=\pm1$ is the common sign factor of both coefficients that remains undetermined. 
\subsection{First low-energy expansion. \label{LE}}
The semiclassical regimes described above are realized  at small $f$ at generic values of parameters
$P$ and $\Lambda$, since in this case the solutions of equation \eqref{ELam}   are well separated from each other. On the other hand, three
low-energy and three crossover regimes take place,
when $\Lambda$ approaches some critical value of the function $\mathcal{E}(z)$, at which two or four solutions of \eqref{ELam} merge in $S_1$. 

The first low-energy regime is realized at $0\le P< P_c(\eta)$ and $\Lambda$ slightly above $\mathcal{E}(z)\big|_{z=1}$. In this case, two saddle points $z_a,z_a^{-1}\in S_1$ shown in Figure \ref{fig:S1} approach the value $z=1$. The perturbative calculation of the energy spectrum
$\Lambda_{\iota, n}$ in this regime is  based on equations \eqref{J12}, \eqref{CA}.
The small-$f$ asymptotics of the integral in \eqref{CA} is determined be the 
contribution of the degenerate saddle point $z=1$.
In order to calculate this contribution, 
we proceed to 
the integration variable $p=\frac{\mathrm{arg}\, z}{2}$ and  replace  the functions  $\mathcal{F}(z,\Lambda)$,  $U_\iota(z)$, and $z^{-i\delta_{\iota,0}/2}$   in \eqref{CA}
by two initial terms  in their Taylor expansions in $p$:
\begin{subequations}\label{smallp}
\begin{align}
&\mathcal{F}(z,\Lambda)=-2 p\,\delta \Lambda+\frac{\epsilon''(0)}{3}p^3+O(p^5),\\\label{Uio}
&{U}_\iota(z)=
{U}_\iota(1)
\left[1+ i \mathfrak{c}_{\iota} p+O(p^2)\right],\\&
z^{-i\delta_{\iota,0}/2}=1-i \delta_{\iota,0}\, p+O(p^2).
\end{align}
\end{subequations}
where $z=\exp(2 i p)$,  $\delta \Lambda=\Lambda-\epsilon(0)$,  and $\epsilon(p)$ is 
determined by \eqref{Energ1}. Then, one obtains  for the saddle point asymptotics of the integral 
$J(\Lambda_\iota)$:
\begin{align}\nonumber
&\frac{J(\Lambda_\iota)}{2 i \,{U}_\iota(1)}\cong \int_{-\infty}^\infty dp\, (1+i\,  \tilde{\mathfrak{c}}_{\iota})
e^{
\frac{i}{f}\left[
- p\,\delta \Lambda+\frac{\epsilon''(0)}{6}p^3
\right]
}=\\
&Y(\delta \Lambda_\iota)- \tilde{\mathfrak{c}}_{\iota}\,f\,Y'(\delta \Lambda_\iota)=
Y(\delta \Lambda_\iota- \tilde{\mathfrak{c}}_{\iota}\,f)+O(f^2),\label{Jas}
\end{align}
where
\begin{align}
&\tilde{\mathfrak{c}}_{\iota}={\mathfrak{c}}_{\iota}-\delta_{\iota,0},\\\nonumber
&Y( x)=\int_{-\infty}^\infty dp\, e^{
\frac{i}{f}\left[
- p\,x+\frac{\epsilon''(0)}{6}p^3
\right]
}=\\
&\left(
\frac{2f}{\epsilon''(0)}
\right)^{1/3} 2\pi\, \mathrm{Ai}\,\left(-f^{-2/3}[2/\epsilon''(0)]^{1/3} x\right),
\end{align}
and $\mathrm{Ai}\,(u)$ is the Airy function.

Equating the right-hand side of \eqref{Jas} to zero, we obtain two initial terms of the 
small-$f$ asymptotical expansion for the discrete set $\{ \delta \Lambda_{\iota, n}\}_{n=1}^\infty$ of allowed values of the parameter  $\delta \Lambda_{\iota}$:
\begin{equation}\label{dLa}
\delta \Lambda_{\iota, n}=f^{2/3} [\epsilon''(0)/2]^{1/3} z_n+ \tilde{\mathfrak{c}}_\iota f+\ldots.
\end{equation}
where $-z_n$ denote the zeros of the Airy function, 
$\mathrm{Ai}\,(-z_n)=0$, and $z_{n+1}>z_n$.

The coefficient $\mathfrak{c}_{\iota}$ in \eqref{Uio} depends, besides the other parameters, on the string tension $f$. Its limiting value at $f=0$ is given by the  relation:
\begin{equation}\label{limc}
\lim_{f\to+0}\mathfrak{c}_{\iota }(f)=\frac{i}{2}\,\partial_p\ln W_\iota(p)\Big|_{p=0} .
\end{equation}
This relation, together with \eqref{Wio} and \eqref{scph}, leads to the following formula  for the coefficient $\tilde{c}_\iota(f)$ in the limit $f\to+0$:
\begin{equation}
\lim_{f\to+0}\tilde{\mathfrak{c}}_{\iota }(f)=a_\iota(P),\label{lima}
\end{equation}
where the scattering length $a_\iota(P)$  is given by \eqref{scL}.

In order to prove  equality \eqref{limc}, let us note that the Bethe-Salpeter equation  \eqref{BSQ5}
degenerates at  $f=0$ and $\Lambda_\iota=\mathcal{E}(1)$ to the form:
\begin{equation}
[\mathcal{E}(z)-\mathcal{E}(1)]\psi_\iota(z)=0.
\end{equation}
Its formal solution satisfying the symmetry relation \eqref{psiR} reads,
\begin{equation}\label{psi0}
\psi_\iota(z)\Big|_{z=\exp(2 i p)}=\pi C\left(
\delta'(p)+\frac{\delta(p)}{2}\, \frac{W_\iota'(0)}{W_\iota(0)}\right),
\end{equation}
with some arbitrary constant $C$. Note, that $W_\iota(0)=-1$, as one can see 
from \eqref{Wio}, \eqref{scph}. Substitution of \eqref{psi0} 
 into \eqref{gpm}  yields the auxiliary functions $g_\pm(z)$:
\begin{equation*}
g_{\iota\pm}(z)=C\left[
\frac{2 i z}{(1-z)^2}+\frac{1}{2}\, \frac{W_\iota'(0)}{W_\iota(0)}\frac{1}{1-z}
\right].
\end{equation*}
The corresponding function $U_\iota(z)$ is given then  by  equation \eqref{U} at $f=0$ and 
$\Lambda_\iota=\mathcal{E}(1)$:
\begin{equation*}
U_\iota(z)=[\mathcal{E}(z)-\mathcal{E}(1)]\,g_{\iota\pm}(z).
\end{equation*}
Expanding this function in  $p$ at $p\to0$, one arrives at the equality
\[
U_\iota(z)\Big|_{z=\exp(2 i p)}=- \frac{i C \epsilon''(0)}{4} \left[1-\frac{p}{2}\frac{W_\iota'(0)}{W_\iota(0)}+O(p^2)\right],
\]
that completes proof  of \eqref{limc}.

Combining \eqref{dLa} with \eqref{lima}, we obtain formula \eqref{les} for the meson dispersion law in the first low-energy
regime. 
\subsection{Second low-energy expansion. \label{LE2}}
In  this Section, we obtain two initial terms in the second  low-energy expansion that describes the meson energy spectra slightly 
above the red dashed curves bounding from below
the regions (III)  in Figure  \ref{fig:RegE}. Our analysis will be restricted to the case of the 
meson momenta in the interval   $P\in (P_c,\pi/2)$.

We start from the Bethe-Salpeter equation in the form \eqref{BSQg}, and 
 simplify it in the vicinity of the points $p=p_m$, $p'=p_m$ at energies  $\tilde{E}_\iota$ close to 
the lower bound  $\epsilon(p_m)$, see  equations   \eqref{epm}, \eqref{pmin}, and Figure \ref{fig:2Lac}.
To this end, we proceed in \eqref{BSQg} to the rescaled energy $\mathfrak{e}_\iota$ 
and momentum  variables $\mathfrak{p}$, $\mathfrak{p}'$ defined in the following way: 
\begin{subequations}\label{rEp}
\begin{align}\label{pE}
&p=p_m+t\, \mathfrak{p},\quad p'=p_m+t\, \mathfrak{p}',\\\label{tE}
&\tilde{E}_\iota=\epsilon(p_m)+t^2 \mathfrak{e}_\iota,
\end{align}
\end{subequations}
where $t=f^{1/3}$ is a small parameter. Expanding the result  in $t$ to the first order, we obtain the reduced integral equation:
\begin{align}\label{BSE}
&\left[\frac{\epsilon''(p_m)}{2}\mathfrak{p}^2-\mathfrak{e}_\iota-t \delta_{\iota,0}+t\frac{\epsilon'''(p_m)}{6}\mathfrak{p}^3\right]
\varphi_\iota(\mathfrak{p})=\\\nonumber
&\dashint_{-\infty}^\infty \frac{d \mathfrak{p}}{\pi}\frac{\varphi_\iota(\mathfrak{p}')}{(\mathfrak{p}'-\mathfrak{p})^2}+
t \frac{W_\iota'(p_m)}{W_\iota(p_m)}\int_{-\infty}^\infty \frac{d \mathfrak{p}}{2\pi}\frac{\varphi_\iota(\mathfrak{p}')}{(\mathfrak{p}'-\mathfrak{p}+i0)},
\end{align}
where  $\mathfrak{p},\mathfrak{p}'\in \mathbb{R}$, 
$\varphi_\iota(\mathfrak{p})=\phi_\iota(p_m+t\, \mathfrak{p})$,   and $\dashint$ denotes the integral in the sense of the principal value.
This singular linear integral equation can be solved using the procedure described in Appendixes \ref{Sem1},  \ref{Sem2}.

We introduce two auxiliary functions $\mathfrak{g}_{\iota+}(\mathfrak{p})$ and $\mathfrak{g}_{\iota-}(\mathfrak{p})$,
which are analytical in the half planes  $\mathrm{Im}\,\mathfrak{p}> 0$ and $\mathrm{Im}\,\mathfrak{p}<0$, respectively:
\begin{equation}
\mathfrak{g}_{\iota\pm}(\mathfrak{p})=\int_{-\infty}^\infty\frac{d \mathfrak{p}'}{2\pi i}\frac{\varphi_\iota(\mathfrak{p}')}{\mathfrak{p}'-\mathfrak{p}}, 
\quad \text{for  } \mathrm{Im}\,\mathfrak{p}\gtrless 0.
\end{equation}
In the regions of their analyticity, these functions decay at large $|\mathfrak{p}|\to\infty$ as $O(|\mathfrak{p}|^{-1})$.
The third auxiliary function $\mathfrak{U}_\iota(\mathfrak{p},t)$ defined by equation 
\begin{equation}\label{Upl}
\mathfrak{U}_\iota(\mathfrak{p},t)=\!\!\Big[-i\partial_{\mathfrak{p}}+\frac{\epsilon''(p_m)}{2}\mathfrak{p}^2-{\mathfrak{e}}_\iota
-t\delta_{\iota,0}+\\ t\frac{\epsilon'''(p_m)}{6}\mathfrak{p}^3\Big]
\mathfrak{g}_{\iota+}(\mathfrak{p}),
\end{equation}
admits due to \eqref{BSE} the alternative representation in terms of the function $\mathfrak{g}_{\iota-}(\mathfrak{p})$:
\begin{align}\label{Umn}
&\mathfrak{U}_\iota(\mathfrak{p},t)=\Big[i\partial_{\mathfrak{p}}+\frac{\epsilon''(p_m)}{2}\mathfrak{p}^2-{\mathfrak{e}}_\iota
-t\delta_{\iota,0}+\\\nonumber
&t\frac{\epsilon'''(p_m)}{6}\mathfrak{p}^3
+i t\frac{W_\iota'(p_m)}{W_\iota(p_m)}
\Big]\mathfrak{g}_{\iota-}(\mathfrak{p}).
\end{align}
It follows from \eqref{Upl}, \eqref{Umn}, that at a fixed $t$, the function $\mathfrak{U}_\iota(\mathfrak{p},t)$ is analytical in $\mathfrak{p}$ in the 
whole complex plane, and increases at $\mathfrak{p}\to\infty$ not faster than $C |\mathfrak{p}|^2$, with some constant $C>0$. 
Therefore, this function is just a second-order polynomial
\begin{equation}\label{Ud}
\mathfrak{U}_\iota(\mathfrak{p},t)=\mathfrak{d}_0(t)+\mathfrak{d}_1(t) \mathfrak{p}+\mathfrak{d}_2(t) \mathfrak{p}^2,
\end{equation}
with coefficients $\mathfrak{d}_{j}(t)$, regularly depending on $t$.

Solving the differential equations \eqref{Upl}, \eqref{Umn} with respect to the functions $\mathfrak{g}_{\iota\pm}(\mathfrak{p})$, 
one obtains:
\begin{align}\label{gpm4}
\mathfrak{g}_{\iota\pm}(\mathfrak{p})=\pm i \int_{\mathfrak{p}_0}^\mathfrak{p} d\mathfrak{p}'\mathfrak{U}_\iota(\mathfrak{p}',t)e^{\pm i[\mathfrak{F}_\pm(\mathfrak{p}',t)-\mathfrak{F}_\pm(\mathfrak{p},t)]},
\end{align}
where 
\begin{align}
&\mathfrak{F}_+(\mathfrak{p},t,{\mathfrak{e}}_\iota)=\mathfrak{F}_0(\mathfrak{p},{\mathfrak{e}}_\iota)
+t\frac{\epsilon'''(p_m)}{24}\mathfrak{p}^4-t\delta_{\iota,0}\mathfrak{p},\\
&\mathfrak{F}_-(\mathfrak{p},t,{\mathfrak{e}}_\iota)=\mathfrak{F}_+(\mathfrak{p},t,{\mathfrak{e}}_\iota)+ it \frac{W_\iota'(p_m)}{W_\iota(p_m)} \,\mathfrak{p},\\
&\mathfrak{F}_0(\mathfrak{p},{\mathfrak{e}}_\iota)=\frac{\epsilon''(p_m)}{6}\mathfrak{p}^3-{\mathfrak{e}}_\iota \,\mathfrak{p}.
\end{align}
The lower  integration limit  $\mathfrak{p}_0$ in the integral  in \eqref{gpm4} must guarantee, that the functions 
$\mathfrak{g}_{\iota\pm}(\mathfrak{p})$ determined by the right-hand side of this equation are analytical at 
$\mathrm{Im}\,\mathfrak{p}\gtrless 0$. Two appropriate choices are $\mathfrak{p}_0=\pm\infty$. 

The uniqueness requirement 
for  the solution of the integral equation \eqref{BSE} leads to the constraints:
\begin{equation}\label{constrU}
 \int_{-\infty}^\infty d\mathfrak{p}\,\mathfrak{U}_\iota(\mathfrak{p},t)e^{\pm i \,\mathfrak{F}_\pm(\mathfrak{p},t)}=0.
\end{equation}
These two constraints allow one to determine the small-$t$ asymptotics of the eigenvalues ${\mathfrak{e}}_{\iota,n}(t)$
of the eigenvalue problem \eqref{BSE} to the linear order in $t$. To this end, let us expand the left-hand side 
of \eqref{constrU} in $t$ to the linear order, using the 
following substitutions for the eigenvalue ${\mathfrak{e}}_{\iota,n}(t)$, and 
 coefficients  $\mathfrak{d}_i(t)$:
\begin{align}\label{be}
&{\mathfrak{e}}_{\iota,n}(t)={\mathfrak{e}}_{\iota, n}^{(0)}+t\, b_ {\iota, n}+O(t^2),\\
&\mathfrak{d}_0(t)=\mathfrak{d}_{00}+t \mathfrak{d}_{01}+O(t^2),\\ \nonumber
&\mathfrak{d}_1(t)=\mathfrak{d}_{10}+t \mathfrak{d}_{11}+O(t^2), \\ \nonumber
&\mathfrak{d}_2(t)=t \mathfrak{d}_{21}+O(t^2).
\end{align}
In the zero order in $t$, we obtain this way from \eqref{constrU} two equations
\begin{equation}
 \int_{-\infty}^\infty d\mathfrak{p}\,(\mathfrak{d}_{00}+\mathfrak{d}_{10}\mathfrak{p}) 
 e^{ \pm i \,\mathfrak{F}_0(\mathfrak{p},{\mathfrak{e}}_{\iota, n}^{(0)})}
 =0,
\end{equation}
which admit two series of solutions.
\begin{enumerate}
\item
{Bose-type} solutions:
\begin{align}
&\mathfrak{d}_{00}=0,\quad \mathfrak{d}_{10}\ne0,\\\label{eodd}
&  \mathfrak{e}_{\iota, n}^{(0)}=\left[\frac{\epsilon''(p_m)}{2}\right]^{1/3}z_{(n+1)/2}',
\end{align}
with odd $n=1,3,5,\ldots$, and $(-z_l')$ being the zeros of the derivative of the  Airy functions, $\mathrm{Ai}'\,(-z_l')=0$. 
\item
{Fermi-type} solutions:
\begin{align}
&\mathfrak{d}_{10}=0,\quad \mathfrak{d}_{00}\ne0,\\\label{eev}
 & \mathfrak{e}_{\iota, n}^{(0)}=\left[\frac{\epsilon''(p_m)}{2}\right]^{1/3}z_{n/2}.
\end{align}
with even $n=2,4,6,\ldots$, and $(-z_l)$ being the zeros of  the  Airy functions, $\mathrm{Ai}\,(-z_l)=0$. 
\end{enumerate}
In order to determine the  coefficient $b_ {\iota, n}$ in \eqref{be}, one should
equate to zero the first-order terms in the expansion of the left-hand side of \eqref{constrU} in $t$.
The final result reads:
\begin{align}\label{b1}
&b_ {\iota, n}=-\delta_{\iota,0}+\frac{i}{2}\frac{W_\iota'(p_m)}{W_\iota(p_m)} =\\\nonumber
&-\frac{1}{2}\partial_p \theta_\iota(P/2+p,P/2-p)\Big|_{p=p_m}.
\end{align}
Equations \eqref{tE}, \eqref{eodd}, \eqref{eev}, \eqref{b1} lead to the 
second  low-energy expansion \eqref{slee} for the meson dispersion law.
Although we have  obtained 
 formula \eqref{slee}  for $P\in(P_c,\pi/2)$, it 
holds in fact in the wider interval of the meson momentum: $P\in(P_c,\pi-P_c)$.

\begin{thebibliography}{62}%
\makeatletter
\providecommand \@ifxundefined [1]{%
 \@ifx{#1\undefined}
}%
\providecommand \@ifnum [1]{%
 \ifnum #1\expandafter \@firstoftwo
 \else \expandafter \@secondoftwo
 \fi
}%
\providecommand \@ifx [1]{%
 \ifx #1\expandafter \@firstoftwo
 \else \expandafter \@secondoftwo
 \fi
}%
\providecommand \natexlab [1]{#1}%
\providecommand \enquote  [1]{``#1''}%
\providecommand \bibnamefont  [1]{#1}%
\providecommand \bibfnamefont [1]{#1}%
\providecommand \citenamefont [1]{#1}%
\providecommand \href@noop [0]{\@secondoftwo}%
\providecommand \href [0]{\begingroup \@sanitize@url \@href}%
\providecommand \@href[1]{\@@startlink{#1}\@@href}%
\providecommand \@@href[1]{\endgroup#1\@@endlink}%
\providecommand \@sanitize@url [0]{\catcode `\\12\catcode `\$12\catcode
  `\&12\catcode `\#12\catcode `\^12\catcode `\_12\catcode `\%12\relax}%
\providecommand \@@startlink[1]{}%
\providecommand \@@endlink[0]{}%
\providecommand \url  [0]{\begingroup\@sanitize@url \@url }%
\providecommand \@url [1]{\endgroup\@href {#1}{\urlprefix }}%
\providecommand \urlprefix  [0]{URL }%
\providecommand \Eprint [0]{\href }%
\providecommand \doibase [0]{http://dx.doi.org/}%
\providecommand \selectlanguage [0]{\@gobble}%
\providecommand \bibinfo  [0]{\@secondoftwo}%
\providecommand \bibfield  [0]{\@secondoftwo}%
\providecommand \translation [1]{[#1]}%
\providecommand \BibitemOpen [0]{}%
\providecommand \bibitemStop [0]{}%
\providecommand \bibitemNoStop [0]{.\EOS\space}%
\providecommand \EOS [0]{\spacefactor3000\relax}%
\providecommand \BibitemShut  [1]{\csname bibitem#1\endcsname}%
\let\auto@bib@innerbib\@empty
\bibitem [{\citenamefont {Narison}(2004)}]{Nar04}%
  \BibitemOpen
  \bibfield  {author} {\bibinfo {author} {\bibfnamefont {S.}~\bibnamefont
  {Narison}},\ }\href {\doibase https://doi.org/10.1017/CBO9780511535000}
  {\emph {\bibinfo {title} {QCD as a Theory of Hadrons}}}\ (\bibinfo
  {publisher} {Cambridge University Press},\ \bibinfo {address} {Cambridge},\
  \bibinfo {year} {2004})\BibitemShut {NoStop}%
\bibitem [{\citenamefont {Coldea}\ \emph {et~al.}(2010)\citenamefont {Coldea},
  \citenamefont {Tennant}, \citenamefont {Wheeler}, \citenamefont {Wawrzynska},
  \citenamefont {Prabhakaran}, \citenamefont {Telling}, \citenamefont
  {Habicht}, \citenamefont {Smeibidl},\ and\ \citenamefont
  {Kiefer}}]{Coldea10}%
  \BibitemOpen
  \bibfield  {author} {\bibinfo {author} {\bibfnamefont {R.}~\bibnamefont
  {Coldea}}, \bibinfo {author} {\bibfnamefont {D.~A.}\ \bibnamefont {Tennant}},
  \bibinfo {author} {\bibfnamefont {E.~M.}\ \bibnamefont {Wheeler}}, \bibinfo
  {author} {\bibfnamefont {E.}~\bibnamefont {Wawrzynska}}, \bibinfo {author}
  {\bibfnamefont {D.}~\bibnamefont {Prabhakaran}}, \bibinfo {author}
  {\bibfnamefont {M.}~\bibnamefont {Telling}}, \bibinfo {author} {\bibfnamefont
  {K.}~\bibnamefont {Habicht}}, \bibinfo {author} {\bibfnamefont
  {P.}~\bibnamefont {Smeibidl}}, \ and\ \bibinfo {author} {\bibfnamefont
  {K.}~\bibnamefont {Kiefer}},\ }\href {\doibase 10.1126/science.1180085}
  {\bibfield  {journal} {\bibinfo  {journal} {Science}\ }\textbf {\bibinfo
  {volume} {327}},\ \bibinfo {pages} {177} (\bibinfo {year}
  {2010})}\BibitemShut {NoStop}%
\bibitem [{\citenamefont {Morris}\ \emph {et~al.}(2014)\citenamefont {Morris},
  \citenamefont {Vald\'es~A.}, \citenamefont {Ghosh}, \citenamefont
  {Koohpayeh}, \citenamefont {Krizan}, \citenamefont {Cava}, \citenamefont
  {Tchernyshyov}, \citenamefont {McQueen},\ and\ \citenamefont
  {Armitage}}]{Mor14}%
  \BibitemOpen
  \bibfield  {author} {\bibinfo {author} {\bibfnamefont {C.~M.}\ \bibnamefont
  {Morris}}, \bibinfo {author} {\bibfnamefont {R.}~\bibnamefont {Vald\'es~A.}},
  \bibinfo {author} {\bibfnamefont {A.}~\bibnamefont {Ghosh}}, \bibinfo
  {author} {\bibfnamefont {S.~M.}\ \bibnamefont {Koohpayeh}}, \bibinfo {author}
  {\bibfnamefont {J.}~\bibnamefont {Krizan}}, \bibinfo {author} {\bibfnamefont
  {R.~J.}\ \bibnamefont {Cava}}, \bibinfo {author} {\bibfnamefont
  {O.}~\bibnamefont {Tchernyshyov}}, \bibinfo {author} {\bibfnamefont {T.~M.}\
  \bibnamefont {McQueen}}, \ and\ \bibinfo {author} {\bibfnamefont {N.~P.}\
  \bibnamefont {Armitage}},\ }\href {\doibase 10.1103/PhysRevLett.112.137403}
  {\bibfield  {journal} {\bibinfo  {journal} {Phys. Rev. Lett.}\ }\textbf
  {\bibinfo {volume} {112}},\ \bibinfo {pages} {137403} (\bibinfo {year}
  {2014})},\ \Eprint {http://arxiv.org/abs/arXiv:1312.4514} {arXiv:1312.4514}
  \BibitemShut {NoStop}%
\bibitem [{\citenamefont {Fava}\ \emph {et~al.}(2020)\citenamefont {Fava},
  \citenamefont {Coldea},\ and\ \citenamefont {Parameswaran}}]{Coldea20}%
  \BibitemOpen
  \bibfield  {author} {\bibinfo {author} {\bibfnamefont {M.}~\bibnamefont
  {Fava}}, \bibinfo {author} {\bibfnamefont {R.}~\bibnamefont {Coldea}}, \ and\
  \bibinfo {author} {\bibfnamefont {S.~A.}\ \bibnamefont {Parameswaran}},\
  }\href {\doibase 10.1073/pnas.2007986117} {\bibfield  {journal} {\bibinfo
  {journal} {Proceedings of the National Academy of Sciences}\ }\textbf
  {\bibinfo {volume} {117}},\ \bibinfo {pages} {25219} (\bibinfo {year}
  {2020})},\ \bibinfo {note} {arXiv:2004.04169v2}\BibitemShut {NoStop}%
\bibitem [{\citenamefont {Sachdev}(1999)}]{Sach99}%
  \BibitemOpen
  \bibfield  {author} {\bibinfo {author} {\bibfnamefont {S.}~\bibnamefont
  {Sachdev}},\ }\href {\doibase https://doi.org/10.1017/CBO9780511973765}
  {\emph {\bibinfo {title} {Quantum Phase Transitions}}}\ (\bibinfo
  {publisher} {Cambridge University Press},\ \bibinfo {address} {Cambridge},\
  \bibinfo {year} {1999})\BibitemShut {NoStop}%
\bibitem [{\citenamefont {Wang}\ \emph {et~al.}(2015)\citenamefont {Wang},
  \citenamefont {Schmidt}, \citenamefont {Bera}, \citenamefont {Islam},
  \citenamefont {Lake}, \citenamefont {Loidl},\ and\ \citenamefont
  {Deisenhofer}}]{Wang15}%
  \BibitemOpen
  \bibfield  {author} {\bibinfo {author} {\bibfnamefont {Z.}~\bibnamefont
  {Wang}}, \bibinfo {author} {\bibfnamefont {M.}~\bibnamefont {Schmidt}},
  \bibinfo {author} {\bibfnamefont {A.~K.}\ \bibnamefont {Bera}}, \bibinfo
  {author} {\bibfnamefont {A.~T. M.~N.}\ \bibnamefont {Islam}}, \bibinfo
  {author} {\bibfnamefont {B.}~\bibnamefont {Lake}}, \bibinfo {author}
  {\bibfnamefont {A.}~\bibnamefont {Loidl}}, \ and\ \bibinfo {author}
  {\bibfnamefont {J.}~\bibnamefont {Deisenhofer}},\ }\href {\doibase
  10.1103/PhysRevB.91.140404} {\bibfield  {journal} {\bibinfo  {journal} {Phys.
  Rev. B}\ }\textbf {\bibinfo {volume} {91}},\ \bibinfo {pages} {140404}
  (\bibinfo {year} {2015})},\ \Eprint {http://arxiv.org/abs/arXiv:1503.06351}
  {arXiv:1503.06351} \BibitemShut {NoStop}%
\bibitem [{\citenamefont {Bera}\ \emph {et~al.}(2017)\citenamefont {Bera},
  \citenamefont {Lake}, \citenamefont {Essler}, \citenamefont {Vanderstraeten},
  \citenamefont {Hubig}, \citenamefont {Schollw\"ock}, \citenamefont {Islam},
  \citenamefont {Schneidewind},\ and\ \citenamefont
  {Quintero-Castro}}]{Bera17}%
  \BibitemOpen
  \bibfield  {author} {\bibinfo {author} {\bibfnamefont {A.~K.}\ \bibnamefont
  {Bera}}, \bibinfo {author} {\bibfnamefont {B.}~\bibnamefont {Lake}}, \bibinfo
  {author} {\bibfnamefont {F.~H.~L.}\ \bibnamefont {Essler}}, \bibinfo {author}
  {\bibfnamefont {L.}~\bibnamefont {Vanderstraeten}}, \bibinfo {author}
  {\bibfnamefont {C.}~\bibnamefont {Hubig}}, \bibinfo {author} {\bibfnamefont
  {U.}~\bibnamefont {Schollw\"ock}}, \bibinfo {author} {\bibfnamefont {A.~T.
  M.~N.}\ \bibnamefont {Islam}}, \bibinfo {author} {\bibfnamefont
  {A.}~\bibnamefont {Schneidewind}}, \ and\ \bibinfo {author} {\bibfnamefont
  {D.~L.}\ \bibnamefont {Quintero-Castro}},\ }\href {\doibase
  10.1103/PhysRevB.96.054423} {\bibfield  {journal} {\bibinfo  {journal} {Phys.
  Rev. B}\ }\textbf {\bibinfo {volume} {96}},\ \bibinfo {pages} {054423}
  (\bibinfo {year} {2017})},\ \Eprint {http://arxiv.org/abs/arXiv:1705.01259}
  {arXiv:1705.01259} \BibitemShut {NoStop}%
\bibitem [{\citenamefont {Grenier}\ \emph {et~al.}(2015)\citenamefont
  {Grenier}, \citenamefont {Petit}, \citenamefont {Simonet}, \citenamefont
  {Can\'evet}, \citenamefont {Regnault}, \citenamefont {Raymond}, \citenamefont
  {Canals}, \citenamefont {Berthier},\ and\ \citenamefont {Lejay}}]{Gr15}%
  \BibitemOpen
  \bibfield  {author} {\bibinfo {author} {\bibfnamefont {B.}~\bibnamefont
  {Grenier}}, \bibinfo {author} {\bibfnamefont {S.}~\bibnamefont {Petit}},
  \bibinfo {author} {\bibfnamefont {V.}~\bibnamefont {Simonet}}, \bibinfo
  {author} {\bibfnamefont {E.}~\bibnamefont {Can\'evet}}, \bibinfo {author}
  {\bibfnamefont {L.-P.}\ \bibnamefont {Regnault}}, \bibinfo {author}
  {\bibfnamefont {S.}~\bibnamefont {Raymond}}, \bibinfo {author} {\bibfnamefont
  {B.}~\bibnamefont {Canals}}, \bibinfo {author} {\bibfnamefont
  {C.}~\bibnamefont {Berthier}}, \ and\ \bibinfo {author} {\bibfnamefont
  {P.}~\bibnamefont {Lejay}},\ }\href {\doibase 10.1103/PhysRevLett.114.017201}
  {\bibfield  {journal} {\bibinfo  {journal} {Phys. Rev. Lett.}\ }\textbf
  {\bibinfo {volume} {114}},\ \bibinfo {pages} {017201} (\bibinfo {year}
  {2015})},\ \Eprint {http://arxiv.org/abs/arXiv:1407.0213} {arXiv:1407.0213}
  \BibitemShut {NoStop}%
\bibitem [{\citenamefont {{Faure}}\ \emph {et~al.}(2018)\citenamefont
  {{Faure}}, \citenamefont {{Takayoshi}}, \citenamefont {{Petit}},
  \citenamefont {{Simonet}}, \citenamefont {{Raymond}}, \citenamefont
  {{Regnault}}, \citenamefont {{Boehm}}, \citenamefont {{White}}, \citenamefont
  {{M{\aa}nsson}}, \citenamefont {{R{\"u}egg}}, \citenamefont {{Lejay}},
  \citenamefont {{Canals}}, \citenamefont {{Lorenz}}, \citenamefont {{Furuya}},
  \citenamefont {{Giamarchi}},\ and\ \citenamefont {{Grenier}}}]{Faur17}%
  \BibitemOpen
  \bibfield  {author} {\bibinfo {author} {\bibfnamefont {Q.}~\bibnamefont
  {{Faure}}}, \bibinfo {author} {\bibfnamefont {S.}~\bibnamefont
  {{Takayoshi}}}, \bibinfo {author} {\bibfnamefont {S.}~\bibnamefont
  {{Petit}}}, \bibinfo {author} {\bibfnamefont {V.}~\bibnamefont {{Simonet}}},
  \bibinfo {author} {\bibfnamefont {S.}~\bibnamefont {{Raymond}}}, \bibinfo
  {author} {\bibfnamefont {L.-P.}\ \bibnamefont {{Regnault}}}, \bibinfo
  {author} {\bibfnamefont {M.}~\bibnamefont {{Boehm}}}, \bibinfo {author}
  {\bibfnamefont {J.~S.}\ \bibnamefont {{White}}}, \bibinfo {author}
  {\bibfnamefont {M.}~\bibnamefont {{M{\aa}nsson}}}, \bibinfo {author}
  {\bibfnamefont {C.}~\bibnamefont {{R{\"u}egg}}}, \bibinfo {author}
  {\bibfnamefont {P.}~\bibnamefont {{Lejay}}}, \bibinfo {author} {\bibfnamefont
  {B.}~\bibnamefont {{Canals}}}, \bibinfo {author} {\bibfnamefont
  {T.}~\bibnamefont {{Lorenz}}}, \bibinfo {author} {\bibfnamefont {S.~C.}\
  \bibnamefont {{Furuya}}}, \bibinfo {author} {\bibfnamefont {T.}~\bibnamefont
  {{Giamarchi}}}, \ and\ \bibinfo {author} {\bibfnamefont {B.}~\bibnamefont
  {{Grenier}}},\ }\href {\doibase https://doi.org/10.1038/s41567-018-0126-8}
  {\bibfield  {journal} {\bibinfo  {journal} {Nature Physics}\ }\textbf
  {\bibinfo {volume} {14}},\ \bibinfo {pages} {716} (\bibinfo {year} {2018})},\
  \Eprint {http://arxiv.org/abs/arXiv:1706.05848} {arXiv:1706.05848}
  \BibitemShut {NoStop}%
\bibitem [{\citenamefont {Wang}\ \emph {et~al.}(2019)\citenamefont {Wang},
  \citenamefont {Schmidt}, \citenamefont {Loidl}, \citenamefont {Wu},
  \citenamefont {Zou}, \citenamefont {Yang}, \citenamefont {Dong},
  \citenamefont {Kohama}, \citenamefont {Kindo}, \citenamefont {Gorbunov},
  \citenamefont {Niesen}, \citenamefont {Breunig}, \citenamefont {Engelmayer},\
  and\ \citenamefont {Lorenz}}]{Wang19}%
  \BibitemOpen
  \bibfield  {author} {\bibinfo {author} {\bibfnamefont {Z.}~\bibnamefont
  {Wang}}, \bibinfo {author} {\bibfnamefont {M.}~\bibnamefont {Schmidt}},
  \bibinfo {author} {\bibfnamefont {A.}~\bibnamefont {Loidl}}, \bibinfo
  {author} {\bibfnamefont {J.}~\bibnamefont {Wu}}, \bibinfo {author}
  {\bibfnamefont {H.}~\bibnamefont {Zou}}, \bibinfo {author} {\bibfnamefont
  {W.}~\bibnamefont {Yang}}, \bibinfo {author} {\bibfnamefont {C.}~\bibnamefont
  {Dong}}, \bibinfo {author} {\bibfnamefont {Y.}~\bibnamefont {Kohama}},
  \bibinfo {author} {\bibfnamefont {K.}~\bibnamefont {Kindo}}, \bibinfo
  {author} {\bibfnamefont {D.~I.}\ \bibnamefont {Gorbunov}}, \bibinfo {author}
  {\bibfnamefont {S.}~\bibnamefont {Niesen}}, \bibinfo {author} {\bibfnamefont
  {O.}~\bibnamefont {Breunig}}, \bibinfo {author} {\bibfnamefont
  {J.}~\bibnamefont {Engelmayer}}, \ and\ \bibinfo {author} {\bibfnamefont
  {T.}~\bibnamefont {Lorenz}},\ }\href {\doibase
  10.1103/PhysRevLett.123.067202} {\bibfield  {journal} {\bibinfo  {journal}
  {Phys. Rev. Lett.}\ }\textbf {\bibinfo {volume} {123}},\ \bibinfo {pages}
  {067202} (\bibinfo {year} {2019})}\BibitemShut {NoStop}%
\bibitem [{\citenamefont {McCoy}\ and\ \citenamefont {Wu}(1978)}]{McCoy78}%
  \BibitemOpen
  \bibfield  {author} {\bibinfo {author} {\bibfnamefont {B.~M.}\ \bibnamefont
  {McCoy}}\ and\ \bibinfo {author} {\bibfnamefont {T.~T.}\ \bibnamefont {Wu}},\
  }\href {\doibase 10.1103/PhysRevD.18.1259} {\bibfield  {journal} {\bibinfo
  {journal} {Phys. Rev. D}\ }\textbf {\bibinfo {volume} {18}},\ \bibinfo
  {pages} {1259} (\bibinfo {year} {1978})}\BibitemShut {NoStop}%
\bibitem [{\citenamefont {Delfino}\ \emph {et~al.}(1996)\citenamefont
  {Delfino}, \citenamefont {Mussardo},\ and\ \citenamefont
  {Simonetti}}]{Del96}%
  \BibitemOpen
  \bibfield  {author} {\bibinfo {author} {\bibfnamefont {G.}~\bibnamefont
  {Delfino}}, \bibinfo {author} {\bibfnamefont {G.}~\bibnamefont {Mussardo}}, \
  and\ \bibinfo {author} {\bibfnamefont {P.}~\bibnamefont {Simonetti}},\ }\href
  {\doibase https://doi.org/10.1016/0550-3213(96)00265-9} {\bibfield  {journal}
  {\bibinfo  {journal} {Nucl. Phys. B}\ }\textbf {\bibinfo {volume} {473}},\
  \bibinfo {pages} {469} (\bibinfo {year} {1996})},\ \Eprint
  {http://arxiv.org/abs/arXiv:hep-th/9603011} {arXiv:hep-th/9603011}
  \BibitemShut {NoStop}%
\bibitem [{\citenamefont {Delfino}\ and\ \citenamefont
  {Mussardo}(1998)}]{DelMus98}%
  \BibitemOpen
  \bibfield  {author} {\bibinfo {author} {\bibfnamefont {G.}~\bibnamefont
  {Delfino}}\ and\ \bibinfo {author} {\bibfnamefont {G.}~\bibnamefont
  {Mussardo}},\ }\href {\doibase https://doi.org/10.1016/S0550-3213(98)00063-7}
  {\bibfield  {journal} {\bibinfo  {journal} {Nucl. Phys. B}\ }\textbf
  {\bibinfo {volume} {516}},\ \bibinfo {pages} {675} (\bibinfo {year}
  {1998})},\ \Eprint {http://arxiv.org/abs/hep-th/9709028} {hep-th/9709028}
  \BibitemShut {NoStop}%
\bibitem [{\citenamefont {Mussardo}(2011)}]{Mus11}%
  \BibitemOpen
  \bibfield  {author} {\bibinfo {author} {\bibfnamefont {G.}~\bibnamefont
  {Mussardo}},\ }\href {http://stacks.iop.org/1742-5468/2011/i=01/a=P01002}
  {\bibfield  {journal} {\bibinfo  {journal} {Journal of Statistical Mechanics:
  Theory and Experiment}\ }\textbf {\bibinfo {volume} {2011}},\ \bibinfo
  {pages} {P01002} (\bibinfo {year} {2011})}\BibitemShut {NoStop}%
\bibitem [{\citenamefont {Shiba}(1980)}]{Shiba80}%
  \BibitemOpen
  \bibfield  {author} {\bibinfo {author} {\bibfnamefont {H.}~\bibnamefont
  {Shiba}},\ }\href {\doibase 10.1143/PTP.64.466} {\bibfield  {journal}
  {\bibinfo  {journal} {Progress of Theoretical Physics}\ }\textbf {\bibinfo
  {volume} {64}},\ \bibinfo {pages} {466} (\bibinfo {year} {1980})}\BibitemShut
  {NoStop}%
\bibitem [{\citenamefont {Fonseca}\ and\ \citenamefont
  {Zamolodchikov}(2003)}]{FonZam2003}%
  \BibitemOpen
  \bibfield  {author} {\bibinfo {author} {\bibfnamefont {P.}~\bibnamefont
  {Fonseca}}\ and\ \bibinfo {author} {\bibfnamefont {A.~B.}\ \bibnamefont
  {Zamolodchikov}},\ }\href {\doibase https://doi.org/10.1023/A:1022147532606}
  {\bibfield  {journal} {\bibinfo  {journal} {J. Stat. Phys.}\ }\textbf
  {\bibinfo {volume} {110}},\ \bibinfo {pages} {527} (\bibinfo {year}
  {2003})},\ \Eprint {http://arxiv.org/abs/arXiv:hep-th/0309228}
  {arXiv:hep-th/0309228} \BibitemShut {NoStop}%
\bibitem [{\citenamefont {Fonseca}\ and\ \citenamefont
  {Zamolodchikov}(2006)}]{FZ06}%
  \BibitemOpen
  \bibfield  {author} {\bibinfo {author} {\bibfnamefont {P.}~\bibnamefont
  {Fonseca}}\ and\ \bibinfo {author} {\bibfnamefont {A.~B.}\ \bibnamefont
  {Zamolodchikov}},\ }\href@noop {} {\enquote {\bibinfo {title} {Ising
  spectroscopy ${{\rm I}} $: Mesons at ${T}<{T}_c$},}\ } (\bibinfo {year}
  {2006}),\ \Eprint {http://arxiv.org/abs/arXiv:hep-th/0612304}
  {arXiv:hep-th/0612304} \BibitemShut {NoStop}%
\bibitem [{\citenamefont {Rutkevich}(2005)}]{Rut05}%
  \BibitemOpen
  \bibfield  {author} {\bibinfo {author} {\bibfnamefont {S.~B.}\ \bibnamefont
  {Rutkevich}},\ }\href {\doibase 10.1103/PhysRevLett.95.250601} {\bibfield
  {journal} {\bibinfo  {journal} {Phys. Rev. Lett.}\ }\textbf {\bibinfo
  {volume} {95}},\ \bibinfo {pages} {250601} (\bibinfo {year} {2005})},\
  \Eprint {http://arxiv.org/abs/arXiv:hep-th/0509149} {arXiv:hep-th/0509149}
  \BibitemShut {NoStop}%
\bibitem [{\citenamefont {Delfino}\ and\ \citenamefont {Grinza}(2008)}]{Del08}%
  \BibitemOpen
  \bibfield  {author} {\bibinfo {author} {\bibfnamefont {G.}~\bibnamefont
  {Delfino}}\ and\ \bibinfo {author} {\bibfnamefont {P.}~\bibnamefont
  {Grinza}},\ }\href {\doibase https://doi.org/10.1016/j.nuclphysb.2007.09.003}
  {\bibfield  {journal} {\bibinfo  {journal} {Nucl. Phys. B}\ }\textbf
  {\bibinfo {volume} {791}},\ \bibinfo {pages} {265} (\bibinfo {year}
  {2008})}\BibitemShut {NoStop}%
\bibitem [{\citenamefont {Mussardo}\ and\ \citenamefont
  {Tak\'acs}(2009)}]{MusTak09}%
  \BibitemOpen
  \bibfield  {author} {\bibinfo {author} {\bibfnamefont {G.}~\bibnamefont
  {Mussardo}}\ and\ \bibinfo {author} {\bibfnamefont {G.}~\bibnamefont
  {Tak\'acs}},\ }\href {\doibase
  https://doi.org/10.1088/1751-8113/42/30/304022} {\bibfield  {journal}
  {\bibinfo  {journal} {J. Phys. A}\ }\textbf {\bibinfo {volume} {42}},\
  \bibinfo {pages} {304022} (\bibinfo {year} {2009})}\BibitemShut {NoStop}%
\bibitem [{\citenamefont {Kormos}\ \emph {et~al.}(2017)\citenamefont {Kormos},
  \citenamefont {Collura}, \citenamefont {Tak{\'a}cs},\ and\ \citenamefont
  {Calabrese}}]{Kor16}%
  \BibitemOpen
  \bibfield  {author} {\bibinfo {author} {\bibfnamefont {M.}~\bibnamefont
  {Kormos}}, \bibinfo {author} {\bibfnamefont {M.}~\bibnamefont {Collura}},
  \bibinfo {author} {\bibfnamefont {G.}~\bibnamefont {Tak{\'a}cs}}, \ and\
  \bibinfo {author} {\bibfnamefont {P.}~\bibnamefont {Calabrese}},\ }\href
  {\doibase https://doi.org/10.1038/nphys3934} {\bibfield  {journal} {\bibinfo
  {journal} {Nat. Phys.}\ }\textbf {\bibinfo {volume} {13}},\ \bibinfo {pages}
  {246} (\bibinfo {year} {2017})},\ \Eprint
  {http://arxiv.org/abs/arXiv:1604.03571} {arXiv:1604.03571} \BibitemShut
  {NoStop}%
\bibitem [{\citenamefont {Robinson}\ \emph {et~al.}(2019)\citenamefont
  {Robinson}, \citenamefont {James},\ and\ \citenamefont {Konik}}]{Rob19}%
  \BibitemOpen
  \bibfield  {author} {\bibinfo {author} {\bibfnamefont {N.~J.}\ \bibnamefont
  {Robinson}}, \bibinfo {author} {\bibfnamefont {A.~J.~A.}\ \bibnamefont
  {James}}, \ and\ \bibinfo {author} {\bibfnamefont {R.~M.}\ \bibnamefont
  {Konik}},\ }\href {\doibase 10.1103/PhysRevB.99.195108} {\bibfield  {journal}
  {\bibinfo  {journal} {Phys. Rev. B}\ }\textbf {\bibinfo {volume} {99}},\
  \bibinfo {pages} {195108} (\bibinfo {year} {2019})}\BibitemShut {NoStop}%
\bibitem [{\citenamefont {Lagnese}\ \emph {et~al.}(2020)\citenamefont
  {Lagnese}, \citenamefont {Surace}, \citenamefont {Kormos},\ and\
  \citenamefont {Calabrese}}]{Lagnese_2020}%
  \BibitemOpen
  \bibfield  {author} {\bibinfo {author} {\bibfnamefont {G.}~\bibnamefont
  {Lagnese}}, \bibinfo {author} {\bibfnamefont {F.~M.}\ \bibnamefont {Surace}},
  \bibinfo {author} {\bibfnamefont {M.}~\bibnamefont {Kormos}}, \ and\ \bibinfo
  {author} {\bibfnamefont {P.}~\bibnamefont {Calabrese}},\ }\href {\doibase
  10.1088/1742-5468/abb368} {\bibfield  {journal} {\bibinfo  {journal} {Journal
  of Statistical Mechanics: Theory and Experiment}\ }\textbf {\bibinfo {volume}
  {2020}},\ \bibinfo {pages} {093106} (\bibinfo {year} {2020})}\BibitemShut
  {NoStop}%
\bibitem [{\citenamefont {Lagnese}\ \emph {et~al.}(2022)\citenamefont
  {Lagnese}, \citenamefont {Surace}, \citenamefont {Kormos},\ and\
  \citenamefont {Calabrese}}]{Lagnese_2022}%
  \BibitemOpen
  \bibfield  {author} {\bibinfo {author} {\bibfnamefont {G.}~\bibnamefont
  {Lagnese}}, \bibinfo {author} {\bibfnamefont {F.~M.}\ \bibnamefont {Surace}},
  \bibinfo {author} {\bibfnamefont {M.}~\bibnamefont {Kormos}}, \ and\ \bibinfo
  {author} {\bibfnamefont {P.}~\bibnamefont {Calabrese}},\ }\href {\doibase
  10.1088/1751-8121/ac5215} {\bibfield  {journal} {\bibinfo  {journal} {Journal
  of Physics A: Mathematical and Theoretical}\ }\textbf {\bibinfo {volume}
  {55}},\ \bibinfo {pages} {124003} (\bibinfo {year} {2022})}\BibitemShut
  {NoStop}%
\bibitem [{\citenamefont {Ramos}\ \emph {et~al.}(2020)\citenamefont {Ramos},
  \citenamefont {Lencs\'es}, \citenamefont {Xavier},\ and\ \citenamefont
  {Pereira}}]{Ram20}%
  \BibitemOpen
  \bibfield  {author} {\bibinfo {author} {\bibfnamefont {F.~B.}\ \bibnamefont
  {Ramos}}, \bibinfo {author} {\bibfnamefont {M.}~\bibnamefont {Lencs\'es}},
  \bibinfo {author} {\bibfnamefont {J.~C.}\ \bibnamefont {Xavier}}, \ and\
  \bibinfo {author} {\bibfnamefont {R.~G.}\ \bibnamefont {Pereira}},\ }\href
  {\doibase 10.1103/PhysRevB.102.014426} {\bibfield  {journal} {\bibinfo
  {journal} {Phys. Rev. B}\ }\textbf {\bibinfo {volume} {102}},\ \bibinfo
  {pages} {014426} (\bibinfo {year} {2020})}\BibitemShut {NoStop}%
\bibitem [{\citenamefont {Lencs\'es}\ \emph {et~al.}(2022)\citenamefont
  {Lencs\'es}, \citenamefont {Mussardo},\ and\ \citenamefont
  {Tak\'acs}}]{Mus22}%
  \BibitemOpen
  \bibfield  {author} {\bibinfo {author} {\bibfnamefont {M.}~\bibnamefont
  {Lencs\'es}}, \bibinfo {author} {\bibfnamefont {G.}~\bibnamefont {Mussardo}},
  \ and\ \bibinfo {author} {\bibfnamefont {G.}~\bibnamefont {Tak\'acs}},\
  }\href {\doibase https://doi.org/10.1016/j.physletb.2022.137008} {\bibfield
  {journal} {\bibinfo  {journal} {Physics Letters B}\ }\textbf {\bibinfo
  {volume} {828}},\ \bibinfo {pages} {137008} (\bibinfo {year} {2022})},\
  \Eprint {http://arxiv.org/abs/arXiv:2111.05360} {arXiv:2111.05360}
  \BibitemShut {NoStop}%
\bibitem [{\citenamefont {Lencs{\'e}s}\ and\ \citenamefont
  {Tak{\'a}cs}(2014)}]{Tak14}%
  \BibitemOpen
  \bibfield  {author} {\bibinfo {author} {\bibfnamefont {M.}~\bibnamefont
  {Lencs{\'e}s}}\ and\ \bibinfo {author} {\bibfnamefont {G.}~\bibnamefont
  {Tak{\'a}cs}},\ }\href {\doibase https://doi.org/10.1007/JHEP09(2014)052}
  {\bibfield  {journal} {\bibinfo  {journal} {Journal of High Energy Physics}\
  }\textbf {\bibinfo {volume} {2014}},\ \bibinfo {pages} {52} (\bibinfo {year}
  {2014})},\ \Eprint {http://arxiv.org/abs/arXiv:1405.3157} {arXiv:1405.3157}
  \BibitemShut {NoStop}%
\bibitem [{\citenamefont {Lencs{\'e}s}\ and\ \citenamefont
  {Tak{\'a}cs}(2015)}]{LT2015}%
  \BibitemOpen
  \bibfield  {author} {\bibinfo {author} {\bibfnamefont {M.}~\bibnamefont
  {Lencs{\'e}s}}\ and\ \bibinfo {author} {\bibfnamefont {G.}~\bibnamefont
  {Tak{\'a}cs}},\ }\href {\doibase https://doi.org/10.1007/JHEP09(2015)146}
  {\bibfield  {journal} {\bibinfo  {journal} {Journal of High Energy Physics}\
  }\textbf {\bibinfo {volume} {2015}},\ \bibinfo {pages} {146} (\bibinfo {year}
  {2015})},\ \Eprint {http://arxiv.org/abs/arXiv:1506.06477} {arXiv:1506.06477}
  \BibitemShut {NoStop}%
\bibitem [{\citenamefont {{Rutkevich}}(2009)}]{Rut09}%
  \BibitemOpen
  \bibfield  {author} {\bibinfo {author} {\bibfnamefont {S.~B.}\ \bibnamefont
  {{Rutkevich}}},\ }\href {\doibase
  https://doi.org/10.1088/1751-8113/42/30/304025} {\bibfield  {journal}
  {\bibinfo  {journal} {J. Phys. A}\ }\textbf {\bibinfo {volume} {42}},\
  \bibinfo {pages} {304025} (\bibinfo {year} {2009})},\ \Eprint
  {http://arxiv.org/abs/0901.1571} {arXiv:0901.1571} \BibitemShut {NoStop}%
\bibitem [{\citenamefont {Rutkevich}(2017)}]{Rut17P}%
  \BibitemOpen
  \bibfield  {author} {\bibinfo {author} {\bibfnamefont {S.~B.}\ \bibnamefont
  {Rutkevich}},\ }\href {\doibase
  https://doi.org/10.1016/j.nuclphysb.2017.08.009} {\bibfield  {journal}
  {\bibinfo  {journal} {Nuclear Physics B}\ }\textbf {\bibinfo {volume}
  {923}},\ \bibinfo {pages} {508 } (\bibinfo {year} {2017})},\ \Eprint
  {http://arxiv.org/abs/arXiv:1706.05281} {arXiv:1706.05281} \BibitemShut
  {NoStop}%
\bibitem [{Note1()}]{Note1}%
  \BibitemOpen
  \bibinfo {note} {In the high-energy physics, the Bethe-Salpeter equation was
  applied to the confinement problem by 't~Hooft \cite {Hooft74}, who
  considered a model for QCD in one space and one time dimension in the limit
  of an infinite number of colours.}\BibitemShut {Stop}%
\bibitem [{\citenamefont {{Rutkevich}}(2008)}]{Rut08a}%
  \BibitemOpen
  \bibfield  {author} {\bibinfo {author} {\bibfnamefont {S.~B.}\ \bibnamefont
  {{Rutkevich}}},\ }\href {\doibase https://doi.org/10.1007/s10955-008-9495-1}
  {\bibfield  {journal} {\bibinfo  {journal} {J. Stat. Phys.}\ }\textbf
  {\bibinfo {volume} {131}},\ \bibinfo {pages} {917} (\bibinfo {year}
  {2008})},\ \Eprint {http://arxiv.org/abs/arXiv:0712.3189v1}
  {arXiv:0712.3189v1} \BibitemShut {NoStop}%
\bibitem [{\citenamefont {{Rutkevich}}(2010)}]{RutP09}%
  \BibitemOpen
  \bibfield  {author} {\bibinfo {author} {\bibfnamefont {S.~B.}\ \bibnamefont
  {{Rutkevich}}},\ }\href {\doibase
  https://doi.org/10.1088/1751-8113/43/23/235004} {\bibfield  {journal}
  {\bibinfo  {journal} {J. Phys. A}\ }\textbf {\bibinfo {volume} {43}},\
  \bibinfo {pages} {235004} (\bibinfo {year} {2010})},\ \Eprint
  {http://arxiv.org/abs/arXiv:0907.3697v2} {arXiv:0907.3697v2} \BibitemShut
  {NoStop}%
\bibitem [{\citenamefont {Rutkevich}(2018)}]{Rut18}%
  \BibitemOpen
  \bibfield  {author} {\bibinfo {author} {\bibfnamefont {S.~B.}\ \bibnamefont
  {Rutkevich}},\ }\href {\doibase https://doi.org/10.1209/0295-5075/121/37001}
  {\bibfield  {journal} {\bibinfo  {journal} {{EPL} (Europhysics Letters)}\
  }\textbf {\bibinfo {volume} {121}},\ \bibinfo {pages} {37001} (\bibinfo
  {year} {2018})}\BibitemShut {NoStop}%
\bibitem [{\citenamefont {Jimbo}\ and\ \citenamefont {Miwa}(1995)}]{Jimbo94}%
  \BibitemOpen
  \bibfield  {author} {\bibinfo {author} {\bibfnamefont {M.}~\bibnamefont
  {Jimbo}}\ and\ \bibinfo {author} {\bibfnamefont {T.}~\bibnamefont {Miwa}},\
  }\href@noop {} {\emph {\bibinfo {title} {Algebraic {A}nalysis of {S}olvable
  {L}attice {M}odels}}},\ \bibinfo {series} {Conference Board of the
  Mathematical Sciences}\ No.~\bibinfo {number} {85}\ (\bibinfo  {publisher}
  {American Mathematical Soc.},\ \bibinfo {year} {1995})\BibitemShut {NoStop}%
\bibitem [{\citenamefont {Orbach}(1958)}]{Orb58}%
  \BibitemOpen
  \bibfield  {author} {\bibinfo {author} {\bibfnamefont {R.}~\bibnamefont
  {Orbach}},\ }\href {\doibase 10.1103/PhysRev.112.309} {\bibfield  {journal}
  {\bibinfo  {journal} {Phys. Rev.}\ }\textbf {\bibinfo {volume} {112}},\
  \bibinfo {pages} {309} (\bibinfo {year} {1958})}\BibitemShut {NoStop}%
\bibitem [{\citenamefont {Takahashi}(2005)}]{Tak09}%
  \BibitemOpen
  \bibfield  {author} {\bibinfo {author} {\bibfnamefont {M.}~\bibnamefont
  {Takahashi}},\ }\href {https://books.google.de/books?id=ubGcM-JCT0IC} {\emph
  {\bibinfo {title} {Thermodynamics of One-Dimensional Solvable Models}}}\
  (\bibinfo  {publisher} {Cambridge University Press},\ \bibinfo {year}
  {2005})\BibitemShut {NoStop}%
\bibitem [{\citenamefont {Zvyagin}(2010)}]{Zv10}%
  \BibitemOpen
  \bibfield  {author} {\bibinfo {author} {\bibfnamefont {A.~A.}\ \bibnamefont
  {Zvyagin}},\ }\href {https://books.google.de/books?id=DJfeYgEACAAJ} {\emph
  {\bibinfo {title} {Quantum Theory of One-Dimensional Spin Systems}}},\
  Kharkov series in physics and mathematics\ (\bibinfo  {publisher} {Cambridge
  Scientific Publishers},\ \bibinfo {year} {2010})\BibitemShut {NoStop}%
\bibitem [{\citenamefont {Dugave}\ \emph {et~al.}(2015)\citenamefont {Dugave},
  \citenamefont {G\"ohmann}, \citenamefont {Kozlowski},\ and\ \citenamefont
  {Suzuki}}]{Dug15}%
  \BibitemOpen
  \bibfield  {author} {\bibinfo {author} {\bibfnamefont {M.}~\bibnamefont
  {Dugave}}, \bibinfo {author} {\bibfnamefont {F.}~\bibnamefont {G\"ohmann}},
  \bibinfo {author} {\bibfnamefont {K.~K.}\ \bibnamefont {Kozlowski}}, \ and\
  \bibinfo {author} {\bibfnamefont {J.}~\bibnamefont {Suzuki}},\ }\href
  {http://stacks.iop.org/1742-5468/2015/i=5/a=P05037} {\bibfield  {journal}
  {\bibinfo  {journal} {Journal of Statistical Mechanics: Theory and
  Experiment}\ }\textbf {\bibinfo {volume} {2015}},\ \bibinfo {pages} {P05037}
  (\bibinfo {year} {2015})},\ \Eprint {http://arxiv.org/abs/arXiv:1412.8217}
  {arXiv:1412.8217} \BibitemShut {NoStop}%
\bibitem [{\citenamefont {Baxter}(1973)}]{Baxter1973}%
  \BibitemOpen
  \bibfield  {author} {\bibinfo {author} {\bibfnamefont {R.~J.}\ \bibnamefont
  {Baxter}},\ }\href {\doibase https://doi.org/10.1007/BF01016845} {\bibfield
  {journal} {\bibinfo  {journal} {Journal of Statistical Physics}\ }\textbf
  {\bibinfo {volume} {9}},\ \bibinfo {pages} {145} (\bibinfo {year}
  {1973})}\BibitemShut {NoStop}%
\bibitem [{\citenamefont {Baxter}(1976)}]{Baxter1976}%
  \BibitemOpen
  \bibfield  {author} {\bibinfo {author} {\bibfnamefont {R.~J.}\ \bibnamefont
  {Baxter}},\ }\href {\doibase https://doi.org/10.1007/BF01020802} {\bibfield
  {journal} {\bibinfo  {journal} {Journal of Statistical Physics}\ }\textbf
  {\bibinfo {volume} {15}},\ \bibinfo {pages} {485} (\bibinfo {year}
  {1976})}\BibitemShut {NoStop}%
\bibitem [{\citenamefont {Izergin}\ \emph {et~al.}(1999)\citenamefont
  {Izergin}, \citenamefont {Kitanine}, \citenamefont {Maillet},\ and\
  \citenamefont {Terras}}]{Iz99}%
  \BibitemOpen
  \bibfield  {author} {\bibinfo {author} {\bibfnamefont {A.~G.}\ \bibnamefont
  {Izergin}}, \bibinfo {author} {\bibfnamefont {N.}~\bibnamefont {Kitanine}},
  \bibinfo {author} {\bibfnamefont {J.~M.}\ \bibnamefont {Maillet}}, \ and\
  \bibinfo {author} {\bibfnamefont {V.}~\bibnamefont {Terras}},\ }\href
  {\doibase https://doi.org/10.1016/S0550-3213(99)00273-4} {\bibfield
  {journal} {\bibinfo  {journal} {Nuclear Physics B}\ }\textbf {\bibinfo
  {volume} {554}},\ \bibinfo {pages} {679 } (\bibinfo {year}
  {1999})}\BibitemShut {NoStop}%
\bibitem [{\citenamefont {Yang}\ and\ \citenamefont
  {Yang}(1966{\natexlab{a}})}]{YY66_1}%
  \BibitemOpen
  \bibfield  {author} {\bibinfo {author} {\bibfnamefont {C.~N.}\ \bibnamefont
  {Yang}}\ and\ \bibinfo {author} {\bibfnamefont {C.~P.}\ \bibnamefont
  {Yang}},\ }\href {\doibase 10.1103/PhysRev.150.321} {\bibfield  {journal}
  {\bibinfo  {journal} {Phys. Rev.}\ }\textbf {\bibinfo {volume} {150}},\
  \bibinfo {pages} {321} (\bibinfo {year} {1966}{\natexlab{a}})}\BibitemShut
  {NoStop}%
\bibitem [{\citenamefont {Yang}\ and\ \citenamefont
  {Yang}(1966{\natexlab{b}})}]{YY66_2}%
  \BibitemOpen
  \bibfield  {author} {\bibinfo {author} {\bibfnamefont {C.~N.}\ \bibnamefont
  {Yang}}\ and\ \bibinfo {author} {\bibfnamefont {C.~P.}\ \bibnamefont
  {Yang}},\ }\href {\doibase 10.1103/PhysRev.150.327} {\bibfield  {journal}
  {\bibinfo  {journal} {Phys. Rev.}\ }\textbf {\bibinfo {volume} {150}},\
  \bibinfo {pages} {327} (\bibinfo {year} {1966}{\natexlab{b}})}\BibitemShut
  {NoStop}%
\bibitem [{\citenamefont {Johnson}\ \emph {et~al.}(1973)\citenamefont
  {Johnson}, \citenamefont {Krinsky},\ and\ \citenamefont {McCoy}}]{McCoy73}%
  \BibitemOpen
  \bibfield  {author} {\bibinfo {author} {\bibfnamefont {J.~D.}\ \bibnamefont
  {Johnson}}, \bibinfo {author} {\bibfnamefont {S.}~\bibnamefont {Krinsky}}, \
  and\ \bibinfo {author} {\bibfnamefont {B.~M.}\ \bibnamefont {McCoy}},\ }\href
  {\doibase 10.1103/PhysRevA.8.2526} {\bibfield  {journal} {\bibinfo  {journal}
  {Phys. Rev. A}\ }\textbf {\bibinfo {volume} {8}},\ \bibinfo {pages} {2526}
  (\bibinfo {year} {1973})}\BibitemShut {NoStop}%
\bibitem [{Note2()}]{Note2}%
  \BibitemOpen
  \bibinfo {note} {The rapidity variable $\alpha $ is simply related with the
  rapidity variable $\lambda $ used previously in \cite {Rut18}: $\alpha =\pi
  -\lambda $. The definition of the rapidity $\alpha $ adopted here has been
  changed in order to harmonise it with notations in the monograph \cite
  {Jimbo94} by Miwa and Jimbo, see equation (7.18) there.}\BibitemShut {Stop}%
\bibitem [{\citenamefont {Zabrodin}(1992)}]{Zabr92}%
  \BibitemOpen
  \bibfield  {author} {\bibinfo {author} {\bibfnamefont {A.}~\bibnamefont
  {Zabrodin}},\ }\href@noop {} {\bibfield  {journal} {\bibinfo  {journal}
  {Modern Physics Letters A}\ }\textbf {\bibinfo {volume} {07}},\ \bibinfo
  {pages} {441} (\bibinfo {year} {1992})}\BibitemShut {NoStop}%
\bibitem [{\citenamefont {Davies}\ \emph {et~al.}(1993)\citenamefont {Davies},
  \citenamefont {Foda}, \citenamefont {Jimbo}, \citenamefont {Miwa},\ and\
  \citenamefont {Nakayashiki}}]{Miwa93}%
  \BibitemOpen
  \bibfield  {author} {\bibinfo {author} {\bibfnamefont {B.}~\bibnamefont
  {Davies}}, \bibinfo {author} {\bibfnamefont {O.}~\bibnamefont {Foda}},
  \bibinfo {author} {\bibfnamefont {M.}~\bibnamefont {Jimbo}}, \bibinfo
  {author} {\bibfnamefont {T.}~\bibnamefont {Miwa}}, \ and\ \bibinfo {author}
  {\bibfnamefont {A.}~\bibnamefont {Nakayashiki}},\ }\href {\doibase
  https://doi.org/10.1007/BF02096750} {\bibfield  {journal} {\bibinfo
  {journal} {Commun. Math. Phys.}\ }\textbf {\bibinfo {volume} {151}},\
  \bibinfo {pages} {89 } (\bibinfo {year} {1993})}\BibitemShut {NoStop}%
\bibitem [{\citenamefont {Bougourzi}\ \emph {et~al.}(1998)\citenamefont
  {Bougourzi}, \citenamefont {Karbach},\ and\ \citenamefont
  {M\"uller}}]{Kar98}%
  \BibitemOpen
  \bibfield  {author} {\bibinfo {author} {\bibfnamefont {A.~H.}\ \bibnamefont
  {Bougourzi}}, \bibinfo {author} {\bibfnamefont {M.}~\bibnamefont {Karbach}},
  \ and\ \bibinfo {author} {\bibfnamefont {G.}~\bibnamefont {M\"uller}},\
  }\href {\doibase 10.1103/PhysRevB.57.11429} {\bibfield  {journal} {\bibinfo
  {journal} {Phys. Rev. B}\ }\textbf {\bibinfo {volume} {57}},\ \bibinfo
  {pages} {11429} (\bibinfo {year} {1998})}\BibitemShut {NoStop}%
\bibitem [{\citenamefont {Caux}\ \emph {et~al.}(2008)\citenamefont {Caux},
  \citenamefont {Mossel},\ and\ \citenamefont {Castillo}}]{Caux_2008}%
  \BibitemOpen
  \bibfield  {author} {\bibinfo {author} {\bibfnamefont {J.-S.}\ \bibnamefont
  {Caux}}, \bibinfo {author} {\bibfnamefont {J.}~\bibnamefont {Mossel}}, \ and\
  \bibinfo {author} {\bibfnamefont {I.~P.}\ \bibnamefont {Castillo}},\ }\href
  {\doibase 10.1088/1742-5468/2008/08/p08006} {\bibfield  {journal} {\bibinfo
  {journal} {Journal of Statistical Mechanics: Theory and Experiment}\ }\textbf
  {\bibinfo {volume} {2008}},\ \bibinfo {pages} {P08006} (\bibinfo {year}
  {2008})}\BibitemShut {NoStop}%
\bibitem [{\citenamefont {Lashkevich}(2002)}]{Lash02}%
  \BibitemOpen
  \bibfield  {author} {\bibinfo {author} {\bibfnamefont {M.}~\bibnamefont
  {Lashkevich}},\ }\href {\doibase
  https://doi.org/10.1016/S0550-3213(01)00598-3} {\bibfield  {journal}
  {\bibinfo  {journal} {Nuclear Physics B}\ }\textbf {\bibinfo {volume}
  {621}},\ \bibinfo {pages} {587 } (\bibinfo {year} {2002})}\BibitemShut
  {NoStop}%
\bibitem [{\citenamefont {Lukyanov}\ and\ \citenamefont
  {Terras}(2003)}]{LukTer03}%
  \BibitemOpen
  \bibfield  {author} {\bibinfo {author} {\bibfnamefont {S.}~\bibnamefont
  {Lukyanov}}\ and\ \bibinfo {author} {\bibfnamefont {V.}~\bibnamefont
  {Terras}},\ }\href {\doibase https://doi.org/10.1016/S0550-3213(02)01141-0}
  {\bibfield  {journal} {\bibinfo  {journal} {Nuclear Physics B}\ }\textbf
  {\bibinfo {volume} {654}},\ \bibinfo {pages} {323 } (\bibinfo {year}
  {2003})}\BibitemShut {NoStop}%
\bibitem [{\citenamefont {Ishimura}\ and\ \citenamefont
  {Shiba}(1980)}]{Shiba_80}%
  \BibitemOpen
  \bibfield  {author} {\bibinfo {author} {\bibfnamefont {N.}~\bibnamefont
  {Ishimura}}\ and\ \bibinfo {author} {\bibfnamefont {H.}~\bibnamefont
  {Shiba}},\ }\href {\doibase https://doi.org/10.1143/PTP.63.743} {\bibfield
  {journal} {\bibinfo  {journal} {Progress of Theoretical Physics}\ }\textbf
  {\bibinfo {volume} {63}},\ \bibinfo {pages} {743} (\bibinfo {year}
  {1980})}\BibitemShut {NoStop}%
\bibitem [{\citenamefont {Castillo}(2020)}]{Cas20}%
  \BibitemOpen
  \bibfield  {author} {\bibinfo {author} {\bibfnamefont {I.~P.}\ \bibnamefont
  {Castillo}},\ }\href@noop {} {\enquote {\bibinfo {title} {The exact
  two-spinon longitudinal dynamical structure factor of the anisotropic {XXZ}
  model},}\ } (\bibinfo {year} {2020}),\ \Eprint
  {http://arxiv.org/abs/arXiv:2005.10729} {arXiv:2005.10729} \BibitemShut
  {NoStop}%
\bibitem [{\citenamefont {Babenko}\ \emph {et~al.}(2021)\citenamefont
  {Babenko}, \citenamefont {G\"ohmann}, \citenamefont {Kozlowski},
  \citenamefont {Sirker},\ and\ \citenamefont {Suzuki}}]{Bab21}%
  \BibitemOpen
  \bibfield  {author} {\bibinfo {author} {\bibfnamefont {C.}~\bibnamefont
  {Babenko}}, \bibinfo {author} {\bibfnamefont {F.}~\bibnamefont {G\"ohmann}},
  \bibinfo {author} {\bibfnamefont {K.~K.}\ \bibnamefont {Kozlowski}}, \bibinfo
  {author} {\bibfnamefont {J.}~\bibnamefont {Sirker}}, \ and\ \bibinfo {author}
  {\bibfnamefont {J.}~\bibnamefont {Suzuki}},\ }\href {\doibase
  10.1103/PhysRevLett.126.210602} {\bibfield  {journal} {\bibinfo  {journal}
  {Phys. Rev. Lett.}\ }\textbf {\bibinfo {volume} {126}},\ \bibinfo {pages}
  {210602} (\bibinfo {year} {2021})}\BibitemShut {NoStop}%
\bibitem [{Note3()}]{Note3}%
  \BibitemOpen
  \bibinfo {note} {Another equivalent possibility \cite {Rut05,Rut08a,FZ06} is
  to remove constraint (\ref {XX}) and to replace the linear potential
  $f\protect \tmspace +\thinmuskip {.1667em}(x_2-x_1)$ in (\ref {Hk}) by
  $f\protect \tmspace +\thinmuskip {.1667em}|x_2-x_1|$.}\BibitemShut {Stop}%
\bibitem [{\citenamefont {Bloch}(1929)}]{Bloch29}%
  \BibitemOpen
  \bibfield  {author} {\bibinfo {author} {\bibfnamefont {F.}~\bibnamefont
  {Bloch}},\ }\href {\doibase https://doi.org/10.1007/BF01339455} {\bibfield
  {journal} {\bibinfo  {journal} {Zeitschrift f\"ur Physik}\ }\textbf {\bibinfo
  {volume} {52}},\ \bibinfo {pages} {555} (\bibinfo {year} {1929})}\BibitemShut
  {NoStop}%
\bibitem [{\citenamefont {Landau}\ and\ \citenamefont {Lifshitz}(1981)}]{LL3}%
  \BibitemOpen
  \bibfield  {author} {\bibinfo {author} {\bibfnamefont {L.}~\bibnamefont
  {Landau}}\ and\ \bibinfo {author} {\bibfnamefont {E.}~\bibnamefont
  {Lifshitz}},\ }\href@noop {} {\emph {\bibinfo {title} {Quantum Mechanics:
  Non-Relativistic Theory}}},\ Course of Theoretical Physics\ (\bibinfo
  {publisher} {Elsevier Science},\ \bibinfo {year} {1981})\BibitemShut
  {NoStop}%
\bibitem [{\citenamefont {Abramowitz}\ and\ \citenamefont
  {Stegun}(1965)}]{AbrSt}%
  \BibitemOpen
  \bibfield  {author} {\bibinfo {author} {\bibfnamefont {M.}~\bibnamefont
  {Abramowitz}}\ and\ \bibinfo {author} {\bibfnamefont {I.~A.}\ \bibnamefont
  {Stegun}},\ }\href@noop {} {\emph {\bibinfo {title} {Handbook of Mathematical
  Functions}}}\ (\bibinfo  {publisher} {Dover},\ \bibinfo {year}
  {1965})\BibitemShut {NoStop}%
\bibitem [{\citenamefont {Muskhelishvili}(1977)}]{Mus}%
  \BibitemOpen
  \bibfield  {author} {\bibinfo {author} {\bibfnamefont {N.~I.}\ \bibnamefont
  {Muskhelishvili}},\ }\href@noop {} {\emph {\bibinfo {title} {Singular
  Integral Equations}}}\ (\bibinfo  {publisher} {Noordhoff International
  Publishing},\ \bibinfo {address} {Leyden},\ \bibinfo {year}
  {1977})\BibitemShut {NoStop}%
\bibitem [{\citenamefont {Smirnov}(1992)}]{Sm92}%
  \BibitemOpen
  \bibfield  {author} {\bibinfo {author} {\bibfnamefont {F.~A.}\ \bibnamefont
  {Smirnov}},\ }\href@noop {} {\emph {\bibinfo {title} {Form-factors in
  completely integrable models of quantum field theory, ({A}dvanced {S}eries in
  {M}athematical {P}hysics, {V}ol. 14)}}}\ (\bibinfo  {publisher} {World
  Scientific},\ \bibinfo {address} {Singapore},\ \bibinfo {year}
  {1992})\BibitemShut {NoStop}%
\bibitem [{\citenamefont {{}'t Hooft}(1974)}]{Hooft74}%
  \BibitemOpen
  \bibfield  {author} {\bibinfo {author} {\bibfnamefont {G.}~\bibnamefont {{}'t
  Hooft}},\ }\href {\doibase https://doi.org/10.1016/0550-3213(74)90088-1}
  {\bibfield  {journal} {\bibinfo  {journal} {Nucl. Phys. B}\ }\textbf
  {\bibinfo {volume} {75}},\ \bibinfo {pages} {461} (\bibinfo {year}
  {1974})}\BibitemShut {NoStop}%
\end{thebibliography}
%

\end{document}